\pdfoutput=1

\documentclass[11pt,twoside,a4paper,cmspaper,final,collab]{cms-tdr}
\def\svnVersion{230896}\def\cmsCernNo{2013-037}\def\cmsCernDate{\today}\def\cmsMessage{Published in the Journal of High Energy Physics as \href{http://dx.doi.org/10.1007/JHEP01(2014)096}{\doi{10.1007/JHEP01(2014)096.}}}

\begin{document}\cmsNoteHeader{HIG-13-023}

\hyphenation{had-ron-i-za-tion}
\hyphenation{cal-or-i-me-ter}
\hyphenation{de-vices}

\RCS$Revision: 222091 $
\RCS$HeadURL: svn+ssh://alverson@svn.cern.ch/reps/tdr2/papers/HIG-13-023/trunk/HIG-13-023.tex $
\RCS$Id: HIG-13-023.tex 222091 2013-12-30 16:43:56Z alverson $
\newcommand{\CLs}{\ensuremath{CL_\mathrm{s}}}
\newcommand{\CLb}{\ensuremath{CL_\mathrm{b}}}
\newcommand{\CLsb}{\ensuremath{CL_\mathrm{s+b}}}

\newcommand{\pipm}{\ensuremath{\pi^{\pm}}}
\newcommand{\pizero}{\ensuremath{\pi^{0}}}
\newcommand{\Hi}{\PH\xspace}
\newcommand{\V}{\ensuremath{\mathrm{V}}}
\newcommand{\W}{\PW}
\newcommand{\Wjets}{\ensuremath{\mathrm{W+jets}}}
\newcommand{\Zjets}{\ensuremath{\mathrm{Z+jets}}}
\newcommand{\Wt}{\ensuremath{\mathrm{Wt}}}
\newcommand{\Wstar}{\ensuremath{\mathrm{W}^{*}}}
\newcommand{\Wparenthesisstar}{\ensuremath{\mathrm{W}^{(*)}}}
\newcommand{\WW}{\PW\PW\xspace}
\newcommand{\Zstar}{\ensuremath{\mathrm{Z}^{*}}}
\newcommand{\Astar}{\ensuremath{\mathrm{\gamma}^{*}}}
\newcommand{\ZZ}{\ensuremath{\Z\Z}}
\newcommand{\WZ}{\ensuremath{\W\Z}}
\newcommand{\VVV}{\ensuremath{\V\V\V}}
\newcommand{\Wgstar}{\ensuremath{\W\gamma}^{*}}
\newcommand{\E}{\Pe}
\newcommand{\Ep}{\Pep}
\newcommand{\Epm}{\ensuremath{\Pe^{\pm}}}
\newcommand{\Emp}{\ensuremath{\Pe^{\mp}}}
\newcommand{\M}{\Pgm}
\newcommand{\Mp}{\Pgmp}
\newcommand{\Mm}{\Pgmm}
\newcommand{\Mpm}{\ensuremath{\mu^{\pm}}}
\newcommand{\Mmp}{\ensuremath{\mu^{\mp}}}
\newcommand{\Tau}{\ensuremath{\tau}}
\newcommand{\Nu}{\ensuremath{\nu}}
\newcommand{\Nubar}{\ensuremath{\overline{\nu}}}
\newcommand{\Lep}{\ensuremath{\ell}}
\newcommand{\Lepp}{\ensuremath{\ell^{+}}}
\newcommand{\Lepm}{\ensuremath{\ell^{-}}}
\newcommand{\Lprime}{\ensuremath{\Lep^{\prime}}}
\newcommand{\Prot}{\Pp}
\newcommand{\Pbar}{\Pap}
\newcommand{\PP}{\Pp\Pp}
\newcommand{\PPbar}{\Pp\Pap}
\newcommand{\qq}{\ensuremath{\mathrm{q}\mathrm{q}}}
\newcommand{\qqbar}{\ensuremath{\mathrm{q}\overline{\mathrm{q}}}}
\newcommand{\Wtb}{\ensuremath{\W\mathrm{t}\mathrm{b}}}
\newcommand{\Top}{\ensuremath{\mathrm{t}}}
\newcommand{\Bot}{\ensuremath{\mathrm{b}}}
\newcommand{\Atop}{\ensuremath{\overline{\mathrm{t}}}}
\newcommand{\Abot}{\ensuremath{\overline{\mathrm{b}}}}
\newcommand{\To}{\ensuremath{\rightarrow}}

\newcommand{\mH}{\ensuremath{m_{\PH}}}
\newcommand{\mHi}{\ensuremath{m_{\PH}}\xspace}
\newcommand{\mW}{\ensuremath{m_{\PW}}}
\newcommand{\mZ}{\ensuremath{m_{\cPZ}}}
\newcommand{\mll}{\ensuremath{m_{\Lep\Lep}}}
\newcommand{\mt}{\ensuremath{m_{\mathrm{T}}}}
\newcommand{\mth}{\ensuremath{m_{\mathrm{T}}}}
\newcommand{\mtlnjj}{\ensuremath{m_{\mathrm{T}}^{\ell\nu 2j}}}
\newcommand{\mr}{\ensuremath{m_{\mathrm{R}}}}
\newcommand{\delphir}{\ensuremath{\Delta\phi_{\mathrm{R}}}}

\newcommand{\pythia} {\textsc{pythia}}
\newcommand{\geant} {\textsc{geant4}}
\newcommand{\herwig} {\textsc{herwig}}
\newcommand{\acermc} {\textsc{acermc}}
\newcommand{\jimmy} {\textsc{jimmy}}
\newcommand{\mcatnlo} {\textsc{mc@nlo}}
\newcommand{\sherpa} {\textsc{sherpa}}
\newcommand{\madgraph} {\textsc{madgraph}}
\newcommand{\mcfm} {\textsc{mcfm}}
\newcommand{\powheg} {\textsc{powheg}}
\newcommand{\Phantom} {\textsc{phantom}}
\newcommand{\fastjet} {\textsc{fastJet}}
\newcommand{\tauola} {\textsc{tauola}}

\newcommand{\Et}{\ensuremath{E_\mathrm{T}}}
\newcommand{\ptveto}{\ensuremath{\pt^\text{veto}}}
\newcommand{\ptl}{\ensuremath{p_\perp^{\Lep}}}
\newcommand{\ptlmax}{\ensuremath{p_{\mathrm{T}}^{\Lep,\text{max}}}}
\newcommand{\ptlmin}{\ensuremath{p_{\mathrm{T}}^{\Lep,\text{min}}}}
\newcommand{\ptll}{\ensuremath{\pt^{\ell\ell}}}
\newcommand{\met}{\ensuremath{\Et^{\text{miss}}}}
\newcommand{\delphill}{\ensuremath{\Delta\phi_{\Lep\Lep}}}
\newcommand{\deletall}{\ensuremath{\Delta\eta_{\Lep\Lep}}}
\newcommand{\delphimetl}{\ensuremath{\Delta\phi(\vec{E}_\mathrm{T}^{\text{miss}},\Lep)}}
\newcommand{\delphillmet}{\ensuremath{\Delta\phi(\Lep\Lep,\vec{E}_\mathrm{T}^{\text{miss}})}}
\newcommand{\delR}{\ensuremath{\Delta R}}
\newcommand{\Eta}{\ensuremath{\eta}}
\newcommand{\GAMMA}{\ensuremath{\gamma}}
\newcommand{\pmet}{\ensuremath{E^{\text{miss}\angle}_\mathrm{T}}}
\newcommand{\vmet}{\ensuremath{\vec{E}_\mathrm{T}}^{\text{miss}}}

\newcommand{\effsig}{\ensuremath{\varepsilon_{\text{bkg}}^{\mathrm{S}}}}
\newcommand{\effnorm}{\ensuremath{\varepsilon_{\text{bkg}}^{\mathrm{N}}}}
\newcommand{\Nsig}{\ensuremath{N_{\text{bkg}}^{\mathrm{S}}}}
\newcommand{\Nnorm}{\ensuremath{N_{\text{bkg}}^{\mathrm{N}}}}

\newcommand{\dyeepm}{\ensuremath{{\Z}/\GAMMA^*{\to \Pep\Pem}}}
\newcommand{\dymmpm}{\ensuremath{{\Z/}\GAMMA^*\to\Pgmp\Pgmm}}
\newcommand{\dyttpm}{\ensuremath{{\Z}/\GAMMA^* \to\tau^+\tau^-}}
\newcommand{\dyllpm}{\ensuremath{{\Z}/\GAMMA^*{\to \ell^+\ell^-}}}
\newcommand{\dyee}{\ensuremath{{\Z}/\GAMMA^*{\to \Pe\Pe}}}
\newcommand{\dymm}{\ensuremath{{\Z/}\GAMMA^*\to\mu\mu}}
\newcommand{\dytt}{\ensuremath{{\Z}/\GAMMA^* \to\tau\tau}}
\newcommand{\dyll}{\ensuremath{{\Z}/\GAMMA^*{\to \ell\ell}}}
\newcommand{\zee}{\ensuremath{{\Z\to \Pep\Pem}}}
\newcommand{\zmm}{\ensuremath{{\Z}\to\Pgmp\Pgmm}}
\newcommand{\ztt}{\ensuremath{{\Z}\to\tau^+\tau^-}}
\newcommand{\zll}{\ensuremath{{\Z\to \ell^+\ell^-}}}
\newcommand{\ppww}{\ensuremath{pp \to \PWp\PWm}}
\newcommand{\wwlnln}{\ensuremath{\PWp\PWm\to \ell^+\nu \ell^-\overline{\nu}}}
\newcommand{\ww}{\ensuremath{\W\W}}
\newcommand{\WWpm}{\ensuremath{\PWp\PWm}}
\newcommand{\Hww}{\Hi\to\WW}
\newcommand{\hww}{\Hi\to\WW}
\newcommand{\hwwnopm}{\Hi\to\ww}
\newcommand{\hwwlnlnnopm}{\ensuremath{\Hi\to \W\W\to \ell \nu \ell \nu}}
\newcommand{\wz}{{\PW\cPZ}}
\newcommand{\zz}{{\cPZ\cPZ}}
\newcommand{\wgamma}{\ensuremath{\W\GAMMA}}
\newcommand{\wjets}{\ensuremath{\PW+}\text{jets}}
\newcommand{\tw}{\ensuremath{\mathrm{t}\W}}
\newcommand{\singletopt}{\ensuremath{t} ($t$-chan)}
\newcommand{\singletops}{\ensuremath{t} ($s$-chan)}

\def\fixme{({\bf FixMe})}
\newcommand{\ee}{\ensuremath{ee}}
\newcommand{\emu}{\ensuremath{e\mu}}
\def\mm{\ensuremath{\mu\mu}}
\newcommand{\spintwopmin}{\ensuremath{2^+_\text{min}}}

\newcommand{\usedLumiSeven}{4.9 \fbinv}
\newcommand{\usedLumiWithSystSeven}{\ensuremath{4.9 \pm 0.1 \fbinv}}
\newcommand{\usedLumi}{19.4 \fbinv}
\newcommand{\usedLumiWithSyst}{\ensuremath{19.4 \pm 0.5 \fbinv}}

\cmsNoteHeader{HIG-13-023} 
\title{Measurement of Higgs boson production and properties in the WW decay
channel with leptonic final states}

\date{\today}

\abstract{
A search for the standard model Higgs boson decaying to a W-boson
pair at the LHC is reported.
The event sample corresponds to an integrated luminosity
of 4.9\fbinv and 19.4\fbinv
collected with the CMS detector in pp collisions at $\sqrt{s} = 7$ and 8\TeV, respectively.
The Higgs boson candidates are selected in events with two or three charged leptons.
An excess of events above background is observed, consistent with
the expectation from the standard model Higgs boson with a mass of around 125\GeV.
The probability to observe an excess equal or larger than the one seen, under the
background-only hypothesis, corresponds to a significance of 4.3 standard deviations
for $m_\PH$ = 125.6\GeV.
The observed signal cross section times the branching fraction to $\PW\PW$ for $m_\PH$ = 125.6\GeV is
$0.72^{+0.20}_{-0.18}$ times the standard model expectation.
The spin-parity $J^P=0^+$ hypothesis is favored against a narrow resonance
with $J^P=2^+$ or $J^P=0^-$ that decays to a $\PW$-boson pair.
This result provides strong evidence for a Higgs-like boson
decaying to a $\PW$-boson pair.
}

\hypersetup{%
pdfauthor={CMS Collaboration},%
pdftitle={Measurement of Higgs boson production and properties in the WW decay
channel with leptonic final states},%
pdfsubject={CMS},%
pdfkeywords={CMS, Higgs physics}}

\maketitle 

\section{Introduction}\label{sec:intro}
The origin of the masses of the fundamental particles is one of the main open questions
in the standard model (SM) of particle physics~\cite{SM1,SM2,SM3}.
Within the SM, the masses of the electroweak vector bosons arise by the spontaneous breaking of
electroweak symmetry by the Higgs field~\cite{Higgs1, Higgs2, Higgs3,Higgs4, Higgs5, Higgs6}.
Precision electroweak data constrain the mass of the SM Higgs boson ($\mHi$) to
be less than 158\GeV at the 95\% confidence level (CL)~\cite{EWK,2005ema}.
The ATLAS and CMS experiments at the Large Hadron Collider (LHC), have reported the discovery
of a new boson with a mass of approximately 125\GeV with a significance of five or more standard
deviations each~\cite{AtlasPaperCombination,CMSPaperCombination,CMSPaperCombinationLong}.
Both observations show consistency with the expected properties of the SM Higgs
boson at that mass.
The CDF and D0 experiments at the Tevatron have also reported evidence for a new particle in the mass range 120--135\GeV with
a significance of up to three standard deviations~\cite{Tevatron2012,Tevatron2013}.
The determination of the properties of the observed boson, such as its couplings to
other particles, mass, and quantum numbers, including spin and parity, is crucial
for establishing the nature of this boson. Some of these properties are measured
using the $\PH \to \PWp\PWm$ decay channel with leptonic final states.

Finding such a signal in the complex environment of a
hadron collider is not straightforward.
A complete reconstruction of all the final-state particles is not possible
because of the presence of neutrinos which are not directly detected.
Kinematic observables such as the opening angle between the two charged
leptons in the transverse plane, the dilepton mass, and
the transverse mass of the system of the two leptons and the neutrinos,
can be used to distinguish not only the Higgs boson signal from background processes
with similar signature~\cite{Barger:1990mn,dittmar}, but also between the SM Higgs
boson hypothesis and other narrow exotic resonances with different spin or parity.
Phenomenological studies of the amplitudes for the decay of a Higgs or an exotic boson
into the $\W\W$ final state demonstrate a good sensitivity to distinguish
between the SM Higgs boson hypothesis (spin-parity $0^+$) and a spin-2 resonance, which couples to
the bosons through minimal couplings, referred to as \spintwopmin~\cite{JCPExpPaper2}.
Some sensitivity has also been shown with this final state to distinguish between
the $0^+$ and the pseudoscalar $0^-$ boson hypotheses.

Searches for the SM Higgs boson in the $\PH \to \WW$ final state at the
LHC have previously been performed using data at $\sqrt{s}= 7\TeV$
by CMS~\cite{HWW2010,HWW2011,WH2011}, excluding the presence of the SM Higgs
boson at the 95\% CL in the mass range 129--270\GeV, and by ATLAS~\cite{AtlasHWW7TeV},
excluding the mass range 133--261\GeV. Using their full dataset at 7
and 8\TeV, ATLAS have reported a $\PH \to \WW$ signal with a statistical
significance of 3.8 standard deviations~\cite{AtlasProperties} as well as
evidence for the spin zero nature of the Higgs boson~\cite{AtlasSpin}.

This paper reports a measurement of the production and properties of
the Higgs boson in the WW decay channel
using the entire dataset collected by the CMS experiment during the
2011 and 2012 LHC running period. Various
production modes, using events with two or three charged leptons ($\ell$),
electrons or muons, are investigated. The small contribution proceeding
through an intermediate $\tau$ lepton is included. For Higgs boson masses
around 125\GeV, the expected branching fraction of the Higgs boson to a pair of
$\W$ bosons is about 22\%. The production modes of the SM Higgs boson targeted by
this analysis are the dominant gluon fusion ($\Pg\Pg\PH$), the vector-boson
fusion (VBF), and the associated production with a $\W$ or $\Z$ boson ($\V\PH$).
The fraction of events from associated production with
a top-quark pair ($\ttbar\PH$) passing the analysis selection is negligible,
and therefore this process is not considered in any of the measurements described in this paper.
The analysis is performed in five exclusive event categories based on the final-state leptons
and jets: $2\ell 2\nu$ + 0/1~jet targeting the $\Pg\Pg\PH$ production,
$2\ell 2\nu$ +2~jets targeting the VBF production, $2\ell 2\nu$ + 2~jets targeting
the $\V\PH$ production, $3\ell 3\nu$ targeting the $\W\PH$ production, and $3\ell \nu$ + 2~jets
targeting the $\Z\PH$ production with one hadronically decaying $\W$ boson. The overall
sensitivity is dominated by the first category while the other categories
probe different production modes of the SM Higgs boson.
The search discussed here is performed for a Higgs boson with mass in the range
110--600\GeV. The search range stops at
$\mHi$ = 200\GeV for the analyses targeting the $\V\PH$ production since
for larger masses the expected $\V\PH$ cross section becomes negligible.
In the dilepton categories, non-resonant $\W\W$ production gives rise to the largest background contribution
while top-quark production is dominant in events with high jet multiplicity.
In the trilepton categories, $\W\Z$ and $\Z\Z$ production are the main background processes.
Because of the large inclusive cross section, the instrumental backgrounds from
$\W$-boson and $\Z$-boson production with associated jets or photons are
also present in the kinematic regions similar to that of the Higgs boson signal.

The paper is organized as follows.
After a brief description of the CMS detector in Section~\ref{sec:detector}
and the data and simulated samples in Section~\ref{sec:samples},
the event reconstruction is detailed in Section~\ref{sec:objects}.
The statistical procedure applied and the uncertainties considered for the interpretation of the results
are explained in Section~\ref{sec:stat},
followed by the description of analysis strategies and performance
for the dilepton categories and trilepton categories in sections~\ref{sec:sel_2l} and~\ref{sec:sel_3l}, respectively.
Finally, the results from the measurements of the Higgs boson production and properties
combining all analysis categories are reported in Section~\ref{sec:combined},
and the summary given in Section~\ref{sec:summary}.

\section{CMS detector}\label{sec:detector}
The CMS detector, described in detail in ref.~\cite{CMSdetector}, is a
multipurpose apparatus designed to study high transverse momentum ($\pt$)
physics processes in proton-proton and heavy-ion collisions.
CMS uses a right-handed coordinate system, with the origin at the
nominal interaction point, the $x$ axis pointing to the center of the
LHC, the $y$ axis pointing upwards, perpendicular to the plane of the
LHC ring, and the $z$ axis along the counterclockwise beam
direction. A superconducting solenoid occupies its central region, providing a magnetic
field of 3.8\unit{T} parallel to the beam direction. Charged-particle
trajectories are measured by the silicon pixel and strip trackers, which
cover a pseudorapidity region of $\abs{\eta} < 2.5$. Here, the pseudorapidity is
defined as $\eta=-\ln{[\tan{(\theta/2)}]}$, where $\theta$ is the polar angle
of the particle trajectory with respect to the direction of the
counterclockwise beam. A crystal electromagnetic calorimeter (ECAL) and
a brass/scintillator hadron calorimeter surround the tracking volume
and cover $\abs{\eta} < 3$. The steel/quartz-fiber Cherenkov hadron forward (HF)
calorimeter extends the coverage to $\abs{\eta} < 5$. The muon system consists of gas-ionization
detectors embedded in the steel flux return yoke outside the solenoid, and
covers $\abs{\eta} < 2.4$. The first level of the CMS trigger system,
composed of custom hardware processors, is designed to select the most
interesting events in less than 4\mus, using information from the
calorimeters and muon detectors. The high-level trigger processor farm
further reduces the event rate to a few hundred Hz before data storage.

\section{Data and simulated samples}\label{sec:samples}
\subsection{Data samples}
The data samples used in this analysis correspond to an integrated luminosity
of $\usedLumiSeven$ at a center-of-mass energy of $\sqrt{s} = 7$\TeV collected
in 2011 and of $\usedLumi$ at $\sqrt{s} = 8$\TeV
collected in 2012. The integrated luminosity is measured using data from the HF system and
the pixel detector~\cite{lumiPAS2011,lumiPAS2012}. The uncertainties in the
integrated luminosity measurement are 2.2\% in 2011 and 2.6\% in 2012.

For the analyses described in this paper, events are triggered by requiring
the presence of one or two high-$\pt$ electrons or muons.
The trigger paths consist of several single-lepton triggers with relatively
tight lepton identification. The trigger thresholds for the electron $\pt$ are
in the range of 17 to 27\GeV, while the muon $\pt$ threshold ranges
from 17 to 24\GeV. The higher thresholds are used for the periods of higher
instantaneous luminosity.
For the dilepton triggers, the minimal $\pt$ of the leading and trailing lepton is 17
and 8\GeV, respectively.
The trigger efficiency for signal events that pass any of the analysis
selections is measured to be larger than 97\% for the SM Higgs boson with
$\mHi \sim 125\GeV$. The trigger efficiency increases with the Higgs boson
mass. This efficiency is measured in data using $\Z \to \ell\ell$ events,
recorded with dedicated triggers~\cite{wzxs}. The uncertainty in the yields
derived from simulation due to the trigger efficiency is about 1\%.

\subsection{The Monte Carlo event generators}
Several Monte Carlo (MC) event generators are used to simulate the signal and
background processes. The simulated samples are used to optimize the event selection,
evaluate selection efficiencies and systematic uncertainties, and
compute expected yields.

Simulated Higgs boson signals from gluon fusion and VBF are generated with
the \POWHEG 1.0 generator~\cite{powheg}. Events for alternative spin and
parity signal hypotheses are produced by a leading-order (LO) matrix element generator,
\textsc{jhugen} 1.0~\cite{JCPExpPaper1,JCPExpPaper2}.
The simulation of associated-production
samples uses the \PYTHIA 6.4 generator~\cite{pythia}.
The mass lineshape of the Higgs boson signal at the generator level is corrected to match the results
presented in refs.~\cite{Passarino:2012ri,Goria:2011wa,Kauer:2012hd,LHCHiggsCrossSectionWorkingGroup:2013tqa}, where the
complex-pole mass scheme for the Higgs boson propagator is used.
The effects on the cross section due to the interference between the SM
Higgs boson signal and the $\Pg\Pg\rightarrow \WW$ background, as computed
in refs.~\cite{Kauer:2012ma,Campbell:2011cu}, are included.
The SM Higgs boson
production cross sections are taken
from~\cite{LHCHiggsCrossSectionWorkingGroup:2011ti,Djouadi:1991tka,Dawson:1990zj,Spira:1995rr,Harlander:2002wh,Anastasiou:2002yz,Tackmann:2011,Ravindran:2003um,Catani:2003zt,Aglietti:2004nj,Degrassi:2004mx,Baglio:2010ae,Actis:2008ug,Anastasiou:2008tj,deFlorian:2009hc,Ciccolini:2007jr,Ciccolini:2007ec,Arnold:2008rz,Brein:2003wg,Ciccolini:2003jy,Djouadi:1997yw,Denner:2011mq,Bredenstein:2006rh,Bredenstein:2006ha}.

The $\W\Z$, $\Z\Z$, $\VVV$ ($\V$ = $\W/\Z$), Drell--Yan (DY) production of $\Z/\gamma^*$, $\Wjets$, and $\qqbar \to\WW$ processes
are generated using the \MADGRAPH 5.1 event generator~\cite{Madgraph},
the $\Pg\Pg \to \WW$ process using the \textsc{gg2ww} 3.1 generator~\cite{ggww}, and the $\ttbar$
and $\tw$ processes are generated with \POWHEG.
The electroweak production of non-resonant $\WW$ + 2~jets process,
which is not part of the inclusive $\WW$ + jets sample,
has been generated using the $\Phantom$ 1.1 event generator~\cite{phantom}
including terms of order
$(\alpha_{EW}^{6})$. As a cross-check, the \MADGRAPH
generator has also been used to generate such events.
All other processes are generated using \PYTHIA.

The set of parton distribution functions (PDF) used is CTEQ6L~\cite{cteq66}
for LO generators, while CT10~\cite{Lai:2010vv} is used for
next-to-leading-order (NLO) generators. All the event generators are interfaced
to \PYTHIA for the showering of partons.
For all processes, the detector response is simulated using a detailed
description of the CMS detector, based on the \GEANTfour
package~\cite{Agostinelli:2002hh}.
Additional simulated pp interactions overlapping with the event of interest
in the same bunch crossing, denoted as pileup events, are added in the
simulated samples to reproduce the pileup distribution measured in data.
The average numbers of pileup events per beam
crossing in the 2011 and 2012 data are approximately 9 and 21, respectively.

The $\dytt$ and $\wgamma^{*}$ background processes are
evaluated with a combination of simulated and data samples.
The $\dytt$ background process is estimated using $\dymm$ events selected in
data, in which the muons are replaced with simulated $\Tau$ decays, thus
providing a more accurate description of the experimental conditions with respect to
the full simulation. The \TAUOLA package~\cite{tauola} is used
in the simulation of $\Tau$ decays to account for $\Tau$-polarization effects.
The uncertainty in the estimation of this background
process is about 10\%.

The \MADGRAPH generator is used to estimate the $\wgamma^{*}$
background contribution from asymmetric virtual photon decays~\cite{wgammastart},
in which one lepton escapes detection. To obtain the
normalization scale of the simulated events, a high-purity control sample of
$\wgamma^{*}$ events with three reconstructed leptons is defined and compared
to the simulation, as described in Appendix~\ref{app:wgstar}. As a result of the
analysis in that control sample, a factor of $1.5\pm0.5$ with
respect to the predicted LO cross section is found.

\subsection{Theoretical uncertainties}
The uncertainties in the signal and background production rates due
to theoretical uncertainties include several components, which are assumed to be
independent: the PDFs and $\alpha_{s}$, the underlying event and parton shower model, the
effect of missing higher-order corrections via variations of the renormalization
and factorization scales, and the corrections for the interference between the signal
and the background $\WW$ production.

The effect on the yields from variations in the choice of PDFs and the value of $\alpha_s$
is considered following the {PDF4LHC} prescription~\cite{Alekhin:2011sk,Botje:2011sn},
using the CT10, NNPDF2.1~\cite{Ball:2011mu}, and
MSTW2008~\cite{Martin:2009iq} PDF sets.
For the gluon-initiated signal processes ($\Pg\Pg\PH$ and $\ttbar\PH$), the PDF
uncertainty is about 8\%, while for the quark-initiated processes (VBF and $\V\PH$)
it is 3--5\%. The PDF uncertainties for background processes are 3--6\%.
These uncertainties are assumed to be correlated among processes with identical
LO initial states, without considering whether or not they are signal or
background processes.

The systematic uncertainties due to the underlying event and parton shower model~\cite{ue_cms_7tev,ue_cms_7tev_dy} are
estimated by comparing samples simulated with different MC event generators.
In particular, for the main signal process, $\Pg\Pg\PH$, the \POWHEG
MC generator, interfaced with
\PYTHIA for the parton shower and hadronization, is compared to the
\MCATNLO 4.0 generator~\cite{MCatNLO}, interfaced with \HERWIG{++}~\cite{herwig} for
the parton shower and hadronization model.
Alternative $\qqbar \to\WW$ samples for dedicated studies are produced
with the \MCATNLO and \POWHEG event generators, and compared
to the default \MADGRAPH, while alternative top-quark samples
are produced with \MADGRAPH and compared to the default \POWHEG sample.

The uncertainties in the yields from missing higher-order corrections are evaluated by independently
varying up and down the factorization and renormalization scales by a factor of two.
The categorization of events based on jet multiplicity
introduces additional uncertainties, mainly driven by the factorization
and renormalization scales, as explained in
refs.~\cite{LHCHiggsCrossSectionWorkingGroup:2011ti,Tackmann:2011,XieThesis}.
These uncertainties range between 10\% and 40\%, depending on the jet category
and production mode. They are calculated using the \MCFM program~\cite{MCFM} for
the VBF and $\V\PH$ signal and the diboson ($\W\Z$ and $\Z\Z$) background processes, while for the $\Pg\Pg\PH$
process the \textsc{hqt} program~\cite{HqT1,HqT2} is used.

The uncertainties associated with the interference effect between the SM Higgs boson
signal and the $\Pg\Pg\rightarrow \WW$ background process is up to 30\% at a Higgs
boson mass of 600\GeV, and becomes negligible for masses below 400\GeV.

\section{Event reconstruction}\label{sec:objects}

A particle-flow algorithm~\cite{PFT-09-001} is used to reconstruct the
observable particles in the event. Clusters of energy deposition measured
by the calorimeters and charged-particle tracks identified in the central tracking system
and the muon detectors are combined to reconstruct individual particles
and to set quality criteria to select and define final-state observables.

For each event, the analyses require two or three high-$\pt$ lepton candidates
(electrons or muons) originating from a single primary vertex. Among
the vertices identified in the event, the vertex with the largest $\sum \pt^2$,
where the sum runs over all tracks associated with the vertex, is chosen as the primary vertex.

Electron candidates are defined by a reconstructed charged-particle track in the
tracking detector pointing to a cluster of energy deposition in the ECAL.
A multivariate~\cite{tmva} approach to identify electrons is employed combining
several measured quantities describing the track quality, the ECAL cluster shapes,
and the compatibility of the measurements from the two detectors.
The electron energy is measured primarily from the ECAL cluster energy.
For low-$\pt$ electrons, a dedicated algorithm combines the momentum of
the track and the ECAL cluster energy, improving the energy resolution~\cite{Chatrchyan:2013dga}.
Muon candidates are identified by signals of charged-particle tracks in the muon system
that are compatible with a track reconstructed in the central tracking system.
The precision of the muon momentum measurement from the curvature of the track
in the magnetic field is ensured by minimum requirements on the
number of hits in the layers of sensors and on the quality of the full track fit.
Uncertainties in the lepton momentum scale and resolution are 0.5--4\% per lepton
depending on the kinematic properties, and the effect on the yields at the
analysis selection level is approximately 2\% for electrons and 1.5\% for muons.

Electrons and muons are required to be isolated to distinguish between prompt leptons
from $\W/\Z$-boson decays and those from QCD production or misidentified leptons,
usually situated inside or near jets of hadrons.
The variable $\delR = \sqrt{(\Delta\eta)^2 + (\Delta\phi)^2}$ is used to measure
the separation between reconstructed objects in the detector, where $\phi$ is the angle
(in radians) of the trajectory of the object in the plane transverse to the direction of the
proton beams.
Isolation criteria are set based on the distribution of low-momentum particles
in the ($\eta,\phi$) region around the leptons.
To remove the contribution from the overlapping pileup interactions in this isolation region,
the charged particles included in the computation of the isolation variable
are required to originate from the lepton vertex.
A correction is applied to the neutral component in the isolation $\delR$ cone based on the average energy density
deposited by the neutral particles from additional interactions~\cite{Cacciari:subtraction}. The correction is measured
in a region of the detector away from the known hard scatter in a control sample.
Electron isolation is characterized by the ratio of
the total transverse momentum of the particles reconstructed in a $\delR =~0.3$ cone
around the electron, excluding the candidate itself,
to the transverse energy of the electron.
Isolated electrons are selected by requiring this ratio to be
below $\sim$10\%. The exact threshold value depends on the electron $\eta$ and $\pt$~\cite{XieThesis,10281_40097}.
For each muon candidate, the scalar sum of the transverse energy of all particles
originating from the primary vertex is reconstructed in
$\delR$ cones of several radii around the muon direction, excluding the
contribution from the muon itself. This information is combined using a
multivariate algorithm that exploits the differential energy deposition in the
isolation region to discriminate between the signal of prompt muons and muons
from hadron decays inside a jet.

Lepton selection efficiencies are determined using $\Z \to \ell\ell$ events~\cite{wzxs}.
Simulated samples are corrected by the difference in the efficiencies found
in data and simulation. The total uncertainty in lepton
efficiencies, that includes effects from reconstruction, trigger, and various
identification criteria, amounts to about 2\% per lepton. The lepton selection
criteria in the 7 and 8\TeV samples were tuned to maintain an efficiency
independent of the instantaneous luminosity.

Jets are reconstructed using the anti-$\mathrm{k_T}$ clustering algorithm~\cite{antikt}
with a distance parameter of 0.5, as implemented in the \textsc{fastjet}
package~\cite{Cacciari:fastjet1,Cacciari:fastjet2}.
A similar correction as for the lepton isolation is applied to account for the contribution
to the jet energy from pileup events. Furthermore, the properties of the hard jets
are modified by particles from pileup interactions. A
combinatorial background arises from low-$\pt$ jets from pileup interactions which get
clustered into high-$\pt$ jets. At $\sqrt{s} = 8\TeV$ the number of pileup events is larger than
at $\sqrt{s} = 7\TeV$ and a multivariate selection is applied to separate jets from the
primary interaction and those reconstructed due to energy deposits associated with
pileup interactions~\cite{jetIdPAS}. The discrimination is based on the differences in the jet shapes,
on the relative multiplicity of charged and neutral components, and on the different
fraction of transverse momentum which is carried by the hardest components.
Within the tracker acceptance the tracks belonging to each jet are also required to be compatible
with the primary vertex. Jet energy corrections are applied as a function
of the jet $\pt$ and $\eta$~\cite{cmsJEC}. The jet energy scale and resolution
gives rise to an uncertainty in the yields of 2\% (5\%) for the low (high) jet
multiplicity events. Jets considered for the event categorization are required
to have $\pt>30\GeV$ and $\abs{\eta}<4.7$. Studies have been performed selecting
$\Zjets$ events and comparing the number of jets distribution as a function of
the number of reconstructed vertices. A rather flat behavior has been found,
which indicates that the effect from pileup interactions is properly mitigated.

Identification of decays of the bottom (b) quark is used to discriminate
the background processes containing top-quark
that subsequently decays to a bottom-quark and a $\W$ boson.
The bottom-quark decay is identified by the presence
of a soft-muon in the event from the semileptonic decay of the bottom-quark
and by bottom-quark jet (b-jet) tagging criteria based on the impact parameter
of the constituent tracks~\cite{btag}. In particular, the Track Counting High
Efficiency algorithm is used with a value greater than 2.1 to assign a given
jet as b-tagged. Soft-muon candidates are defined without isolation requirements
and are required to have $\pt >$ 3\GeV.
The set of veto criteria retain about 95\% of the light-quark jets, while rejecting
about 70\% of the b-jets. The performance of b-jet identification
for light-quark jets is verified in $\dyll$ candidate events, and is found to
be consistent between data and simulation within 1\% for the events with up to
one jet and within 3\% for the events with two central jets.

The missing transverse energy vector $\vmet$ is defined as
the negative vector sum of the transverse momenta of all reconstructed
particles (charged or neutral) in the event, with $\met = \abs{\vmet}$.
For the dilepton analyses, a \textit{projected}~$\met$ variable is defined
as the component of $\vmet$ transverse
to the nearest lepton if the lepton is situated
within the azimuthal angular window of $\pm \pi/2$ from
the $\vmet$ direction, or the $\met$ itself otherwise.
A selection using this observable efficiently rejects
$\dytt$ background events, in which the $\vmet$ is preferentially aligned with leptons,
as well as $\dyll$ events with mismeasured $\vmet$ associated with poorly
reconstructed leptons or jets.
Since the $\vmet$ resolution is degraded by pileup,
the minimum of two projected~$\met$ variables is
used ($\pmet$): one constructed from all identified particles
(full $\met$), and another constructed from the charged particles only
(track $\met$).
The uncertainty in the resolution of the $\vmet$ measurement is
approximately 10\%, which is estimated from $\Z \to \ell\ell$ events with the same lepton
selection applied as in the rest of the analysis. Randomly smearing the measured
$\vmet$ by one standard deviation gives rise to a 2\% variation in the estimation
of signal yields after the full selection for all analyses.

\section{Statistical procedure}\label{sec:stat}
The statistical methodology used to interpret subsets of data selected for the $\PH\to\WW$ analyses
and to combine the results from the independent categories
has been developed by the ATLAS and CMS collaborations in the context of the LHC Higgs Combination Group.
A general description of the methodology can be found in refs.~\cite{LHC-HCG,Chatrchyan:2012tx}.
Results presented in this paper also make use of asymptotic formulae from ref.~\cite{Cowan:2010st}
and recent updates available in the \textsc{RooStats} package~\cite{RooStats}.

Several quantities are defined to compare the observation in data with the expectation for
the analyses: upper limits on the production cross section
of the $\PH\to\WW$ process with and without the presence
of the observed new boson;
a significance, or a p-value, characterizing the probability of
background fluctuations to reproduce an observed excess;
signal strengths ($\sigma/\sigma_\mathrm{SM}$) that quantify
the compatibility of the sizes of the observed excess with the SM signal expectation;
and results from a test of two independent signal hypotheses,
namely a SM-like Higgs boson with spin $0^+$ with respect to
a $\spintwopmin$~resonance or a pseudoscalar $0^{-}$ boson.
The modified frequentist method, $\mathrm{CL_s}$~\cite{Read1,junkcls},
is used to define the exclusion limits.
A description of the statistical formulae defining these
quantities is found in ref.~\cite{LHC-HCG,CMSPaperCombination}.

The number of events in each category and in each bin of the discriminant distributions
used to extract the signal is modeled as a Poisson random variable,
whose mean value is the sum of the contributions from the processes under consideration.
Systematic uncertainties are represented by individual nuisance parameters
with log-normal distributions. An exception is applied to the $\qqbar \to \W\W$
normalization in the 0-jet and 1-jet dilepton shape-based fit analyses, described in Section~\ref{sec:hww2l2n_01j},
which is an unconstrained parameter in the fit. The uncertainties affect the overall
normalization of the signal and backgrounds
as well as the shape of the predictions across the distribution of the observables.
Correlation between systematic uncertainties in different categories and final states
are taken into account. In particular, the main sources of correlated systematic
uncertainties are those in the experimental measurements
such as the integrated luminosity, the lepton and trigger selection efficiencies, the lepton momentum
scale, the jet energy scale and missing transverse energy resolution (Section~\ref{sec:objects}), and
the theoretical uncertainties affecting the signal and background processes (Section~\ref{sec:samples}).
Uncertainties in the background
normalizations or background model parameters from control regions
(sections~\ref{sec:sel_2l} and~\ref{sec:sel_3l}) and uncertainties of statistical nature are uncorrelated.
A summary of the systematic uncertainties is shown in Table~\ref{tab:systww},
with focus on the 0-jet and 1-jet dilepton categories.

\begin{table}[htbp]
\centering
\topcaption{Summary of systematic uncertainties relative to the yields (in \%) from various signal
and background processes. Precise values depend on the final state, jet category, and data taking period.
The values listed in the table apply to the 0-jet and 1-jet dilepton categories.
The horizontal bar (---) indicates that the corresponding uncertainty is not applicable.
The jet categorization uncertainty originates from the uncertainties in the
renormalization and factorization scales that change the fraction of events
in each jet category. The systematic uncertainty from the same source is considered
fully correlated across all relevant processes listed.\label{tab:systww}}
\resizebox{\textwidth}{!}{
\begin{tabular}{lcccccccc}
\hline\hline
\multirow{2}{*}{Source} & $\mathrm{H}\to$         & $\qqbar \to$ & $\Pg\Pg \to$  & Non-$\Z$ resonant & $\ttbar+\tw$ & $\dyll$ & $\Wjets$ & $\V\gamma^{(*)}$    \\
                        &  $\WW$  & $\WW$    & $\WW$       & $\WZ/\ZZ$         &     &         &             &                                 \\
\hline
Luminosity                               &2.2--2.6& --- & --- & 2.2--2.6& --- & --- & --- & 2.2--2.6  \\
Lepton efficiency                        & 3.5   & 3.5 & 3.5 & 3.5    & --- & --- & --- &  3.5     \\
Lepton momentum scale                    & 2.0   & 2.0 & 2.0 & 2.0    & --- & --- & --- &  2.0     \\
$\vmet$ resolution                       & 2.0   & 2.0 & 2.0 & 2.0    & --- & --- & --- &  1.0     \\
Jet counting categorization              & 7--20  & --- & 5.5 & 5.5    & --- & --- & --- &  5.5     \\
Signal cross section                     & 5--15  & --- & --- & ---    & --- & --- & --- &  ---     \\
$\qqbar \to WW$ normalization            & ---   &  10 & --- & ---    & --- & --- & --- &  ---     \\
$\Pg\Pg \to WW$ normalization            & ---   & --- &  30 & ---    & --- & --- & --- &  ---     \\
$\WZ/\ZZ$ cross section                  & ---   & --- & --- & 4.0    & --- & --- & --- &  ---     \\
$\ttbar+\tw$ normalization                       & ---   & --- & --- & ---    &  20 & --- & --- &  ---     \\
$\dyll$ normalization			 & ---   & --- & --- & ---    & --- &  40 & --- &  ---     \\
$\Wjets$ normalization                   & ---   & --- & --- & ---    & --- & --- &  36 &  ---     \\
MC statistics                            & 1.0   & 1.0 & 1.0 & 4.0    & 5.0 &  20 &  20 &   20     \\
\hline
\end{tabular}
}
\end{table}

\section{Final states with two charged leptons}\label{sec:sel_2l}

The $\PH \to \WW \to 2\ell 2\nu$ decay features a signature with two
isolated, high-$\pt$, charged leptons and moderate $\met$.
After all selection criteria are applied, the contribution from other Higgs boson
decay channels is negligible.
Kinematic distributions of the decay products exhibit
the characteristic properties of the parent boson.
The three main observables are: the azimuthal opening angle between
the two leptons ($\delphill$), which is correlated to the spin of the Higgs boson;
the dilepton mass ($\mll$), which is one of the most discriminating kinematic variables
for a Higgs boson with low mass, especially against the $\dyll$ background;
and the transverse mass ($\mth$) of the final state objects,
which scales with the Higgs boson mass.
The transverse mass is defined as $\mth^2 = 2 \ptll \met \big(1-\cos\delphillmet\big)$,
where $\ptll$ is the dilepton transverse momentum and
$\delphillmet$ is the azimuthal angle between the dilepton momentum and $\vmet$.

\subsection{\texorpdfstring{$\W\W$}{WW}~~selection and background rejection}

To increase the sensitivity to the SM Higgs boson signal,
events are categorized into lepton pairs of same flavor
(two electrons or two muons, $\Pe\Pe/\mu\mu$) and of different flavor
(one electron and one muon, $\Pe\mu$),
and according to jet multiplicities in zero (0-jet), one (1-jet), and two or more jet (2-jet) categories, where
the jets are selected as described in Section~\ref{sec:objects}.
Splitting the events into categories that differ in signal and background composition
imposes additional constraints on the backgrounds
and defines regions with high signal purity.

The Higgs boson signal events in 0-jet and 1-jet categories are mostly produced by the gluon fusion process.
These categories have relatively high yield and purity and allow measurements of the Higgs boson properties.
The 2-jet category is further separated into events
with a characteristic signature of VBF production with two energetic forward-backward jets
and heavily suppressed additional hadronic activity due to the lack of color flow between the parent quarks,
and those with a VH signature in which two central jets originate from the vector boson decay.
While the sensitivity of the 2-jet category is limited with the current dataset,
the two sub-categories explore specific production modes.
A summary of the selection requirements and analysis approach, as well as the most important
  background processes in the dilepton categories
is shown in Table~\ref{tab:summary_2l}.

\begin{table}[htbp]
\centering
\topcaption{A summary of the selection requirements and analysis approach, as well as the most important
background processes in the dilepton categories.
The same-flavor final states make use of a counting analysis approach in all categories.}
\resizebox{\textwidth}{!}
{
\begin{tabular} {lccc}
\hline\hline
 & Zero-jet and one-jet {$\Pg\Pg\PH$}~~tag & Two-jet VBF tag & Two-jet {$\V\PH$}~~tag \\
\hline
Number of jets & $= 0/1$ & $\geq 2$ & $\geq 2$ \\
\hline
\multirow{2}{*}{Default analysis} & binned shape-based ($\Pe\mu$) & binned shape-based ($\Pe\mu$)  & \multirow{2}{*}{counting} \\
                                  &counting ($\Pe\Pe$, $\mu\mu$)  & counting ($\Pe\Pe$, $\mu\mu$)  &                           \\
\multirow{2}{*}{Alternative analyses} & parametric shape-based & \multirow{2}{*}{counting} & \multirow{2}{*}{binned shape-based} \\
                                      & counting               &                           &                                     \\
VBF tagging & --- & applied & vetoed \\
Main backgrounds & $\W\W$, top-quark, $\Wjets$, $\W\gamma^{(*)}$ & $\W\W$, top-quark & $\W\W$, top-quark \\
\hline
  \end{tabular}
}
\label{tab:summary_2l}
\end{table}

For all jet multiplicity categories, candidate events are composed of exactly two oppositely charged
leptons with $\pt >20\GeV$ for the leading lepton ($\ptlmax$)
and $\pt >10\GeV$ for the trailing lepton ($\ptlmin$).
Events with additional leptons are analyzed separately, as described in Section~\ref{sec:sel_3l}.
The electrons and muons considered in the analysis
include a small contribution from decays via intermediate $\Pgt$ leptons.
The $\pmet$ variable is required to be above 20\GeV. The analysis is
restricted to the kinematic region with
$\mll > 12\GeV$, $\ptll > 30\GeV$, and $\mth > 30\GeV$,
where the signal-to-background ratio is
high and the background content is correctly described.

The main background processes from non-resonant $\WW$ production and
from top-quark production, including top-quark pair ($\ttbar$) and
single-top-quark (mainly $\tw$) processes, are estimated using data.
Instrumental backgrounds arising from
misi\-denti\-fi\-ed (``non-prompt") leptons in $\W$+jets production and
mismeasurement of $\vmet$ in $\Z/\gamma^*$+jets events are also estimated from data.
Contributions from $\wgamma$, $\wgamma^*$, and other sub-dominant diboson ($\W\Z$ and $\Z\Z$)
and triboson ($\VVV$, $\V$ = $\W/\Z$) production
processes are estimated partly from simulated samples,
see Section~\ref{sec:samples}. The $\wgamma^*$
cross section is measured from data, as described in Appendix~\ref{app:wgstar}.
The shapes of the discriminant variables used in the signal extraction
for the $\wgamma$ process are obtained from data, as explained in Appendix~\ref{app:wgamma}.

The non-prompt lepton background, originating from leptonic decays of heavy quarks,
hadrons misi\-denti\-fi\-ed as leptons, and electrons from photon conversions
in $\Wjets$ and QCD multijet production,
is suppressed by the identification and isolation requirements
on electrons and muons, as described in Section~\ref{sec:objects}.
The remaining contribution from the non-prompt lepton background is estimated directly from data.
A control sample is defined by
one lepton that passes the standard lepton selection criteria and another
lepton candidate that fails the criteria, but passes a looser selection,
resulting in a sample of ``pass-fail" lepton pairs.
The efficiency, $\epsilon_\text{pass}$, for a jet that satisfies the loose lepton
requirements to pass the standard selection is determined using an independent sample
dominated by events with non-prompt leptons from QCD multijet processes.
This efficiency, parameterized as a function of $\pt$ and $\eta$ of the lepton, is then
used to weight the events in the pass-fail sample by \mbox{$\epsilon_\text{pass}$/(1 - $\epsilon_\text{pass}$)},
to obtain the estimated contribution from the non-prompt lepton background in the signal region.
The systematic uncertainties from the determination of $\epsilon_\text{pass}$
dominate the overall uncertainty of this method.
The systematic uncertainty has two sources: the dependence of
$\epsilon_\text{pass}$ on the sample composition, and the method. The first source is estimated
by modifying the jet $\pt$ threshold in the QCD multijet sample, which modifies the jet sample
composition. The uncertainty in the method is obtained from a closure test,
where $\epsilon_\text{pass}$ is derived from simulated QCD multijet events and applied
to simulated samples to predict the number of background events.
The total uncertainty in $\epsilon_\text{pass}$, including the statistical
precision of the control sample, is of the order of 40\%.
Validation of the estimate of this background using lepton pairs with the same charge is described
in Section~\ref{sec:hww2l2n_01j}.

The Drell--Yan $\Z/\gamma^*$ production is the largest source of
same-flavor lepton pair production because of its large production
cross section and the finite resolution of the $\vmet$ measurement.
In order to suppress this background, a few additional selection requirements are applied in the same-flavor final states.
The resonant component of the Drell--Yan production is rejected by requiring $\mll$ to be more than 15\GeV away
from the $\Z$ boson mass. To suppress the remaining off-peak contribution, in the 8\TeV sample,
a dedicated multivariate selection combining $\met$ and kinematic and topological variables is used.
In the 7\TeV sample the amount of pileup interactions is smaller on average and
a selection based on a set of simple kinematic variables is adopted.
The $\ptlmin$ and $\mll$ thresholds are raised to 15\GeV and 20\GeV respectively,
and the selection based on $\pmet$ is applied progressively tighter
as a function of the number of reconstructed vertices,
$N_\text{vtx}$, $\pmet$ $>$ (37 + $N_\text{vtx}/2$)\GeV. This requirement is
chosen to obtain a background efficiency nearly constant as a function
of $N_\text{vtx}$.
Events in which the direction of the dilepton momentum and that of the most energetic
jet with $p_T >15$\GeV have an angular difference in the transverse plane greater than 165 degrees are rejected.
For the 2-jet category, the dominant source of $\vmet$ is the mismeasurement of the hadronic recoil
and the best performance in terms of signal-to-background separation is obtained by simply requiring $\MET > 45\GeV$
and the azimuthal separation of the dilepton and dijet momenta to be $\Delta\phi(\ell\ell,jj) <$ 165 degrees.
These selection requirements effectively reduce the Drell--Yan background by three orders of magnitude,
while retaining more than 50\% of the signal.
The $\Z/\gamma^*\to \Pe\Pe/\mu\mu$ contribution to the analysis in the same-flavor final states is obtained by
normalizing the Drell--Yan background to data in the region within ${\pm}7.5\GeV$
of the $\Z$ boson mass after flavor symmetric contributions from other processes are
subtracted using $\Pe\mu$ events.
The extrapolation to the signal region is performed using the simulation
together with a cross-check using data. A more detailed explanation of the
Drell--Yan background estimation is given in Appendix~\ref{app:dyll}. The largest
uncertainty in the estimate arises from the dependence of this extrapolation factor
on $\met$ and the multivariate Drell--Yan discriminant, and is about 20 to 50\%.
The contribution of this background is also evaluated with an alternative method
using $\gamma$ + jets events, which provides results consistent with the
primary method. The $\Z$ boson and the photon exhibit similar
kinematic properties at high $\pt$ and the hadronic recoil is similar in the two cases, and therefore
a $\gamma$ + jets sample is suitable to estimate the Drell--Yan background.

To suppress the background from top-quark production,
events that are top-tagged are rejected based on soft-muon and b-jet
identification (Section~\ref{sec:objects}).
The reduction of the top-quark background is about 50\% in the 0-jet category and above 80\%
for events with at least one jet with $\pt >30\GeV$.
The top-quark background contribution in the analysis is estimated using
top-tagged events ($N_\text{tagged}$).
The top-tagging efficiency ($\epsilon_\text{top-tagged}$) is measured in a control sample
dominated by $\ttbar$ and $\tw$ events, which is selected by requiring one jet to be b-tagged.
The number of top-quark background events ($N_\text{not-tagged}$) expected in the signal region is estimated as:
$N_\text{not-tagged} = N_\text{tagged} \times (1-\epsilon_\text{top-tagged})/\epsilon_\text{top-tagged}$.
Background contributions from other sources are subtracted from the top-tagged sample.
The total uncertainty in $N_\text{not-tagged}$ amounts to about 20\% in the 0-jet, 5\% in the 1-jet,
and 30--40\% in the 2-jet category. Additional selection requirements in the 2-jet category limit the precision of the control sample.
A more detailed explanation of the top-quark background estimation is given in Appendix~\ref{app:top}.

The criteria described above define the $\WW$ selection. The remaining
data sample is dominated by non-resonant $\WW$ events, in particular in the 0-jet category.
The normalization of the $\WW$ background is obtained from the data 0-jet and 1-jet categories.
The procedure depends on the analysis strategy being pursued, as described in
Section~\ref{sec:hww01j_stragegy}.
In the counting analysis, the $\WW$ contribution is normalized to data after subtracting
backgrounds from other sources in the signal-free region of high dilepton mass, $\mll > 100\GeV$,
for $\mH \leq 200\GeV$. For the higher Higgs boson mass hypotheses and in
the 2-jet category, the control region for $\WW$ production is contaminated by
the signal together with other backgrounds. In this case the $\WW$ background prediction
is obtained from simulation and the theoretical uncertainty is
20--30\% for the $\V\PH$ and the VBF selection requirements. Both shape and normalization of
the $\W\W$ background in the $\Pe\mu$ final state for the 0-jet and 1-jet categories
are determined from a fit to data, as described in Section~\ref{sec:hww2l2n_01j}.
Studies to validate the fitting procedure are also summarized in that section.

A summary of the estimation of the background processes in the dilepton
categories is shown in Table~\ref{tab:summary_2l_bkg}.

\begin{table}[htbp]
\centering
\topcaption{Summary of the estimation of the background processes in dilepton
categories in cases where data events are used to estimate either the
normalization or the shape of the discriminant variables. A brief description of
the control/template sample is given. The $\W\W$ estimation in the 2-jet category
is purely from simulation.}
\resizebox{\textwidth}{!}{
\begin{tabular} {lccl}
  \hline\hline
 Process   & Normalization & Shape & Control/template sample \\
\hline
$\W\W$     & data & simulation & events at high $\mll$ and $\mt$ \\
Top-quark  & data & simulation & top-tagged events \\
$\Wjets$   & data & data & events with loosely identified leptons \\
$\W\gamma$ & simulation & data & events with an identified photon \\
$\W\gamma^{*}$ & data & simulation & $\W\gamma^{*} \to 3\mu$ sample \\
$\dymm$ \& $\dyee$ & data & simulation & events at low $\met$ \\
$\dytt$ & data & data & $\tau$ embedded sample \\
\hline
\end{tabular}
 }
\label{tab:summary_2l_bkg}
\end{table}

\begin{figure}[htbp]
\begin{center}
{\includegraphics[width=0.45\textwidth]{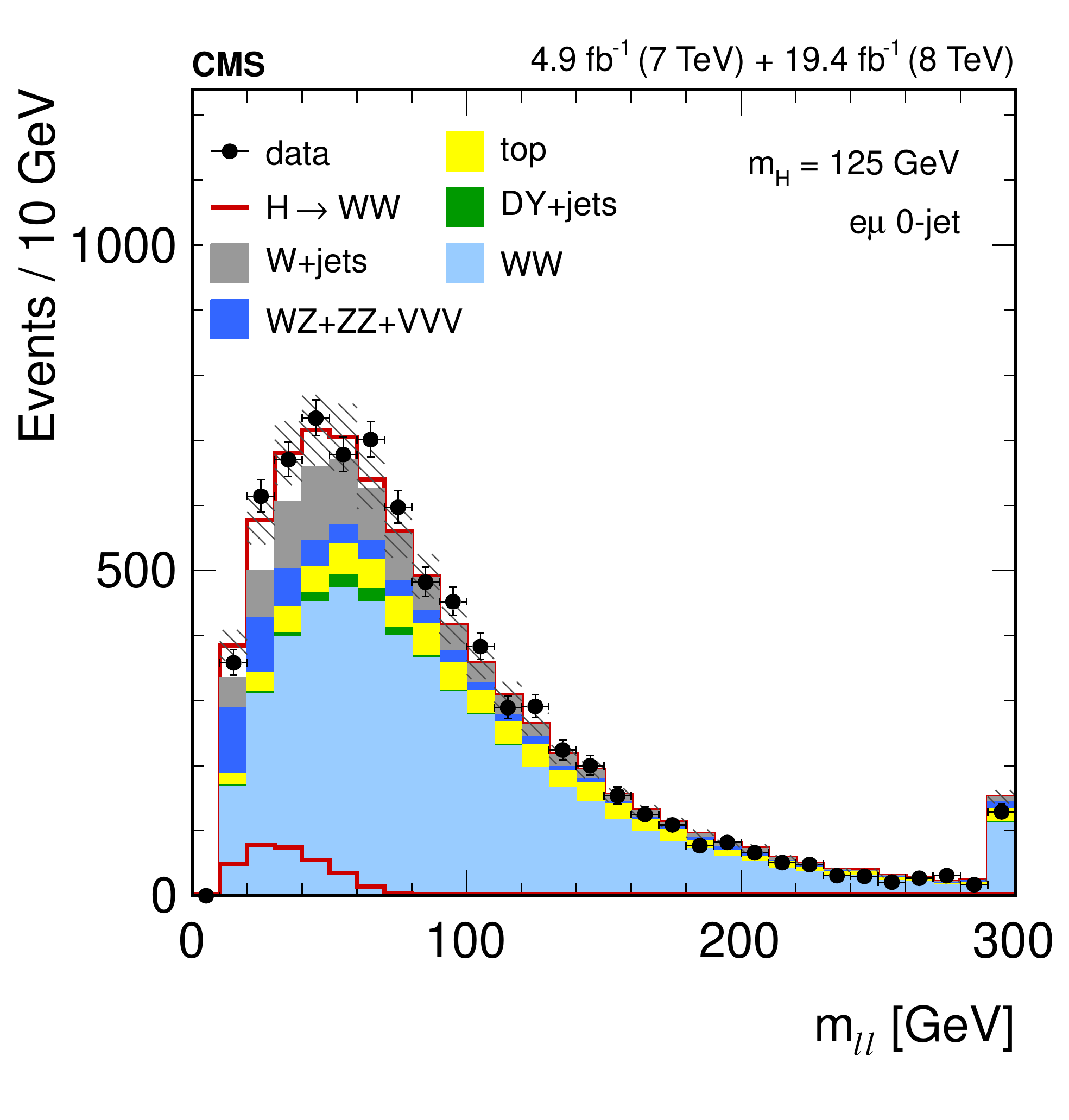}}
{\includegraphics[width=0.45\textwidth]{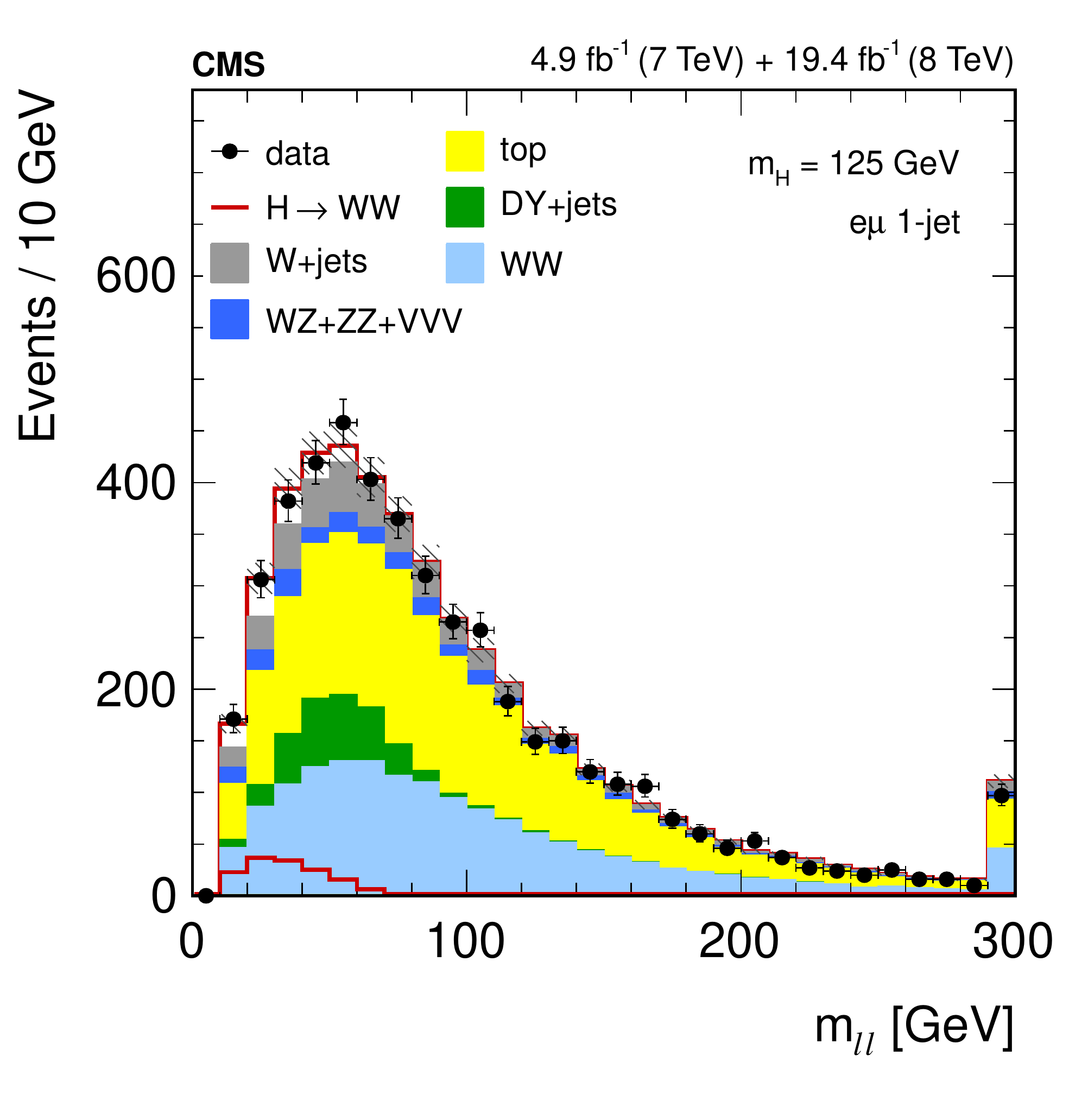}}
\caption{Distributions of the dilepton invariant mass
in the 0-jet category (left), and in the 1-jet category (right), in the $\Pe\mu$ final state
for the main backgrounds (stacked histograms), and for a SM Higgs boson signal with $\mH= 125\GeV$
(superimposed and stacked open histogram) at the $\WW$ selection level.
The last bin of the histograms includes overflows.
\label{fig:wwlevel_01jet}}
\end{center}
\end{figure}

\begin{figure}[htbp]
\begin{center}
{\includegraphics[width=0.45\textwidth]{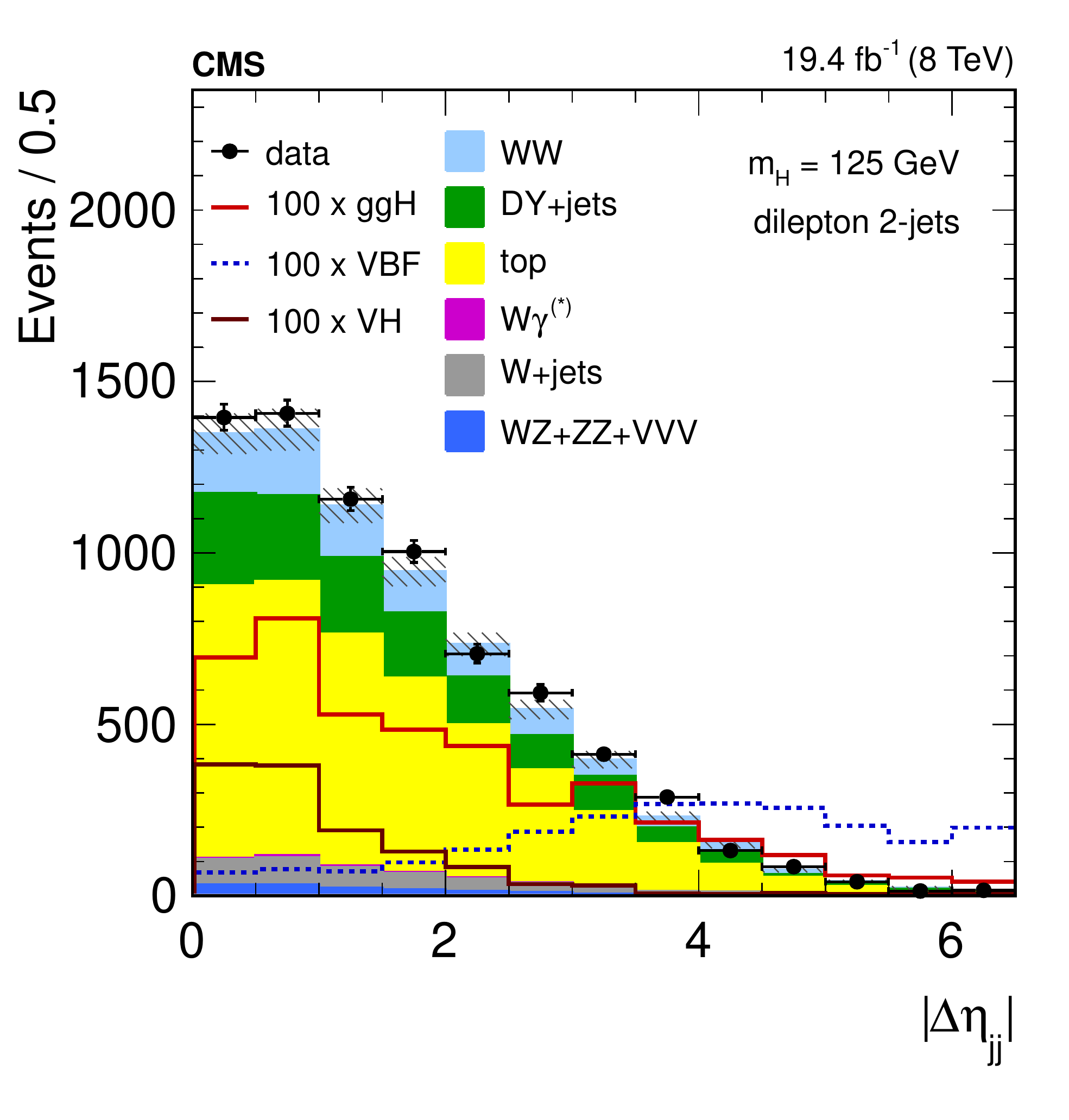}}
{\includegraphics[width=0.45\textwidth]{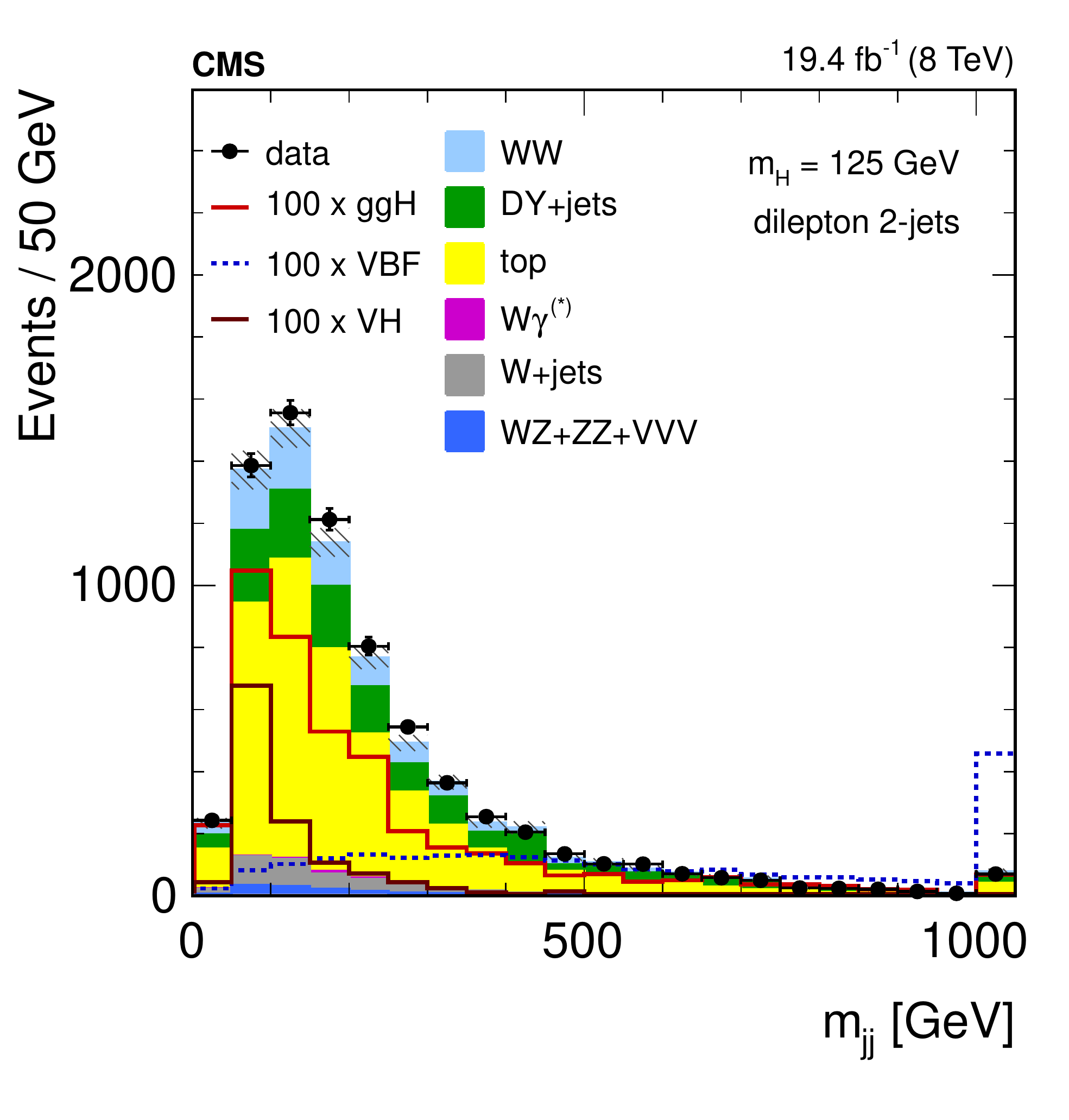}}
\caption{Distributions of the pseudorapidity separation between two highest $\pt$ jets (left)
and the dijet invariant mass (right) in the 2-jet category for the main backgrounds (stacked histograms), and for a SM Higgs boson signal with $\mH= 125\GeV$
(superimposed histogram) at the $\WW$ selection level. The signal contributions are multiplied by 100. All three final states,
$\Pe\Pe$, $\mu\mu$, and $\Pe\mu$, are included.
The last bin of the histograms includes overflows.
\label{fig:wwlevel_2jet}
}
\end{center}
\end{figure}

The $\mll$ distributions after the $\WW$ selection in the $\Pe\mu$ final state
for the 0-jet and 1-jet categories are shown in Fig.~\ref{fig:wwlevel_01jet},
together with the expectation for a SM Higgs boson with $\mH=125\GeV$.
The clear difference in the shape between the $\hwwnopm$ and the non-resonant $\ww$
processes for $m_{\ell\ell}$ is mainly due to the spin-0 nature of the SM Higgs boson.
For a SM Higgs boson with $\mH=125\GeV$, an excess of events with respect to the backgrounds
is expected at low $m_{\ell\ell}$.
For the 2-jet category, the dijet variables which are used to distinguish $\V\PH$
production from VBF production are shown in Fig.~\ref{fig:wwlevel_2jet}.
Control regions in a similar kinematic topology are studied to cross-check the background
normalization and distribution.

\subsection{The zero-jet and one-jet \texorpdfstring{$\Pg\Pg\PH$}{gluon gluon H}~~tag}
\label{sec:hww2l2n_01j}
The analysis in this category provides good sensitivity to
identify Higgs boson production, and to test the spin-0 hypothesis
against the spin-2 hypothesis.
The majority of the SM Higgs boson events originate from the gluon fusion process,
and the event selection relies entirely on the Higgs boson decay signature
of two leptons and $\met$.

While the dominant background is the non-resonant $\WW$ production, a
relatively small contamination from $\Wjets$ and $\wgamma^{(*)}$ production nevertheless
contributes sizeably to the total uncertainty in the measurements
since these processes are less precisely known and can mimic the signal topology.
Separating the analysis in lepton flavor pairs
isolates the most sensitive $\Pe\mu$ final state from the $\Pe\Pe/\mu\mu$ final states,
which have additional background contributions from processes with a $\dyll$ decay.
Splitting the sample into jet multiplicity categories with zero and one
jet distinguishes the kinematic region dominated by top-quark background (1-jet category)
which has jets from bottom-quark fragmentation, as shown in Fig.~\ref{fig:wwlevel_01jet}.

\subsubsection{Analysis strategy}\label{sec:hww01j_stragegy}

To enhance the sensitivity to a Higgs boson signal,
a counting analysis is performed in each final state and category
using a selection optimized for each $\mH$ hypothesis considered.
In addition, a two-dimensional shape analysis is also pursued for
the different-flavor final state only. In this case, a binned template fit is performed
using the most sensitive variables to the presence of signal.
This shape-based analysis is more sensitive than the counting analysis to the presence of a Higgs
boson, as shown in Section~\ref{sec:hww01j_results}, and is used as the default
analysis for the $\Pe\mu$ final state. The counting analysis is
used as the default analysis for the $\Pe\Pe/\mu\mu$ final states, for which modeling
of the $\Z/\gamma^*$ background template is challenging.
Furthermore, an unbinned parametric fit is pursued using alternative variables
and a selection suitable for the measurement of the Higgs boson mass in
the different-flavor final state. The mass measurement using the parametric fit
and the test of spin hypotheses using a binned template fit are performed in
the $\Pe\mu$ final state.

\textbf{Binned template fit in the different-flavor final states}

Kinematic variables such as $\mll$ and $\mth$ are independent quantities
that effectively discriminate the signal against most of the backgrounds
in the dilepton analysis in the 0-jet and 1-jet categories.

The binned fit is performed using template histograms that are obtained from the signal and background models
at the level of the $\WW$ selection.
For the Higgs boson mass hypotheses up to $\mH$ = 250\GeV the template ranges are
12\GeV $< \mll <$ 200\GeV and 60\GeV $< \mth <$ 280\GeV.
For mass hypotheses above 250\GeV the template ranges are 12\GeV $< \mll <$ 600\GeV
and 80\GeV $< \mth <$ 600\GeV, and a higher leading-lepton $\pt$ threshold of
$\ptlmax >$ 50\GeV is required.
The templates have 9 bins in $\mll$ and 14 bins in $\mth$.
The bin widths vary within the given range,
and are optimized to achieve good separation between the SM Higgs boson
signal and backgrounds, as well as between the two spin hypotheses, while retaining adequate
template statistics for all processes in the bins.

\begin{figure}[htb]
\begin{center}
\includegraphics[width=0.45\textwidth]{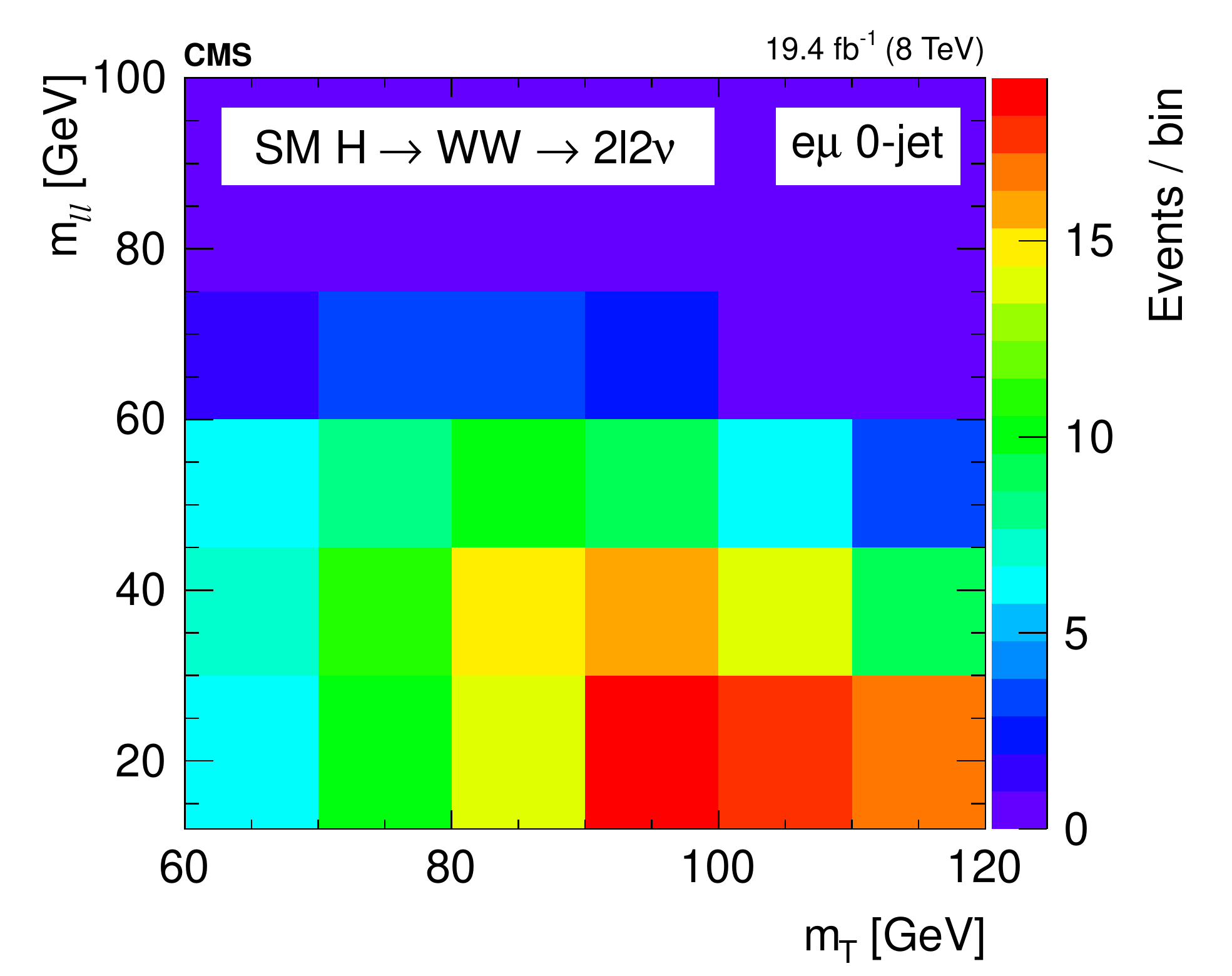}
\includegraphics[width=0.45\textwidth]{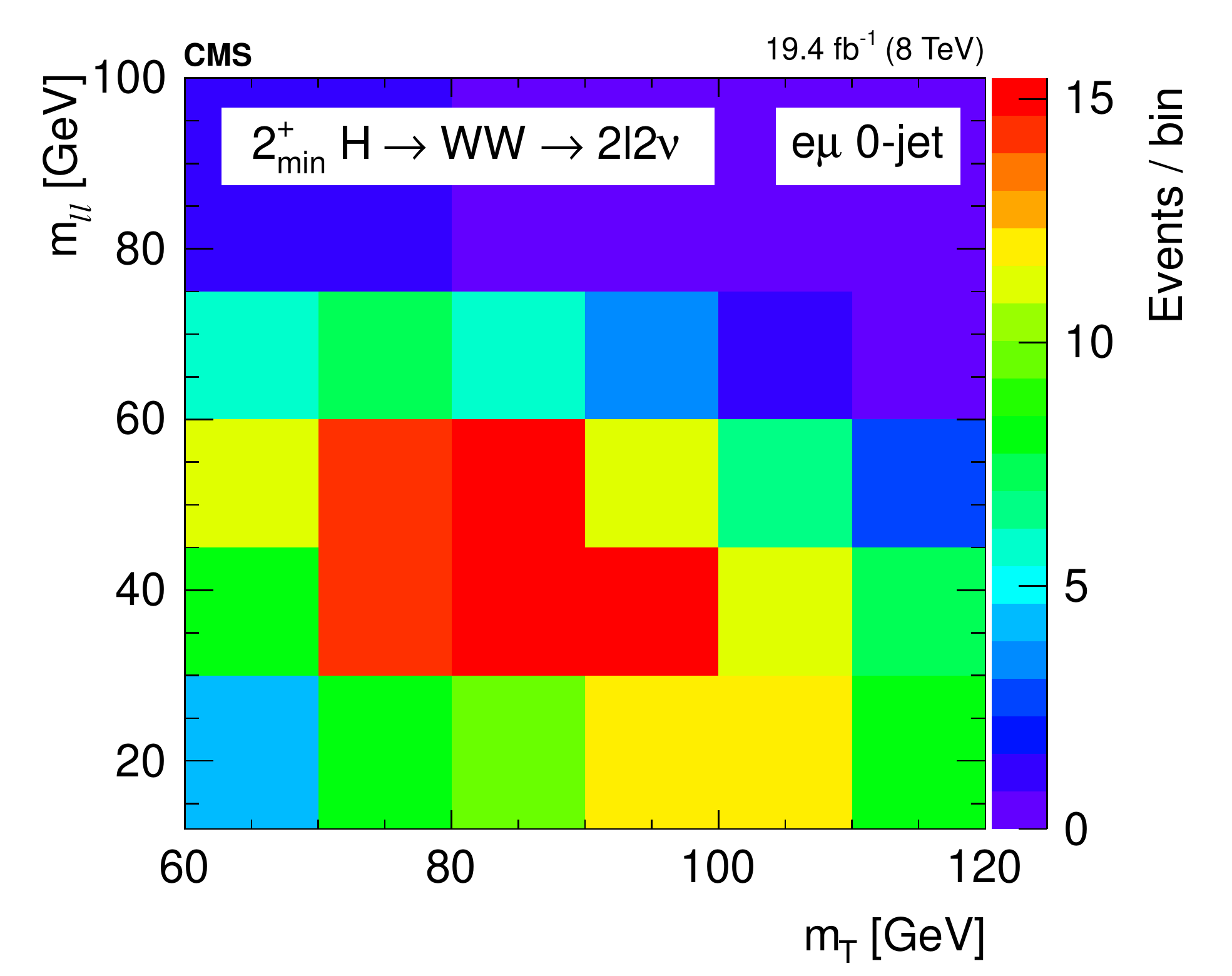} \\
\includegraphics[width=0.45\textwidth]{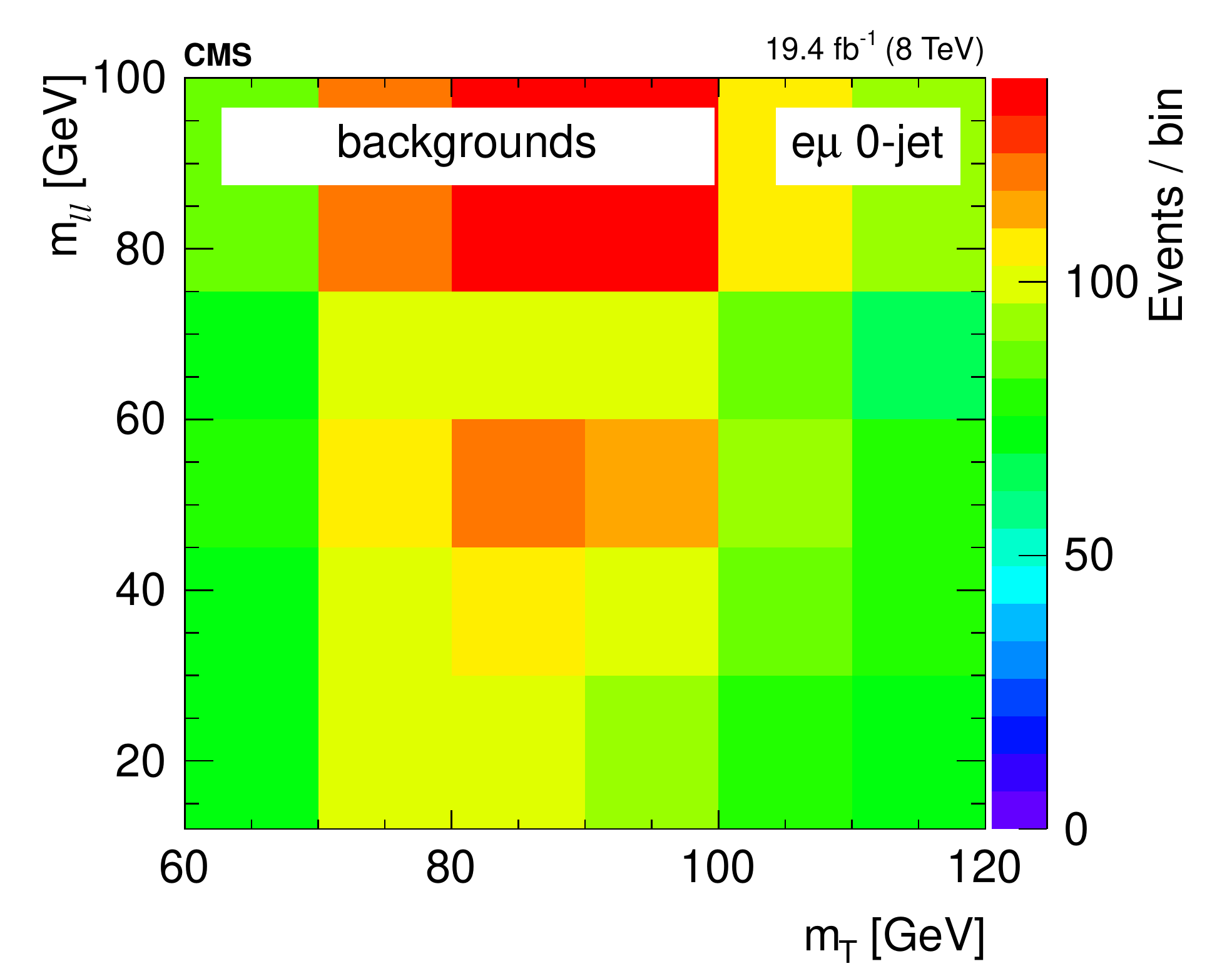}
\includegraphics[width=0.45\textwidth]{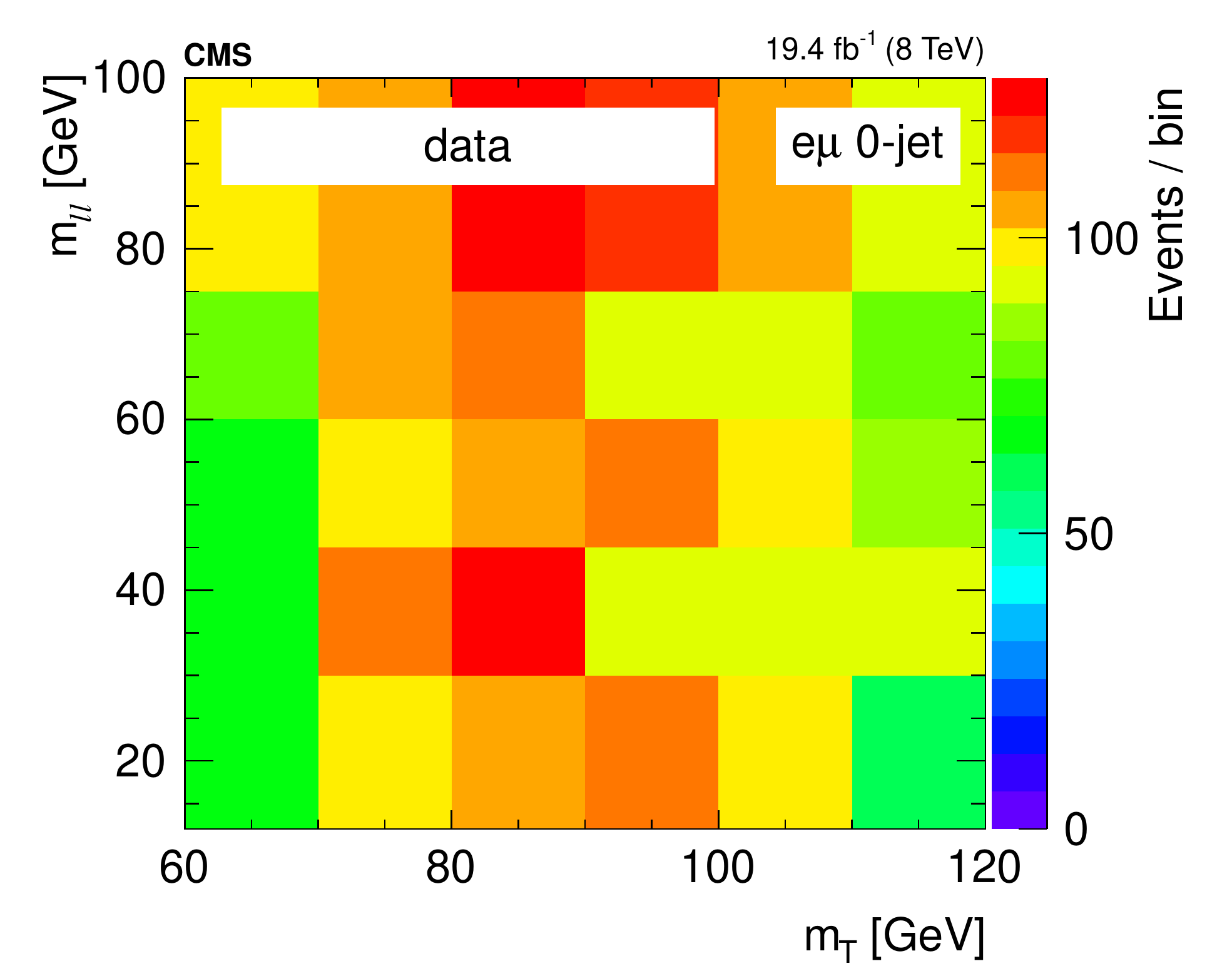}
       \caption{Two-dimensional ($\mth$, $\mll$) distributions for 8\TeV data in the 0-jet category for
       the $\mH = 125\GeV$ SM Higgs boson signal hypothesis (top left), the \spintwopmin hypothesis (top right),
       the background processes (bottom left), and the data (bottom right). The distributions are
       restricted to the signal region expected for a low mass Higgs boson, that is: $\mll$ [12--100]\GeV
       and $\mt$ [60--120]\GeV.}
\label{fig:hww01j_2D_0j}
\end{center}
\end{figure}

\begin{figure}[htb]
\begin{center}
\includegraphics[width=0.45\textwidth]{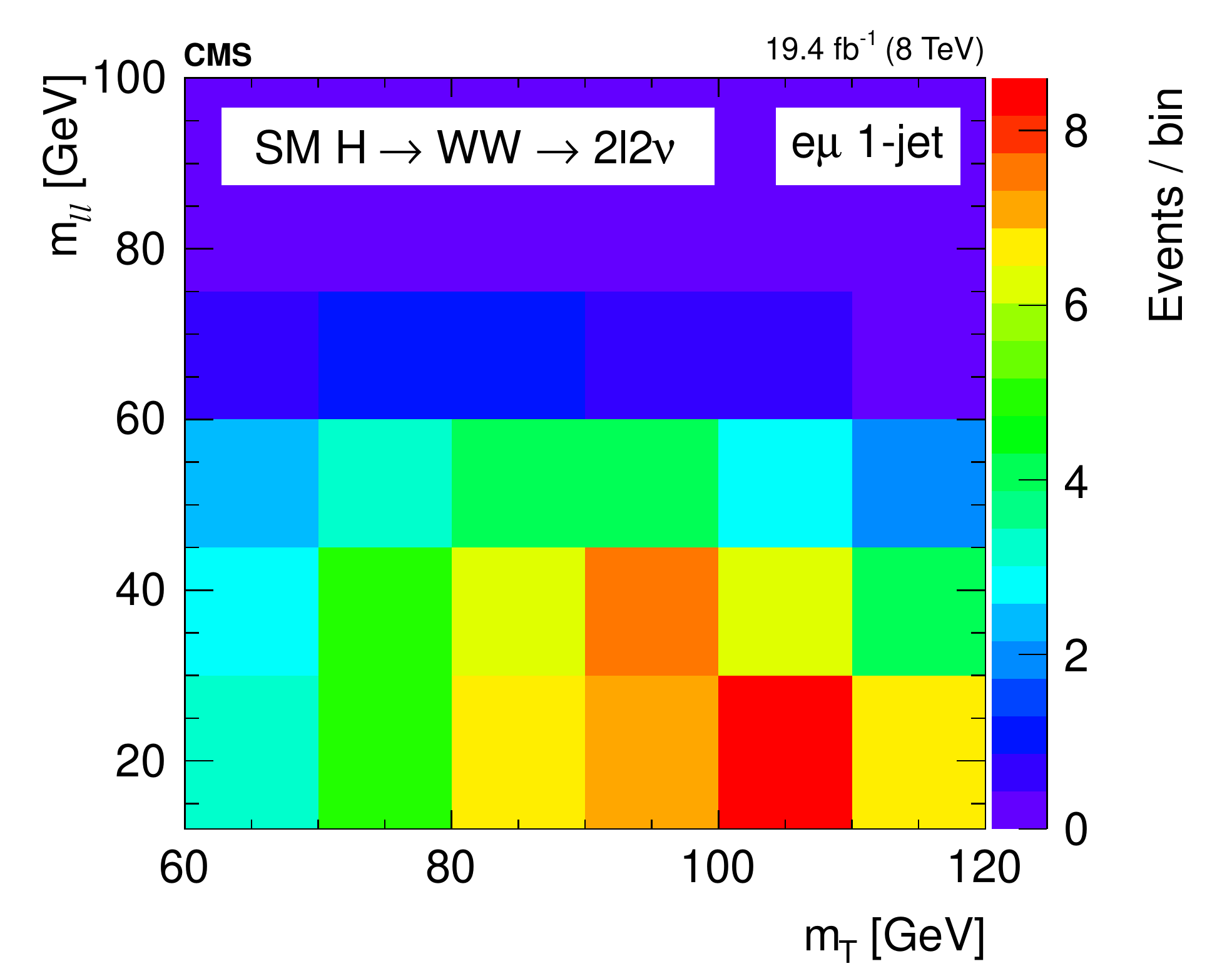}
\includegraphics[width=0.45\textwidth]{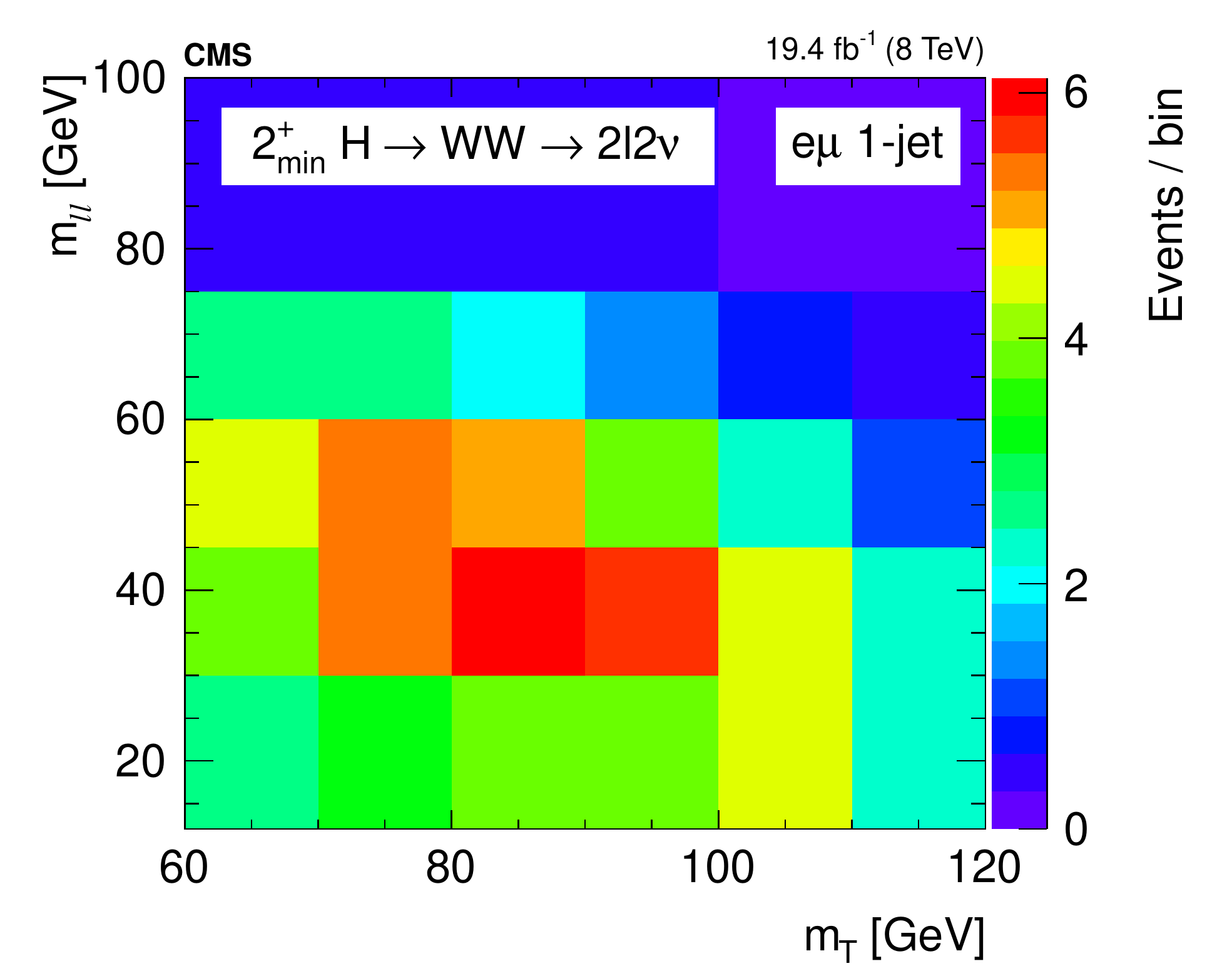} \\
\includegraphics[width=0.45\textwidth]{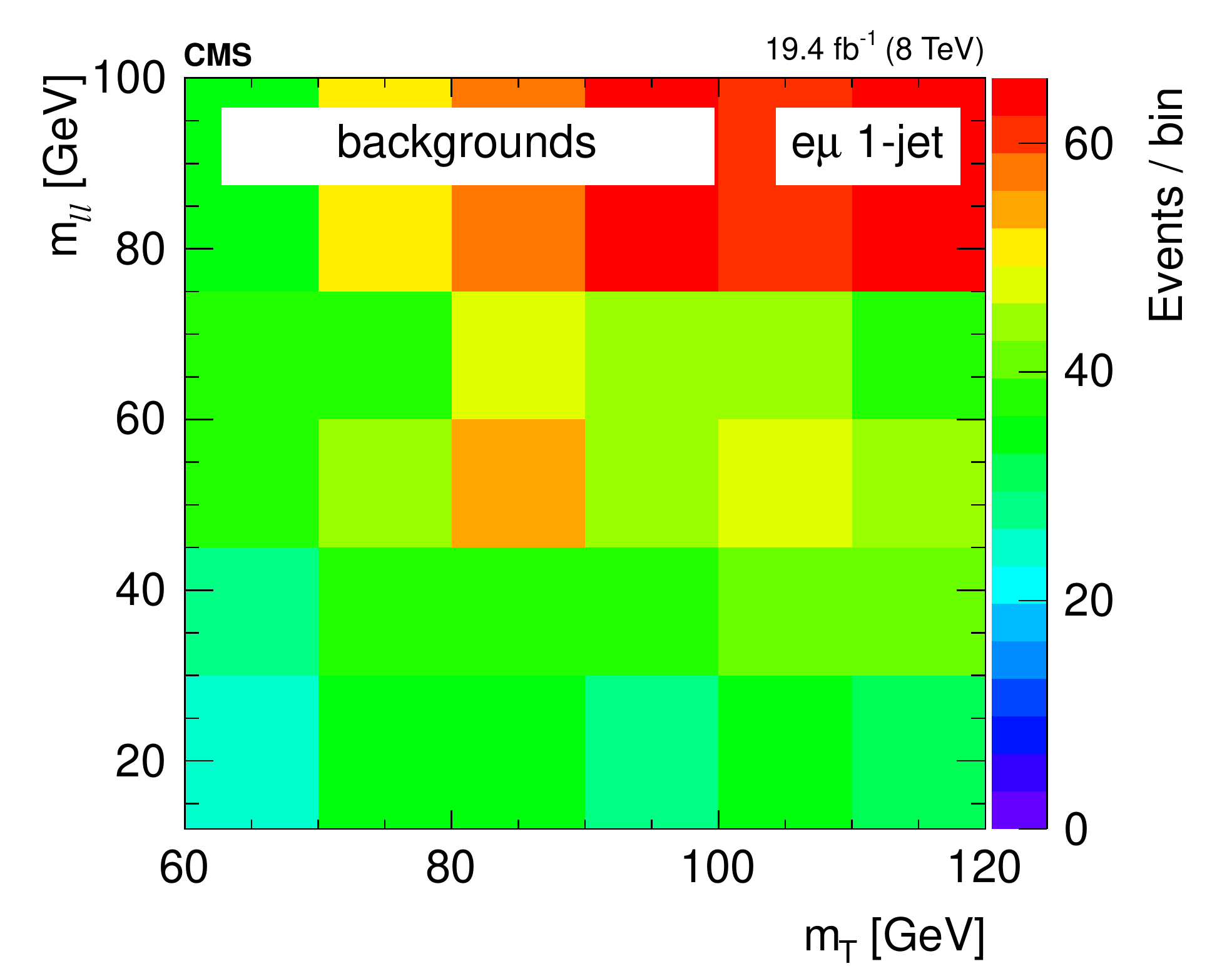}
\includegraphics[width=0.45\textwidth]{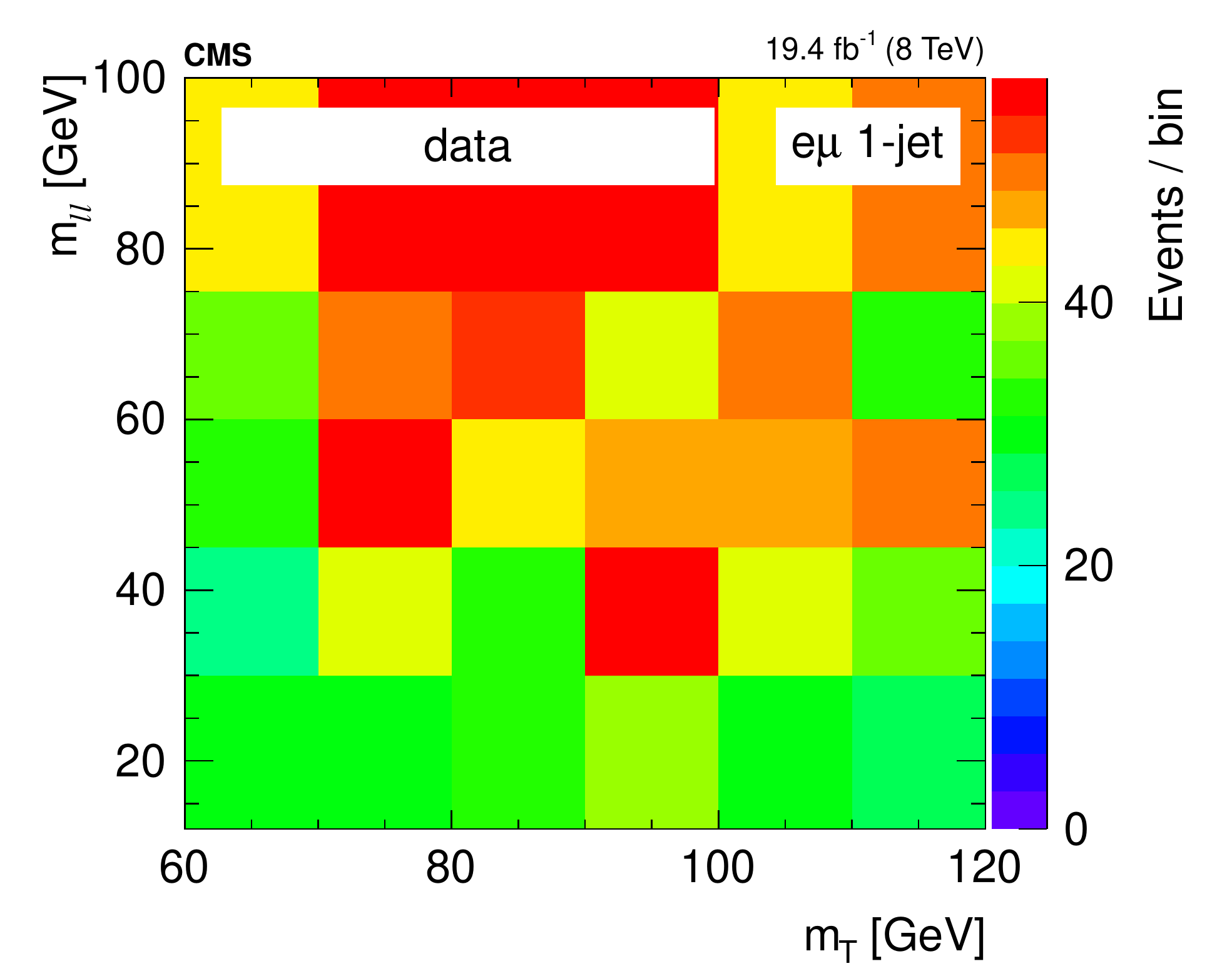}
       \caption{Two-dimensional ($\mth$, $\mll$) distributions in the 1-jet category for
       the $\mH = 125\GeV$ SM Higgs boson signal hypothesis (top left), the \spintwopmin hypothesis (top right),
       the background processes (bottom left), and the data (bottom right). The distributions are
       restricted to the signal region expected for a low mass Higgs boson, that is: $\mll$ [12--100]\GeV
       and $\mt$ [60--120]\GeV.}
\label{fig:hww01j_2D_1j}
\end{center}
\end{figure}

The signal and background templates, as well as the distribution observed in data,
are shown in Fig.~\ref{fig:hww01j_2D_0j} for the 0-jet category and in
Fig.~\ref{fig:hww01j_2D_1j} for the 1-jet category for the $8\TeV$ analysis. The distributions
are restricted to the signal region expected for a low mass Higgs boson, that is:
$\mll$ [12--100]\GeV and $\mt$ [60--120]\GeV.
The distribution of the two variables and the correlation between them are distinct for
the Higgs boson signal and the backgrounds, and clearly separates the two spin hypotheses.
Pseudo-experiments have been performed to assess the stability of the ($\mll$, $\mth$)
template fit method by randomly varying the expected signal and background yields according
to the Poisson statistics and to the spread of the systematic uncertainties, as discussed below.

\textbf{Unbinned parametric fit in the different-flavor final states}

A dedicated analysis to probe the Higgs boson mass
is performed using a two-dimensional parametric maximum
likelihood fit to variables computed in the estimated decay frame of
the Higgs boson candidate, the so-called ``razor frame"~\cite{razor}.
One of the two variables is an estimator of the Higgs boson mass and the other is the
opening angle of the two charged leptons in the razor frame. This
analysis is performed for the Higgs boson mass range 115--180\GeV.

The razor mass variable is based on the generic process of pair
production of heavy particles, each decaying to an unseen particle
plus jets or leptons that are reconstructed in the detector. The
application of this technique in SUSY analyses with hadronic and
leptonic final states has been extensively studied~\cite{PhysRevLett.111.081802}.

Given the presence of the two neutrinos in the final state, the
longitudinal and transverse boosts of the Higgs boson candidate cannot
be determined. The razor frame is an approximation of the Higgs boson rest frame,
defined unambiguously from measured quantities in the laboratory
frame. A longitudinal boost to an intermediate frame, where the
visible energies are written in terms of an overall scale that is
invariant under longitudinal boosts, is defined as:
\begin{equation*}
\label{eq:betaRstar}
\beta^{R^*}_L \equiv \frac{p_z^{{\ell}_1}+p_z^{{\ell}_2}}{E_{{\ell}_1}+E_{{\ell}_2}}
\; ,
\end{equation*}
where $p_z^{\ell_i}$ is the component along the $z$ axis of the four-momentum and
$E_{\ell_i}$ is the energy of the $i$th lepton.
In order to also account for the recoil of the Higgs boson candidate
when produced in association with jets, a transverse boost is further applied,
estimated with the measured $\vmet$. In the razor frame, an invariant quantity that serves as
per-event estimator of the mass scale of the decaying Higgs boson candidate is defined as:

\begin{equation*}
\label{eq:mr}
\mr = \sqrt{\dfrac{1}{2} \left[ \mll^2 - \vmet\cdot\vec{p}_\mathrm{T}^{\ell\ell}
   + \sqrt{(\mll^2 + (\pt^{\ell\ell})^{2}) (\mll^2 + (E_T^{\text{miss}})^{2})}\right]}
\;.
\end{equation*}

This variable has a resolution of around 15\% for a Higgs boson with
$\mHi =125\GeV$, regardless of the jet multiplicity. The distribution
of the $\mr$ variable is parameterized with a relatively simple function with a
linear dependence on the Higgs boson mass, enabling an unbinned fit to
data and a smooth interpolation between mass hypotheses.

\begin{figure}[htbp]
  \centering
    \includegraphics[width=0.45\textwidth]{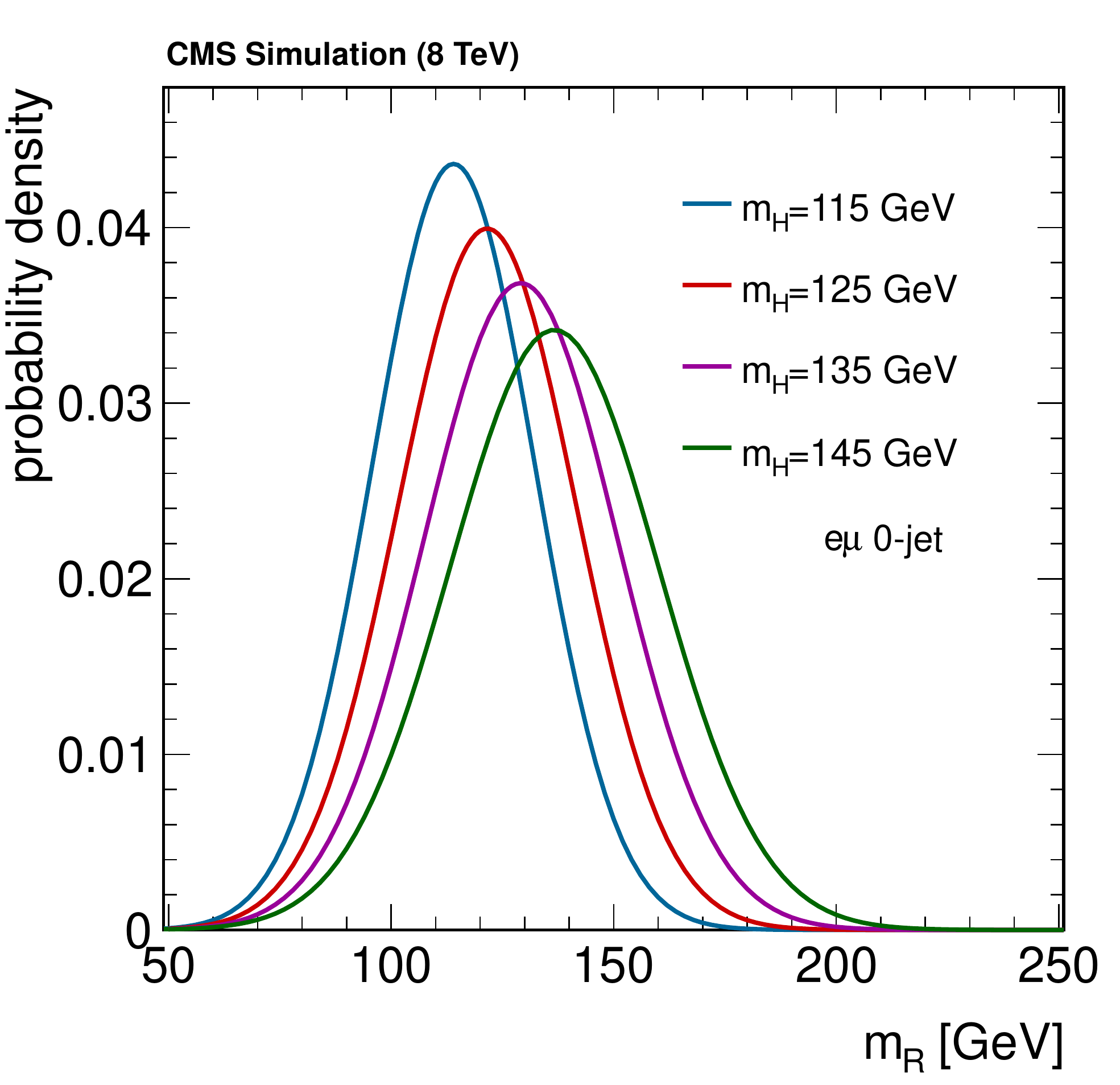}
    \includegraphics[width=0.45\textwidth]{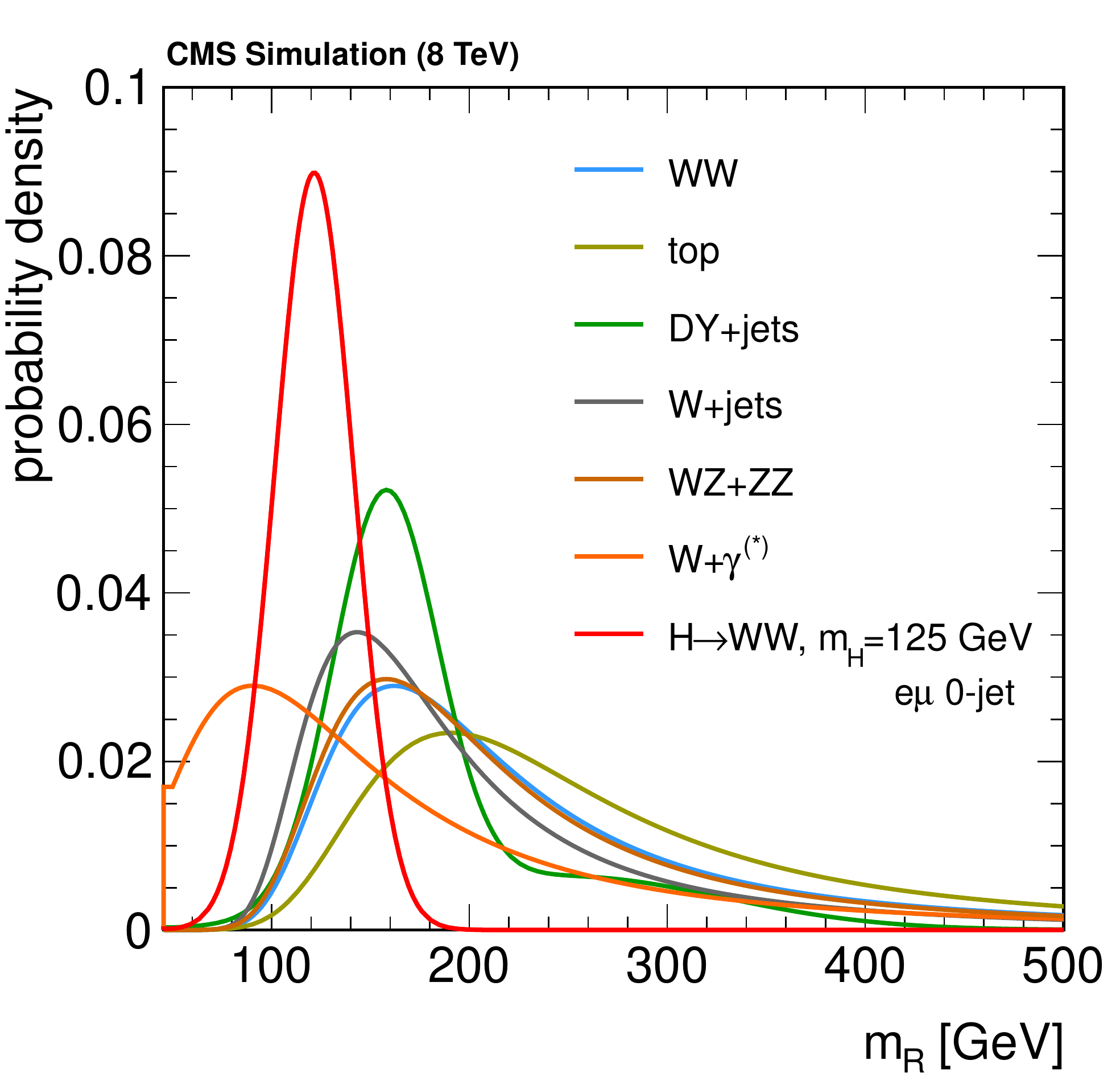}
    \caption{Evolution of $\mr$ distribution with Higgs boson mass hypotheses (left),
    and distribution of $\mr$ for signal and different backgrounds (right), all normalized to unity,
    for the 0-jet category in the $\Pe\mu$ final state.\label{fig:hww01j_mrevolution}}
\end{figure}

The parameterized distributions of the $\mr$ variable for different signal
mass hypotheses and backgrounds are shown in
Fig.~\ref{fig:hww01j_mrevolution}. The functional form of the
Higgs boson signal in $\mr$ is described by the convolution of a Breit--Wigner
function, centered on the expected $\mHi$ and with a width equal to the
expected Higgs boson width, and a Crystal Ball function~\cite{Gaiser}
to describe the resolution of the Gaussian core and the tail. For the
Higgs boson mass hypotheses considered in this analysis, the
theoretical width of the SM Higgs boson is negligible
with respect to the experimental resolution.

The $\mr$ distribution for the majority of the backgrounds is
described with a Landau function~\cite{Landau:1944if}, except for
the $\Z\to\tau\tau$ process which is modeled with a double Gaussian
function. The parametric fit is carried out in bins of
$\delphir$, which is the azimuthal separation between the two
leptons computed in the same reference frame as $\mr$. The two
variables are largely uncorrelated in the decay of the Higgs boson,
while the distributions for backgrounds are correlated.
A total of 10 bins in $\delphir$ are used with finer (coarser)
bin widths at smaller (larger) value of $\delphir$.

A selection tighter than that of the ($\mth$, $\mll$) template fits is chosen for this
analysis by applying $\pt^{\ell\ell} >$ 45\GeV and $\mth >$
80\GeV. The reason for the tighter selection is to
reject a larger fraction of the $\Wjets$ and $\wgamma^{(*)}$
background processes, which otherwise show a maximum at $\mr \sim
125\GeV$ because of kinematic requirements.
The upper bounds on $\mll$ and $\mth$ that are used for the ($\mth$, $\mll$) template fits
are removed.
The range of
50\GeV $< \mr <$ 500\GeV, which contains almost 100\% of the signal,
is used for the fit.

All the theoretical and experimental systematic uncertainties are
taken into account in the parametric fit. The shape uncertainties are
estimated by refitting the distribution produced with the systematic
variation for each source. The parametric fit to the
($\mr$, $\delphir$) distribution has been validated using pseudo-experiments
and the results show no bias in the measurement of the signal
and background yields neither for the 0-jet nor for the 1-jet category.

\textbf{Counting analysis}

A simple counting experiment is performed as a basic cross-check for all categories,
and as default approach for the same-flavor $\Pe\Pe/\mu\mu$ final states.
A tighter selection is applied to increase the signal-to-background ratio
using kinematic variables that characterize the Higgs boson final state.
The minimum requirement on dilepton $\pt$ is raised to $\ptll >$ 45\GeV,
and a series of selections are applied based on the lepton momenta ($\ptlmax$ and $\ptlmin$),
$\mll$, the azimuthal separation between the two leptons ($\delphill$), and $\mth$. The threshold values are
optimized for each Higgs boson mass hypothesis.
Table~\ref{tab:hww01j_cbselection} summarizes the selection requirements used
in the counting analysis for a few representative mass points.

\begin{table}[htbp]
\centering
\topcaption{Event selection requirements for the counting analysis in 0-jet and 1-jet categories.
For the 2-jet categories the lower threshold on $\mth$ is set at 30\GeV.}
\label{tab:hww01j_cbselection}
{
  \begin{tabular} {ccccccc}
  \hline\hline
\mHi [\GeVns{}] & $\ptlmax$ [\GeVns{}] & $\ptlmin$ [\GeVns{}] & $\mll$ [\GeVns{}]   & $\delphill$ [$^\circ$] & $\mth$ [\GeVns{}] \\
\hline
    120 & $>$20  & $>$10     & $>$40  & $<$115 & [80,120] \\
    125 & $>$23  & $>$10     & $>$43  & $<$100 & [80,123] \\
    130 & $>$25  & $>$10     & $>$45  & $<$90  & [80,125] \\
    160 & $>$30  & $>$25     & $>$50  & $<$60  & [90,160] \\
    200 & $>$40  & $>$25     & $>$90  & $<$100 & [120,200] \\
    400 & $>$90  & $>$25     & $>$300 & $<$175 & [120,400] \\
    600 & $>$140 & $>$25     & $>$500 & $<$175 & [120,600] \\
  \hline
  \end{tabular}
}
\end{table}

\subsubsection{Results}\label{sec:hww01j_results}

The data yields and the expected yields for the SM Higgs boson signal and various backgrounds
in each of the jet categories lepton-flavor final states are listed in
tables~\ref{tab:hww01j_yields7tev} and~\ref{tab:hww01j_yields8tev}
for the counting analysis for representative Higgs boson mass hypotheses up to $\mH$ = 600\GeV,
and for the selection used for the shape-based analyses.
For a SM Higgs boson with $\mH = 125\GeV$, a couple of hundred signal events are expected in total,
and the purity of the counting analysis selection is around 20\% in the most sensitive $\Pe\mu$ final state.
The looser selection used for the shape-based analyses recovers a large fraction
of the signal events, and also accommodates background-dominated regions
allowing the fit to impose constraints on the background contributions.

\begin{table}[htbp]
\centering
\topcaption{Signal prediction, observed number of events in data, and background estimates
for $\sqrt{s} = 7\TeV$ after applying the requirements used
for the $\Hww$ counting analysis
and for the shape-based analyses ($\Pe\mu$ final state only).
The combination of statistical uncertainties with experimental and theoretical systematic uncertainties is reported.
The $\dyll$ process includes the $\Pe\Pe$, $\mu\mu$ and $\tau\tau$ final states. The shape-based
selections correspond to the $\mHi = 125\GeV$ selection.}
\label{tab:hww01j_yields7tev}
\resizebox{\textwidth}{!}
{\setlength{\extrarowheight}{1pt}
\begin{tabular} {lccccccccc}
  \hline\hline
$\mH$ [\GeVns{}] & \multirow{2}{*}{$\Pg\Pg\PH$} & \multirow{2}{*}{VBF+$\V\PH$} & \multirow{2}{*}{Data} & \multirow{2}{*} {All bkg.} & \multirow{2}{*}{$\WW$}       & $\WZ+\ZZ$ & \multirow{2}{*}{$\ttbar+\tw$} & \multirow{2}{*}{$\Wjets$} & \multirow{2}{*}{$\wgamma^{(*)}$}\\
(shape)  &                               &                              &		       &			  &   & + $\dyll$  &  		    &				&				  \\
\hline
\multicolumn{10}{c}{7\TeV $\Pe\mu$ final state, 0-jet category} \\
\hline
 $120$ & $12.1\pm2.6$ & $0.15\pm0.01$ & $85$ & $83.1\pm7.7$ & $62.1\pm6.5$ & $1.78\pm0.40$ & $3.39\pm0.83$ & $9.7\pm2.8$ & $6.0\pm2.9$ \\
 $125$ & $20.1\pm4.3$ & $0.19\pm0.03$ & $105$ & $99.0\pm9.0$ & $75.4\pm7.8$ & $2.07\pm0.41$ & $4.2\pm1.0$ & $10.8\pm3.1$ & $6.5\pm3.0$ \\
 $130$ & $32.1\pm6.9$ & $0.42\pm0.04$ & $112$ & $109.6\pm9.9$ & $84.3\pm8.7$ & $2.20\pm0.42$ & $5.0\pm1.2$ & $11.8\pm3.4$ & $6.4\pm3.0$ \\
 $160$ & $73\pm16$ & $0.98\pm0.09$ & $59$ & $53.4\pm5.0$ & $44.8\pm4.6$ & $0.68\pm0.08$ & $4.1\pm1.0$ & $2.6\pm1.1$ & $1.2\pm1.0$ \\
 $200$ & $28.3\pm6.4$ & $0.49\pm0.04$ & $85$ & $86.6\pm7.9$ & $71.3\pm7.4$ & $1.13\pm0.12$ & $11.1\pm2.5$ & $2.9\pm1.2$ & $0.14\pm0.16$ \\
 $400$ & $11.0\pm3.0$ & $0.16\pm0.02$ & $58$ & $63.0\pm5.9$ & $40.0\pm4.3$ & $0.92\pm0.10$ & $17.4\pm3.9$ & $3.3\pm1.3$ & $1.36\pm0.72$ \\
 $600$ & $2.2\pm1.0$ & $0.07\pm0.01$ & $16$ & $18.7\pm1.9$ & $11.7\pm1.3$ & $0.27\pm0.04$ & $5.3\pm1.2$ & $1.07\pm0.54$ & $0.30\pm0.25$ \\
 ($\mth$, $\mll$)    & $50\pm10$ & $0.44\pm0.03$ & $1207$ & $1193\pm50$ & $861\pm12$ & $22.7\pm1.2$ & $91\pm20$ & $150\pm39$ & $68\pm20$ \\
 ($\mr$, $\delphir$) & $30.8\pm8.3$  & $1.4\pm0.1$ & $765$  & $769\pm35$  & $570\pm20$  &  $0.3\pm0.1$ &  $81\pm27$   & $61.0\pm9.2$   & $11.9\pm1.1$ \\
\hline
\multicolumn{10}{c}{7\TeV $\Pe\Pe$/$\mu\mu$ final state, 0-jet category} \\
\hline
 $120$ & $5.0\pm1.1$ & $0.06\pm0.01$ & $48$ & $50.0\pm5.2$ & $35.4\pm3.8$ & $9.7\pm3.5$ & $1.44\pm0.41$ & $2.9\pm1.0$ & $0.64\pm0.39$ \\
 $125$ & $10.0\pm2.2$ & $0.07\pm0.01$ & $66$ & $64.1\pm6.7$ & $46.6\pm4.9$ & $11.4\pm4.4$ & $1.97\pm0.52$ & $3.1\pm1.1$ & $0.94\pm0.53$ \\
 $130$ & $16.2\pm3.5$ & $0.19\pm0.02$ & $78$ & $71.9\pm7.4$ & $54.7\pm5.7$ & $9.7\pm4.3$ & $2.54\pm0.65$ & $4.0\pm1.4$ & $0.94\pm0.53$ \\
 $160$ & $59\pm13$ & $0.74\pm0.07$ & $50$ & $45.8\pm5.4$ & $37.5\pm3.9$ & $3.9\pm3.5$ & $3.31\pm0.82$ & $0.52\pm0.52$ & $0.58\pm0.37$ \\
 $200$ & $24.0\pm5.4$ & $0.43\pm0.04$ & $70$ & $68.2\pm6.3$ & $55.5\pm5.8$ & $4.5\pm1.8$ & $6.9\pm1.6$ & $1.33\pm0.78$ & --- \\
 $400$ & $8.8\pm2.4$ & $0.12\pm0.01$ & $45$ & $46.8\pm4.2$ & $29.5\pm3.2$ & $3.57\pm0.35$ & $11.1\pm2.5$ & $2.5\pm1.0$ & $0.16\pm0.17$ \\
 $600$ & $1.59\pm0.72$ & $0.05\pm0.01$ & $13$ & $12.1\pm1.2$ & $6.57\pm0.79$ & $1.14\pm0.14$ & $3.26\pm0.79$ & $1.12\pm0.53$ & --- \\
\hline
\multicolumn{10}{c}{7\TeV $\Pe\mu$ final state, 1-jet category} \\
\hline
 $120$ & $4.7\pm1.5$ & $0.51\pm0.05$ & $44$ & $36.8\pm3.6$ & $16.3\pm2.8$ & $2.05\pm0.41$ & $11.10\pm0.90$ & $6.2\pm1.9$ & $1.04\pm0.58$ \\
 $125$ & $7.0\pm2.3$ & $0.86\pm0.09$ & $53$ & $44.8\pm4.3$ & $20.1\pm3.4$ & $2.37\pm0.42$ & $13.9\pm1.1$ & $6.3\pm2.0$ & $2.0\pm1.2$ \\
 $130$ & $11.3\pm3.8$ & $1.37\pm0.13$ & $64$ & $50.1\pm4.7$ & $22.6\pm3.8$ & $2.56\pm0.43$ & $15.9\pm1.2$ & $6.8\pm2.1$ & $2.2\pm1.2$ \\
 $160$ & $33\pm11$ & $4.10\pm0.40$ & $32$ & $35.1\pm3.3$ & $18.0\pm3.0$ & $1.10\pm0.12$ & $14.1\pm1.1$ & $1.59\pm0.79$ & $0.29\pm0.24$ \\
 $200$ & $13.7\pm4.1$ & $2.40\pm0.23$ & $49$ & $65.6\pm5.8$ & $31.0\pm5.2$ & $1.28\pm0.14$ & $31.1\pm2.2$ & $2.20\pm0.98$ & $0.04\pm0.04$ \\
 $400$ & $7.6\pm2.3$ & $0.74\pm0.07$ & $60$ & $71.8\pm5.6$ & $31.0\pm4.7$ & $2.07\pm0.69$ & $34.1\pm2.4$ & $4.3\pm1.6$ & $0.31\pm0.25$ \\
 $600$ & $1.94\pm0.82$ & $0.32\pm0.03$ & $19$ & $24.3\pm2.2$ & $10.8\pm1.7$ & $1.36\pm0.68$ & $9.75\pm0.80$ & $2.23\pm0.88$ & $0.16\pm0.17$ \\
 ($\mth$, $\mll$)    & $17.1\pm5.5$ & $2.09\pm0.12$ & $589$ & $573\pm22$ & $249.9\pm4.0$ & $26.4\pm1.4$ & $226\pm14$ & $60\pm16$ & $10.1\pm2.8$ \\
 ($\mr$, $\delphir$) & $15.1\pm4.3$   & $3.41\pm0.21$  & $457$  & $518\pm45$  & $239.0\pm8.6$   &  $0.9\pm0.3$ &  $211\pm44$  & $39.4\pm5.9$   & $3.31\pm0.32$  \\
\hline
\multicolumn{10}{c}{7\TeV $\Pe\Pe$/$\mu\mu$ final state, 1-jet category} \\
\hline
 $120$ & $1.51\pm0.50$ & $0.19\pm0.02$ & $22$ & $23.8\pm3.6$ & $7.6\pm1.3$ & $10.3\pm3.2$ & $4.87\pm0.47$ & $0.65\pm0.48$ & $0.31\pm0.26$ \\
 $125$ & $2.64\pm0.89$ & $0.38\pm0.04$ & $31$ & $28.1\pm4.5$ & $10.1\pm1.7$ & $10.5\pm4.1$ & $6.34\pm0.57$ & $0.88\pm0.55$ & $0.31\pm0.26$ \\
 $130$ & $5.2\pm1.7$ & $0.60\pm0.06$ & $35$ & $31.7\pm4.5$ & $11.7\pm2.0$ & $10.7\pm3.9$ & $7.39\pm0.64$ & $1.60\pm0.75$ & $0.31\pm0.26$ \\
 $160$ & $24.3\pm7.7$ & $2.89\pm0.28$ & $47$ & $34.5\pm4.6$ & $13.0\pm2.2$ & $9.5\pm3.8$ & $10.20\pm0.85$ & $1.64\pm0.93$ & $0.15\pm0.16$ \\
 $200$ & $9.8\pm3.0$ & $1.58\pm0.15$ & $56$ & $60.6\pm6.6$ & $21.9\pm3.7$ & $15.9\pm5.1$ & $20.6\pm1.5$ & $2.2\pm1.1$ & --- \\
 $400$ & $5.3\pm1.6$ & $0.51\pm0.05$ & $65$ & $46.2\pm4.2$ & $17.6\pm2.7$ & $7.1\pm2.7$ & $19.8\pm1.4$ & $1.69\pm0.80$ & --- \\
 $600$ & $1.27\pm0.54$ & $0.20\pm0.02$ & $16$ & $12.4\pm1.2$ & $5.67\pm0.92$ & $0.74\pm0.09$ & $4.94\pm0.46$ & $1.02\pm0.51$ & --- \\
\hline
  \end{tabular}
}
\end{table}

\begin{table}[htb]
\centering
  \topcaption{
  Signal prediction, observed number of events in data, and background estimates
  for $\sqrt{s} = 8$\TeV after applying the requirements used
  for the $\Hww$ counting analysis
  and for the shape-based analyses ($\Pe\mu$ final state only).
  The combination of statistical uncertainties with experimental and theoretical systematic uncertainties is reported.
  The $\dyll$ process includes the $\Pe\Pe$, $\mu\mu$ and $\tau\tau$ final states. The shape-based
  selections correspond to the $\mHi = 125\GeV$ selection.}
   \label{tab:hww01j_yields8tev}
\resizebox{\textwidth}{!}
{
\setlength{\extrarowheight}{1pt}
\begin{tabular} {lccccccccc}
  \hline\hline
$\mH$ [\GeVns{}] & \multirow{2}{*}{$\Pg\Pg\PH$} & \multirow{2}{*}{VBF+$\V\PH$} & \multirow{2}{*}{Data} & \multirow{2}{*} {All bkg.} &  \multirow{2}{*}{$\WW$}	 & $\WZ+\ZZ$ & \multirow{2}{*}{$\ttbar+\tw$} & \multirow{2}{*}{$\Wjets$} & \multirow{2}{*}{$\wgamma^{(*)}$}\\
(shape) &                               &                              &		       &			    &   & + $\dyll$  &  		    &				&				  \\
\hline
\multicolumn{10}{c}{8\TeV $\Pe\mu$ final state 0-jet category} \\
\hline
 $120$ & $51\pm11$ & $1.35\pm0.14$ & $414$ & $347\pm28$ & $246\pm23$ & $9.16\pm0.77$ & $15.8\pm3.5$ & $40\pm10$ & $36\pm12$ \\
 $125$ & $88\pm19$ & $2.19\pm0.22$ & $506$ & $429\pm34$ & $310\pm29$ & $11.4\pm1.0$ & $19.9\pm4.3$ & $48\pm13$ & $39\pm13$ \\
 $130$ & $133\pm28$ & $2.97\pm0.28$ & $567$ & $473\pm37$ & $346\pm32$ & $12.3\pm1.1$ & $21.9\pm4.6$ & $50\pm13$ & $42\pm13$ \\
 $160$ & $370\pm80$ & $8.75\pm0.71$ & $285$ & $239\pm19$ & $196\pm18$ & $5.94\pm0.61$ & $24.9\pm5.4$ & $5.9\pm2.0$ & $6.3\pm3.5$ \\
 $200$ & $150\pm33$ & $3.91\pm0.33$ & $471$ & $394\pm32$ & $318\pm30$ & $10.6\pm1.0$ & $55\pm11$ & $7.0\pm2.5$ & $3.8\pm2.5$ \\
 $400$ & $62\pm17$ & $1.24\pm0.12$ & $306$ & $326\pm29$ & $209\pm22$ & $9.9\pm1.1$ & $92\pm18$ & $9.4\pm3.6$ & $5.2\pm3.1$ \\
 $600$ & $12.8\pm5.8$ & $0.63\pm0.06$ & $95$ & $108\pm10$ & $66.3\pm7.2$ & $4.04\pm0.52$ & $30.2\pm6.4$ & $3.4\pm1.4$ & $3.9\pm2.8$ \\
 ($\mth$, $\mll$)    & $227\pm46$ & $10.27\pm0.41$ & $5747$ & $5760\pm210$ & $4185\pm63$ & $178.3\pm9.5$ & $500\pm96$ & $620\pm160$ & $282\pm76$ \\
 ($\mr$, $\delphir$) & $180\pm49$ & $8.11\pm0.72$  & $3751$ & $3460\pm80$ & $2518\pm62$ &  $71\pm11$ &  $398\pm27$  & $279\pm42$ & $47.0\pm4.6$ \\
\hline
\multicolumn{10}{c}{8\TeV $\Pe\Pe$/$\mu\mu$ final state 0-jet category} \\
\hline
 $120$ & $30.4\pm6.6$ & $0.69\pm0.10$ & $340$ & $289\pm30$ & $158\pm15$ & $92\pm25$ & $7.0\pm1.7$ & $23.7\pm6.4$ & $7.7\pm3.3$ \\
 $125$ & $55\pm12$ & $1.10\pm0.14$ & $423$ & $361\pm37$ & $207\pm19$ & $106\pm31$ & $9.4\pm2.2$ & $29.0\pm7.8$ & $9.3\pm3.8$ \\
 $130$ & $85\pm18$ & $1.81\pm0.21$ & $455$ & $410\pm42$ & $239\pm22$ & $119\pm34$ & $11.2\pm2.5$ & $30.5\pm8.1$ & $10.7\pm4.1$ \\
 $160$ & $319\pm69$ & $6.78\pm0.58$ & $258$ & $214\pm19$ & $164\pm15$ & $28.5\pm9.7$ & $14.0\pm3.2$ & $5.7\pm1.9$ & $1.72\pm0.92$ \\
 $200$ & $120\pm27$ & $3.31\pm0.28$ & $389$ & $351\pm27$ & $260\pm24$ & $39.7\pm8.0$ & $41.9\pm8.7$ & $7.0\pm2.3$ & $2.9\pm1.3$ \\
 $400$ & $53\pm15$ & $0.97\pm0.09$ & $290$ & $314\pm34$ & $182\pm19$ & $52\pm24$ & $72\pm14$ & $6.8\pm2.6$ & $1.28\pm0.87$ \\
 $600$ & $11.1\pm5.0$ & $0.52\pm0.05$ & $94$ & $92.7\pm8.2$ & $60.1\pm6.6$ & $7.46\pm0.75$ & $21.8\pm4.7$ & $2.7\pm1.2$ & $0.52\pm0.54$ \\
\hline
\multicolumn{10}{c}{8\TeV $\Pe\mu$ final state 1-jet category} \\
\hline
 $120$ & $20.0\pm6.5$ & $4.02\pm0.33$ & $182$ & $173\pm12$ & $65.7\pm8.7$ & $10.56\pm0.96$ & $63.3\pm4.0$ & $22.4\pm6.0$ & $10.7\pm4.5$ \\
 $125$ & $37\pm12$ & $6.53\pm0.53$ & $228$ & $209\pm14$ & $80\pm11$ & $12.9\pm1.2$ & $79.2\pm4.6$ & $25.9\pm6.9$ & $11.2\pm4.6$ \\
 $130$ & $51\pm17$ & $9.60\pm0.79$ & $262$ & $233\pm15$ & $90\pm12$ & $13.9\pm1.3$ & $90.4\pm3.7$ & $27.8\pm7.4$ & $11.4\pm4.6$ \\
 $160$ & $180\pm57$ & $30.6\pm2.5$ & $226$ & $174\pm11$ & $73.3\pm9.6$ & $7.98\pm0.83$ & $83.2\pm4.7$ & $8.7\pm2.8$ & $1.07\pm0.69$ \\
 $200$ & $78\pm23$ & $15.2\pm1.3$ & $421$ & $346\pm19$ & $130\pm17$ & $11.7\pm1.2$ & $188.2\pm8.4$ & $13.6\pm4.0$ & $2.9\pm2.4$ \\
 $400$ & $42\pm13$ & $4.39\pm0.44$ & $363$ & $379\pm23$ & $134\pm20$ & $12.8\pm1.2$ & $213.4\pm9.1$ & $17.5\pm5.5$ & $1.41\pm0.92$ \\
 $600$ & $11.2\pm4.7$ & $2.08\pm0.21$ & $112$ & $130.4\pm9.3$ & $50.4\pm7.7$ & $5.47\pm0.61$ & $65.0\pm4.2$ & $9.1\pm3.0$ & $0.44\pm0.47$ \\
 ($\mth$, $\mll$)    & $88\pm28$ & $19.83\pm0.81$ & $3281$ & $3242\pm90$ & $1268\pm21$ & $193\pm11$ & $1443\pm46$ & $283\pm72$ & $55\pm14$ \\
 ($\mr$, $\delphir$) & $91\pm26$  & $20.4\pm1.7$ & $2536$ & $2400\pm83$ & $792\pm28$  &  $1.9\pm0.6$ &  $1260\pm70$ & $222\pm33$ & $13.21\pm1.33$ \\
\hline
\multicolumn{10}{c}{8\TeV $\Pe\Pe$/$\mu\mu$ final state 1-jet category} \\
\hline
 $120$ & $8.2\pm2.7$ & $1.65\pm0.16$ & $110$ & $90.1\pm7.3$ & $31.0\pm4.2$ & $19.0\pm4.8$ & $30.7\pm2.6$ & $6.0\pm1.9$ & $3.3\pm1.7$ \\
 $125$ & $15.8\pm5.1$ & $3.09\pm0.28$ & $141$ & $111.9\pm8.6$ & $39.9\pm5.4$ & $21.2\pm5.4$ & $40.8\pm3.1$ & $6.6\pm2.0$ & $3.3\pm1.7$ \\
 $130$ & $23.4\pm7.8$ & $4.74\pm0.42$ & $168$ & $125.1\pm9.4$ & $45.7\pm6.1$ & $21.4\pm5.6$ & $47.0\pm3.4$ & $8.0\pm2.4$ & $2.9\pm1.6$ \\
 $160$ & $103\pm33$ & $16.8\pm1.5$ & $134$ & $113.8\pm8.2$ & $46.8\pm6.2$ & $13.8\pm3.9$ & $48.0\pm3.2$ & $3.9\pm1.5$ & $1.3\pm1.0$ \\
 $200$ & $48\pm14$ & $8.57\pm0.77$ & $263$ & $240\pm14$ & $86\pm11$ & $27.5\pm5.9$ & $120.6\pm6.3$ & $6.2\pm2.0$ & --- \\
 $400$ & $29.5\pm8.9$ & $2.96\pm0.30$ & $215$ & $236\pm21$ & $75\pm11$ & $33\pm17$ & $122.1\pm6.0$ & $4.9\pm1.7$ & $1.08\pm0.88$ \\
 $600$ & $7.1\pm3.0$ & $1.29\pm0.13$ & $63$ & $63.5\pm4.8$ & $26.6\pm4.1$ & $4.21\pm0.53$ & $31.0\pm2.2$ & $1.71\pm0.79$ & --- \\
\hline
  \end{tabular}
  }
\end{table}

The overall signal efficiency uncertainty is estimated to be about 20\%
and is dominated by the theoretical uncertainty due to missing
higher-order corrections and PDF uncertainties. The total uncertainty in
the background estimations in the signal region is about 15\%,
dominated by the statistical uncertainty in the number of observed events in the
background control regions and the theoretical uncertainties affecting the non-resonant $\WW$
production. A summary of the systematic uncertainties is given in Table~\ref{tab:systww}.
The obtained $\W\W$ continuum normalization uncertainty is between 3\% and 12\%
depending on the jet category and center-of-mass energy.

\begin{figure}[htb]
\begin{center}
\includegraphics[width=0.48\textwidth]{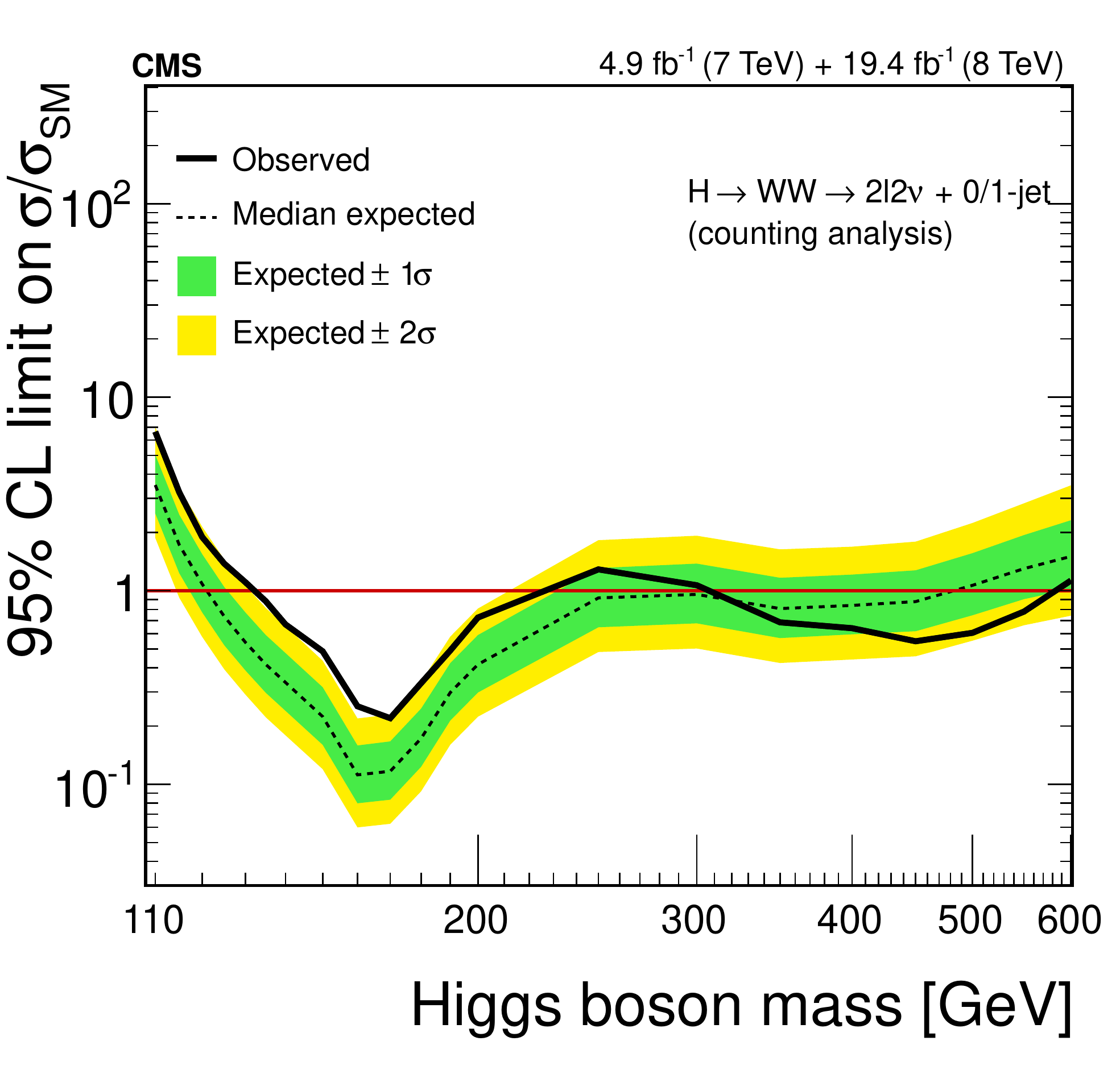}
\includegraphics[width=0.48\textwidth]{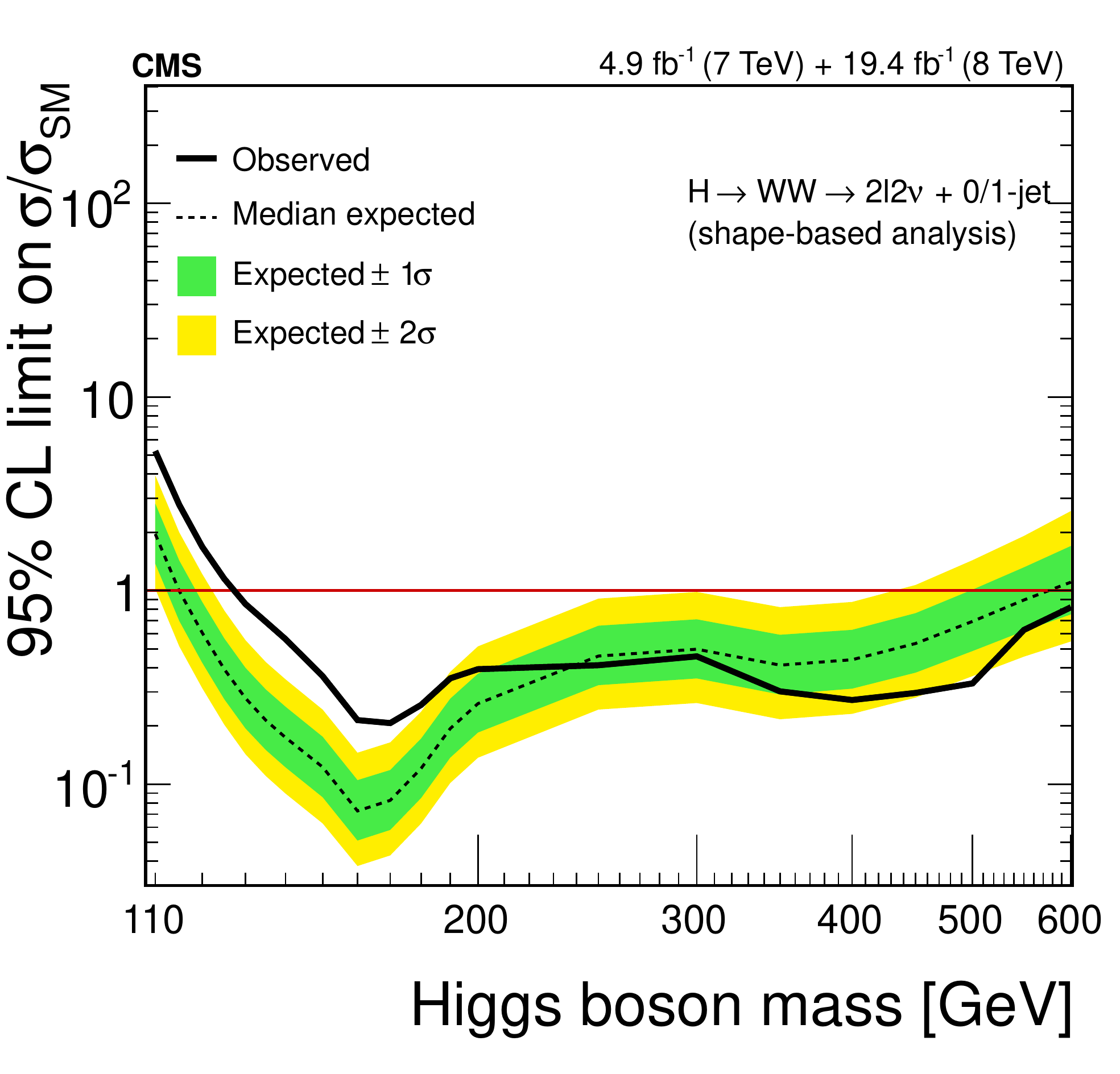}
\caption{
Expected and observed 95\% CL upper limits on the $\PH \to \WW$ production cross section
relative to the SM Higgs boson expectation
using the counting analysis (left) and the shape-based template fit
approach (right) in the 0-jet and 1-jet categories.
The shape-based analysis results use a binned template fit to ($\mth$, $\mll$)
for the $\Pe\mu$ final state, combined with the counting analysis results
for the $\Pe\Pe/\mu\mu$ final states.
  }
\label{fig:hww01j_lim}
\end{center}
\end{figure}

Given the expected number of signal and background events, the
sensitivity is limited by the systematic uncertainties for the counting analysis.
The additional information from the distributions of the kinematic variables
enables a significant improvement over the counting analysis.
Expected and observed 95\% CL upper limits on the production cross section
of the $\PH\to\WW$ process
relative to the SM prediction are shown in Fig.~\ref{fig:hww01j_lim},
for counting and shape-based analyses.
An excess of events is observed for low Higgs boson mass hypotheses,
which makes the observed limits weaker than expected.

\begin{figure}[htbp]
  \centering
     \includegraphics[width=0.45\textwidth]{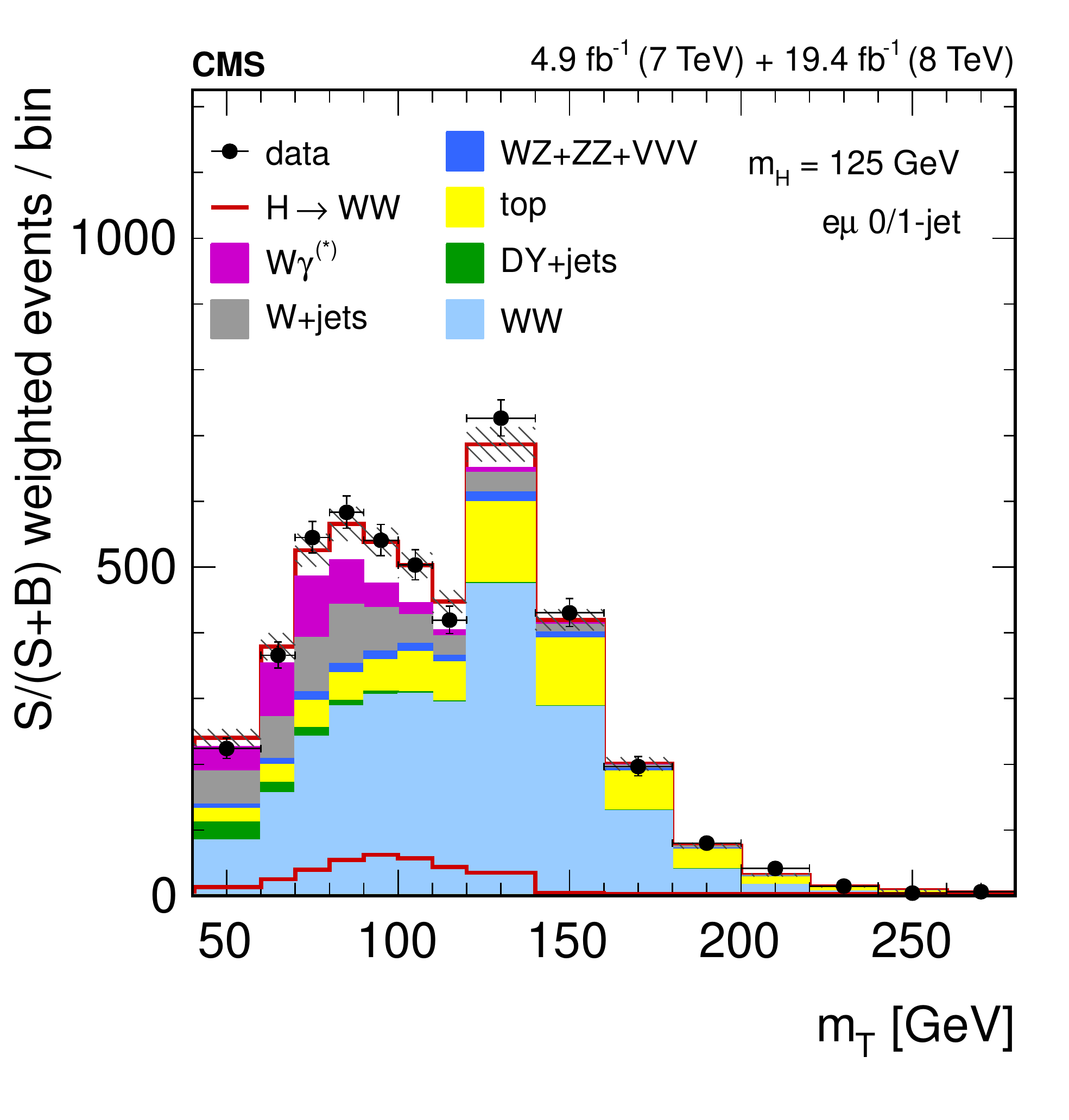}
     \includegraphics[width=0.45\textwidth]{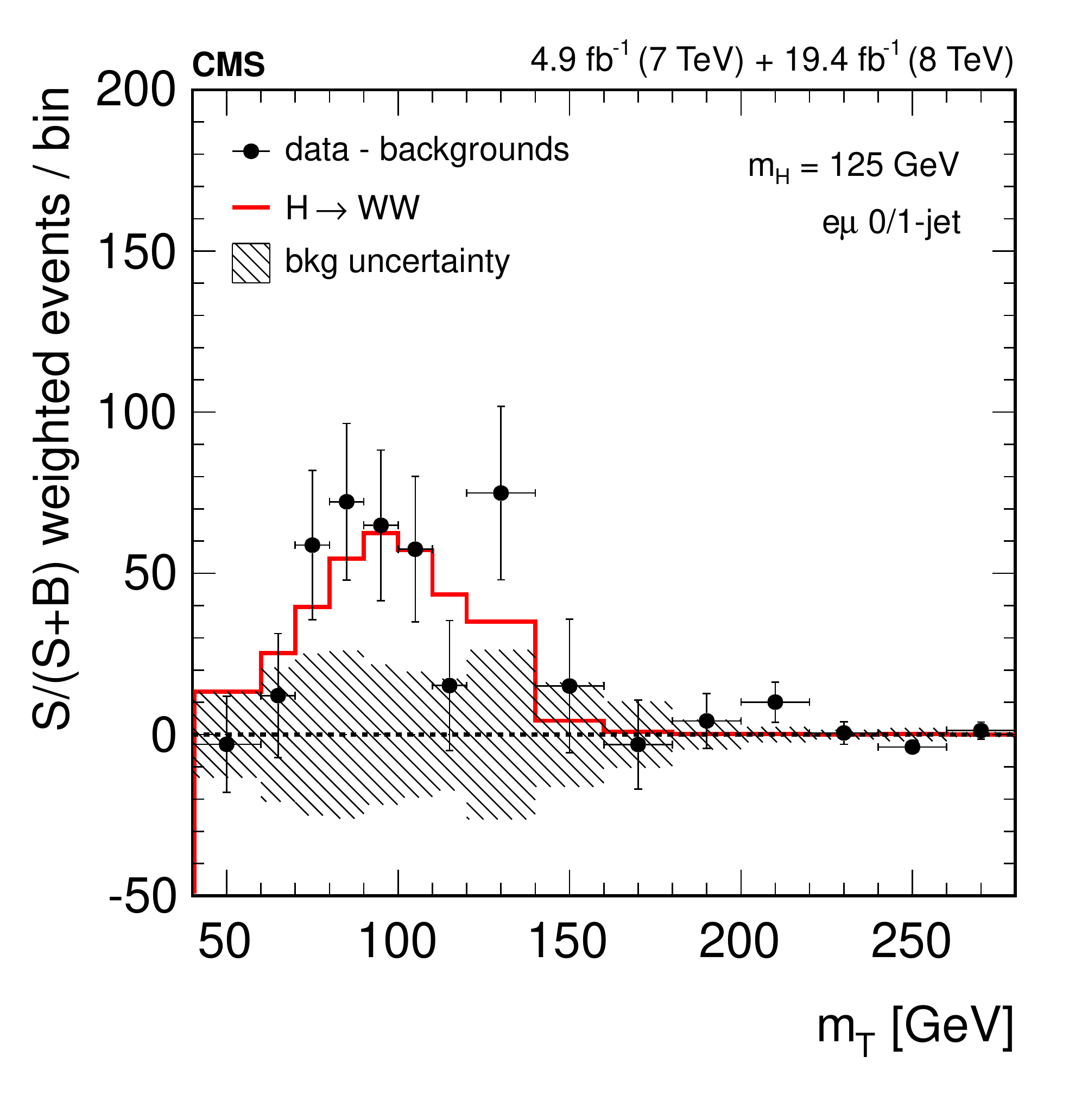}
    \caption{
         The $\mth$ distribution in the $\Pe\mu$ final state for the 0-jet and 1-jet categories combined
	 for observed data superimposed on signal + background events
	 and separately for the signal events alone (left)
	 and background-subtracted data with best-fit signal component (right).
         The signal and background processes are normalized to the result of
	 the template fit to the ($\mth$, $\mll$) distribution
	 and weighted according to the observed S/(S+B) ratio in each bin of
         the $\mll$ distribution integrating over the $\mth$ variable.
         To better visualize a peak structure, an extended $\mth$ range including $\mth$=[40,60]\GeV is shown, with the normalization
         of signal and background events extrapolated from the fit result.
         \label{fig:hww01j_postfit_mth}
	    }
\vspace{5mm}
  \centering
     \includegraphics[width=0.45\textwidth]{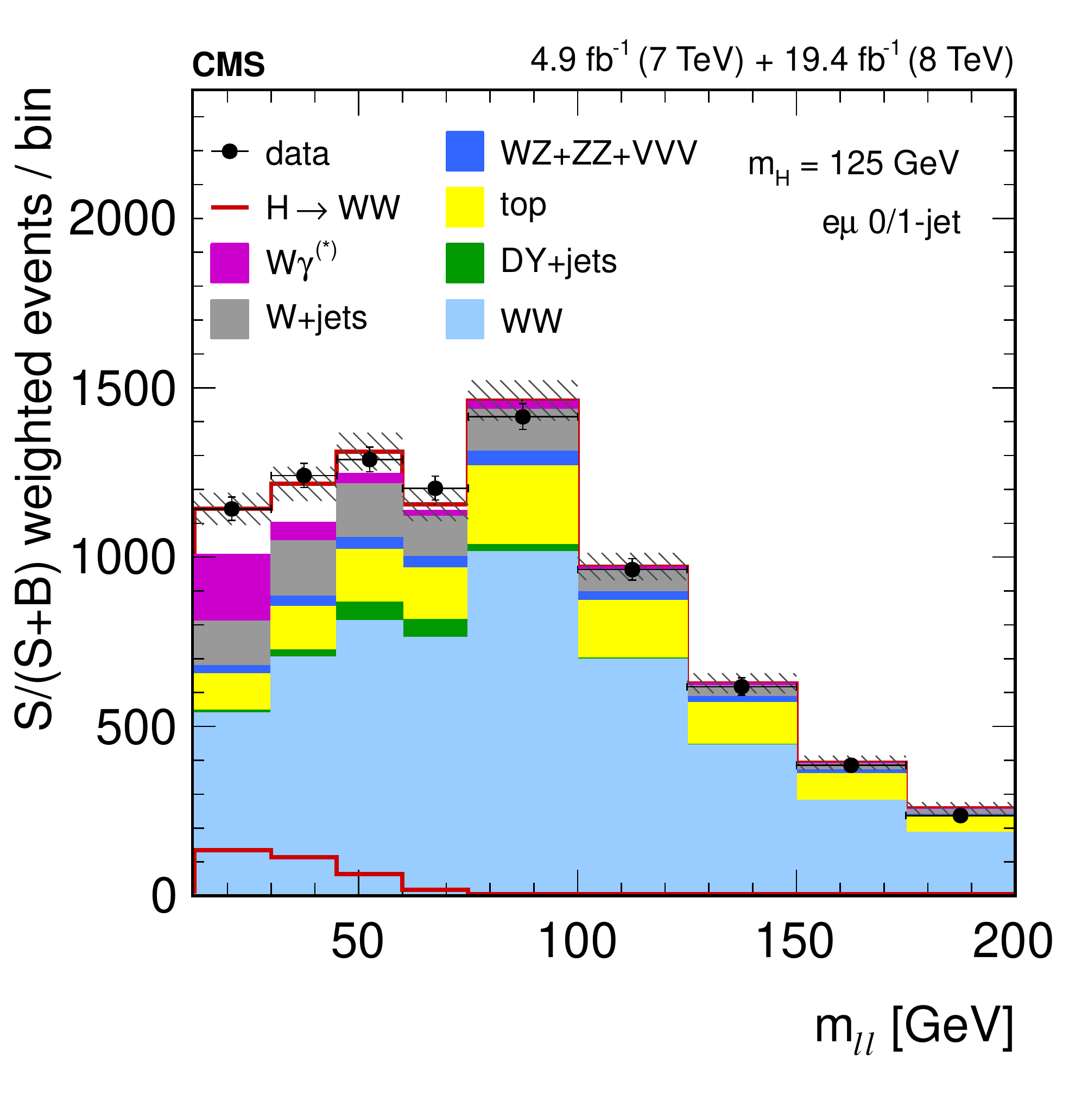}
     \includegraphics[width=0.45\textwidth]{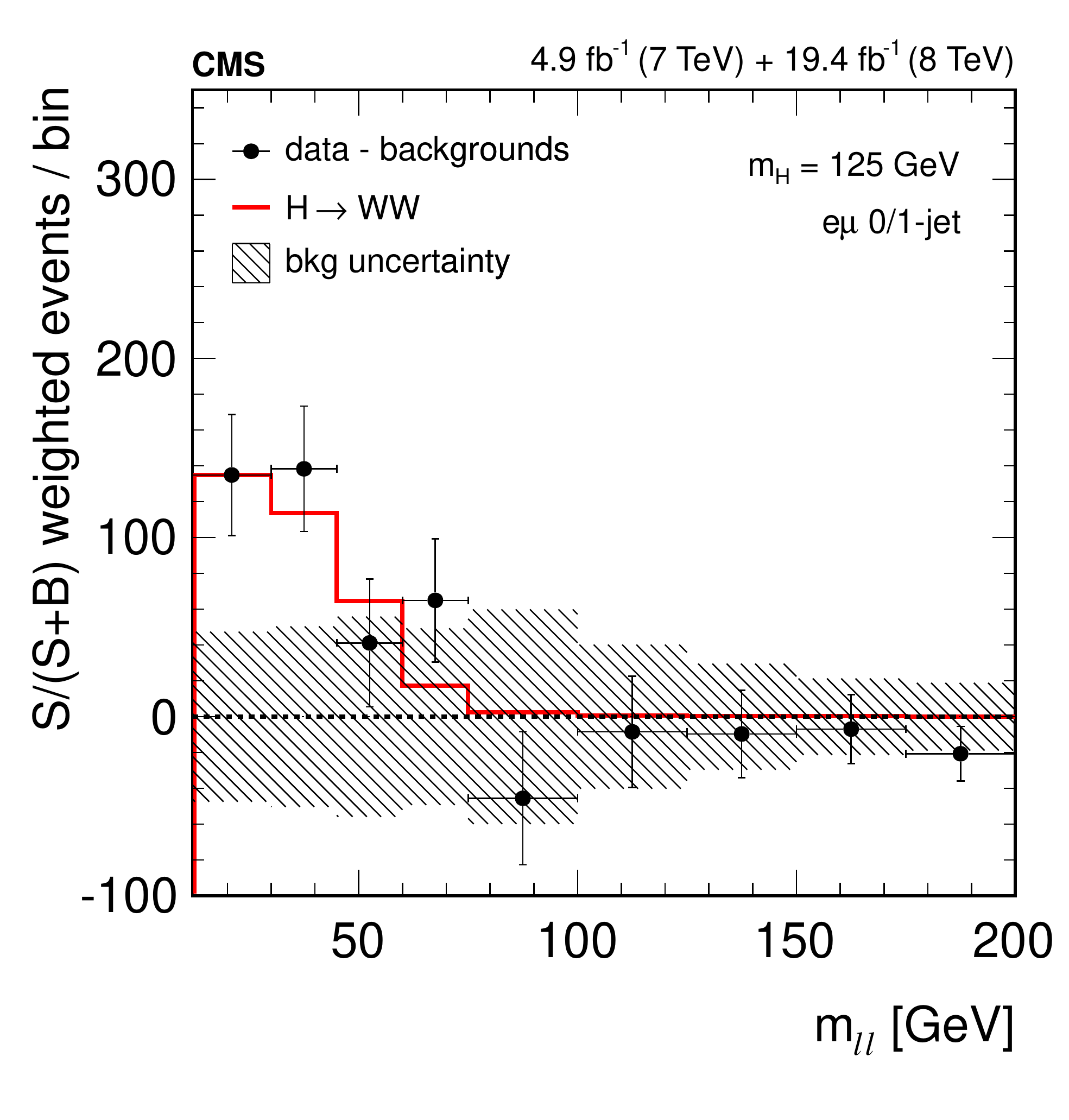}
    \caption{
         The $\mll$ distribution in the $\Pe\mu$ final state for the 0-jet and 1-jet categories combined
	 for observed data superimposed on signal + background events,
	 and separately for the signal events alone (left)
	 and background-subtracted data with best-fit signal component (right).
         The signal and background processes are normalized to the result of
	 the template fit to the ($\mth$, $\mll$) distribution
	 and weighted according to the observed S/(S+B) ratio in each bin of
         the $\mth$ distribution integrating over the $\mll$ variable.
         \label{fig:hww01j_postfit_mll}
	    }
\end{figure}

After the template fit to the ($\mth$, $\mll$) distribution, the
observed signal events as a function of $\mth$ and $\mll$ are shown in
Figures~\ref{fig:hww01j_postfit_mth} and~\ref{fig:hww01j_postfit_mll},
respectively. In these figures, each process is normalized to the fit
result and weighted using the other variable. This means for the
$\mth$ distribution, the $\mll$ distribution is used to compute the
ratio of the fitted signal (S) to the sum of signal and background
(S+B) in each bin of the $\mll$ distribution integrating over the
$\mth$ variable. Since the $\mth$ and $\mll$ variables are essentially
uncorrelated, the procedure allows to show unbiased background
subtracted data distributions. The observed distributions show good
agreement with the expected SM Higgs boson distributions.

\begin{figure}[htbp]
  \centering
     \includegraphics[width=0.45\textwidth]{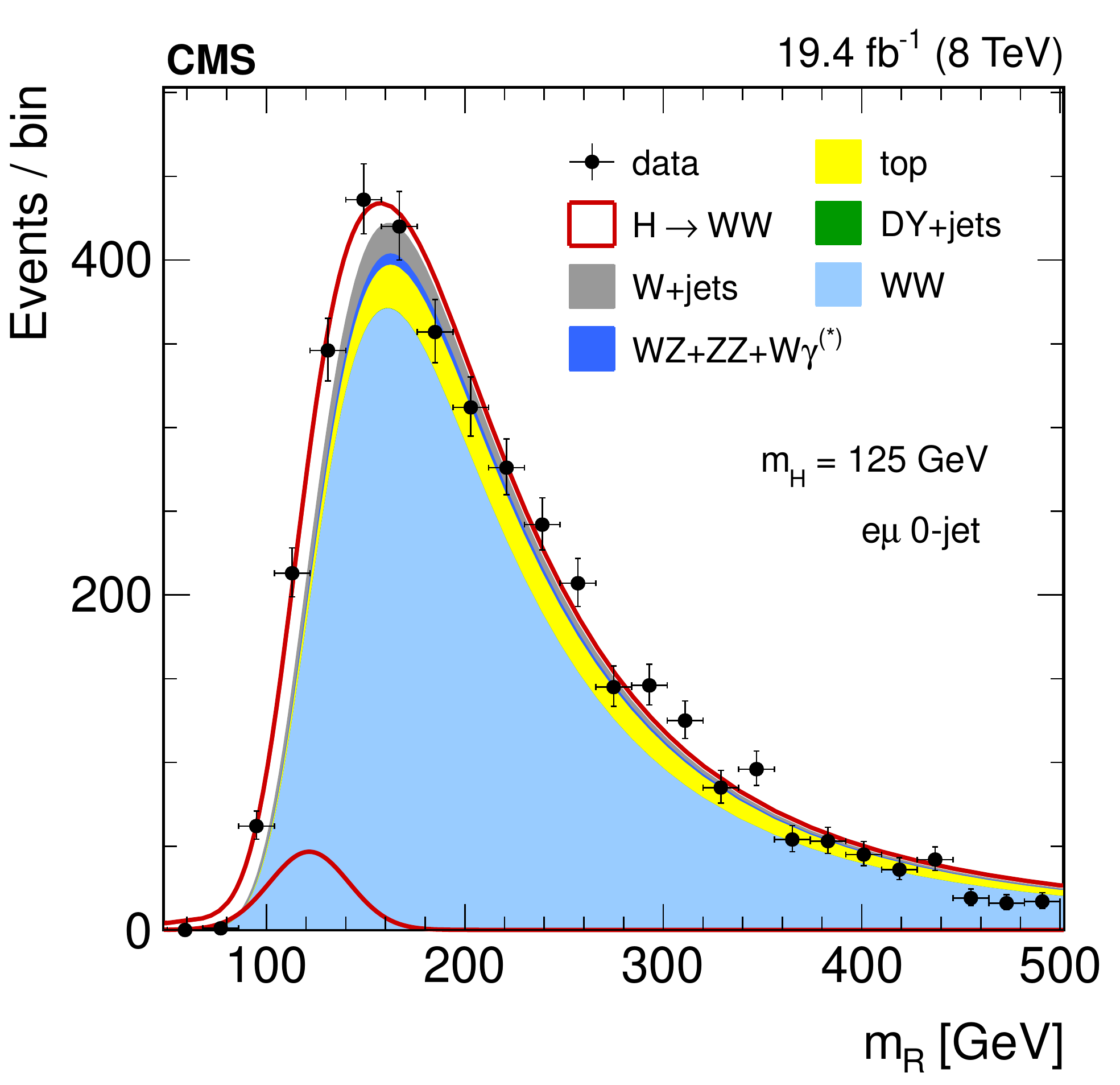}
     \includegraphics[width=0.45\textwidth]{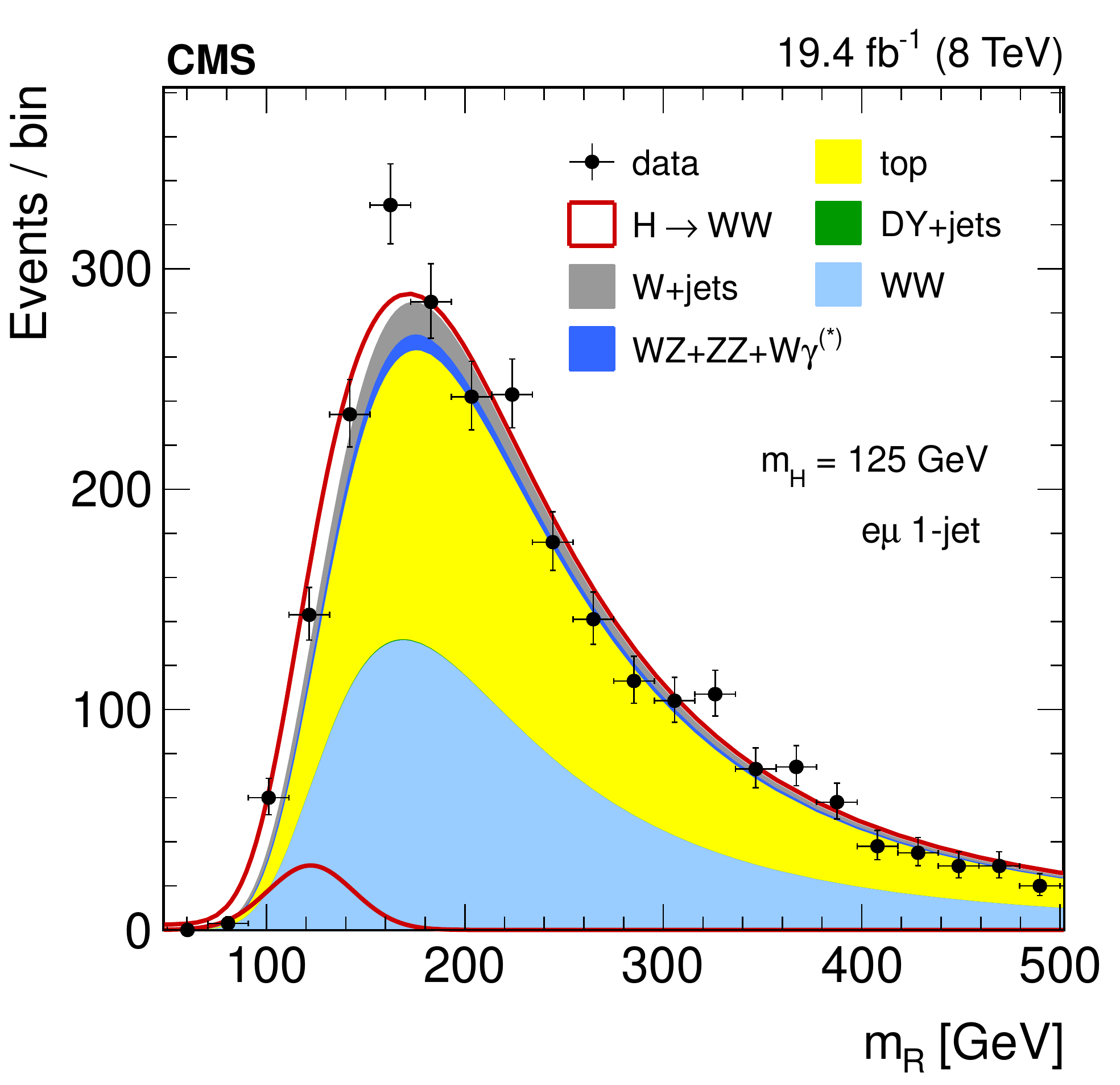} \\
    \caption{Distributions of $\mr$ showing the composition of signal
    and backgrounds, superimposed on the signal events alone,
    in the $\Pe\mu$ final state for the 0-jet (left)
    and 1-jet (right) categories for $\sqrt{s} = 8\TeV$.
    The signal and background processes are normalized to
    the result of the parametric fit to the ($\mr$, $\delphir$) distribution.
    \label{fig:hww01j_postfit_mr-dphi}}
\vspace{5mm}
  \centering
     \includegraphics[width=0.45\textwidth]{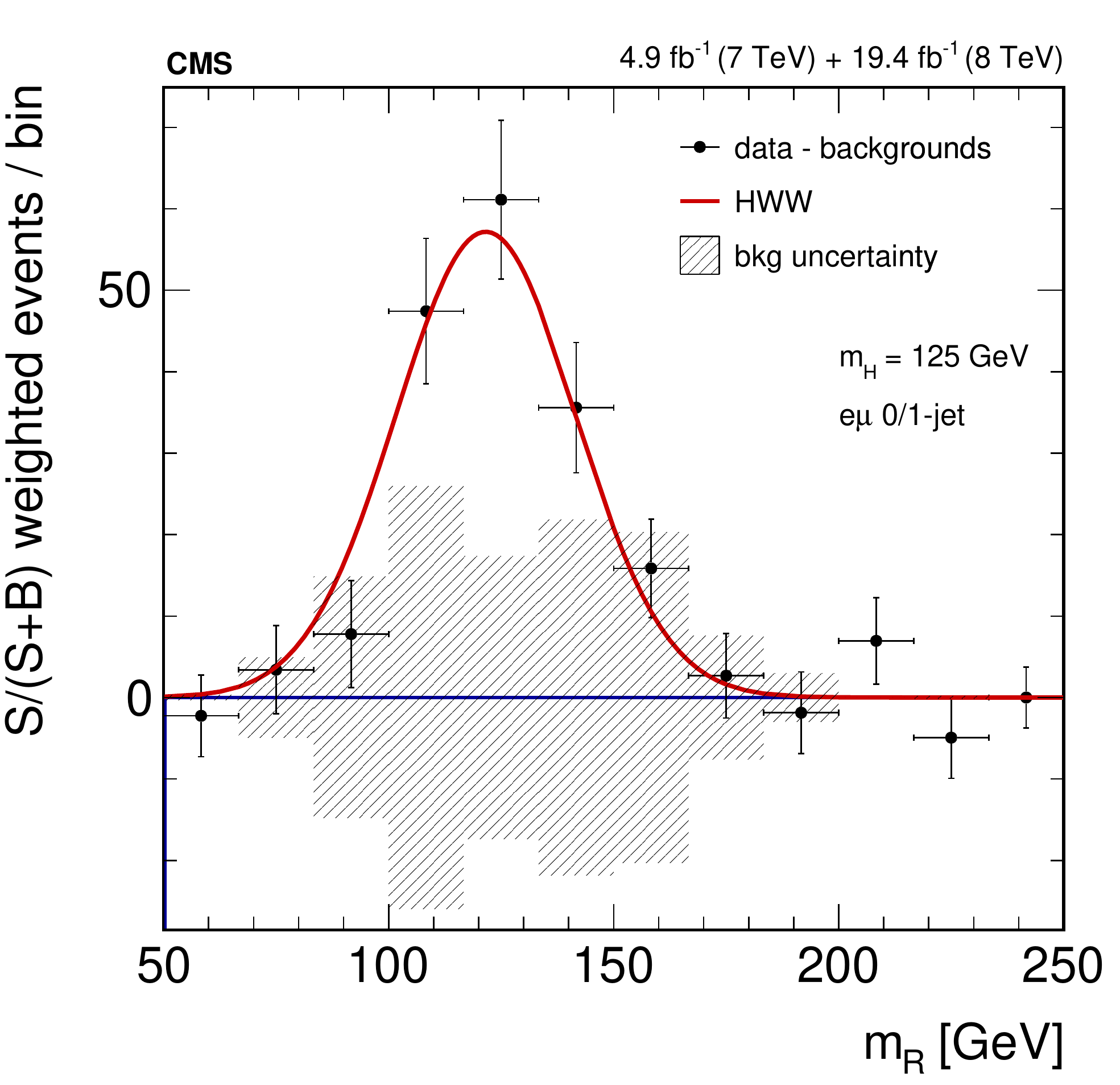}
     \includegraphics[width=0.45\textwidth]{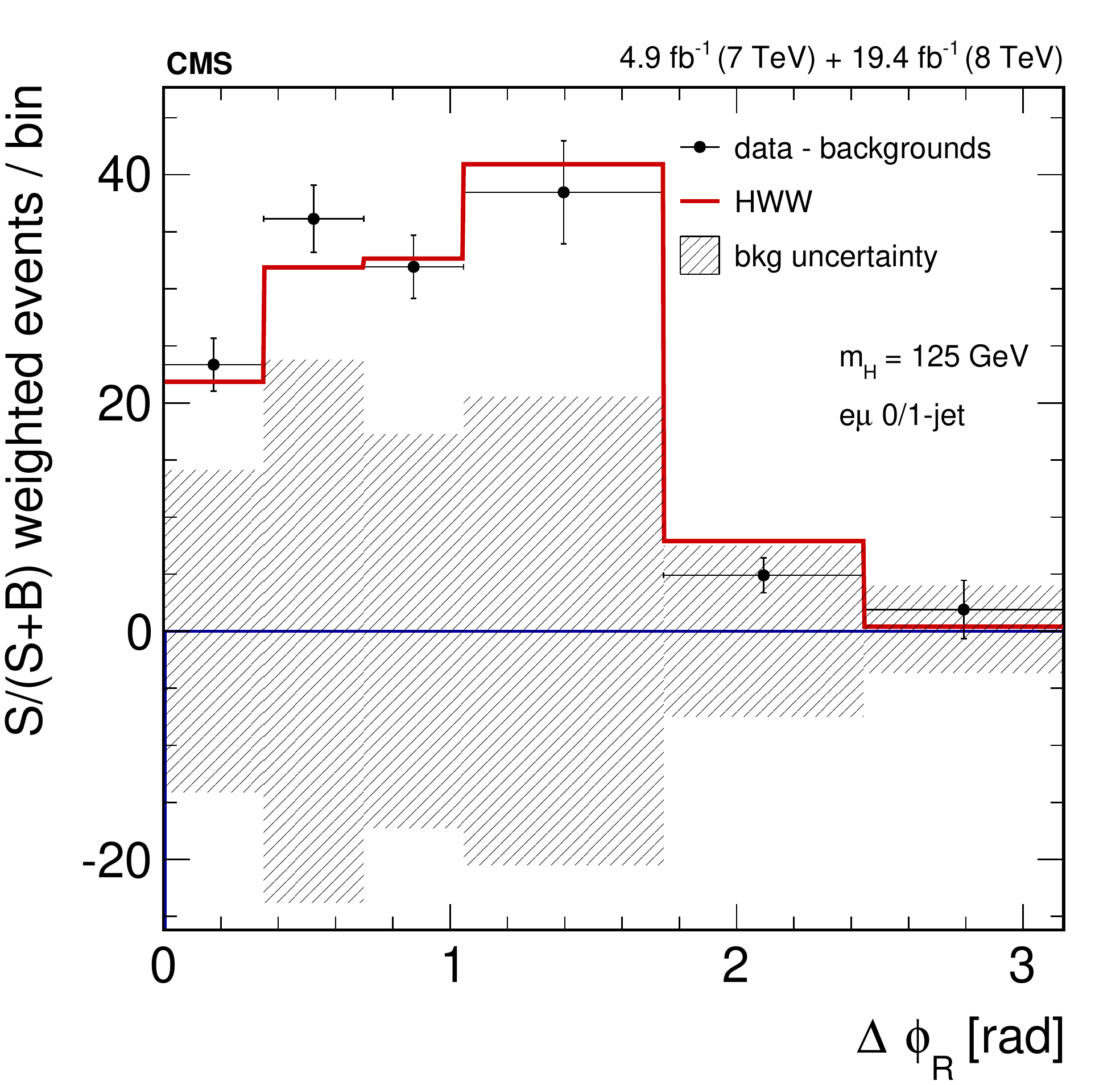} \\
    \caption{
    The background-subtracted data distribution for $\mr$ (left)
    and $\Delta\phi_R$ (right) with the best-fit superimposed for
    the 0-jet and 1-jet categories combined for $\sqrt{s} = $ 7 and
    $8\TeV$. The signal and background processes are normalized to
    the result of the parametric fit to the ($\mr$, $\delphir$) distribution.
    The events are weighted according to the
    observed S/(S+B) ratio of the second variable.\label{fig:hww01j_bkgsub_mr-dphi}}
\end{figure}

Similarly, the fit results for the parametric approach using the ($\mr$, $\delphir$) distribution
are shown in Figures~\ref{fig:hww01j_postfit_mr-dphi} and~\ref{fig:hww01j_bkgsub_mr-dphi}.
The fit projection of the $\mr$ variable integrated over
$\delphir$ is shown superimposed to the data distribution.
The background-subtracted data distributions are shown weighted by the
S/(S+B) ratio using the same weighting method previously described.

The expected and observed results for the $\PH\to\WW\to2\ell2\nu$ analyses in the 0/1-jet bin
are summarized in Table~\ref{tab:hww01j_results125}.
The upper limits on the $\PH \to \WW$ production cross section are slightly
higher than the SM expectation. The observed significance is 4.0 standard deviations for
the default shape-based analysis for $\mHi = 125\GeV$ using a template fit to
the ($\mth$, $\mll$) distribution and the expected significance is 5.2 standard deviations.
The best-fit signal strength, $\sigma/\sigma_\mathrm{SM}$,
which is the ratio of the measured  $\PH \to \WW$ signal yield
to the expectation for a SM Higgs boson is 0.76 $\pm$ 0.21.

\begin{table}[htbp]
\centering
\topcaption{
A summary of the expected and observed 95\% CL upper limits on  the $\PH \to \WW$ production cross section
relative to the SM prediction, the significances for the background-only
hypothesis to account for the excess in units of standard deviations (sd),
and the best-fit signal strength $\sigma/\sigma_\mathrm{SM}$,
the ratio of measured signal yield to the expected yield at $\mH = 125\GeV$ for the
0-jet and 1-jet categories. The $\Pe\mu$ and $\Pe\Pe/\mu\mu$ final states are combined
for these results.
The shape-based analysis results using a binned template fit or a parametric fit
for the $\Pe\mu$ final state are combined with counting analysis results
for the $\Pe\Pe/\mu\mu$ final states. The binned template fit to ($\mth$, $\mll$) is used to obtain the
default results.
}
\label{tab:hww01j_results125}
\resizebox{\textwidth}{!}
{
\begin{tabular}{lccc}
\hline\hline
0/1-jet analysis & 95\% CL limits on $\sigma/\sigma_\mathrm{SM}$ & Significance & $\sigma/\sigma_\mathrm{SM}$ \\
 $\mHi = 125\GeV$ & expected / observed & expected / observed & observed \\ \hline
($\mth$, $\mll$) template fit (default)	    & 0.4 / 1.2 & 5.2 / 4.0 sd & 0.76 $\pm$ 0.21 \\
($\mr$, $\delphir$) parametric fit             & 0.5 / 1.4 & 5.0 / 4.0 sd & 0.88 $\pm$ 0.25 \\
Counting analysis                        & 0.7 / 1.4 & 2.7 / 2.0 sd & 0.72 $\pm$ 0.37 \\ \hline
\end{tabular}
}
\end{table}

\textbf{Validation of the template fits}

The two-dimensional fit procedure has been extensively validated through pseudo-experiments
and fits in data control regions. The former are used to validate the fit
under known input conditions, while the latter are used
to check the accuracy of background templates and the model of correlations
between systematic uncertainties.

Assuming the SM expectation, the fit performance has been evaluated
with pseudo-experiments in terms of process normalizations
and nuisance parameters, both under default conditions and in the
presence of input biases, which correspond to $\pm$1 standard deviation on
either normalization or shape of the most important backgrounds. Fit results are
very stable and in most cases the signal yield is determined with no significant
bias. The largest deviation is observed for input bias applied on the $\Wjets$ background
normalization, with an average shift no larger than 10\% which is more than three
times smaller than the uncertainty in the signal yield. All nuisance parameter values
and uncertainties resulting from the fit performed on data are compatible with expectations from
pseudo-experiments. The most constrained parameters
are related to the $\WW$ (and, secondarily, top-quark) background, as the fit can gauge
it from a large signal-free region. It is therefore crucial to verify with data that
the $\WW$ correlation model is correct.

\begin{figure}[htbp]
\begin{center}
\includegraphics[width=0.45\linewidth]{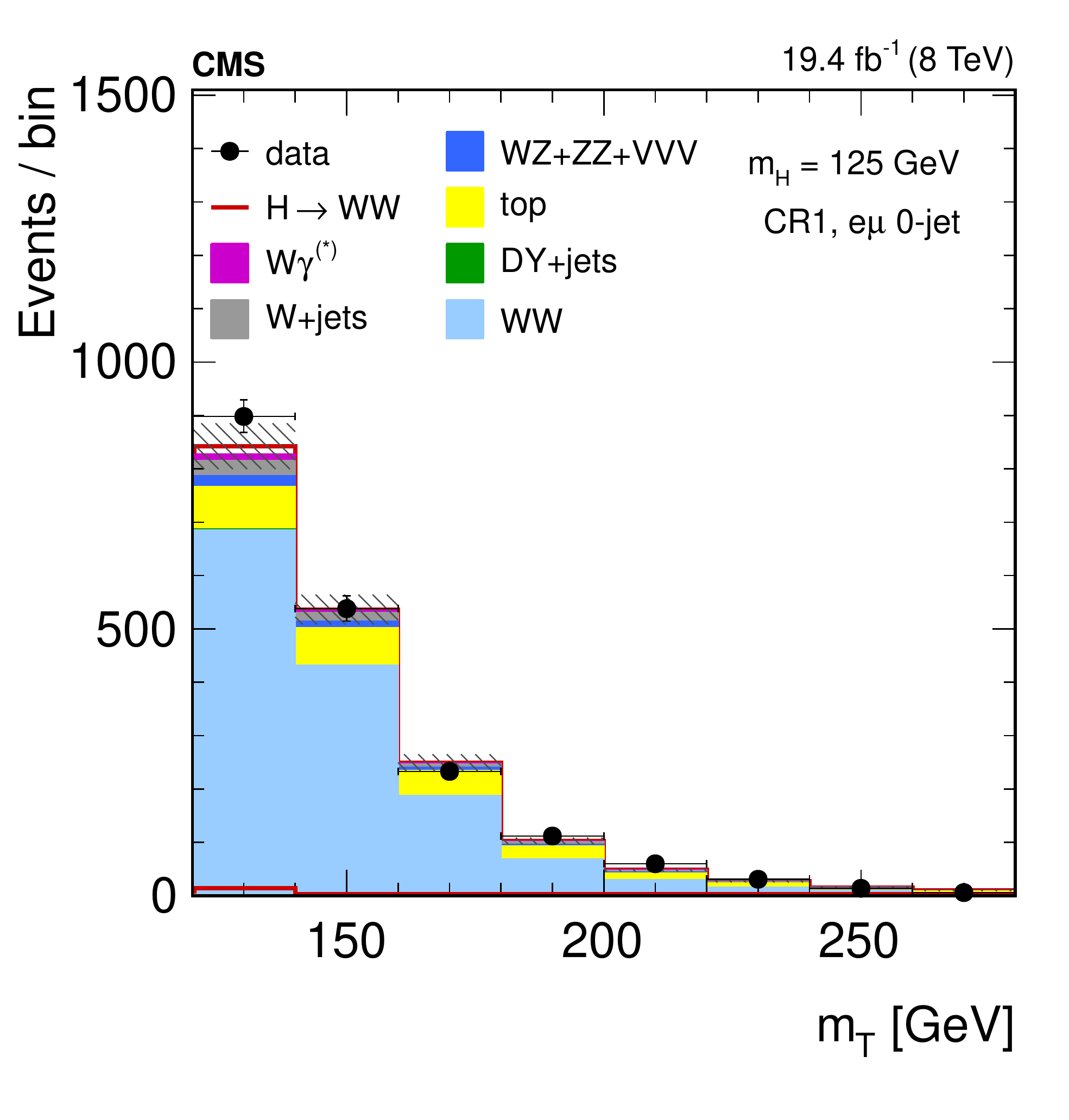}
\includegraphics[width=0.45\linewidth]{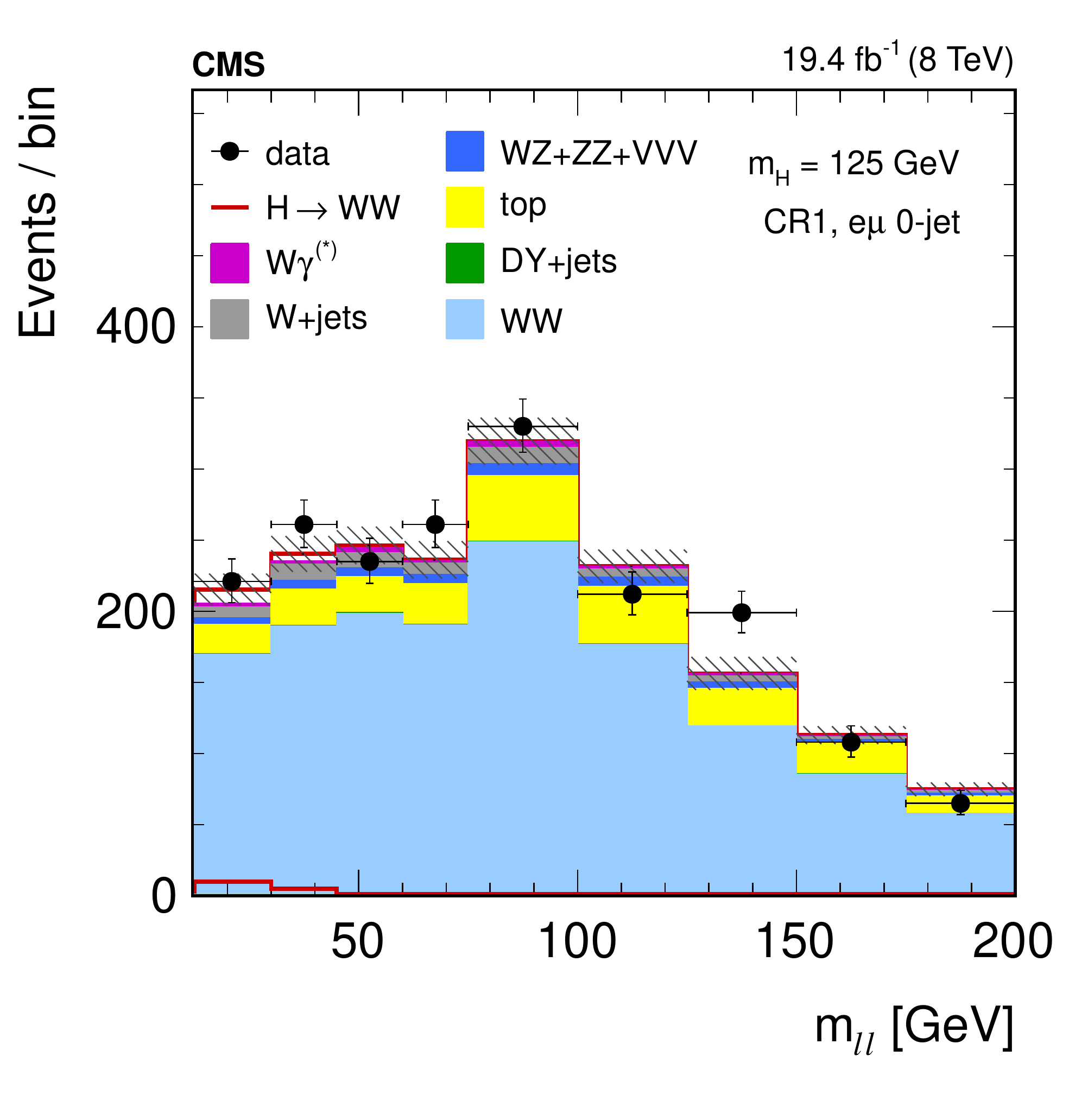}
\includegraphics[width=0.45\linewidth]{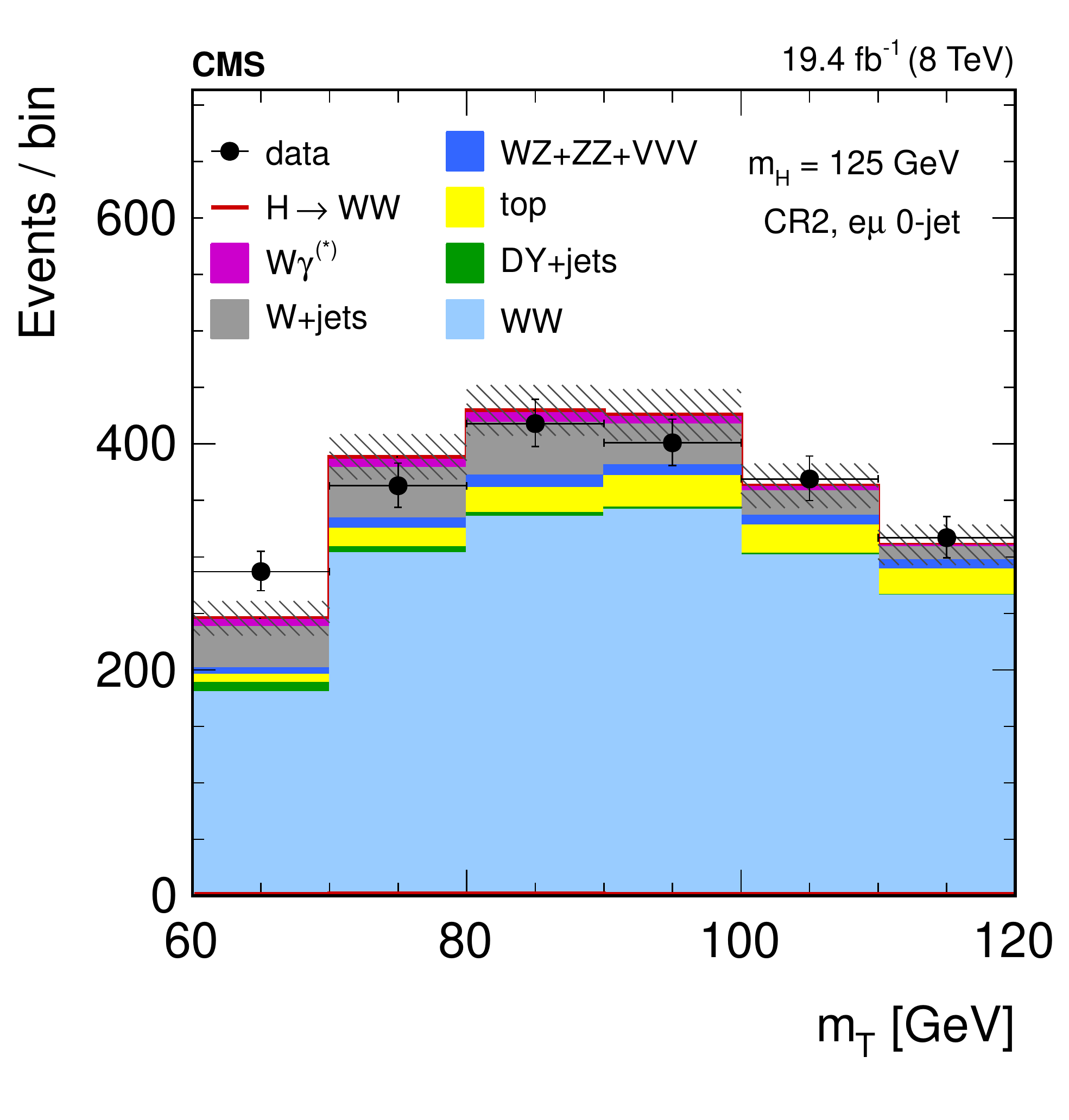}
\includegraphics[width=0.45\linewidth]{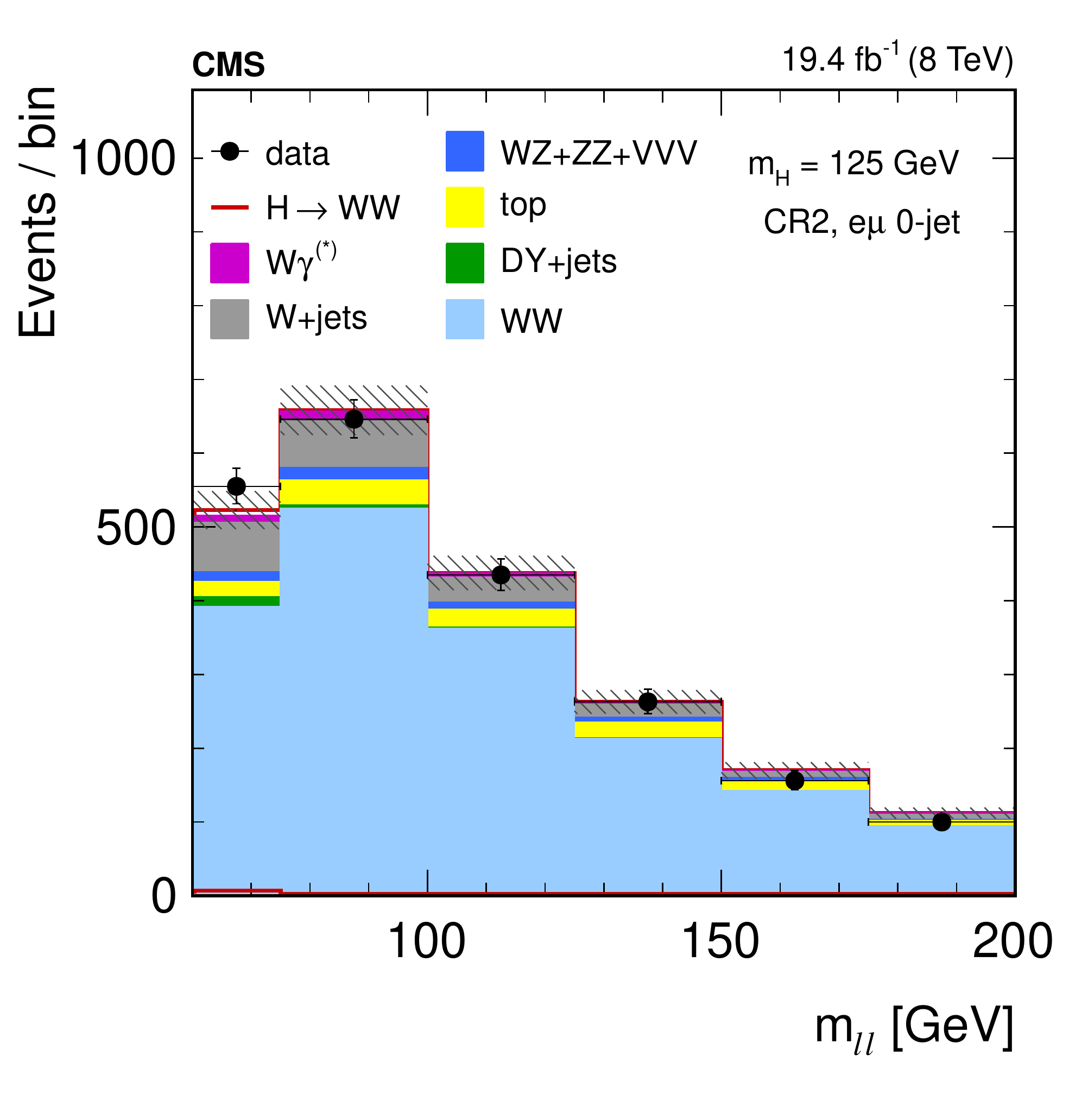}
\caption{Distributions of $\mt$ (left) and $\mll$ (right) extrapolated to the control regions CR1 (top) and
CR2 (bottom) in the 0-jet bin category, after fitting the other control region.}
\label{fig:mt_mll_CR}
\end{center}
\end{figure}

For the purpose of checking the $\W\W$ model a dedicated test is developed. First, the signal-free
$\WW$ control sample is separated into two non-overlapping regions with a similar number of events.
Then, each region is fitted separately. In this fit, only the $\WW$ background is allowed to change.
In order to avoid fluctuations due to non-$\WW$ components, all other processes are fixed to the
values obtained in the fit performed in the full range. The first region (CR1, high $\mt$) is defined by
requiring 120\GeV $< \mt <$ 280\GeV and 12\GeV $< \mll <$ 200\GeV, while the second region (CR2, high $\mll$) is defined
by requiring 60\GeV $< \mt <$ 120\GeV and 60\GeV $< \mll <$ 200\GeV.
The $\WW$ normalization and shape obtained from the fit
in one region are extrapolated to the other region and compared
to data.
Figure~\ref{fig:mt_mll_CR} shows the $\mt$ and $\mll$ distributions in the control
regions CR1 and CR2 using fit results from the other control region.
The uncertainty band is evaluated from pseudo-experiments.
In each bin of the two-dimensional distribution, the uncertainty in the background processes is obtained from
the fit in the full range. All distributions show generally good agreement with data, indicating that
the $\WW$ fit model is not biased.

\begin{figure}[htbp]
\begin{center}
{\includegraphics[width=0.45\textwidth]{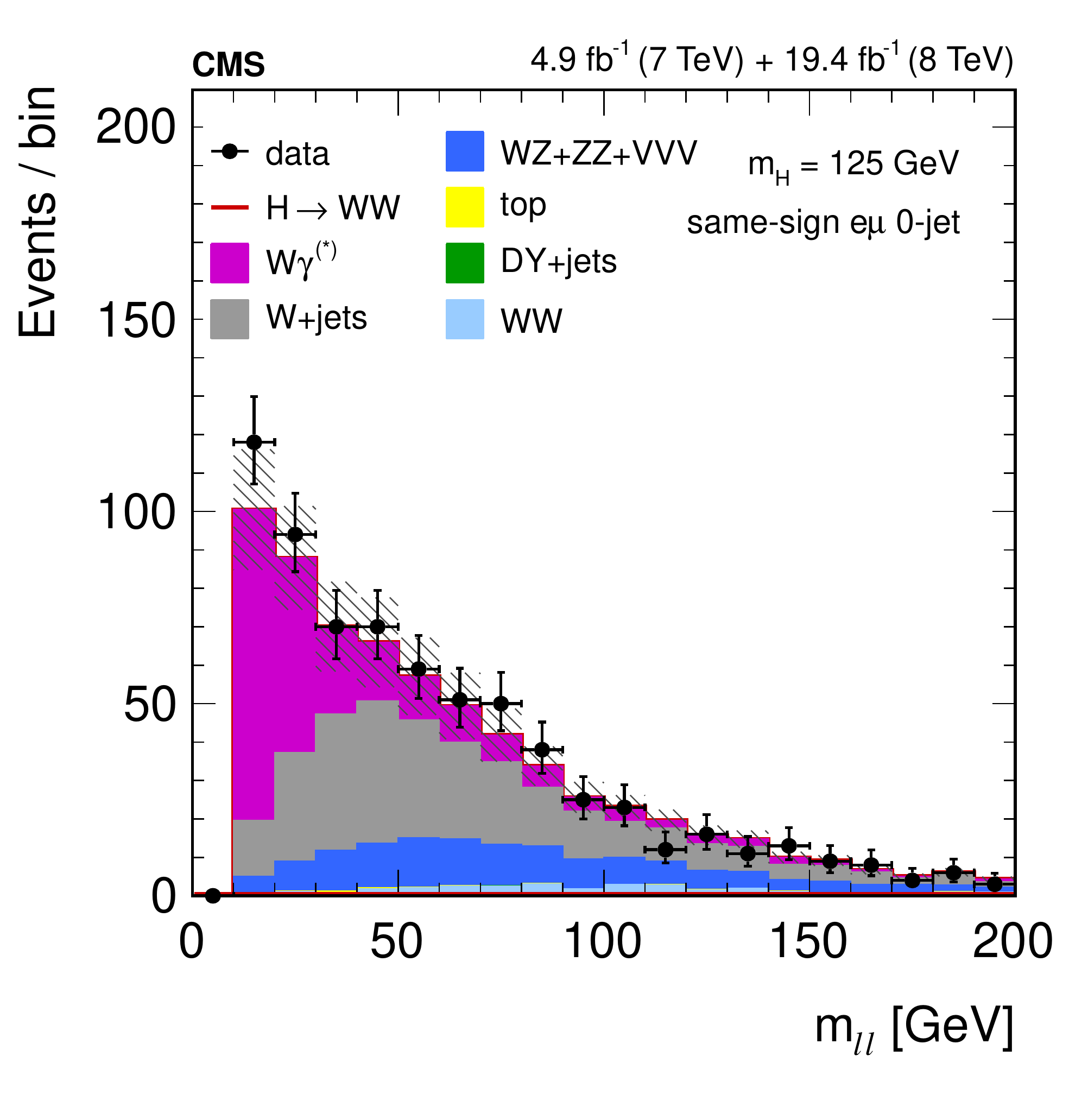}}
{\includegraphics[width=0.45\textwidth]{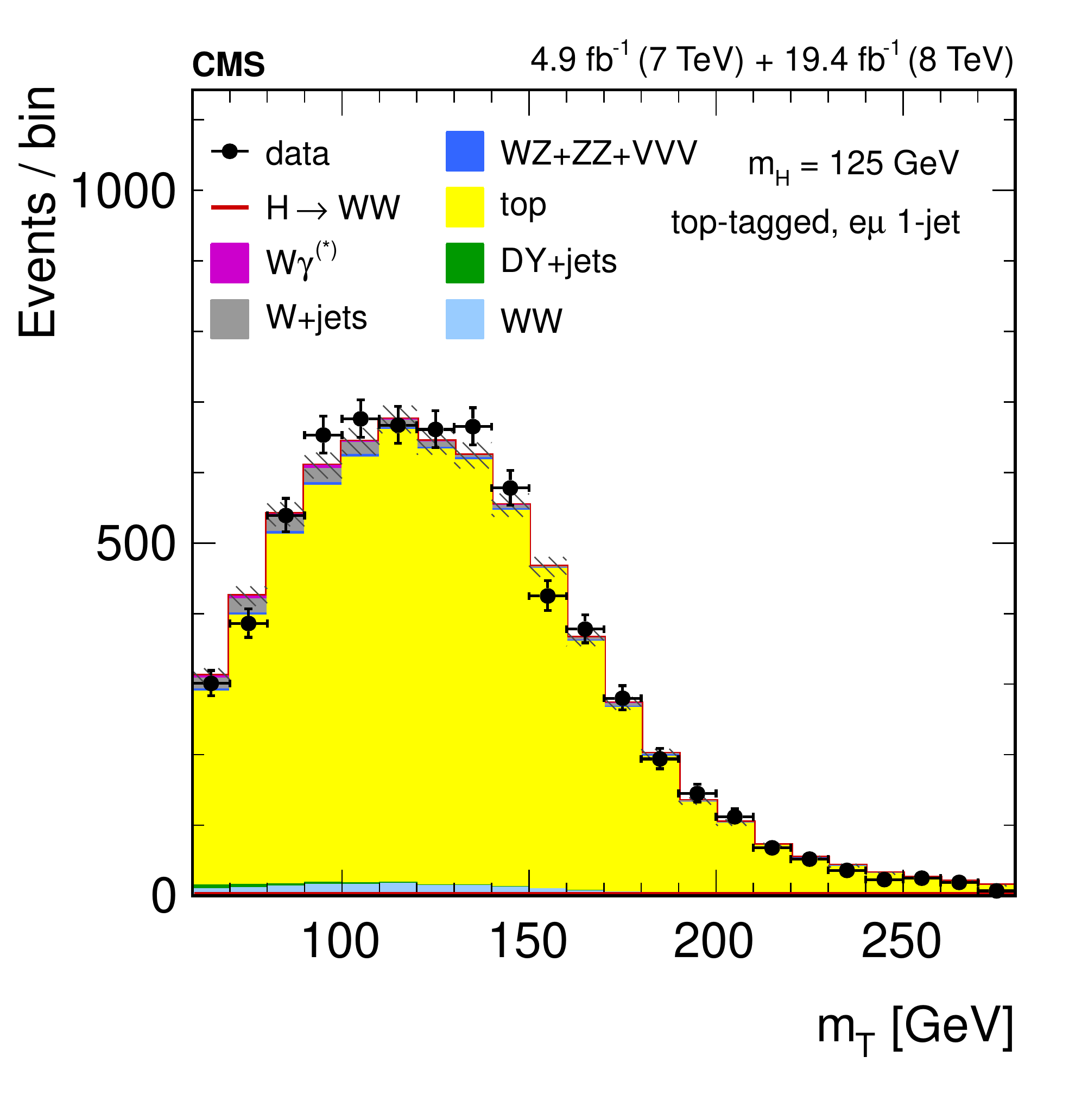}} \\
\caption{Distributions of the dilepton mass (left) in the same-charge dilepton control region in the 0-jet category
and the transverse mass (right) in the top-tagged control region in the 1-jet category of
the $\Pe\mu$ final state.
}
\label{fig:wwlevel_sstopcontrol}
\end{center}
\end{figure}

Fits are performed in two types of control samples, one defined by b-tagged jets
and the other by two leptons with the same charge. The first
sample is dominated by top-quark processes, while the second sample is dominated by the $\Wjets$ and
$\wgamma^{(*)}$ processes. In both cases the background yields agree with the
expectations and no signal component is found.
Distributions of the discriminating variables in some of these
control regions are shown in Fig.~\ref{fig:wwlevel_sstopcontrol}.

In summary, the templates for all main backgrounds ($\WW$, $\ttbar+\tw$, $\Wjets$, and
$\wgamma^{(*)}$) have been tested in dedicated control regions with data.
Both the fit procedure and the background estimations are found to be very robust.

Finally, the template shape for the dominant $\qqbar\to\WW$ background process has been
cross-checked by replacing the template histogram obtained from the default generator
by another one and rederiving the shape uncertainty templates that are allowed to vary
in the fit. Table~\ref{tab:sig_mu_nH125_test} summarizes the results of this procedure
using \MADGRAPH (a priori default used in the analysis),
\MCATNLO, and \POWHEG. The signal significance, and the
best-fit signal strength are found to be consistent with one another for the
three different $\qqbar\to\WW$ template models tested.

\begin{table}[htbp]
\centering
\topcaption{A summary of the expected and observed 95\% CL upper limits on the $\PH \to \WW$ production cross section
relative to the SM prediction, significances in units of standard deviations (sd), and
the best-fit value of $\sigma/\sigma_\mathrm{SM}$ for the SM Higgs boson with a mass of 125\GeV for
the 0-jet and 1-jet categories using the template fit to ($\mth$, $\mll$), where three different
generators have been used to model the $\qqbar \to \WW$ background process.\label{tab:sig_mu_nH125_test}}
{
\begin{tabular}{lccc}
\hline\hline
$\qqbar \to \W\W$  & 95\% CL limits on $\sigma/\sigma_\mathrm{SM}$ & Significance & $\sigma/\sigma_\mathrm{SM}$ \\
  generator    & expected / observed & expected / observed & observed \\
\hline
\MADGRAPH (default)   & 0.4 / 1.2 & 5.2 / 4.0 sd & 0.76 $\pm$ 0.21 \\
\MCATNLO               & 0.4 / 1.2 & 5.3 / 4.2 sd & 0.82 $\pm$ 0.24 \\
\POWHEG               & 0.4 / 1.2 & 5.1 / 3.9 sd & 0.74 $\pm$ 0.21 \\
\hline
\end{tabular}
}
\end{table}

\subsection{The two-jet VBF tag}
\label{sec:hww2l2n_2j}
The second-largest production mode for the SM Higgs boson is through VBF,
for which the cross section is approximately an order of magnitude smaller than that of the gluon fusion process.
In this process two vector bosons are radiated from initial-state quarks and produce
a Higgs boson at tree level. In the scattering process, the two initial-state
partons may scatter at a polar angle from the beam axis large enough to
be detected as additional jets in the signal events. Furthermore, these two jets,
being remnants of the incoming proton beams, feature the distinct signature of
having high momentum and large separation in pseudorapidity, hence sizeable
invariant mass, with an absence of additional hadronic activity in the central rapidity region
due to the lack of color exchange between the parent quarks.
By exploiting this specific signature, VBF searches typically
have a good signal-to-background ratio.
In this analysis the signal-to-background ratio approaches one
after all the selection criteria are applied.

To select events with the characteristics of the VBF process, the
two highest $\pt$ jets in the event are required
to have pseudorapidity separation of $\abs{\Delta \eta_{jj}} >$ 3.5
and to form an invariant mass $m_{jj} >$ 500\GeV.
Events with an additional jet situated
in the pseudorapidity range between the two leading jets are rejected.
Both leptons are also required to be within the pseudorapidity
region defined by the two highest $\pt$ jets.

\subsubsection{Analysis strategy}

Given the small event yield for the 2-jet category with VBF tag with the currently available
datasets, the signal extraction uses a template fit to a single kinematic variable
with appropriately-sized bins. The dilepton mass, $\mll$, has been chosen for its
simple definition and discrimination power, and also because the hadronic information
is already extensively used in the event selection.
The counting analysis is pursued for the same-flavor category,
and also used as a cross-check of the shape-based approach for the different-flavor final state.

Since the fit to data uses only the $\mll$ distribution, the events are preselected
to satisfy $\mth$ smaller than the Higgs boson mass of the given hypothesis.
For Higgs boson mass hypotheses of 250\GeV and above,
$\ptlmax$ is required to be greater than 50\GeV.
The $\mll$ template has 14 bins for the 8\TeV sample and 10 bins for the 7\TeV sample,
covering the range from 12\GeV to 600\GeV.

For the counting analysis, the same requirements as the 0-jet and 1-jet analyses
are applied, as summarized in Table~\ref{tab:hww01j_cbselection},
except for the lower $\mth$ threshold which is kept at 30\GeV for all Higgs
boson mass hypotheses. The results of the same-flavor counting analysis
are combined with the results of the different-flavor shape analysis to provide the result for this category.

\subsubsection{Results}

The data yields and the expected yields for the SM Higgs boson signal and various backgrounds
in each of the lepton-flavor final states for the VBF analysis are listed in
tables~\ref{tab:vbf_yields7tev} and~\ref{tab:vbf_yields8tev}, for several representative Higgs boson mass hypotheses.
For a Higgs boson with $\mH = 125\GeV$, a few signal events are expected to be observed
with a signal-to-background ratio of about one.
The contribution to the VBF selection from gluon fusion Higgs boson production after
all selection requirements is approximately 20\% of the total signal
yield~\cite{10281_40097}.
\begin{table}[htbp]
\centering
\topcaption{
Signal prediction, observed number of events in data, and background estimates
for $\sqrt{s}=7\TeV$ after applying the $\hww$ VBF tag counting analysis selection requirements
and the requirements used for the shape-based approach ($\Pe\mu$ final state only).
The combined statistical, experimental, and theoretical systematic uncertainties are reported.
The $\dyll$ process includes the dimuon, dielectron and ditau final state.
The $\mathrm{VZ}$ background denotes the contributions from $\WZ$ and $\ZZ$ processes.
}
\label{tab:vbf_yields7tev}
{
\scriptsize
\setlength{\extrarowheight}{1pt}
\begin{tabular} {ccccccccc}
\hline\hline
\multirow{2}{*}{$\mH$ [\GeVns{}]}  & \multirow{2}{*}{$\Pg\Pg\PH$} & \multirow{2}{*}{VBF+$\V\PH$} & \multirow{2}{*}{Data} & \multirow{2}{*} {All bkg.} & \multirow{2}{*}{$\WW$} & $\mathrm{VZ}+\wgamma^{(*)}$ & \multirow{2}{*}{$\ttbar+\tw$} & \multirow{2}{*}{$\Wjets$} \\
                        &                               &                              &		       &			    &			     & + $\dyll$  &		&					      \\
\hline
\multicolumn{9}{c}{7\TeV $\Pe\mu$ final state, 2-jets category, VBF tag} \\
\hline
 $120$         & $0.07\pm0.03$ & $0.44\pm0.06$ & $0$ & $0.50\pm0.20$ & $0.08\pm0.03$ & $0.15\pm0.14$ & $0.16\pm0.07$ & $0.10\pm0.09$ \\
 $125$         & $0.12\pm0.04$ & $0.73\pm0.10$ & $0$ & $0.66\pm0.23$ & $0.12\pm0.05$ & $0.15\pm0.15$ & $0.20\pm0.08$ & $0.19\pm0.14$ \\
 $130$         & $0.13\pm0.05$ & $1.05\pm0.14$ & $0$ & $0.76\pm0.24$ & $0.18\pm0.08$ & $0.17\pm0.15$ & $0.22\pm0.09$ & $0.19\pm0.14$ \\
 $160$         & $0.63\pm0.21$ & $3.01\pm0.40$ & $0$ & $0.46\pm0.13$ & $0.17\pm0.07$ & $0.02\pm0.01$ & $0.27\pm0.11$ &       ---       \\
 $200$         & $0.47\pm0.14$ & $2.42\pm0.32$ & $2$ & $1.73\pm0.42$ & $0.58\pm0.22$ & $0.07\pm0.02$ & $0.84\pm0.31$ & $0.24\pm0.18$ \\
 $400$         & $0.34\pm0.11$ & $0.87\pm0.11$ & $4$ & $2.03\pm0.54$ & $0.82\pm0.36$ & $0.05\pm0.02$ & $1.00\pm0.37$ & $0.16\pm0.14$ \\
 $600$         & $0.11\pm0.04$ & $0.31\pm0.04$ & $1$ & $0.73\pm0.22$ & $0.35\pm0.16$ & $0.03\pm0.01$ & $0.27\pm0.11$ & $0.08\pm0.10$ \\
 $125$ (shape) & $0.19\pm0.09$ & $1.05\pm0.13$ &  4 & $5.81\pm0.96$  & $0.92\pm0.28$ & $0.08\pm0.01$ & $3.47\pm0.87$ & $0.57\pm0.24$ \\
\hline
\multicolumn{9}{c}{7\TeV $\Pe\Pe$/$\mu\mu$ final state, 2-jets category, VBF tag} \\
\hline
 $120$ & $0.04\pm0.02$ & $0.14\pm0.02$ & $1$ & $0.97\pm1.02$ & $0.08\pm0.05$ & $0.77\pm1.02$ & $0.13\pm0.06$ &       ---       \\
 $125$ & $0.02\pm0.01$ & $0.26\pm0.04$ & $1$ & $1.9\pm2.1$   & $0.10\pm0.07$ & $1.6\pm2.1$   & $0.14\pm0.06$ &       ---       \\
 $130$ & $0.10\pm0.04$ & $0.42\pm0.06$ & $1$ & $1.8\pm1.9$   & $0.14\pm0.08$ & $1.5\pm1.9$   & $0.16\pm0.07$ &       ---       \\
 $160$ & $0.46\pm0.16$ & $1.87\pm0.25$ & $1$ & $0.57\pm0.34$ & $0.22\pm0.11$ & $0.20\pm0.31$ & $0.15\pm0.06$ &       ---       \\
 $200$ & $0.21\pm0.07$ & $1.29\pm0.17$ & $2$ & $2.4\pm2.1$   & $0.42\pm0.17$ & $1.4\pm2.0$   & $0.44\pm0.18$ & $0.16\pm0.14$ \\
 $400$ & $0.18\pm0.06$ & $0.46\pm0.06$ & $1$ & $0.58\pm0.16$ & $0.24\pm0.11$ & $0.01\pm0.01$ & $0.33\pm0.12$ &       ---       \\
 $600$ & $0.06\pm0.02$ & $0.18\pm0.02$ & $0$ & $0.24\pm0.09$ & $0.10\pm0.04$ & $0.01\pm0.01$ & $0.14\pm0.07$ &       ---       \\
\hline
\end{tabular}
}
\end{table}

\begin{table}[htbp]
\centering
\caption{
Signal prediction, observed number of events in data, and background estimates
for $\sqrt{s}=8\TeV$ after applying the $\hww$ VBF tag counting analysis selection requirements
and the requirements used for the shape-based approach ($\Pe\mu$ final state only).
The combination of statistical uncertainties with experimental and theoretical systematic uncertainties is reported.
The $\dyll$ process includes the dimuon, dielectron and ditau final state.
The $\mathrm{VZ}$ background denotes the contributions from $\WZ$ and $\ZZ$ processes.
}
\label{tab:vbf_yields8tev}
 {
 \ifthenelse{\boolean{cms@external}}{\scriptsize}{\scriptsize}
\setlength{\extrarowheight}{1pt}
\begin{tabular} {ccccccccc}
  \hline\hline
\multirow{2}{*}{$\mH$ [\GeVns{}]}  & \multirow{2}{*}{$\Pg\Pg\PH$} & \multirow{2}{*}{VBF+$\V\PH$} & \multirow{2}{*}{Data} & \multirow{2}{*} {All bkg.} & \multirow{2}{*}{$\WW$} & $\mathrm{VZ}+\wgamma^{(*)}$ & \multirow{2}{*}{$\ttbar+\tw$} & \multirow{2}{*}{$\Wjets$} \\
                        &                               &                              &		       &			    &                        & + $\dyll$  &  		    &						  \\
\hline
\multicolumn{9}{c}{8\TeV $\Pe\mu$ final state, 2-jets category, VBF tag} \\
\hline
 $120$         & $0.43\pm0.18$ & $2.06\pm0.28$  & $2$  & $3.34\pm0.55$  & $0.75\pm0.22$ & $0.36\pm0.12$ & $1.75\pm0.42$  & $0.48\pm0.26$ \\
 $125$         & $0.89\pm0.35$ & $3.41\pm0.47$  & $2$  & $4.38\pm0.81$  & $0.86\pm0.24$ & $0.49\pm0.14$ & $2.67\pm0.73$  & $0.36\pm0.22$ \\
 $130$         & $1.55\pm0.54$ & $5.24\pm0.73$  & $5$  & $4.87\pm0.84$  & $1.20\pm0.30$ & $0.56\pm0.15$ & $2.74\pm0.74$  & $0.36\pm0.22$ \\
 $160$         & $3.5\pm1.1$   & $14.8\pm2.0$   & $3$  & $3.98\pm0.78$  & $1.21\pm0.29$ & $0.22\pm0.10$ & $2.55\pm0.71$  &       ---       \\
 $200$         & $2.60\pm0.74$ & $12.0\pm1.6$   & $10$ & $11.2\pm1.8$   & $2.96\pm0.57$ & $0.64\pm0.17$ & $7.2\pm1.6$    & $0.39\pm0.31$ \\
 $400$         & $1.82\pm0.55$ & $4.11\pm0.57$  & $9$  & $12.1\pm2.1$   & $4.3\pm1.3$   & $0.47\pm0.14$ & $7.0\pm1.6$    & $0.30\pm0.23$ \\
 $600$         & $0.57\pm0.23$ & $1.70\pm0.23$  & $3$  & $4.8\pm1.2$    & $2.02\pm0.65$ & $0.12\pm0.07$ & $2.4\pm1.0$    & $0.29\pm0.19$ \\
 $125$ (shape) & $1.39\pm0.62$ & $4.80\pm0.61$  & 24   & $24.8\pm3.2$   & $4.5\pm1.3$   & $0.48\pm0.08$ & $14.0\pm2.8$   & $2.45\pm0.57$ \\
\hline
\multicolumn{9}{c}{8\TeV $\Pe\Pe$/$\mu\mu$ final state, 2-jets category, VBF tag} \\
\hline
 $120$ & $0.29\pm0.13$ & $1.23\pm0.17$  & $11$ & $6.4\pm1.9$  & $0.52\pm0.16$ & $4.1\pm1.8$   & $1.12\pm0.31$ & $0.66\pm0.38$ \\
 $125$ & $0.32\pm0.15$ & $1.91\pm0.27$  & $12$ & $6.6\pm2.0$  & $0.56\pm0.17$ & $4.2\pm1.9$   & $1.17\pm0.31$ & $0.66\pm0.38$ \\
 $130$ & $0.77\pm0.29$ & $2.99\pm0.42$  & $12$ & $6.3\pm2.0$  & $0.56\pm0.17$ & $3.8\pm1.9$   & $1.26\pm0.33$ & $0.65\pm0.38$ \\
 $160$ & $1.62\pm0.58$ & $10.2\pm1.4$   & $7$  & $5.4\pm2.9$  & $0.62\pm0.18$ & $3.4\pm2.8$   & $1.36\pm0.35$ & $0.09\pm0.08$ \\
 $200$ & $1.25\pm0.39$ & $6.61\pm0.92$  & $13$ & $10.2\pm2.5$ & $1.58\pm0.35$ & $5.2\pm2.4$   & $2.97\pm0.64$ & $0.47\pm0.31$ \\
 $400$ & $1.25\pm0.39$ & $3.03\pm0.42$  & $13$ & $8.1\pm1.6$  & $1.99\pm0.63$ & $0.10\pm0.03$ & $5.8\pm1.5$   & $0.19\pm0.21$ \\
 $600$ & $0.42\pm0.17$ & $1.43\pm0.20$  & $2$  & $3.6\pm1.0$  & $0.95\pm0.32$ & $0.06\pm0.03$ & $2.47\pm0.98$ & $0.14\pm0.12$ \\
\hline
  \end{tabular}
  }
\end{table}

\begin{figure}[htbp]
\centering
  \includegraphics[width=0.45\textwidth]{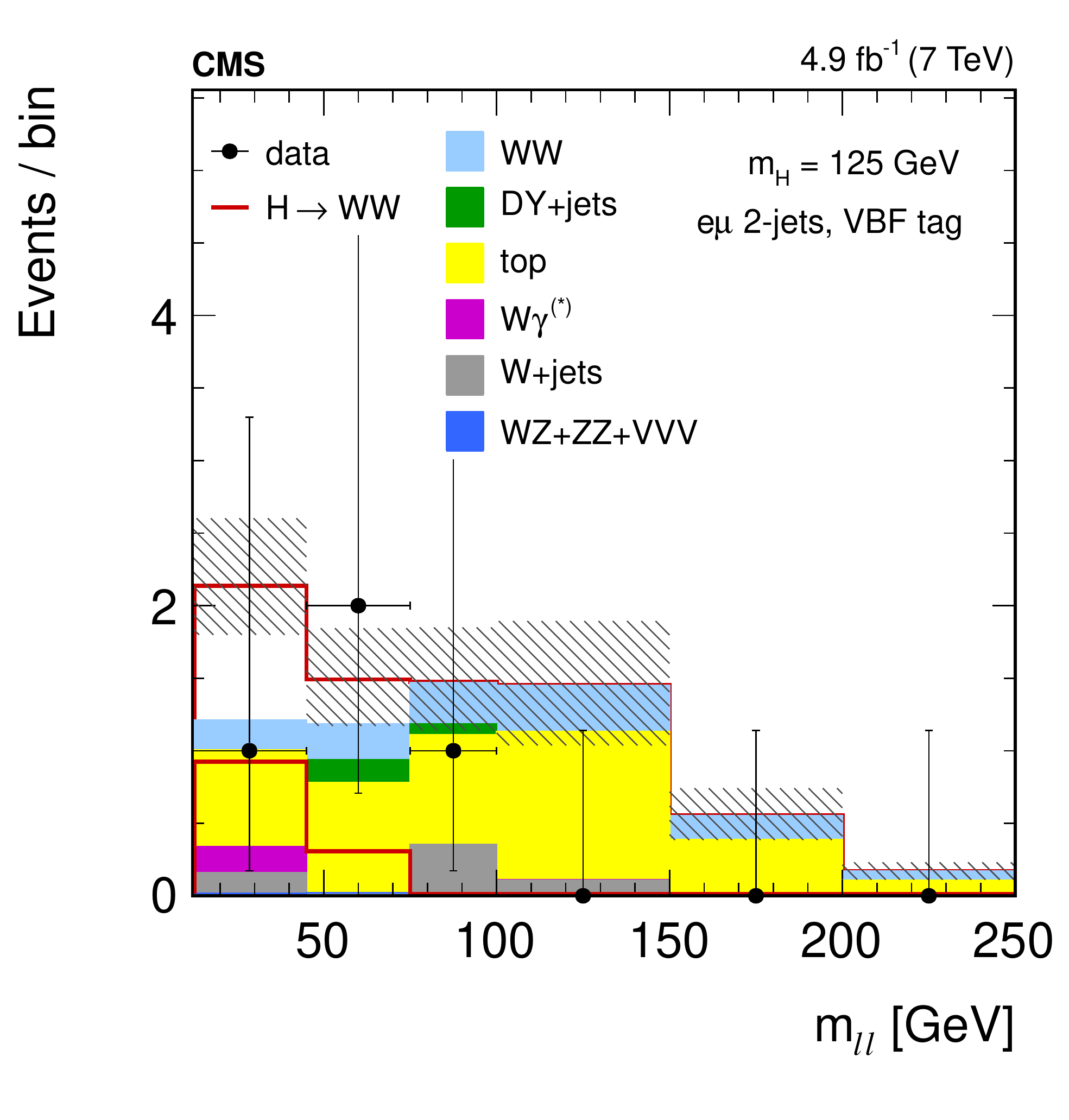}
  \includegraphics[width=0.45\textwidth]{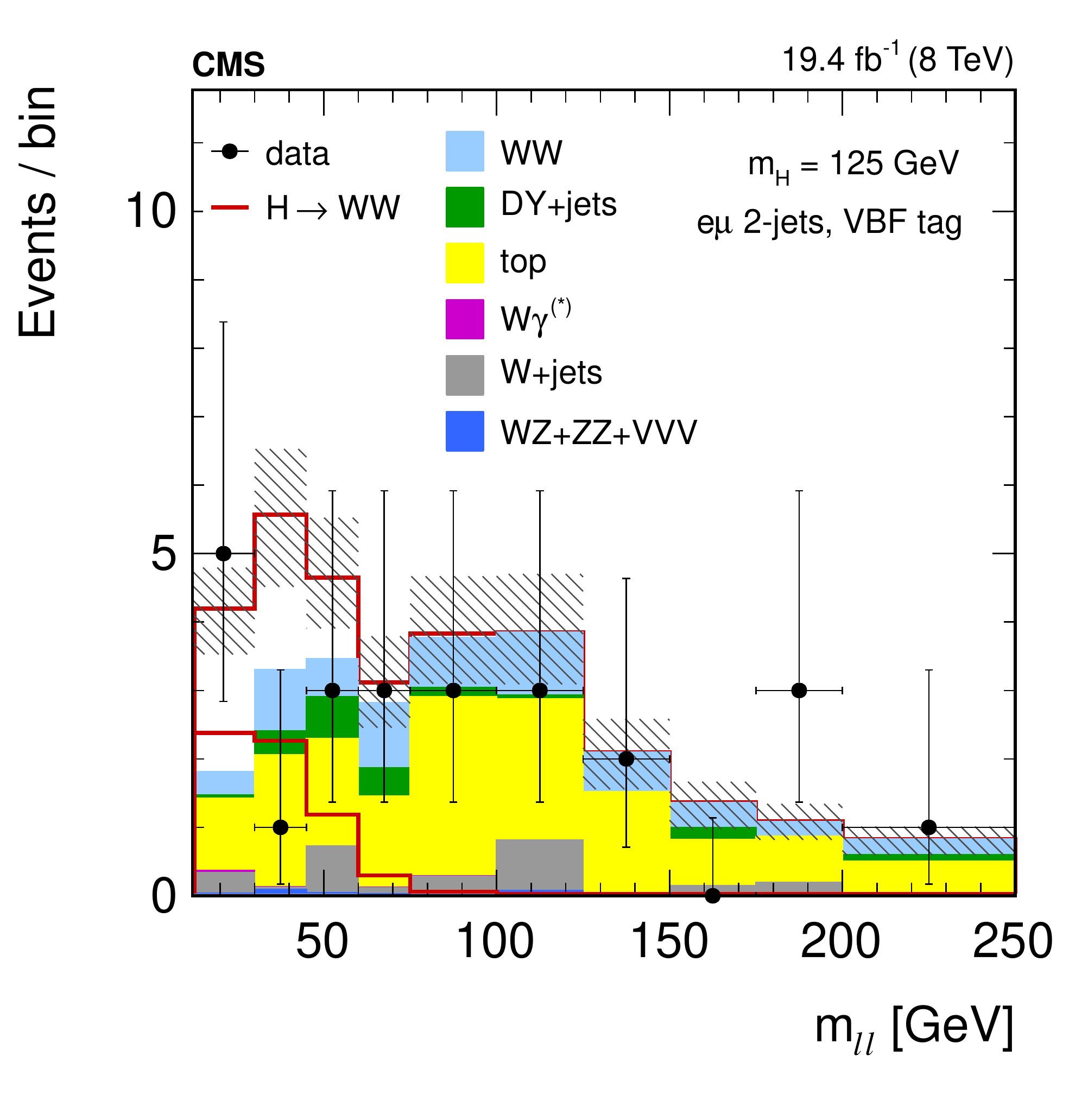}
    \caption{The $\mll$ distributions for the data and background predictions for 7
   \TeV (left) and 8\TeV (right) analyses in the different-flavor final state
    for the 2-jet category with VBF tag. Selection criteria correspond to a Higgs boson mass of
    125\GeV for the shape-based analysis. The uncertainty bands correspond to the
    sum of the statistical and systematic uncertainties in the background processes.
    The expected contribution for a Higgs boson signal with $\mHi = 125\GeV$ (red open histogram)
    is also shown, both separately and stacked with the background histograms. For
    illustration purposes the region between 250 and 600\GeV is not shown in the
    figures, but is used in the measurement.}
    \label{fig:mll_df_shapebased}
\end{figure}

\begin{figure}[htb]
  \begin{center}
 \includegraphics[width=0.48\textwidth]{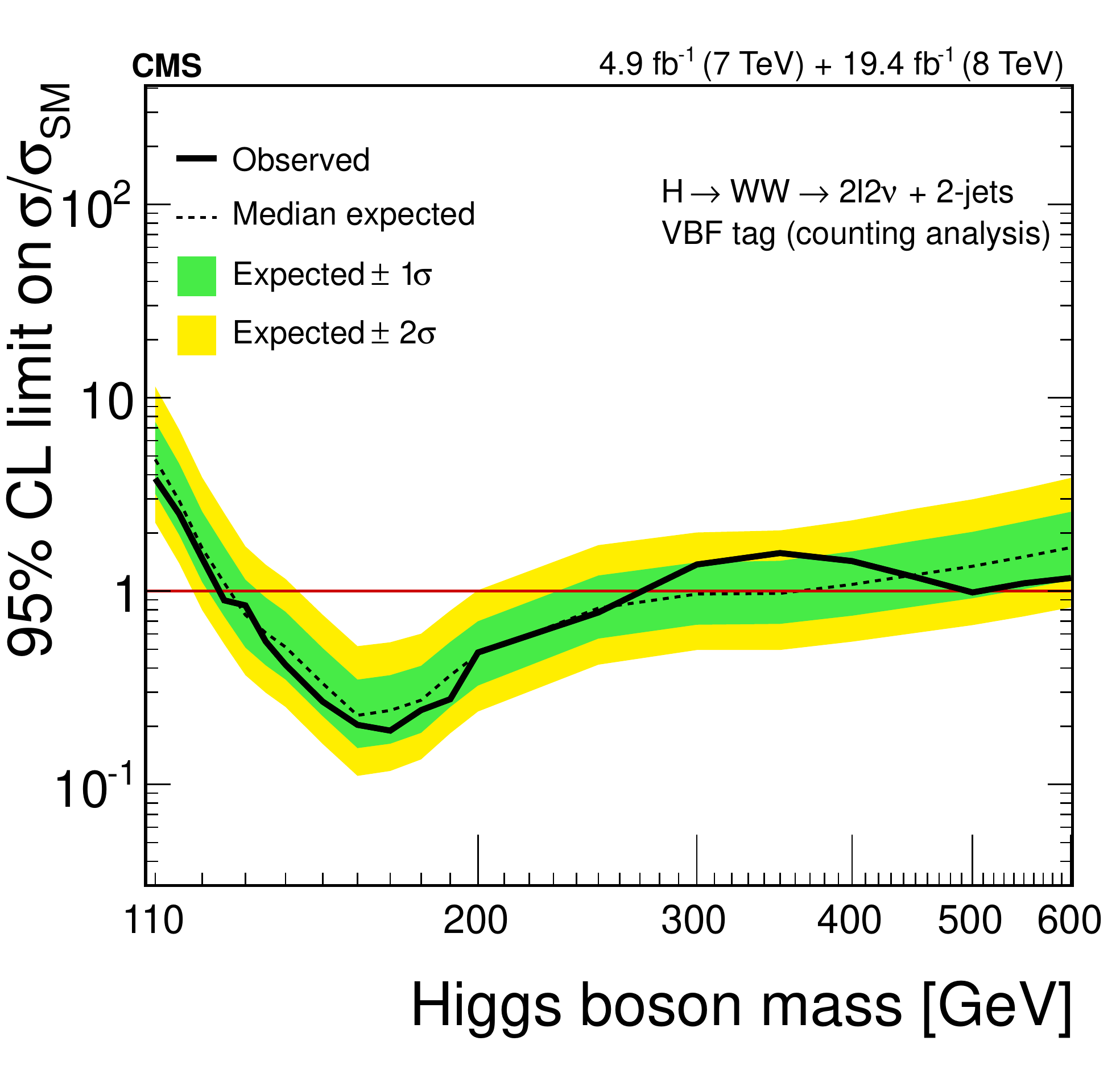}
 \includegraphics[width=0.48\textwidth]{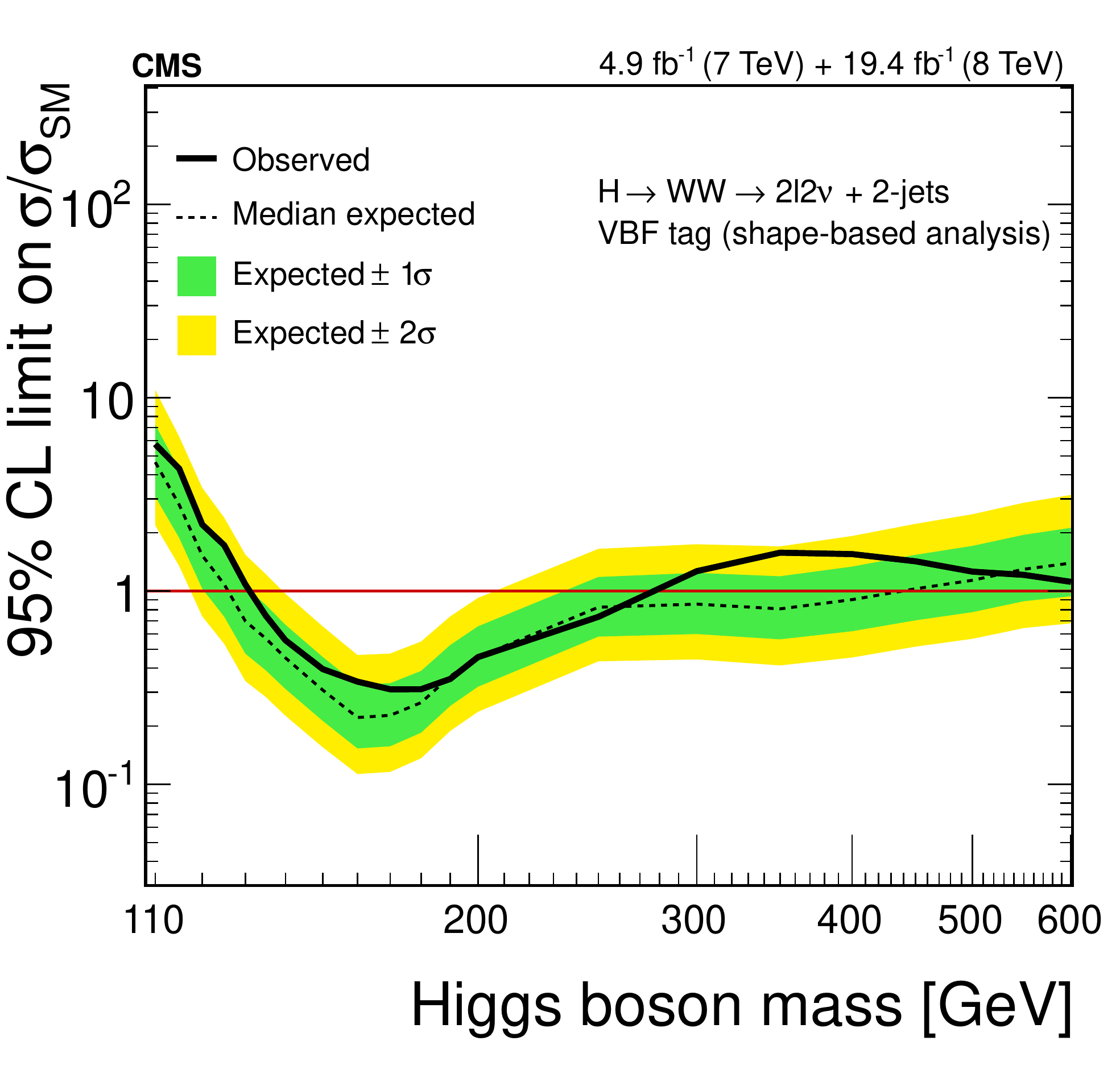}
    \caption{Expected and observed 95\% CL upper
       limits on the $\PH \to \WW$ production cross section
       relative to the SM Higgs boson expectation using the counting analysis (left),
       and shape-based template fit approach (right) in the 2-jet category with VBF tag.
The shape-based analysis results use the one-dimensional binned template fit to $\mll$ distribution
for the $\Pe\mu$ final state, combined with counting analysis inputs
for the $\Pe\Pe/\mu\mu$ final states.
       }
    \label{fig:xsLim_VBF}
  \end{center}
\end{figure}

Figure~\ref{fig:mll_df_shapebased} shows the comparison of $\mll$ between the prediction and
the data for a Higgs boson mass of 125\GeV after the selection for the shape-based analysis.
The 95\% CL observed and median expected upper limits on the production cross section of the $\PH\to\WW$ process are shown
in Fig.~\ref{fig:xsLim_VBF}.
Limits are reported for both counting and shape-based analyses.
The observed (expected) signal significance for the shape-based approach
is 1.3 (2.1)~standard deviations for a SM Higgs boson with mass of $125\GeV$. The observed
signal strength for this mass is $\sigma/\sigma_\mathrm{SM} = 0.62^{+0.58}_{-0.47}$.
A summary of the results for $\mH = 125\GeV$ is shown in Table~\ref{tab:vbf_results125}.

\begin{table}[htbp]
\centering
\topcaption{
A summary of the expected and observed
95\% CL upper limits on the $\PH \to \WW$ production cross section relative to the SM prediction,
the significances for the background-only hypothesis to account for the excess in units of standard deviations (sd),
and the best-fit $\sigma/\sigma_\mathrm{SM}$
at $\mH = 125\GeV$ in the VBF analysis.
The shape-based analysis results use the one-dimensional binned template fit to $\mll$ distribution
for the $\Pe\mu$ final state, combined with counting analysis results
for the $\Pe\Pe/\mu\mu$ final states. The difference in the observed results
between the two analyses is due to the large statistical fluctuations in
the currently available data sample.
}
\label{tab:vbf_results125}
\begin{tabular}{lccc}
\hline\hline
VBF analysis & 95\% CL limits on $\sigma/\sigma_\mathrm{SM}$ & Significance & $\sigma/\sigma_\mathrm{SM}$ \\
 $\mHi = 125\GeV$ & expected / observed & expected / observed & observed \\
\hline
Shape-based (default) & 1.1 / 1.7 & 2.1 / 1.3 sd &  $0.62^{+0.58}_{-0.47}$ \\
Counting analysis     & 1.1 / 0.9 & 2.0 /  ---     &  $-0.35^{+0.43}_{-0.45}$ \\ \hline
\end{tabular}
\end{table}

\subsection{The two-jet \texorpdfstring{$\V\PH$}{VH}~~tag}
\label{sec:vhwwqq2l2n}
The analysis of the associated production of a SM Higgs boson with a $\W$ or a $\Z$ boson
in the dilepton final state selects events with two centrally produced ($\abs{\eta}<2.5$) jets
from the decay of the associated vector boson.
The dijet invariant mass is required to be consistent with the parent boson mass,
i.e. in the range 65\GeV $< m_{jj} <$ 105\GeV,
and the pseudorapidity separation between the two jets within $\abs{\Delta\eta_{jj}} < $ 1.5.
These requirements ensure no overlap of this selection with the VBF analysis for which
a pair of forward-backward jets is required. Additionally,
for $\mHi$ $<$ ($\geq$) 180\GeV, events are required to have 60 (70)\GeV $<$ $\mth$ $<$ $\mH$.

\subsubsection{Analysis strategy}

The default analysis in the dilepton 2-jets category with $\V\PH$ tag is
performed using a counting analysis approach because this category is
statistically limited for the current datasets and the expected signal yield
is relatively small. Further $\mHi$-dependent selections are applied
to suppress top-quark processes, $\dyll$, and WW contamination
based on $\mll$ and angular separation between the two leptons ($\Delta R_{\ell\ell}$).
The lower threshold on $\mll$ is raised to $\mll > 20\GeV$ for $\mH > 135\GeV$,
and the upper bound is $\mll < 60\GeV$ for $\mH< 180\GeV$ and $\mll < 80\GeV$ for the higher Higgs boson masses.
The maximum $\Delta R_{\ell\ell}$ requirement varies between 1.5 and 2.0 from the lowest to
the highest mass hypotheses tested.

As demonstrated for other analyses previously described, the sensitivity to
the Higgs boson signal in this category is expected to gain from a fit to a
kinematic distribution, especially when the integrated luminosity increases.
The method has been tested in the $\Pe\mu$ final state using the invariant
mass of the dilepton system. The selection that is used for
the counting analysis is simplified with $\mll< 200\GeV$ and
$\Delta R_{\ell\ell}< 2.5$ for the shape-based analysis. A total of 9 bins in
$\mll$ have been defined between the lower threshold and 200\GeV.

\subsubsection{Results}

The data yields and the expected yields for the Higgs boson signal and various backgrounds
in each of the categories for the $\V\PH$ analysis
are listed in tables~\ref{tab:vhwwqq2l2n_yields_7tev} and~\ref{tab:vhwwqq2l2n_yields_8tev}.
For a Higgs boson with $\mH = 125\GeV$, a few signal events are expected
with a signal-to-background ratio of approximately 8\%.
Among the selected signal events, the contribution of the associated production mode is
$\sim$40\%, and the majority of the remaining signal originates from gluon fusion process.

\begin{table}[htbp]
\centering
\topcaption{
Signal prediction, observed number of events in data, and background estimates
at $\sqrt{s}=7\TeV$ in the $\V\PH$ counting analysis.
The combination of statistical uncertainties with experimental and theoretical systematic uncertainties is reported.}
\label{tab:vhwwqq2l2n_yields_7tev}
{
\scriptsize
\setlength{\extrarowheight}{1pt}
\begin{tabular} {ccccccccc}
  \hline\hline
\multirow{2}{*}{$\mH$ [\GeVns{}]}  & \multirow{2}{*}{$\Pg\Pg\PH$} & \multirow{2}{*}{VBF+$\V\PH$} & \multirow{2}{*}{Data} & \multirow{2}{*} {All bkg.} & \multirow{2}{*}{$\WW$} & $\WZ+\ZZ$ & \multirow{2}{*}{$\ttbar+\tw$} & \multirow{2}{*}{$\Wjets$} \\
                        &                               &                              &		       &			    &			     & $+\dyll$  &			&			     	    \\
\hline
\multicolumn{9}{c}{7\TeV $\Pe\mu$ final state, 2-jets category, $\V\PH$ tag} \\
\hline
 $120$ & $0.20\pm0.07$ & $0.22\pm0.04$ & $4$ & $6.6\pm1.3$ & $1.66\pm0.40$ & $0.67\pm0.21$ & $1.49\pm0.90$ & $1.12\pm0.52$ \\
 $125$ & $0.34\pm0.11$ & $0.42\pm0.06$ & $4$ & $7.1\pm1.4$ & $1.80\pm0.43$ & $0.67\pm0.21$ & $1.9\pm1.1$ & $1.12\pm0.52$   \\
 $130$ & $0.44\pm0.15$ & $0.42\pm0.06$ & $5$ & $7.9\pm1.7$ & $2.01\pm0.47$ & $0.68\pm0.21$ & $2.4\pm1.4$ & $1.17\pm0.53$   \\
 $160$ & $1.78\pm0.59$ & $0.95\pm0.12$ & $11$ & $9.7\pm1.5$ & $3.02\pm0.69$ & $0.73\pm0.21$ & $3.2\pm1.1$ & $1.12\pm0.47$  \\
 $200$ & $0.89\pm0.30$ & $0.48\pm0.06$ & $12$ & $10.5\pm1.5$ & $3.42\pm0.78$ & $0.55\pm0.15$ & $3.9\pm1.1$ & $0.98\pm0.41$ \\
\hline
\multicolumn{9}{c}{7\TeV $\Pe\Pe$/$\mu\mu$ final state, 2-jets category, $\V\PH$ tag} \\
\hline
 $120$ & $0.05\pm0.02$ & $0.04\pm0.01$ & $2$ & $5.8\pm1.3$ & $0.59\pm0.16$ & $1.29\pm0.33$ & $3.9\pm1.3$ & $0.06\pm0.05$ \\
 $125$ & $0.12\pm0.04$ & $0.11\pm0.03$ & $2$ & $7.5\pm1.8$ & $0.65\pm0.18$ & $1.62\pm0.44$ & $5.2\pm1.7$ & $0.06\pm0.05$ \\
 $130$ & $0.20\pm0.07$ & $0.15\pm0.03$ & $3$ & $8.9\pm2.0$ & $0.85\pm0.22$ & $2.23\pm0.67$ & $5.8\pm1.9$ & $0.04\pm0.03$ \\
 $160$ & $0.89\pm0.31$ & $0.56\pm0.08$ & $5$ & $12.2\pm2.7$ & $1.45\pm0.35$ & $2.95\pm0.83$ & $7.8\pm2.6$ & --- \\
 $190$ & $0.62\pm0.21$ & $0.33\pm0.05$ & $6$ & $13.3\pm2.8$ & $1.81\pm0.43$ & $3.39\pm0.86$ & $8.1\pm2.7$ & --- \\
\hline
\end{tabular}
}
\end{table}

\begin{table}[htbp]
\centering
  \topcaption{
  Signal prediction, observed number of events in data, and background estimates
  at $\sqrt{s}=8\TeV$ in the $\V\PH$ counting and shape-based analyses.
  The combination of statistical uncertainties with experimental and theoretical systematic uncertainties is reported.}
   \label{tab:vhwwqq2l2n_yields_8tev}
 {
\scriptsize
\setlength{\extrarowheight}{1pt}
\begin{tabular} {ccccccccc}
  \hline\hline
\multirow{2}{*}{$\mH$ [\GeVns{}]}  & \multirow{2}{*}{$\Pg\Pg\PH$} & \multirow{2}{*}{VBF+$\V\PH$} & \multirow{2}{*}{Data} & \multirow{2}{*} {All bkg.} & \multirow{2}{*}{$\WW$} & $\WZ+\ZZ$ & \multirow{2}{*}{$\ttbar+\tw$} & \multirow{2}{*}{$\Wjets$} \\
                        &                               &                              &		       &			    &			     & $+\dyll$  &			&			     	    \\
\hline
\multicolumn{9}{c}{8\TeV $\Pe\mu$ final state, 2-jets category, $\V\PH$ tag} \\
\hline
 $120$ & $1.67\pm0.57$ & $1.23\pm0.18$ & $51$ & $40.8\pm5.0$ & $8.3\pm1.9$ & $2.22\pm0.37$ & $22.1\pm4.3$ & $6.1\pm1.3$       \\
 $125$ & $2.32\pm0.79$ & $1.87\pm0.25$ & $55$ & $42.8\pm5.1$ & $9.2\pm2.1$ & $2.31\pm0.37$ & $23.0\pm4.4$ & $6.2\pm1.3$       \\
 $130$ & $2.76\pm0.94$ & $2.86\pm0.37$ & $58$ & $45.5\pm5.5$ & $9.8\pm2.3$ & $2.42\pm0.38$ & $24.5\pm4.7$ & $6.7\pm1.5$       \\
 $160$ & $11.2\pm3.7$ & $6.97\pm0.75$ & $93$ & $79.6\pm9.9$ & $15.7\pm3.5$ & $3.24\pm0.44$ & $47.8\pm8.9$ & $10.8\pm2.3$      \\
 $200$ & $8.0\pm2.6$ & $3.91\pm0.39$ & $126$ & $106\pm13$ & $23.6\pm5.3$ & $4.92\pm0.68$ & $60\pm11$ & $14.9\pm3.1$           \\
 $125$ (shape) &  $2.86\pm0.92$ & $2.30\pm0.18$ & $136$ & $129\pm15$ & $28.3\pm6.2$ & $8.2\pm1.3$ & $67\pm13$ & $23.9\pm4.8$  \\
\hline
\multicolumn{9}{c}{8\TeV $\Pe\Pe$/$\mu\mu$ final state, 2-jets category, $\V\PH$ tag} \\
\hline
 $120$ & $0.76\pm0.27$ & $0.85\pm0.14$ & $74$ & $76.6\pm7.2$ & $5.5\pm1.3$ & $48.9\pm6.1$ & $13.6\pm3.1$ & $7.6\pm1.6$ \\
 $125$ & $1.75\pm0.60$ & $0.94\pm0.16$ & $79$ & $81.0\pm7.2$ & $6.3\pm1.5$ & $51.0\pm5.9$ & $14.4\pm3.2$ & $8.3\pm1.8$ \\
 $130$ & $2.13\pm0.74$ & $1.69\pm0.25$ & $83$ & $88.0\pm7.5$ & $7.1\pm1.7$ & $55.8\pm6.2$ & $15.6\pm3.5$ & $8.6\pm1.8$ \\
 $160$ & $8.9\pm3.0$ & $5.06\pm0.58$ & $96$ & $100\pm11$ & $12.7\pm2.8$ & $42.8\pm8.3$ & $33.5\pm6.4$ & $10.5\pm2.2$   \\
 $200$ & $4.4\pm1.5$ & $2.35\pm0.25$ & $131$ & $134\pm13$ & $18.8\pm4.2$ & $52.0\pm7.9$ & $49.6\pm9.5$ & $12.0\pm2.5$  \\
\hline
  \end{tabular}
  }
\end{table}

\begin{figure}[htbp]
\begin{center}
{\includegraphics[width=0.45\textwidth]{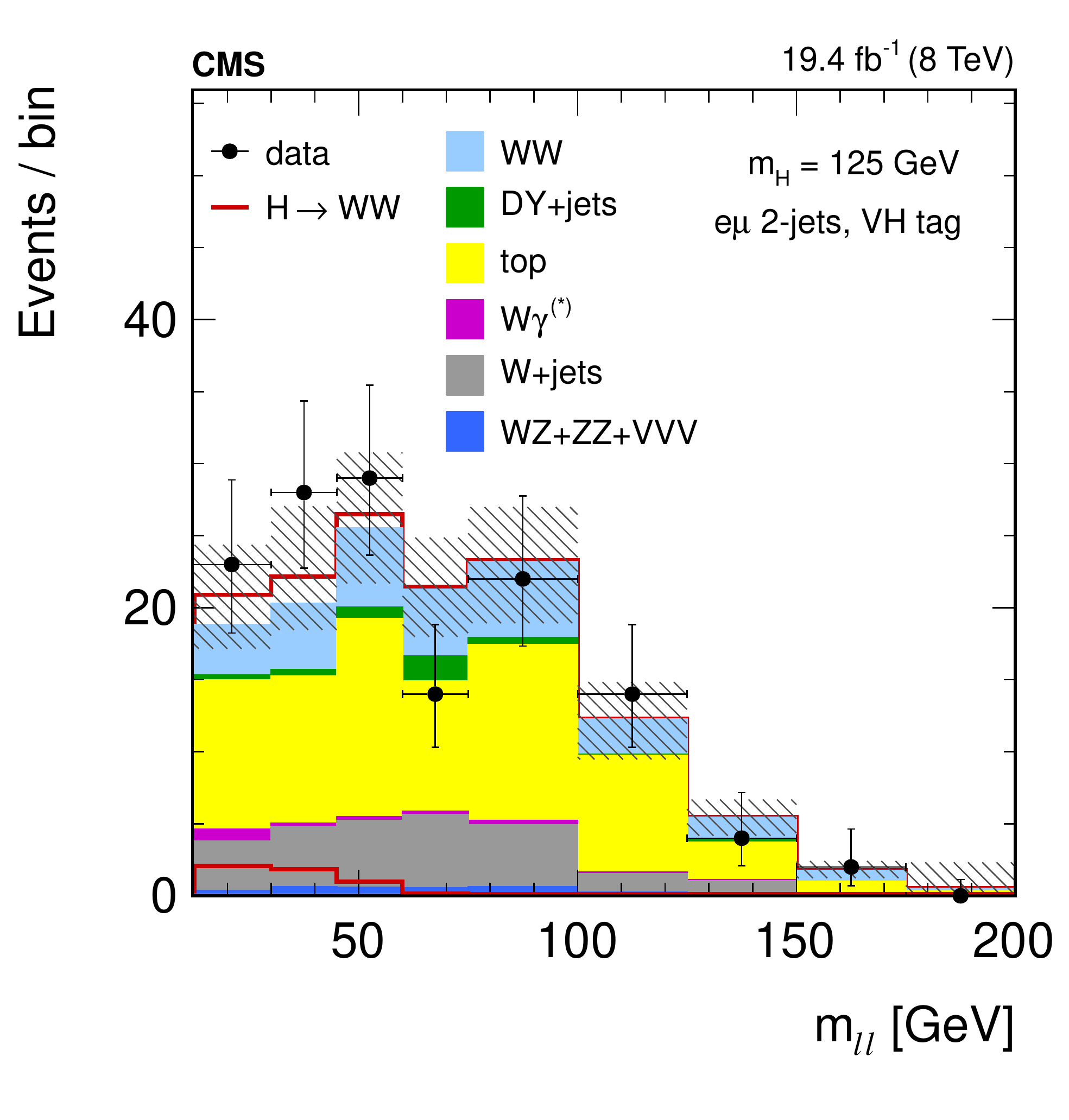}}
\caption{The $\mll$ distribution for $\mH = 125\GeV$ used as input to the template fit
in the $\Pe\mu$ final state for the $\V\PH$ analysis after the corresponding selection at $\sqrt{s}= 8\TeV$.}
\label{fig:histo_mll_aftershapecuts_osof_8tev}
\end{center}
\end{figure}

The $\mll$ distribution at $\sqrt{s}= 8\TeV$ used as an input to the template fit
in the $\Pe\mu$ final state after the corresponding selection for $\mH = 125\GeV$
is shown in Fig.~\ref{fig:histo_mll_aftershapecuts_osof_8tev}.
The shape-based analysis has been tested and compared with the default counting analysis.
No shape-based
analysis was developed at $\sqrt{s}= 7\TeV$ because of very limited statistics.

\begin{figure}[htb]
  \begin{center}
 \includegraphics[width=0.48\textwidth]{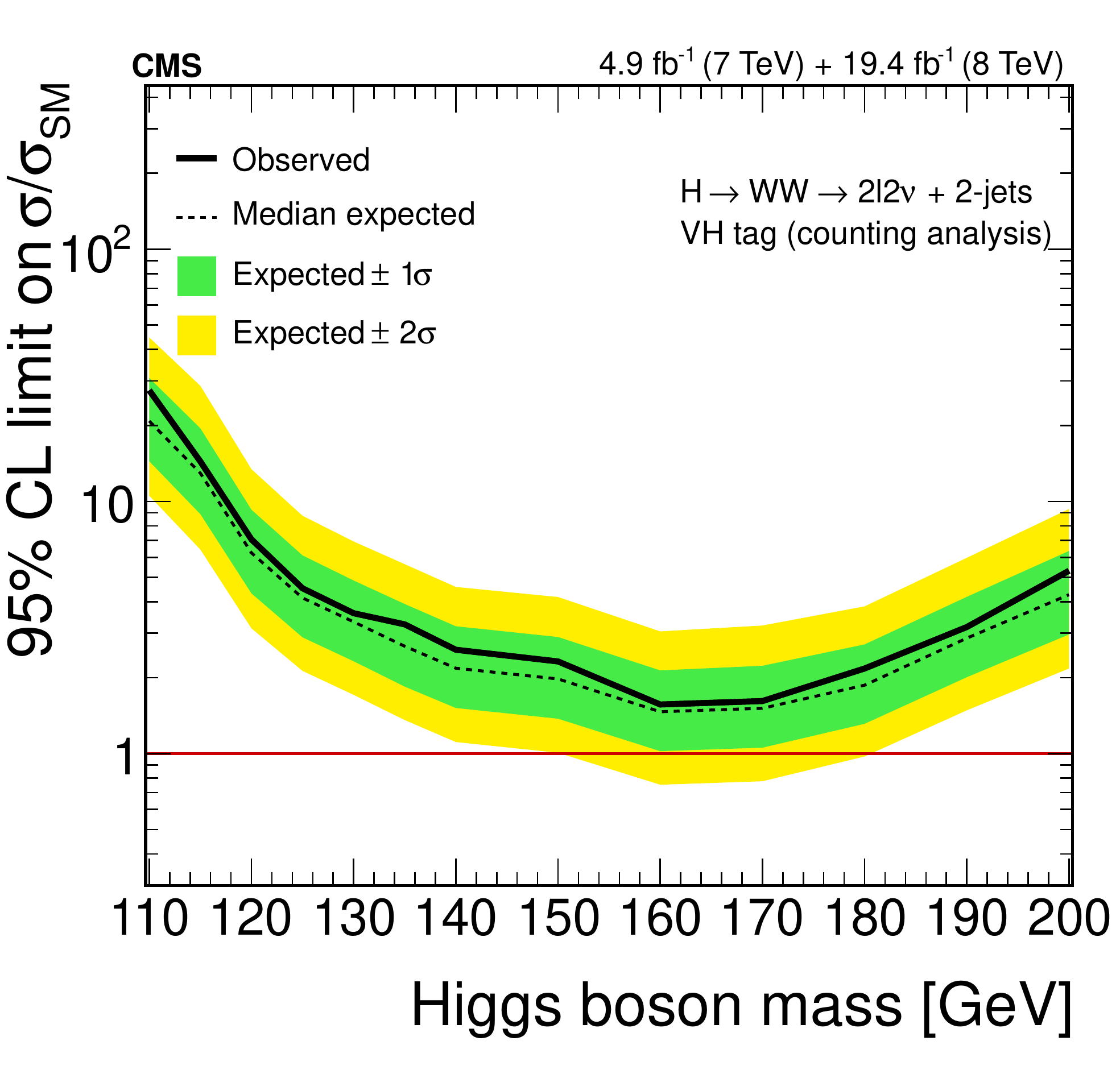}
 \includegraphics[width=0.48\textwidth]{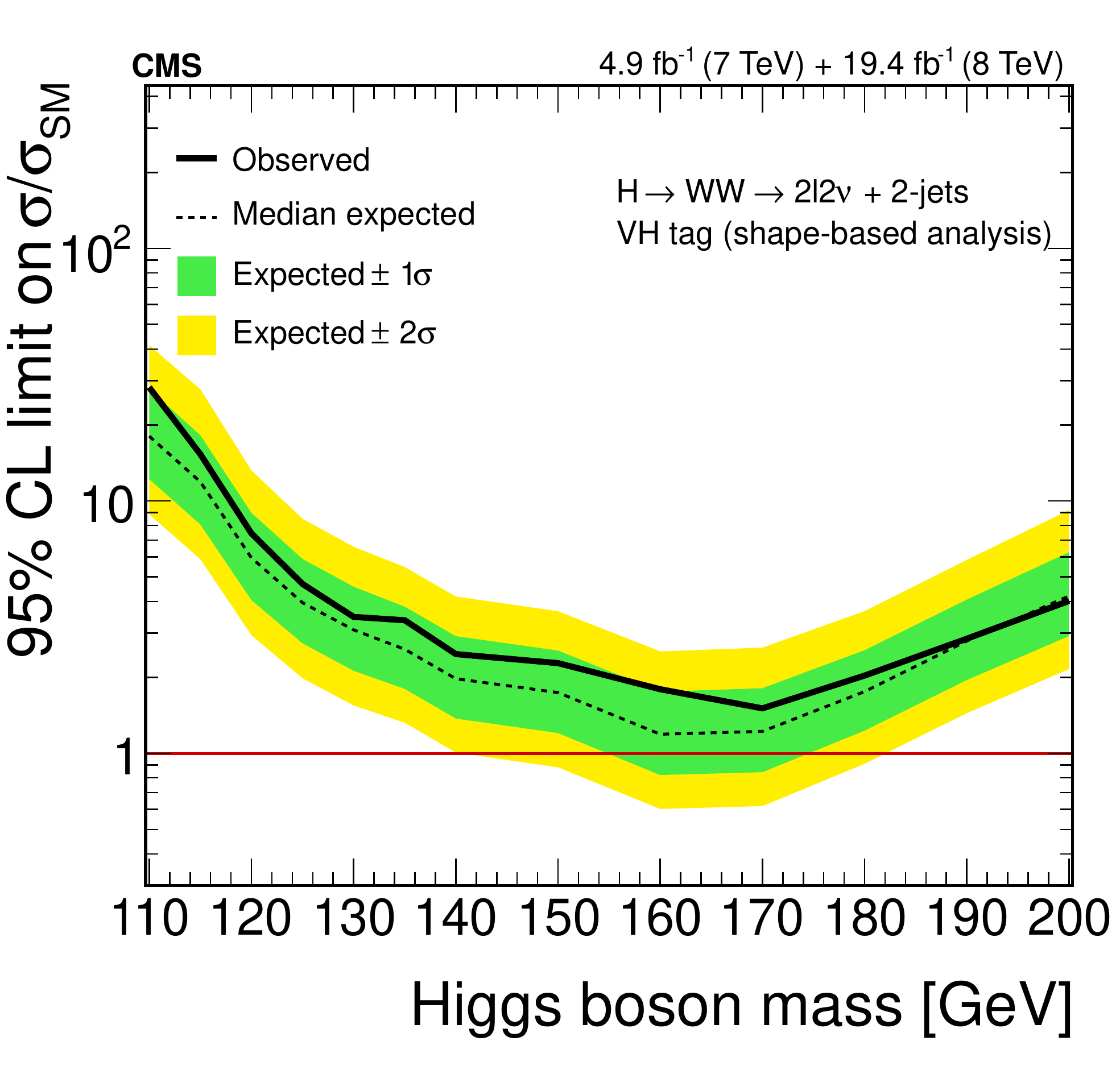}
    \caption{Expected and observed 95\% CL upper
       limits on the $\PH \to \WW$ production cross section
       relative to the SM Higgs boson expectation using the counting analysis (left),
       and the shape-based template fit approach (right) in the $\V\PH$ category.
The shape-based analysis results use the one-dimensional binned template fit to the $\mll$ distribution
for the $\Pe\mu$ final state, combined with counting analysis results
for the $\Pe\Pe/\mu\mu$ final states.
       }
    \label{fig:xsLim_vhwwqq2l2n}
  \end{center}
\end{figure}

The 95\% CL observed and median expected upper limits on the production cross section of the $\PH\to\WW$ process are shown
in Fig.~\ref{fig:xsLim_vhwwqq2l2n}.
Limits are reported for both counting and shape-based analyses. For the latter,
the different-flavor final states are combined with the same-flavor counting analysis.

The expected and observed results for the $\V\PH$ analysis
are summarized in Table~\ref{tab:vhwwqq2l2n_results125}.
The upper limit on the $\PH \to \WW$ production cross section using this category
is about five times the SM expectation, and the observed (expected) significance of the signal
is 0.2 (0.6) standard deviations.

\begin{table}[htbp]
\centering
\topcaption{
A summary of the expected and observed
95\% CL upper limits on the $\PH \to \WW$ production cross section relative to the SM prediction,
the significances for the background-only hypothesis to account for the excess
in units of standard deviations (sd),
and the best-fit $\sigma/\sigma_\mathrm{SM}$
at $\mH = 125\GeV$ for the $\V\PH$ analyses.
The shape-based analysis results use the one-dimensional binned template fit to the $\mll$ distribution
for the $\Pe\mu$ final state, combined with counting analysis results
for the $\Pe\Pe/\mu\mu$ final states.
}
\label{tab:vhwwqq2l2n_results125}
\begin{tabular}{lccc}
\hline\hline
$\V\PH$ analysis & 95\% CL limits on $\sigma/\sigma_\mathrm{SM}$ & Significance & $\sigma/\sigma_\mathrm{SM}$ \\
 $\mHi = 125\GeV$ & expected / observed & expected / observed & observed \\
\hline
Counting analysis (default) & 4.1 / 4.5 & 0.6 / 0.2 sd & $0.40^{+2.03}_{-1.93}$ \\
Shape-based                 & 4.0 / 4.7 & 0.6 / 0.4 sd & $0.73^{+2.04}_{-1.85}$ \\ \hline
\end{tabular}
\end{table}

\section{Final states with three charged leptons}\label{sec:sel_3l}
Events with exactly three identified charged leptons also provide sensitivity to the
$\V\PH$ production mode. Three charged-lepton candidates with total charge
equal to $\pm$1 are required, with $\pt >$20\GeV for the leading lepton and
$\pt >$10\GeV for the other leptons. Events with any further identified lepton
passing the selection criteria defined in Section~\ref{sec:objects} and $\pt >$10
\GeV are rejected. Two analyses have been developed for
this topology. The first analysis selects triboson ($\VVV$, $\V$ = $\W/\Z$) candidates in which all bosons
decay leptonically, yielding  an experimental signature of three isolated
high-$\pt$ leptons, moderate $\met$, and little hadronic activity. The second
analysis requires one opposite-sign same-flavor lepton pair compatible with
a $\Z$ boson decay and two jets compatible with a hadronic $\W$-boson decay,
making the analysis sensitive to $\Z\PH$ production.
A brief summary of the analyses in the trilepton categories is shown in Table~\ref{tab:summary_3l}.

\begin{table}[htbp]
\centering
\topcaption{A summary of the selection requirements and analysis approach, as well as the most important
background processes in the trilepton categories.
The same-flavor final states make use of a counting analysis approach in all categories.}
\label{tab:summary_3l}
 {\small
\begin{tabular} {lcc}
  \hline\hline
 & {$\W\PH \to 3\ell 3\nu$}~~category & {$\Z\PH \to 3\ell \nu \text{+ 2 jets}$}~~category \\
\hline
Number of jets & $=$0 & $\geq$2 \\
Default analysis      & \multicolumn{2}{c}{binned shape-based} \\
Alternative analysis  & \multicolumn{2}{c}{counting} \\
Main backgrounds      & \multicolumn{2}{c}{$\W\Z$, non-prompt leptons} \\
\hline
  \end{tabular}
  }
\end{table}

\subsection{The \texorpdfstring{$\W\PH \to 3\ell 3\nu$}{WH to 3 leptons 3 neutrinos}~~category}
\label{sec:wh3l}
\subsubsection{Analysis strategy}
Signal candidates in this category are split into two final states to improve the
sensitivity: all events that have lepton pairs with the opposite charge and the
same flavor are classified as OSSF final state, all others have
lepton pairs with the same charge and the same flavor, and are classified as SSSF final state. While 1/4 of
the events are selected in the SSSF final state, the expected background is rather
small since physics processes leading to this final state have small cross sections.

To remove the remaining $\Zjets$ background events, the minimum of full $\met$ and
track $\met$ (min-MET) is required to be above 40 (30)\GeV in the
OSSF (SSSF) final state. Since the $\met$ resolution is degraded by pileup,
the minimum of the two variables increases the background rejection for a
given signal efficiency. For this analysis, $\pmet$ is not used
since having three leptons in the event degrades the performance of such variable.
To further suppress the top-quark background, events
are rejected if there is at least one jet with $\pt >$ 40\GeV, or if the
event is top-tagged as described in Section~\ref{sec:objects}.
The $\W\Z \to 3\ell\nu$ background is largely reduced by requiring that all
the OSSF lepton pairs have a dilepton mass at least 25\GeV away from the $\Z$ mass peak.
To reject the $\V\gamma^{(*)}$ background, the dilepton mass of all
opposite-sign lepton pairs is required to be greater than 12\GeV.
In addition to all the above requirements, the signal region is defined by requiring
that the smallest dilepton mass $\mll$ is less than 100\GeV,
and that the smallest distance between the opposite-sign leptons
$\Delta R_{\ell^+\ell^-}$ is less than 2.

Finally, a shape-based analysis is carried out as the main analysis because of its
superior sensitivity with respect to the counting analysis. In this analysis
the requirement on $\Delta R_{\ell^+\ell^-}$ is not applied, and instead that variable is used
as the discriminant. Tests have shown this variable to provide the best
discrimination between signal and background events, both in terms of expected
limits and of expected significance.

\subsubsection{Background estimation}
There are five main background processes in this category: $\W\Z \to 3\ell\nu$, $\Z\Z \to 4\ell$,
tribosons, $\Z\gamma$, and processes with non-prompt leptons. The first four
contributions are estimated from simulation, with corrections from data control
samples, while the non-prompt lepton background is solely evaluated from data.

The $\W\Z \to 3\ell\nu$ decay is the main background in the analysis.
The overall normalization is taken from data using
trilepton events, where one of the same-flavor opposite-sign lepton pairs
has a mass less than 15\GeV away from the $\Z$ boson mass peak. All other
selection requirements are applied, except the $\Delta R_{\ell^+\ell^-}$ and the
upper $\mll$ requirements. The sample is completely dominated by this process, and for
$\mHi = 125\GeV$ less than one signal event is expected in that region.
The uncertainty in the normalization, which mainly arises from the statistics
of the control sample, is 5--10\%.

The $ZZ \to 4\ell$ background is reduced by the $\met$ requirement and
the veto of events containing a fourth lepton. The prediction from
the simulation for this process is used without any further correction. The
triboson background processes are also estimated with simulation.

The $\Z\gamma$ background is normalized in data using events in which
the trilepton mass is compatible with the $\Z$ mass. The number of
selected events for this background after the $\met$ requirements
is very small. A normalization uncertainty of 30\% is assigned from
studies in events with $m_{3\ell}$ compatible with $m_{\Z}$.

The non-prompt lepton backgrounds are estimated as explained in
Section~\ref{sec:sel_2l}, with the only difference that the
contributions are derived from a control
sample in data in which two leptons pass the
standard criteria and the third one does not, but satisfies a
relaxed set of requirements (loose selection), resulting in a ``two-pass and one-fail" sample.
The efficiency for a jet that satisfies the loose lepton selection to pass the tight
selection, $\epsilon_\text{pass}$, is determined using an independent dataset
dominated by non-prompt leptons from multijet events.
Finally, a scale factor of $0.78 \pm 0.31$ is obtained by comparing the prediction from
this method and a trilepton data sample in which a
b-tagged jet is required. This last sample is heavily enriched in top-quark processes
and allows to calibrate the background prediction. The
systematic uncertainty from the efficiency determination
dominates the overall uncertainty of this method, which is estimated to be 40\%.

A summary of the estimation of the background processes in the $\W\PH \to 3\ell 3\nu$
category in cases where data events are used to estimate either the
normalization or the shape of the discriminant variables is shown in
Table~\ref{tab:summary_3l_bkg}.

\begin{table}[htbp]
\centering
\topcaption{Summary of the estimation of the background processes in the $\W\PH \to 3\ell 3\nu$ category
in cases where data events are used to estimate either the normalization or the shape of the
discriminant variables. A brief description of the control/template sample is given.}
\label{tab:summary_3l_bkg}
{\small
\begin{tabular} {lccl}
  \hline\hline
 Process   & Normalization & Shape & Control/template sample \\
\hline
$\W\Z$     & data & simulation & events with $\mll$ close to $m_{\Z}$  \\
$\Z\gamma$   & data & simulation & events with $m_{3\Lep}$ close to $m_{\Z}$ \\
Non-prompt leptons   & data & data & events with loosely identified leptons \\
\hline
  \end{tabular}
  }
\end{table}

\subsubsection{Results}
\begin{figure}[htbp]
\begin{center}
\includegraphics[width=0.45\linewidth]{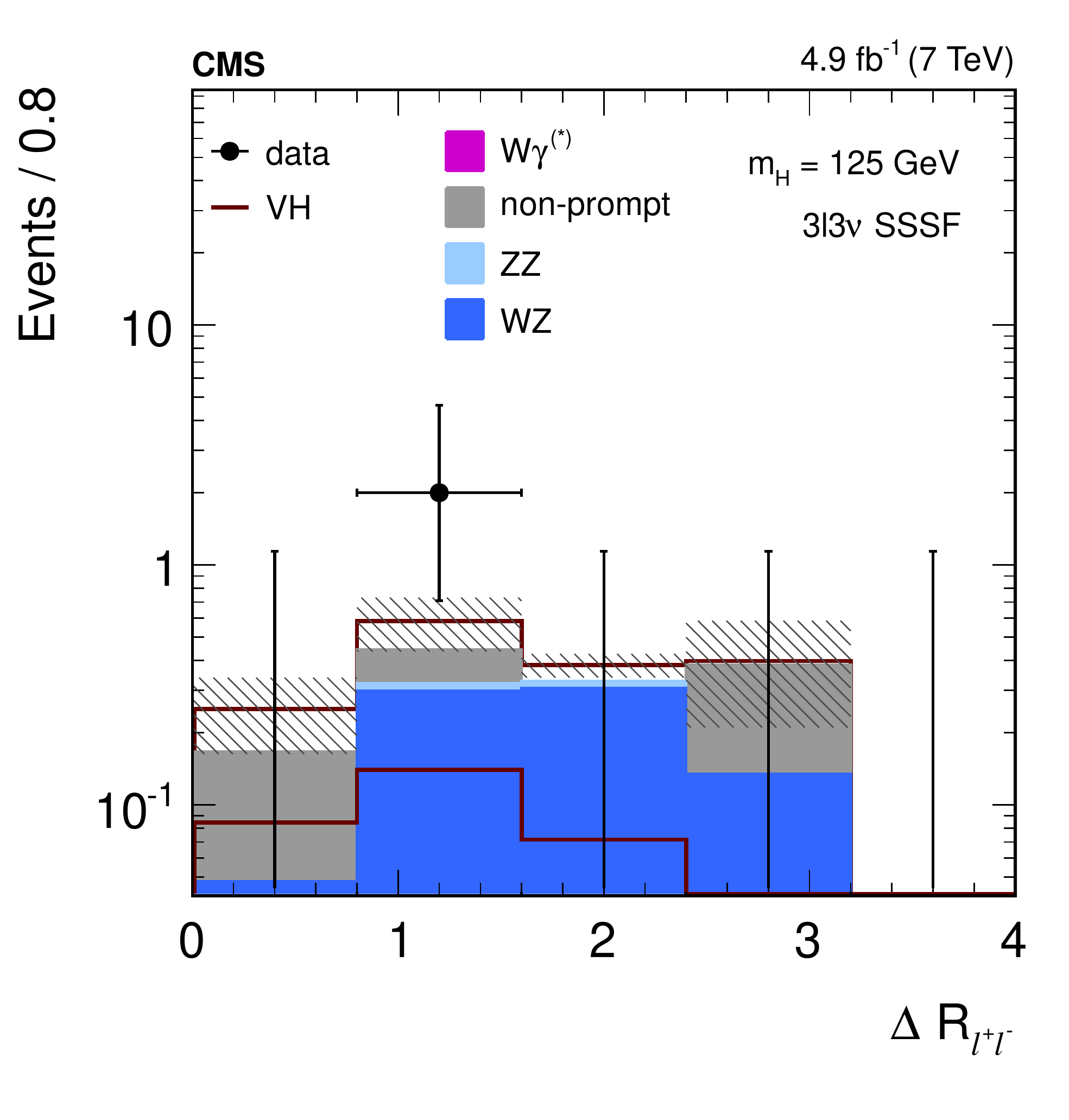}
\includegraphics[width=0.45\linewidth]{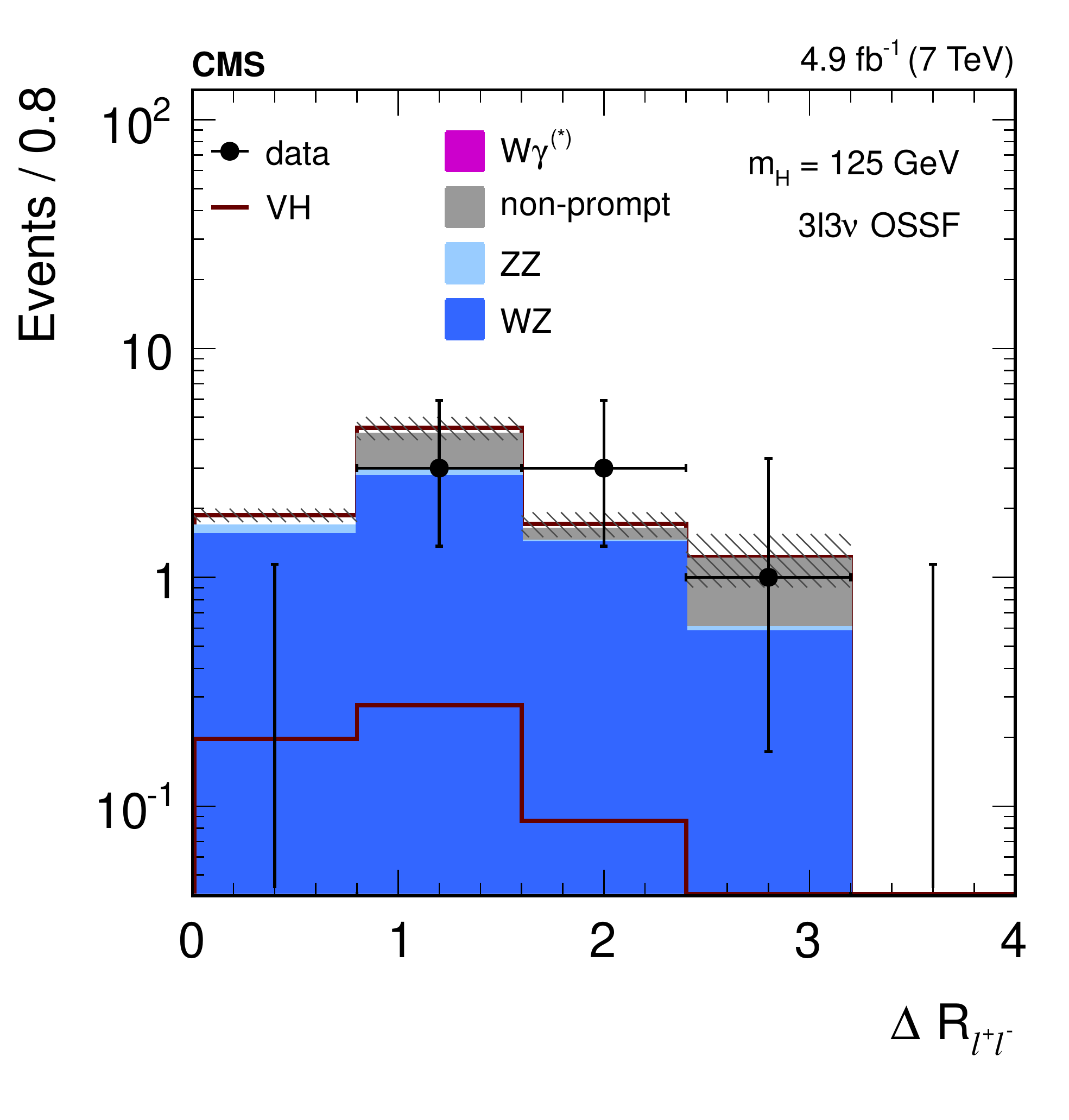}
\includegraphics[width=0.45\linewidth]{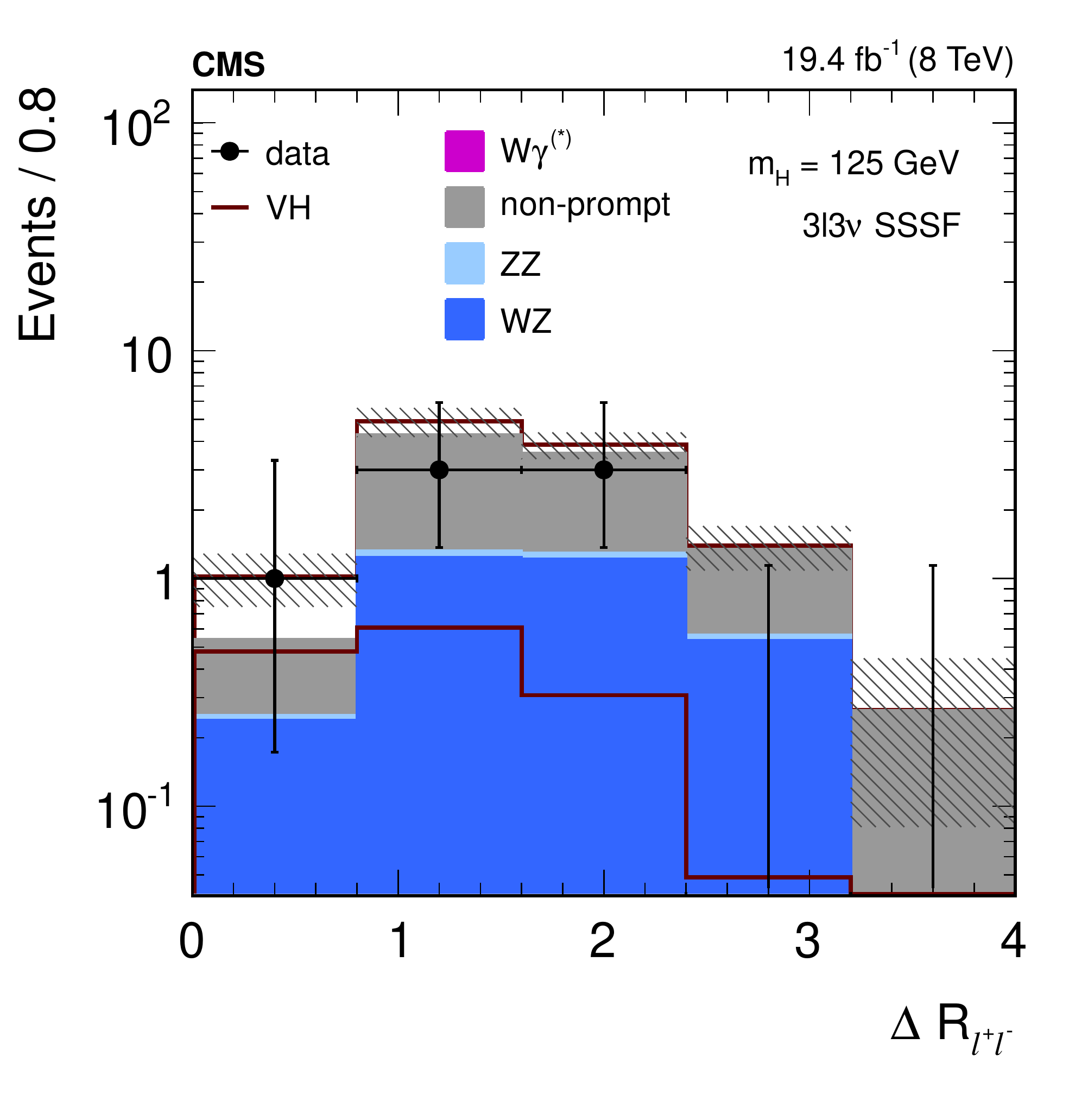}
\includegraphics[width=0.45\linewidth]{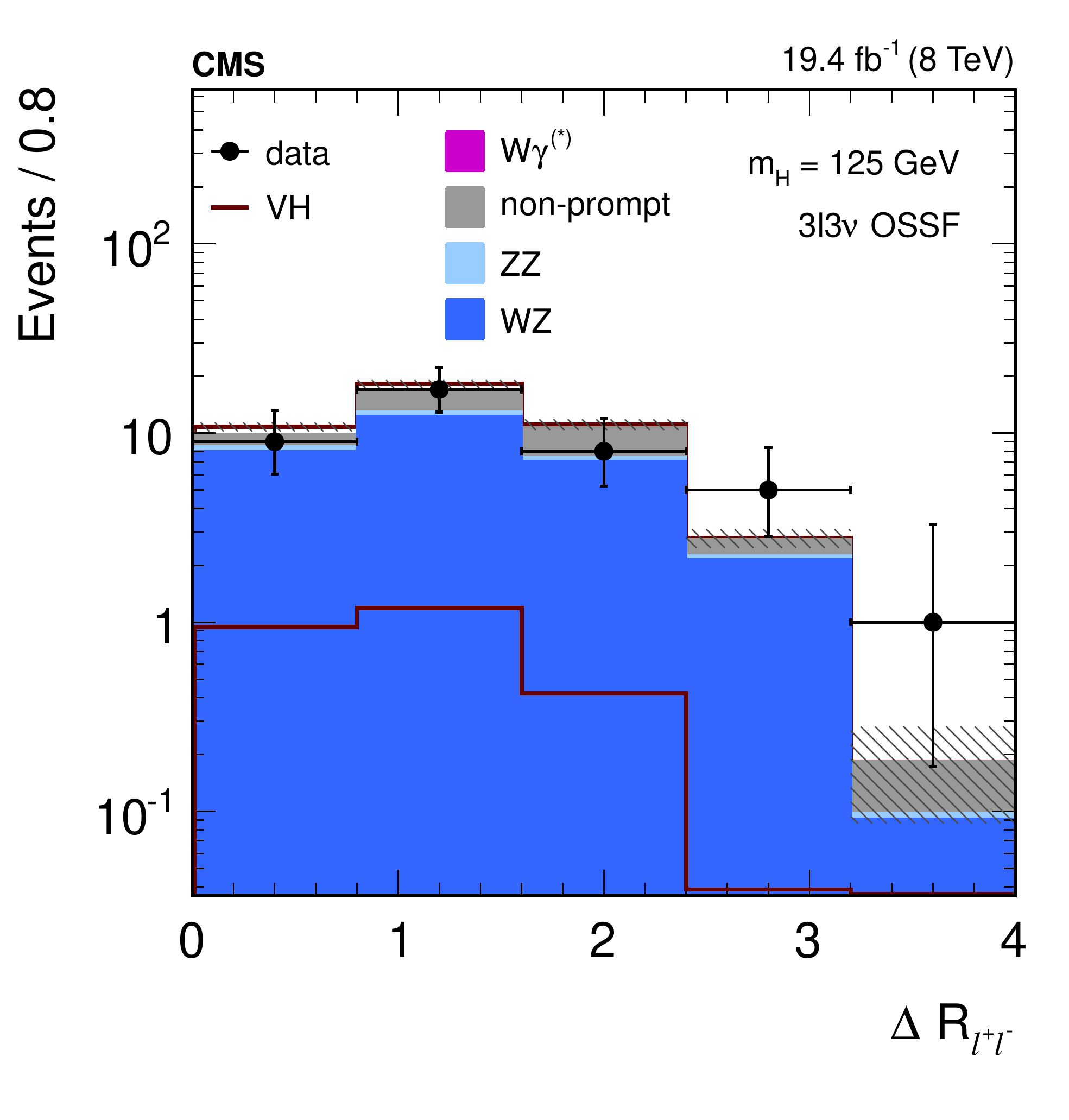}
\caption{The $\Delta R_{\ell^+\ell^-}$ distribution, after applying all other requirements for the
$\W\PH \to 3\ell 3\nu$ analysis, in the SSSF final state at
7\TeV (top left), the OSSF final state at 7\TeV (top right), the SSSF final state at
8\TeV (bottom left), and the OSSF final state at 8\TeV (bottom right). The legend entry
labeled as ``non-prompt" is the combination of the backgrounds from $\Zjets$ and top-quark decays.}
\label{fig:histo_drmin_afterallothercuts}
\end{center}
\end{figure}

The observed number of data events and the expected number of signal
and background events at different stages of the analysis are shown in
Table~\ref{tab:whselection_all}. The signal contribution from $\W\PH$ production
with $\PH\to\tau\tau$ decay to the total number of expected
Higgs boson events decreases from 55\% to 10\% in the mass range 110--130\GeV,
and it is about 15\% for $\mHi=125\GeV$.
The $\Delta R_{\ell^+\ell^-}$ distributions are shown in
Fig.~\ref{fig:histo_drmin_afterallothercuts}.

\begin{table}[htbp]
\centering
\topcaption{
Signal prediction for the SM Higgs boson with $\mH = 125\GeV$, number of observed events in data, and
estimated background at different stages of the $\W\PH \to 3\ell 3\nu$ analysis. Only statistical
uncertainties in the yields are reported in the first four rows of the selection stages,
while all systematic uncertainties are considered in the last row.
The column labeled as ``non-prompt'' is the combination of the backgrounds from $\Zjets$ and top-quark decays.
$\Z\Z$, $\V\gamma^{(*)}$, and triboson processes are not reported separately since
since they constitute a small fraction of the total background.
The 3-lepton selection stage also includes the $\mll~>~12\GeV$ requirement.}
 \label{tab:whselection_all}
\resizebox{\textwidth}{!}
{
  \begin{tabular} {ccccccc}
\hline\hline
   \multirow{2}{*}{Selection stage}    & $\W\PH$          & $\W\PH$     & \multirow{2}{*}{Data} & \multirow{2}{*}{All bkg.} & \multirow{2}{*}{$\WZ$} & \multirow{2}{*}{Non-prompt} \\
                             & $\PH\to\tau\tau$ & $\PH\to\WW$ & & & & \\
  \hline
 & \multicolumn{6}{c}{7\TeV SSSF final state, $\W\PH \to 3\ell 3\nu$~category} \\
\hline
 3 lepton requirement     	     & 0.16 $\pm$  0.02 &    0.42 $\pm$  0.01 &   12 &   12.2  $\pm$  1.3  & 	1.95 $\pm$  0.10 &    9.9  $\pm$  1.3  \\
 Min-MET $>$ 30\GeV  	     & 0.09 $\pm$  0.01 &    0.31 $\pm$  0.01 &    9 &    8.5  $\pm$  1.1  & 	1.29 $\pm$  0.08 &    7.1  $\pm$  1.1  \\
 $Z$ removal			     & 0.09 $\pm$  0.01 &    0.31 $\pm$  0.01 &    9 &    8.5  $\pm$  1.1  & 	1.29 $\pm$  0.08 &    7.1  $\pm$  1.1  \\
 Top-quark veto			     & 0.07 $\pm$  0.01 &    0.24 $\pm$  0.01 &    2 &    1.90 $\pm$  0.44 & 	0.82 $\pm$  0.06 &    1.04 $\pm$  0.43 \\
 $\Delta R_{\ell^+\ell^-}$ \& $\mll$ & 0.04 $\pm$  0.01 &    0.22 $\pm$  0.03 &    2 &    0.79 $\pm$  0.20 & 	0.53 $\pm$  0.07 &    0.23 $\pm$  0.19 \\
\hline
 & \multicolumn{6}{c}{7\TeV OSSF final state, $\W\PH \to 3\ell 3\nu$~category} \\
\hline
 3 lepton requirement     	     & 0.52 $\pm$  0.03 &    1.32 $\pm$  0.01 &  869 &  863    $\pm$ 12    &  475.2  $\pm$  1.5  &  233.9  $\pm$  6.8  \\
 Min-MET $>$ 40\GeV  	     & 0.23 $\pm$  0.02 &    0.81 $\pm$  0.01 &  234 &  238.5  $\pm$  2.5  &  207.3  $\pm$  1.0  &   22.8  $\pm$  2.3  \\
 $Z$ removal			     & 0.14 $\pm$  0.02 &    0.61 $\pm$  0.01 &   25 &   25.7  $\pm$  1.5  &   13.62 $\pm$  0.26 &   11.4  $\pm$  1.5 \\
 Top-quark veto			     & 0.10 $\pm$  0.01 &    0.48 $\pm$  0.01 &    8 &    9.76 $\pm$  0.66 & 	7.34 $\pm$  0.19 &    1.96 $\pm$  0.63 \\
 $\Delta R_{\ell^+\ell^-}$ \& $\mll$ & 0.07 $\pm$  0.01 &    0.45 $\pm$  0.05 &    5 &    6.51 $\pm$  0.84 & 	4.96 $\pm$  0.48 &    1.18 $\pm$  0.69 \\
\hline
 & \multicolumn{6}{c}{8\TeV SSSF final state, $\W\PH \to 3\ell 3\nu$~category} \\
\hline
 3 lepton requirement     	     & 0.72 $\pm$  0.08 &    1.64 $\pm$  0.21 &   71 &   83.7  $\pm$  3.0  & 	7.88 $\pm$  0.30 &   66.8  $\pm$  2.9  \\
 Min-MET $>$ 30\GeV  	     & 0.41 $\pm$  0.06 &    1.21 $\pm$  0.18 &   43 &   60.2  $\pm$  2.5  & 	5.16 $\pm$  0.24 &   48.4  $\pm$  2.5  \\
 $Z$ removal			     & 0.41 $\pm$  0.06 &    1.21 $\pm$  0.18 &   43 &   60.2  $\pm$  2.5  & 	5.16 $\pm$  0.24 &   48.4  $\pm$  2.5  \\
 Top-quark veto			     & 0.29 $\pm$  0.05 &    1.02 $\pm$  0.17 &    7 &   10.41 $\pm$  0.97 & 	2.84 $\pm$  0.18 &    6.60 $\pm$  0.95 \\
 $\Delta R_{\ell^+\ell^-}$ \& $\mll$ & 0.23 $\pm$  0.05 &    1.00 $\pm$  0.20 &    6 &    6.9  $\pm$  2.0  & 	1.71 $\pm$  0.16 &    4.6  $\pm$  2.0  \\
\hline
 & \multicolumn{6}{c}{8\TeV OSSF final state, $\W\PH \to 3\ell 3\nu$~category} \\
\hline
 3 lepton requirement     	     & 1.95 $\pm$  0.12 &    6.08 $\pm$  0.41 & 4340 & 4224    $\pm$ 21    & 2042.7  $\pm$  4.8  & 1369.0  $\pm$ 13    \\
 Min-MET $>$ 40\GeV  	     & 0.91 $\pm$  0.09 &    3.47 $\pm$  0.30 & 1137 & 1140.9  $\pm$  6.0  &  900.0  $\pm$  3.2  &  149.9  $\pm$  4.9  \\
 $Z$ removal			     & 0.56 $\pm$  0.07 &    2.69 $\pm$  0.27 &  153 &  155.3  $\pm$  3.4  &   59.1  $\pm$  0.8  &   79.9  $\pm$  3.3  \\
 Top-quark veto			     & 0.35 $\pm$  0.05 &    2.14 $\pm$  0.23 &   45 &   47.7  $\pm$  1.3  &   34.9  $\pm$  0.6  &    9.6  $\pm$  1.2  \\
 $\Delta R_{\ell^+\ell^-}$ \& $\mll$ & 0.30 $\pm$  0.06 &    2.10 $\pm$  0.34 &   33 &   33.2  $\pm$  3.4  &   24.0  $\pm$  1.4  &    7.2  $\pm$  3.1  \\
\hline
  \end{tabular}
}
\end{table}

\begin{figure}[htb]
\begin{center}
   \includegraphics[width=0.48\textwidth]{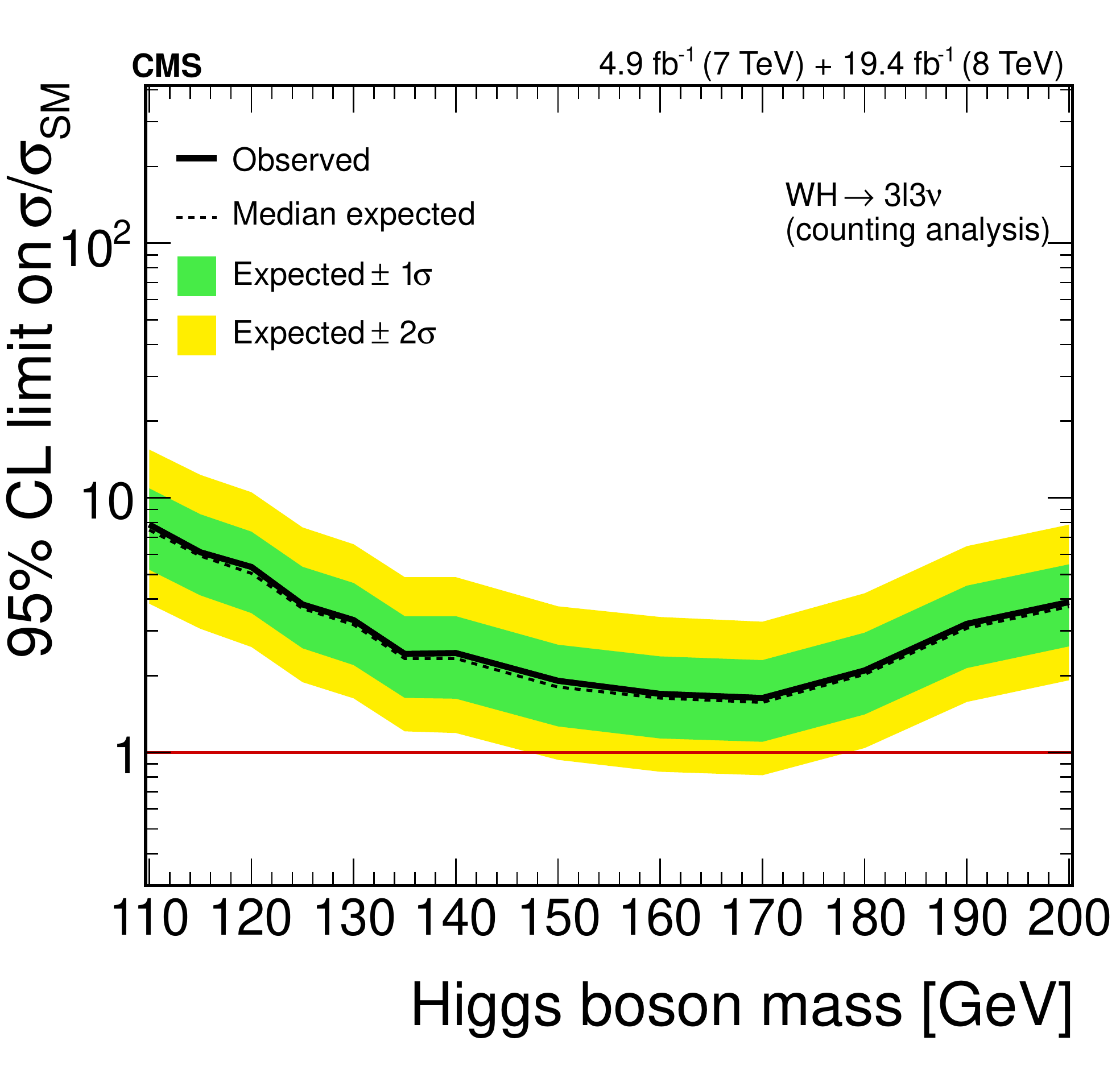}
   \includegraphics[width=0.48\textwidth]{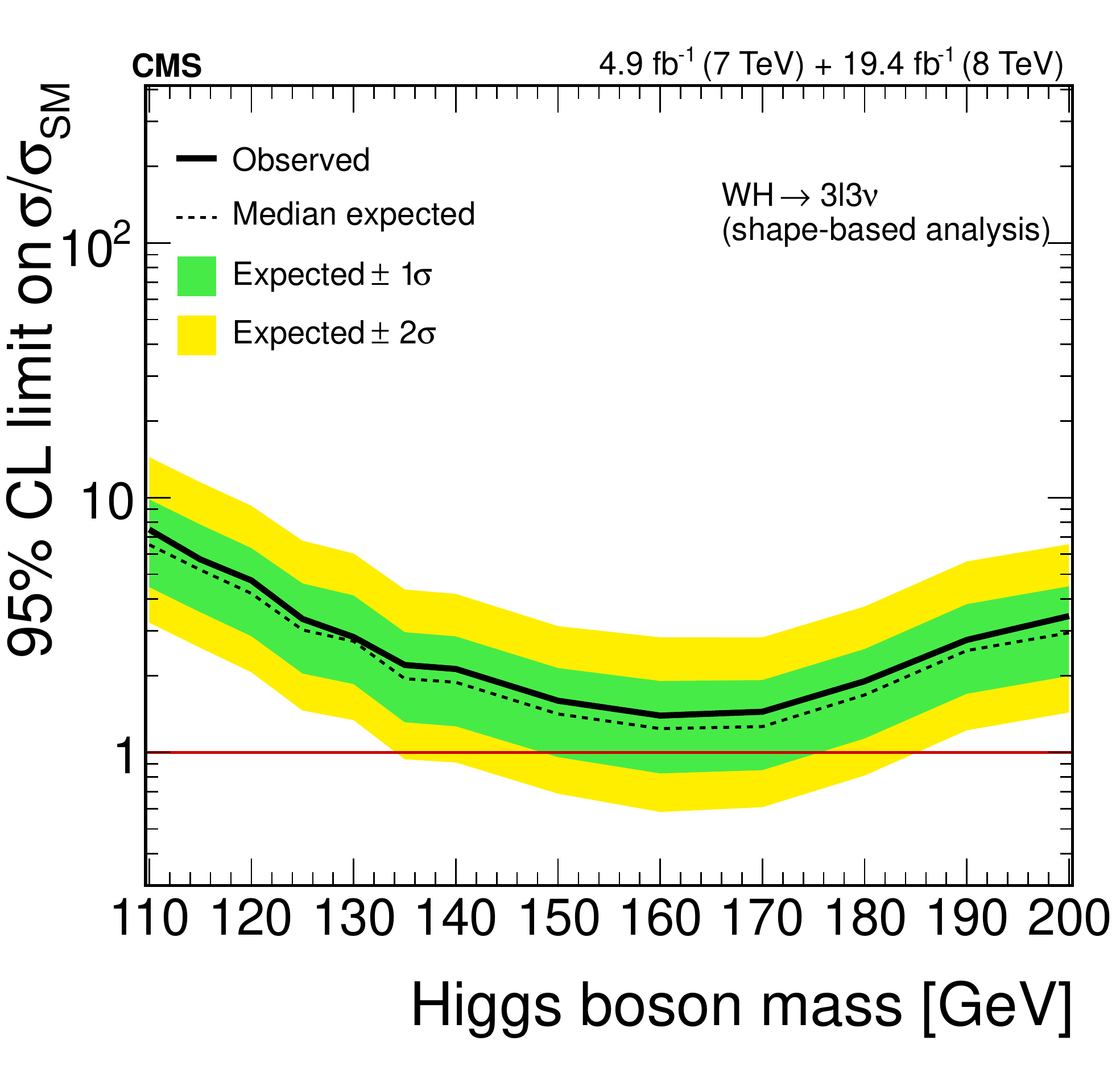}
   \caption{Expected and observed 95\% CL upper limits on the signal production cross section
relative to the SM Higgs boson expectation
using the counting analysis (left) and the shape-based template fit
approach (right) in the $\W\PH \to 3\ell 3\nu$ category.}
   \label{fig:combined_wh3l_from110to200_logx0_logy1}
\end{center}
\end{figure}

No significant excess of events is observed with respect to the background
prediction, and the 95\% CL upper limits are calculated for
the production cross section of the $\W\PH \to 3\ell 3\nu$ process
with respect to the SM Higgs boson expectation.
The expected and observed upper limits are shown in Fig.~\ref{fig:combined_wh3l_from110to200_logx0_logy1}.
Since the analysis is independent of $\mHi$, and the shape of the
$\Delta R_{\ell^+\ell^-}$ distribution has a mild dependence on $\mHi$,
smooth changes are expected for different Higgs boson mass hypotheses.
The observed (expected) upper limit at the 95\% CL is 3.8 (3.7)
times larger than the SM expectation for $\mHi=125\GeV$ for the counting
analysis. For the shape-based analysis, the observed
(expected) upper limit at the 95\% CL is 3.3 (3.0)
times larger than the SM expectation for $\mHi=125\GeV$. A summary of the
results for $\mH = 125\GeV$ is shown in Table~\ref{tab:vh3l_results125}.

\begin{table}[htbp]
\centering
\topcaption{
A summary of the expected and observed
95\% CL upper limits on the signal production cross section relative to the SM prediction,
the significances for the background-only hypothesis to account for the excess
in units of standard deviations (sd),
and the best-fit $\sigma/\sigma_\mathrm{SM}$
at $\mH= 125\GeV$ for the $\W\PH \to 3\ell 3\nu$ category.
}
\label{tab:vh3l_results125}
\begin{tabular}{lccc}
\hline\hline
$\W\PH \to 3\ell 3\nu$ analysis & 95\% CL limits on $\sigma/\sigma_\mathrm{SM}$ & Significance & $\sigma/\sigma_\mathrm{SM}$ \\
 $\mHi = 125\GeV$               & expected / observed & expected / observed & observed \\
\hline
Shape-based (default) & 3.0 / 3.3 & 0.7 / 0.5 sd & $0.57^{+1.28}_{-0.97}$ \\
Counting analysis     & 3.7 / 3.8 & 0.6 / 0.2 sd & $0.37^{+1.65}_{-1.52}$ \\ \hline
\end{tabular}
\end{table}

\subsection{The \texorpdfstring{$\Z\PH \to 3\ell \nu \text{+2 jets}$}{ZH to 3 leptons neutrino + 2 jets}~~category}
\label{sec:zh3lqq}
\subsubsection{Analysis strategy}

To select $\Z\PH$ events, the first step is to identify the leptonic decay of the $\Z$ boson.
Events are required to have one pair of opposite-sign
same-flavor leptons for which $\abs{m_{\ell\ell}-m_{\Z}} < 15\GeV$. If there
is more than one possible combination, the pair with an invariant mass closest
to the $\Z$ mass is chosen. To reject the $\V\gamma^{(*)}$ background, the
dilepton mass of all opposite-sign lepton pairs is required to be greater
than 12\GeV. To reject possible contributions from $\Z$ bosons decaying to
$4\ell$, with one of the leptons not identified, the invariant mass of the
system of the three leptons is required to be
$\abs{m_{\ell\ell\ell}-m_{\Z}}\geq 10\GeV$. As one of the $\W$ bosons in this category
decays hadronically, events are required to have at least two jets.
The requirements described above define the preselection.
The transverse component of the leptonically decaying $\W$ boson is
reconstructed from the remaining lepton, that is not used to reconstruct
the $\Z$ boson, and $\met$.
Events are further required to have the transverse mass $\mt$
of the leptonically decaying $\W$ boson to be less than 85\GeV,
where $\mt^{\ell\nu}$ is defined as
$\mt^{\ell\nu} = \sqrt{ (p_{\mathrm{T},l}+p_{\mathrm{T},\nu})^2 - (p_{\mathrm{x},l}+p_{\mathrm{x},\nu})^2 - (p_{\mathrm{y},l}+p_{\mathrm{y},\nu})^2}$,
where the transverse momentum components of the neutrino are approximated
by the transverse components of $\vmet$. Furthermore,
the invariant mass of the jet pair is required to be compatible with a W decay:
$\abs{m_{jj} - m_{\PW}} \leq 60\GeV$. The angle $\Delta\phi(\ell\nu,jj)$ between
the system of the lepton  and the neutrino, approximated by $\vmet$, and
the system of the two jets in the transverse plane must be smaller than 1.8~radians.
The selection criteria have been optimized for the best $S/\sqrt{B}$ using
simulated samples for a SM Higgs boson signal with $\mHi=125\GeV$.

The criteria listed above comprise the selection for both a counting and a shape-based
analysis in this category. For the shape-based analysis,
which achieves better expected sensitivity than the counting analysis, the transverse mass of
the Higgs boson is reconstructed using the two jets, the $\vmet$ and the
lepton from the $\W$ boson decay, $\mtlnjj$  = $\sqrt{ (\sum \pt)^2 - (\sum p_{\mathrm{x}})^2 - (\sum p_{\mathrm{y}})^2}$,
where in each sum, all the final-state objects from the Higgs boson decay are included.
Therefore $\sum \pt$ is given by $\sum \pt = p_{\mathrm{T}, \ell} + p_{\mathrm{T}, \nu} + p_{\mathrm{T}, j1} + p_{\mathrm{T}, j2}$,
and similarly for $\sum p_{\mathrm{x}}$ and $\sum p_{\mathrm{y}}$.
For the counting analysis,  $\mtlnjj$ is also used with the mass-dependent selection
requirements presented in Table~\ref{tab:zh_cuts}.

\begin{table}[htbp]
\centering
\topcaption{Mass-dependent set of requirements on $\mtlnjj$  used in the
$\Z\PH \to 3\ell \nu~\text{+ 2 jets}$ counting analysis.}
\label{tab:zh_cuts}
\begin{tabular} {cc}
\hline \hline
 $\mHi$ range [\GeVns{}] & Threshold [\GeVns{}] \\
  \hline

$\mHi \leq 135$        &  $\mtlnjj < 140$ \\
$135 < \mHi \leq 160$  &  $\mtlnjj < 170$ \\
$160 < \mHi \leq 170$  &  $\mtlnjj < 180$ \\
$\mHi > 170$           &  --- \\
 \hline
  \end{tabular}
\end{table}

\subsubsection{Background estimation}

Four main background processes are present in the sample after full selection:
$\W\Z$, $\Z\Z$, tribosons, and processes involving non-prompt leptons.
The first three contributions are estimated from simulated samples, while the
last one is evaluated from data. Unlike in the case of the
$\W\PH \to 3\ell 3\nu$ category, the contribution from
$\PH \to \tau\tau$ is negligible in this category.

The non-prompt lepton background processes are estimated
as explained in Section~\ref{sec:wh3l}. This
kind of background arises predominantly from $\Zjets$ production, a small
contribution from top-quark production, and negligible contributions
from other processes.

\subsubsection{Results}
\begin{figure}[htbp]
\begin{center}
\includegraphics[width=0.45\linewidth]{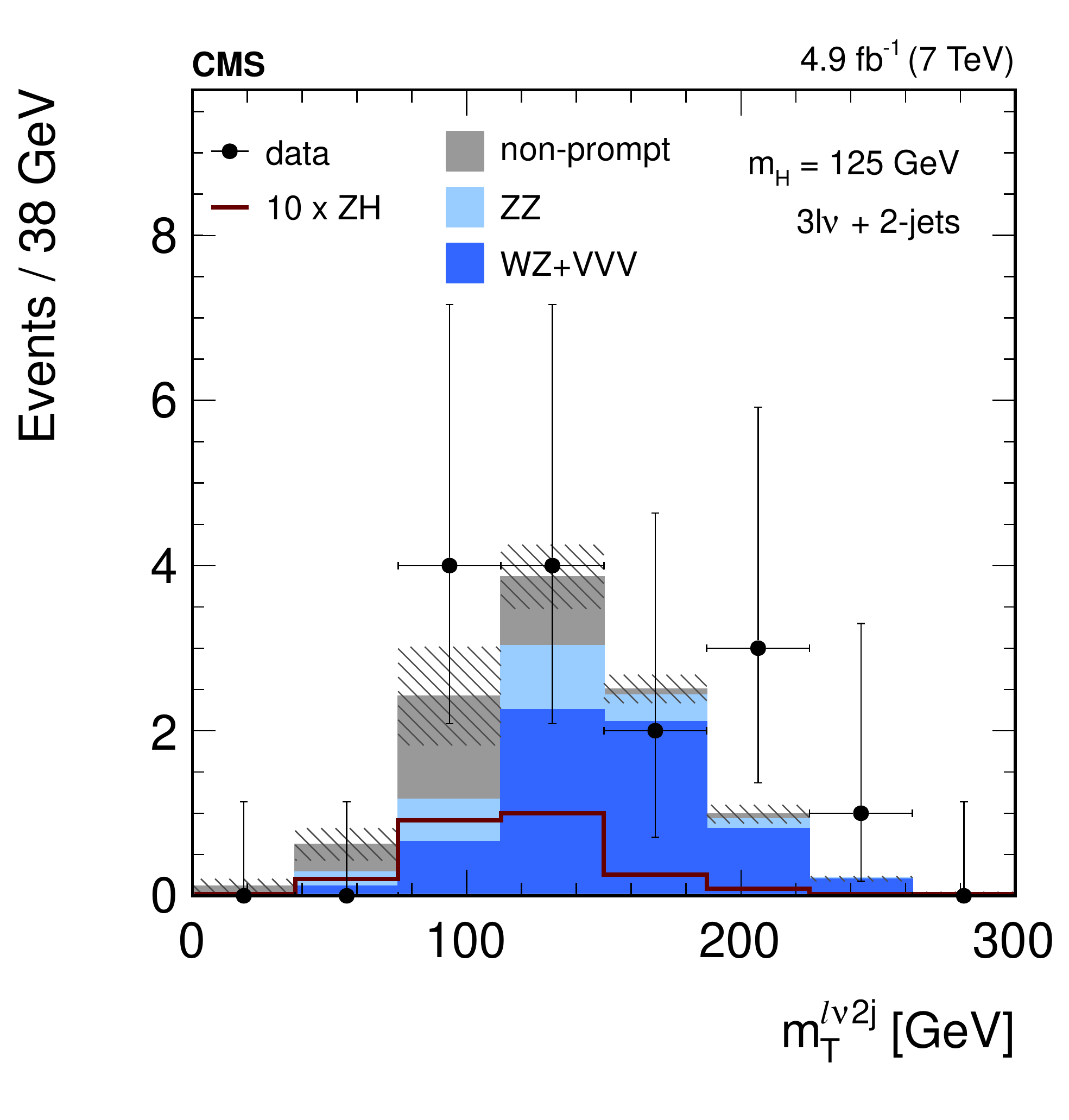}
\includegraphics[width=0.45\linewidth]{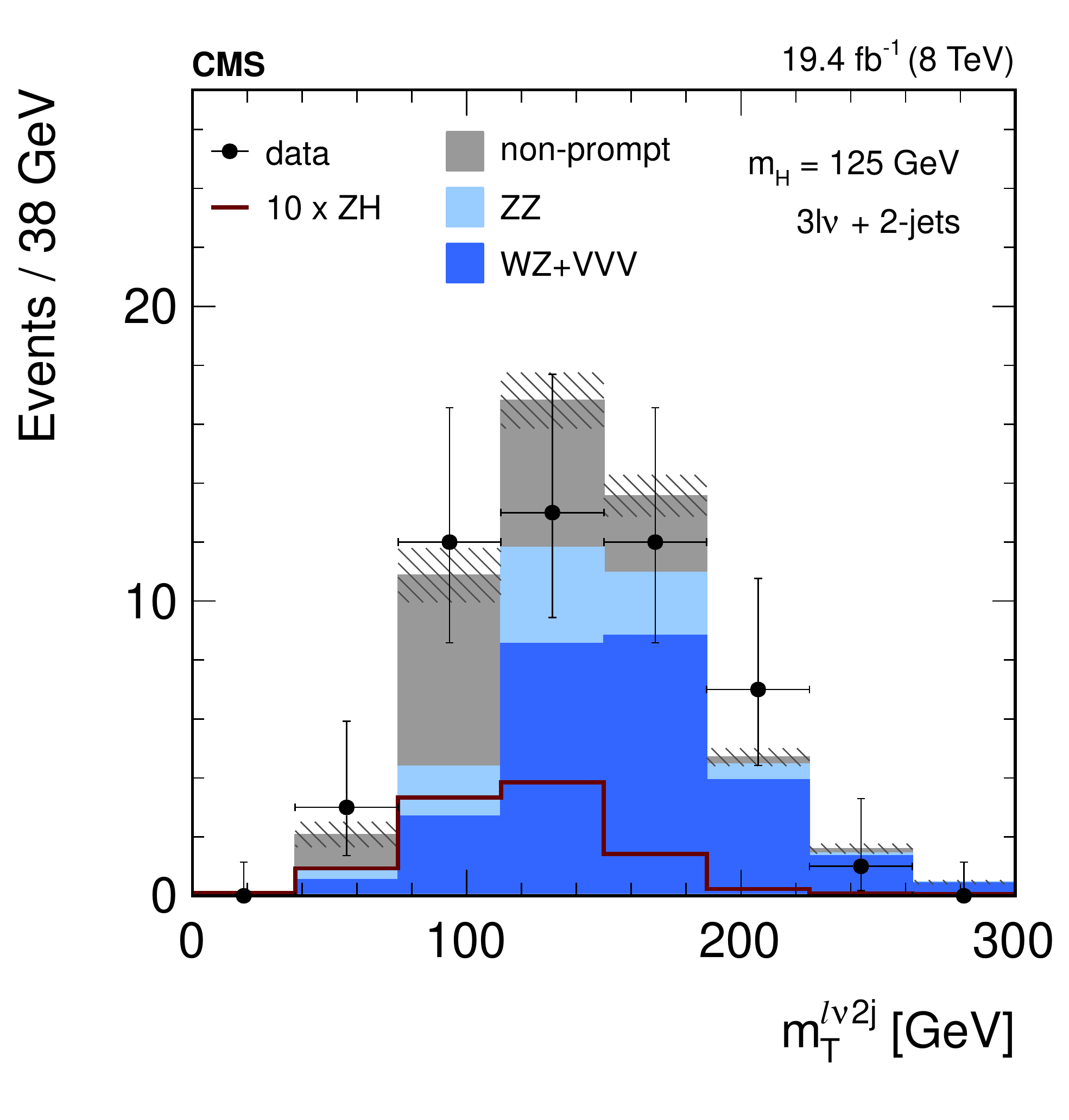}
\caption{The $\mtlnjj$ distribution after all other requirements for the
$\Z\PH \to 3\ell \nu~\text{+ 2 jets}$ analysis at
7\TeV (left), and at 8\TeV (right). The signal yield
(red open histogram) is multiplied by 10 with respect to the SM expectation.
The legend entry labeled as
``non-prompt'' is the combination of the backgrounds from $\Zjets$ and top-quark decays.
}
\label{fig:histo_mh_afterallothercuts}
\end{center}
\end{figure}

The observed number of events and the expected number of signal
and background events at different stages of the shape-based analysis are shown in
Table~\ref{tab:zhselection_all}. The $\mtlnjj$ distributions are shown in
Fig.~\ref{fig:histo_mh_afterallothercuts}.
The final number of events for the counting analysis for four different $\mHi$ values
at 7 and 8\TeV are presented in Table~\ref{tab:zhselection_cut}.

\begin{table}[htbp]
\centering
\topcaption{
Expected signal, number of observed events in data, and estimated background at different stages of the
$\Z\PH \to 3\ell \nu~\text{+ 2 jets}$ shape-based analysis assuming a Higgs boson mass
of 125\GeV.
Only statistical uncertainties in the yields are reported in the first three rows of
the selection stages, while all systematic uncertainties are considered in the last one.
The legend entry labeled as ``non-prompt'' refers to the combination of the backgrounds
from $\Zjets$ and top-quark decays.}
 \label{tab:zhselection_all}
  {\small
  \begin{tabular} {ccccccc}
\hline\hline
\multirow{2}{*}{Selection stage}    & $\Z\PH$ & \multirow{2}{*}{Data} & \multirow{2}{*}{All bkg.} & \multirow{2}{*}{$\WZ+\V\V\V$} & \multirow{2}{*}{Non-prompt} & \multirow{2}{*}{$\cPZ\cPZ$}  \\
& $\PH \to \W\W$ & & & & & \\
  \hline
 & \multicolumn{6}{c}{$7\TeV$ $\Z\PH \to 3\ell \nu~\text{+ 2 jets}$ category} \\
\hline
 Preselection & 0.52 $\pm$ 0.02 & 86 & 93 $\pm$ 2 & 62.1 $\pm$ 0.5 & 21 $\pm$ 2 & 10.0 $\pm$ 0.3\\
$\mt$ & 0.49 $\pm$ 0.01 & 74 & 78 $\pm$ 2 & 50.4 $\pm$ 0.5 & 18 $\pm$ 2 & 9.5 $\pm$ 0.3 \\
$m_{jj}$ & 0.34 $\pm$ 0.01 & 33 & 34 $\pm$ 1 & 20.4 $\pm$ 0.3 & 8 $\pm$ 1 & 5.2 $\pm$ 0.2 \\
$\Delta\phi(l\nu,jj)$ & 0.25 $\pm$ 0.01 & 14 & 10.8 $\pm$ 0.6 & 6.3 $\pm$ 0.2 & 2.6 $\pm$ 0.6 & 1.9 $\pm$ 0.1 \\
 \hline
& \multicolumn{6}{c}{$8\TeV$ $\Z\PH \to 3\ell \nu~\text{+ 2 jets}$ category} \\
\hline
 Preselection & 2.24 $\pm$ 0.06 & 493 & 426 $\pm$ 5 & 263 $\pm$ 2 & 113 $\pm$ 4 & 50.0 $\pm$ 0.2  \\
\mt & 2.08 $\pm$ 0.06 & 386 & 352 $\pm$ 4 & 206 $\pm$ 1 & 101 $\pm$ 4 & 45.3 $\pm$ 0.2  \\
$m_{jj}$ & 1.35 $\pm$ 0.05 & 171 & 150 $\pm$ 3 & 87 $\pm$ 1 & 41 $\pm$ 3 & 22.0 $\pm$ 0.1  \\
$\Delta\phi(l\nu,jj)$ & 0.99 $\pm$ 0.04 & 48 & 50 $\pm$ 4 & 26.7 $\pm$ 2.0 & 15.7 $\pm$ 3.5 & 8.1 $\pm$ 0.4  \\
\hline
  \end{tabular}
}
\end{table}

\begin{table}[htbp]
\centering
\topcaption{
Expected signal, number of observed events in data, and estimated background
for typical Higgs boson signal mass hypotheses used in the counting
$\Z\PH \to 3\ell \nu~\text{+ 2 jets}$ analysis. Statistical and systematic uncertainties in the yields.
Statistical and systematic uncertainties in the yields are reported.}
\label{tab:zhselection_cut}
{\small
  \begin{tabular} {cccc}
\hline\hline
 $\mHi$ [\GeVns{}]  & $\Z\PH, \PH \to \W\W$ & Data & All bkg.  \\
  \hline
 & \multicolumn{3}{c}{$7\TeV$ $\Z\PH \to 3\ell \nu~\text{+ 2 jets}$ category} \\
\hline
125 & 0.20 $\pm$ 0.01 & 7  & 5.9 $\pm$ 0.6 \\
150 & 0.71 $\pm$ 0.03 & 10 & 8.7 $\pm$ 0.6 \\
170 & 0.75 $\pm$ 0.03 & 10 & 9.2 $\pm$ 0.6 \\
190 & 0.41 $\pm$ 0.02 & 14 & 10.8 $\pm$ 0.6 \\
 \hline
& \multicolumn{3}{c}{$8\TeV$ $\Z\PH \to 3\ell \nu~\text{+ 2 jets}$ category} \\
\hline
125 & 0.8 $\pm$ 0.1 & 26 & 25 $\pm$ 3 \\
150 & 2.6 $\pm$ 0.1 & 34 & 38 $\pm$ 3 \\
170 & 2.8 $\pm$ 0.1 & 37 & 41 $\pm$ 4 \\
190 & 2.1 $\pm$ 0.1 & 49 & 50 $\pm$ 4 \\
\hline
  \end{tabular}
}
\end{table}

\begin{figure}[htb]
\begin{center}
   \includegraphics[width=0.48\textwidth]{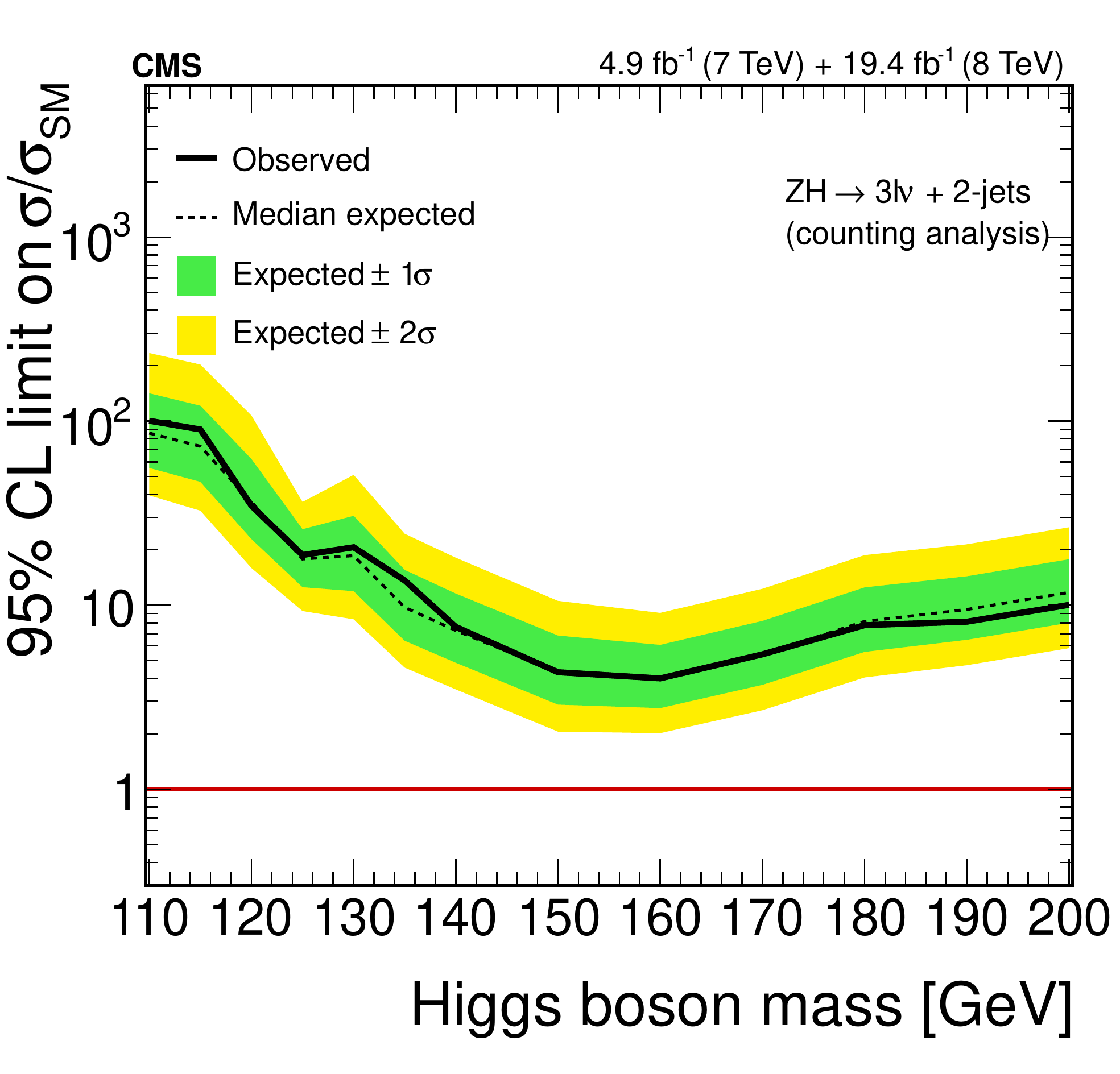}
   \includegraphics[width=0.48\textwidth]{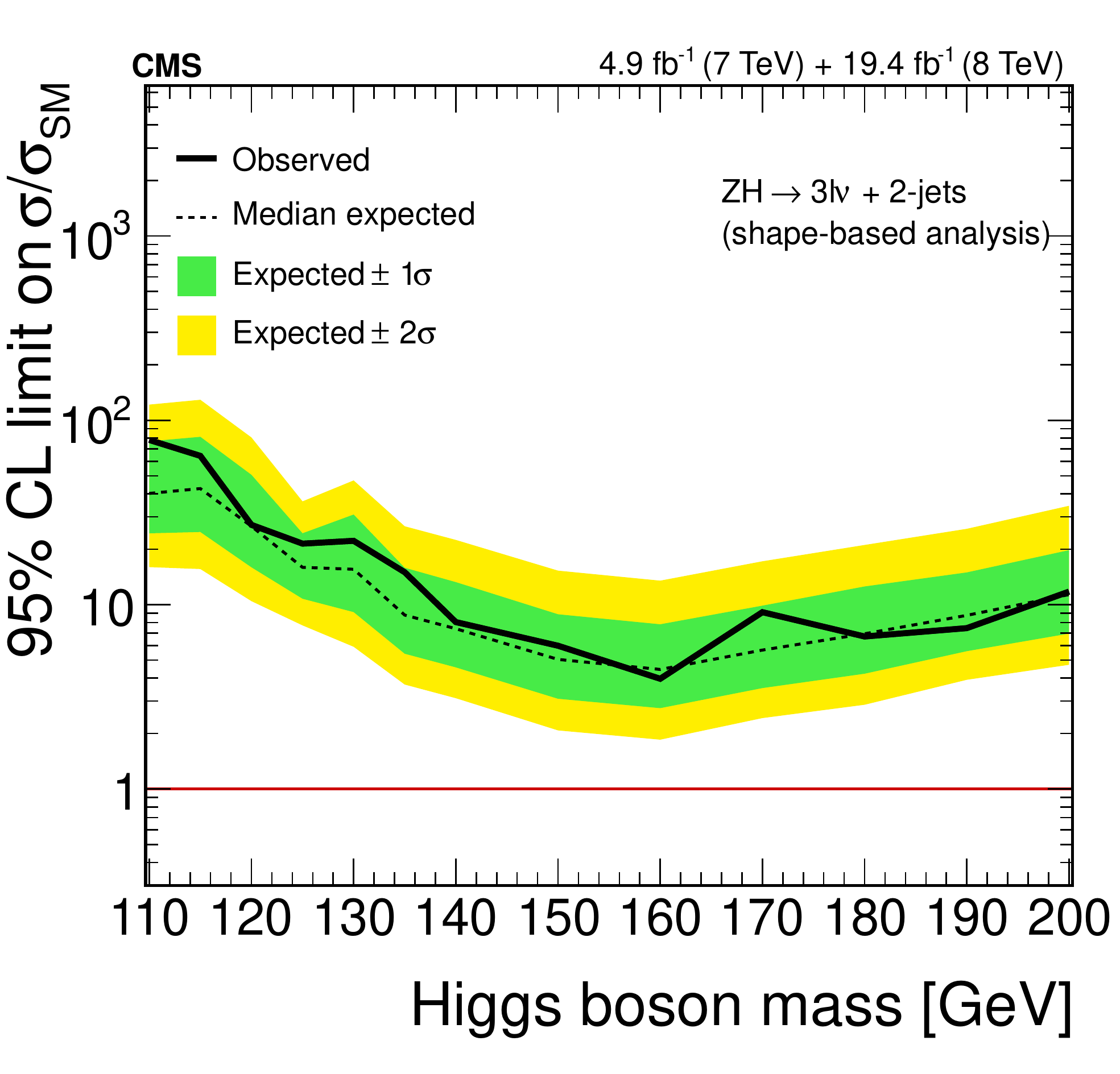}
   \caption{Expected and observed 95\% CL upper limits on the signal production cross section
relative to the SM Higgs boson expectation
using the counting analysis (left) and the shape-based template fit
approach (right) in the $\Z\PH \to 3\ell \nu~\text{+ 2 jets}$ category. }
   \label{fig:combined_zh3l_from110to200_logx0_logy1}
\end{center}
\end{figure}

No significant excess of events is observed with respect to the background
prediction, and the 95\% CL upper limits are calculated for
the production cross section of the $\Z\PH \to 3\ell \nu~\text{+ 2 jets}$ process
with respect to the SM Higgs boson expectation.
Four final states are taken as inputs to the combination:
$\Pe\Pe\Pe$, $\Pe\Pe\mu$, $\mu\mu\Pe$, and $\mu\mu\mu$. These four final states contain
approximately 18\%, 23\%, 24\%, and 35\% of events in the selected sample, respectively.
The upper limits at the 95\% CL for both counting and shape-based analyses are
shown in Fig.~\ref{fig:combined_zh3l_from110to200_logx0_logy1}.
The observed (expected) upper limit at the 95\% CL is 18.7 (17.8)
times larger than the SM expectation for $\mHi=125\GeV$ for the counting
analysis. For the shape-based analysis, the observed
(expected) upper limit at the 95\% CL is 21.4 (15.9)
times larger than the SM expectation for $\mHi=125\GeV$.

\section{Combined results}\label{sec:combined}
In this section, the combined results obtained using all the individual
search categories described in sections~\ref{sec:sel_2l}
and~\ref{sec:sel_3l} are presented. The reference analysis for each
individual search category, selected on the basis of the expected signal
sensitivity, is used in the combination. A summary of the
expected signal production mode fractions for the reference analyses
for a SM Higgs boson with a mass of 125.6\GeV at $\sqrt{s} = 8\TeV$
is shown in Table~\ref{tab:prod_modes_frac}, together with the total
number of expected $\PH \to \W\W$ events at $\sqrt{s} = 7$ and 8\TeV.
The statistical methodology used in this combination is briefly described in Section~\ref{sec:stat}.
The Higgs boson mass hypothesis chosen to evaluate the measurements is $\mHi = 125.6\GeV$, which
corresponds to the mass measurement of the observed boson from the $\PH \to \Z\Z \to 4\ell$ decay
channel~\cite{PAPER-HIG-13-002}. It is important to emphasize that there is a relatively
weak dependence for these analyses on the Higgs boson mass.

\begin{table}[htbp]
\centering
\topcaption{Summary of the expected signal production modes fractions for the
reference analyses for a SM Higgs boson with a mass of 125.6\GeV at
$\sqrt{s} = 8\TeV$. The total number of $\PH \to \W\W$ events is
also reported at $\sqrt{s} = 7$ and 8\TeV.
The shape-based analysis for the 0-jet and 1-jet categories
in the different-flavor final state correspond to
the template fit to the ($\mth$, $\mll$) distribution.
\label{tab:prod_modes_frac}}
{\footnotesize
\begin{tabular}{lccccc}
\hline\hline
\multirow{2}{*}{Category}  & \multirow{2}{*}{$\Pg\Pg\PH$ (\%)} & \multirow{2}{*}{VBF (\%)} & \multirow{2}{*}{$\V\PH$ (\%)} & \multicolumn{2}{c}{Total $\PH \to \W\W$ yield}\\
                           &                                   &                           &                               & $\sqrt{s} = 7\TeV$ & $\sqrt{s} = 8\TeV$ \\
\hline
\multicolumn{6}{c}{Two-lepton analyses} \\
\hline
0-jet different-flavor (shape-based)               &  95.7 &   1.2 &   3.1  & 52.6 & 245   \\
0-jet same-flavor (counting)                       &  98.1 &   0.9 &   1.0  & 10.4 &  58.5 \\
1-jet different-flavor (shape-based)               &  81.6 &  10.3 &   8.1  & 19.8 & 111   \\
1-jet same-flavor (counting)                       &  83.6 &  11.2 &   5.2  &  3.1 &  19.6 \\
2-jet VBF tag different-flavor (shape-based)       &  22.3 &  77.7 &   0.0  &  1.3 &   6.4 \\
2-jet VBF tag same-flavor (counting)               &  14.2 &  85.8 &   0.0  &  0.3 &   2.3 \\
2-jet $\V\PH$ tag different-flavor (counting)      &  55.5 &   4.7 &  39.8  &  0.8 &   4.3 \\
2-jet $\V\PH$ tag same-flavor (counting)           &  65.1 &   4.1 &  30.8  &  0.2 &   2.8 \\ \hline
\multicolumn{6}{c}{Three-lepton analyses} \\
\hline
$\W\PH \to 3\ell 3\nu$ (shape-based)               &   0.0 &   0.0 & 100.0  &  0.7 &   3.8 \\ $\Z\PH \to 3\ell \nu \text{2 jets}$ (shape-based)   &   0.0 &   0.0 & 100.0  &  0.3 &   1.0 \\ \hline
\end{tabular}
}
\end{table}

\subsection{Signal strength}

\begin{figure}[htbp]
\begin{center}
\includegraphics[width=0.49\textwidth]{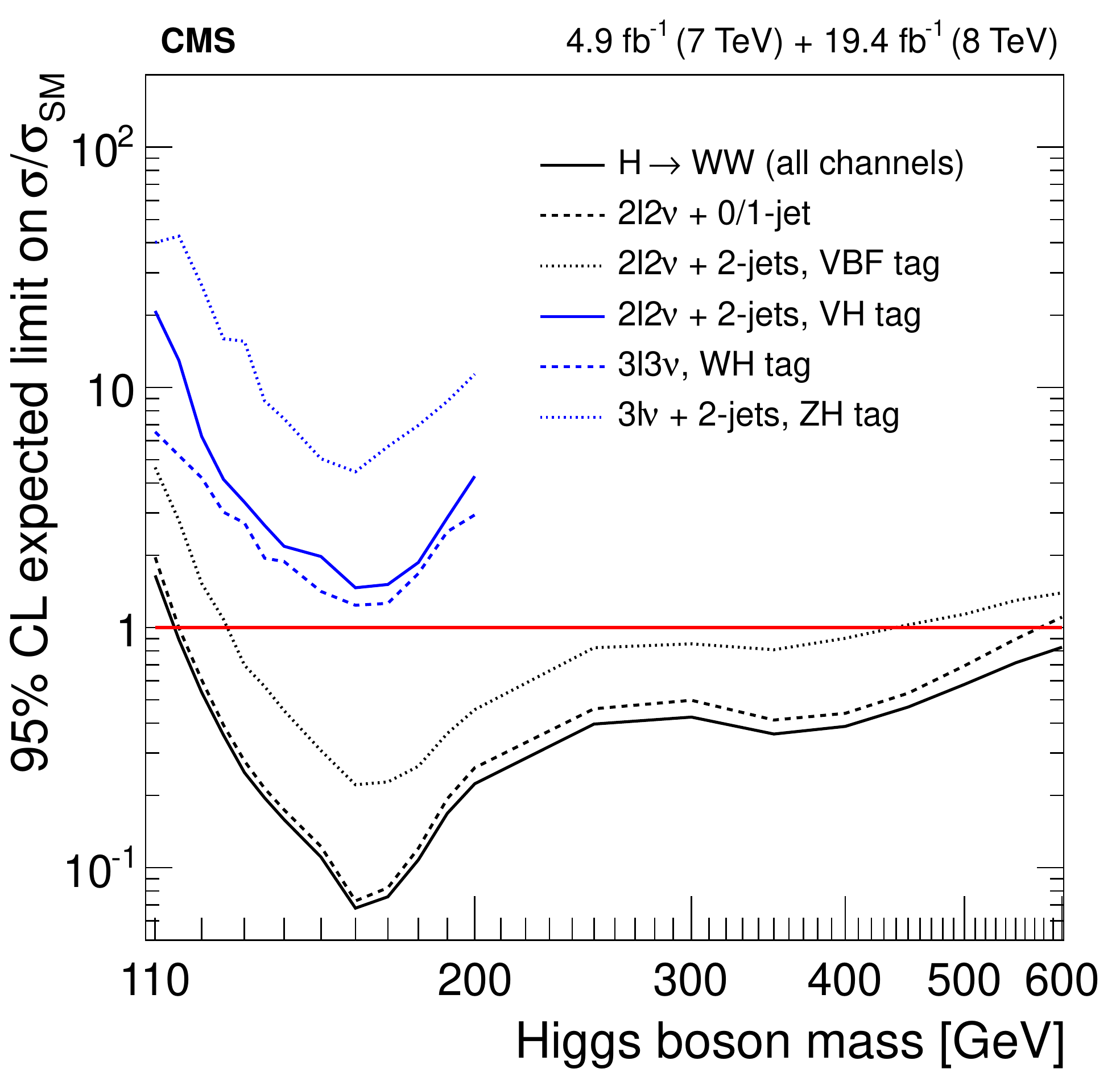}\\
\includegraphics[width=0.49\textwidth]{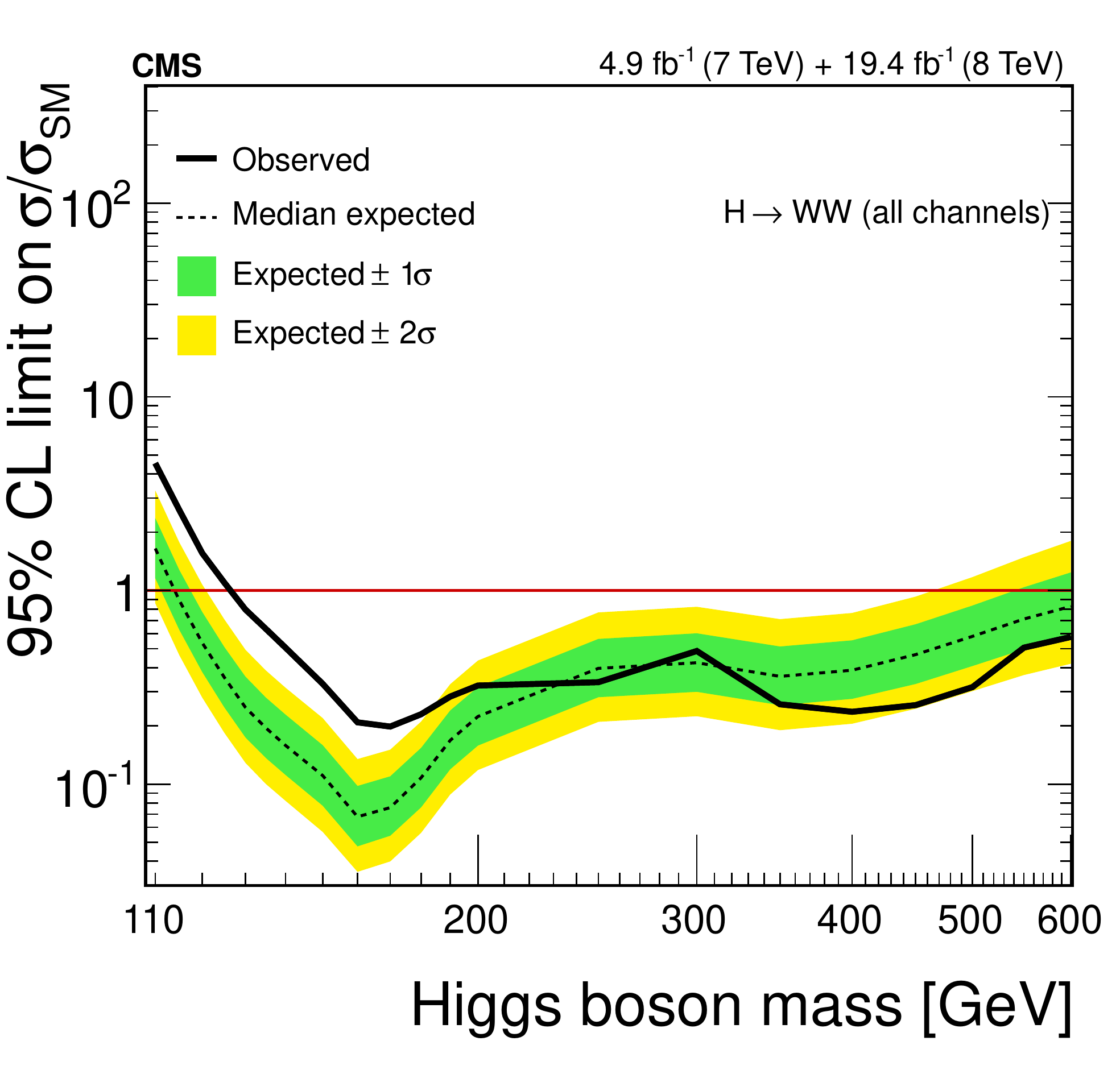}
\includegraphics[width=0.49\textwidth]{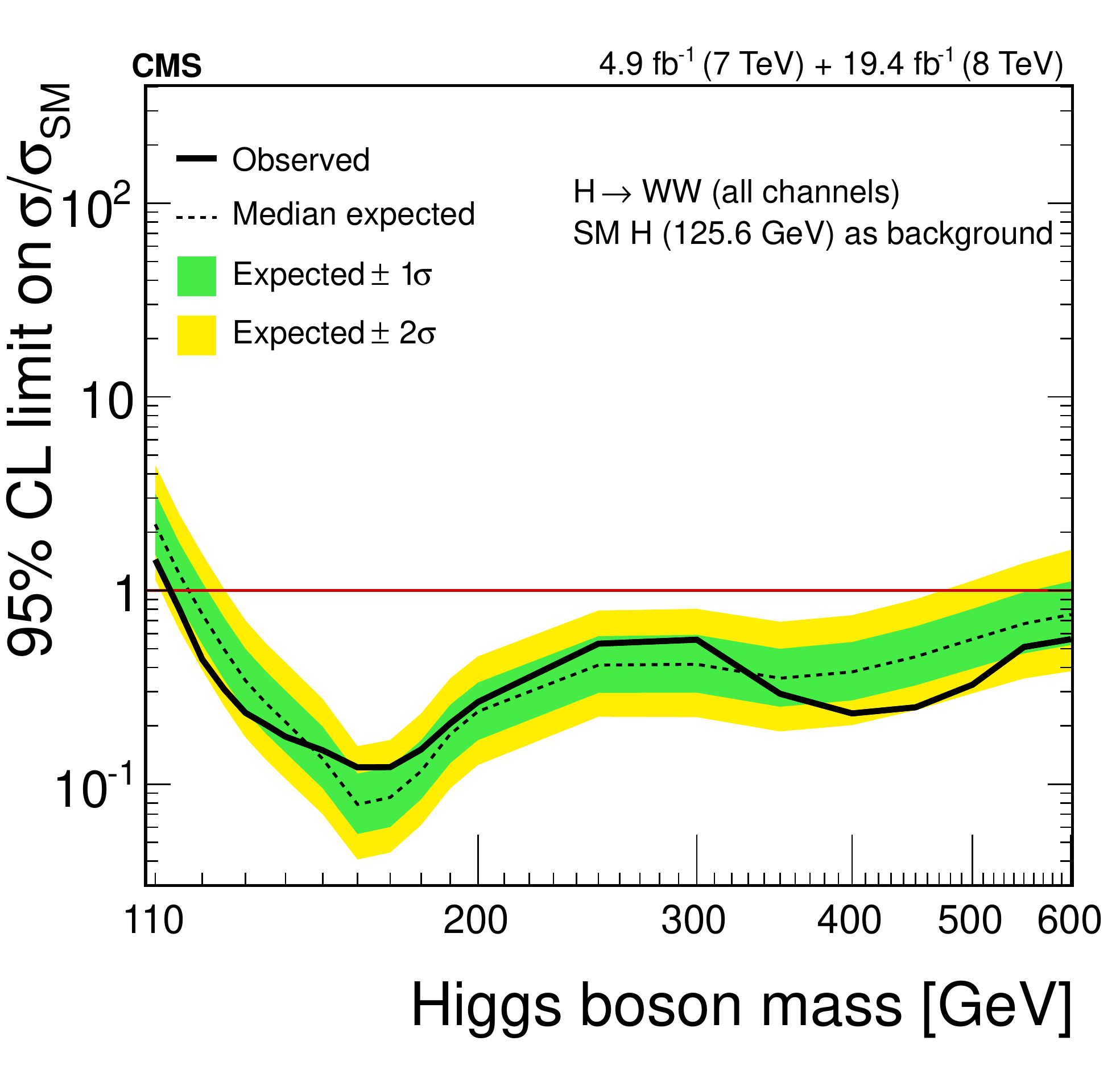}
\caption{
Expected 95\% CL upper limits on the $\PH \to \WW$ production cross section relative
to the SM expectation, shown as a function of the SM Higgs boson mass hypothesis, individually for each search
category considered in the combination, and the combined result from all categories (top).
Expected and observed results are shown with no assumptions on the presence of a Higgs boson (bottom left) and
considering the SM Higgs boson with $\mHi = 125.6\GeV$ as part of the background processes (bottom right). As
expected, the excess observed on the bottom left distribution is reduced on the bottom right by considering
the SM Higgs boson with $\mHi = 125.6\GeV$ as part of the background processes.
  }
\label{fig:hww_limits}
\end{center}
\end{figure}

The expected 95\% CL upper limits on the production cross section of the $\PH\to\WW$ process
with respect to the SM prediction for each category considered
in the combination and the combined result are shown
in Fig.~\ref{fig:hww_limits} (top)
for the Higgs boson mass range \mbox{110--600\GeV.}
Exclusion limits beyond 600\GeV deserve a specific
study and are not addressed in this paper.
The combined observed and expected 95\% CL upper limits on the production cross section
of the $\PH\to\WW$ process with respect to the SM prediction are shown in Fig.~\ref{fig:hww_limits} (bottom).
Results are shown in two ways: without assumptions on the presence of a
SM Higgs boson and considering the SM Higgs boson with $\mHi = 125.6\GeV$ as
part of the background processes. In the first case, an excess of events
is observed for low $\mHi$ hypothesis, which makes the
observed limits much weaker than the expected ones. In particular, the
observed (expected) 95\% CL upper limit on the $\PH \to \WW$ production cross section
with respect to the SM prediction at $\mHi = 125.6\GeV$ is 1.1 (0.3).
The combination of all categories excludes a SM Higgs boson in the
mass range 127--600\GeV at the 95\% CL, while the expected exclusion range
for the background-only hypothesis is 115--600\GeV. In the second case,
to search for another excess, the 95\% CL upper limits are obtained including the SM Higgs
boson with $\mHi = 125.6\GeV$ as a background process, and no significant
excess is found anywhere.
Additional Higgs bosons with SM-like properties are excluded
in the mass range 114--600\GeV at the 95\% confidence level when
assuming that a SM Higgs boson with $\mHi = 125.6\GeV$
is present in the data.

\begin{figure}[htbp]
\begin{center}
\includegraphics[width=0.49\textwidth]{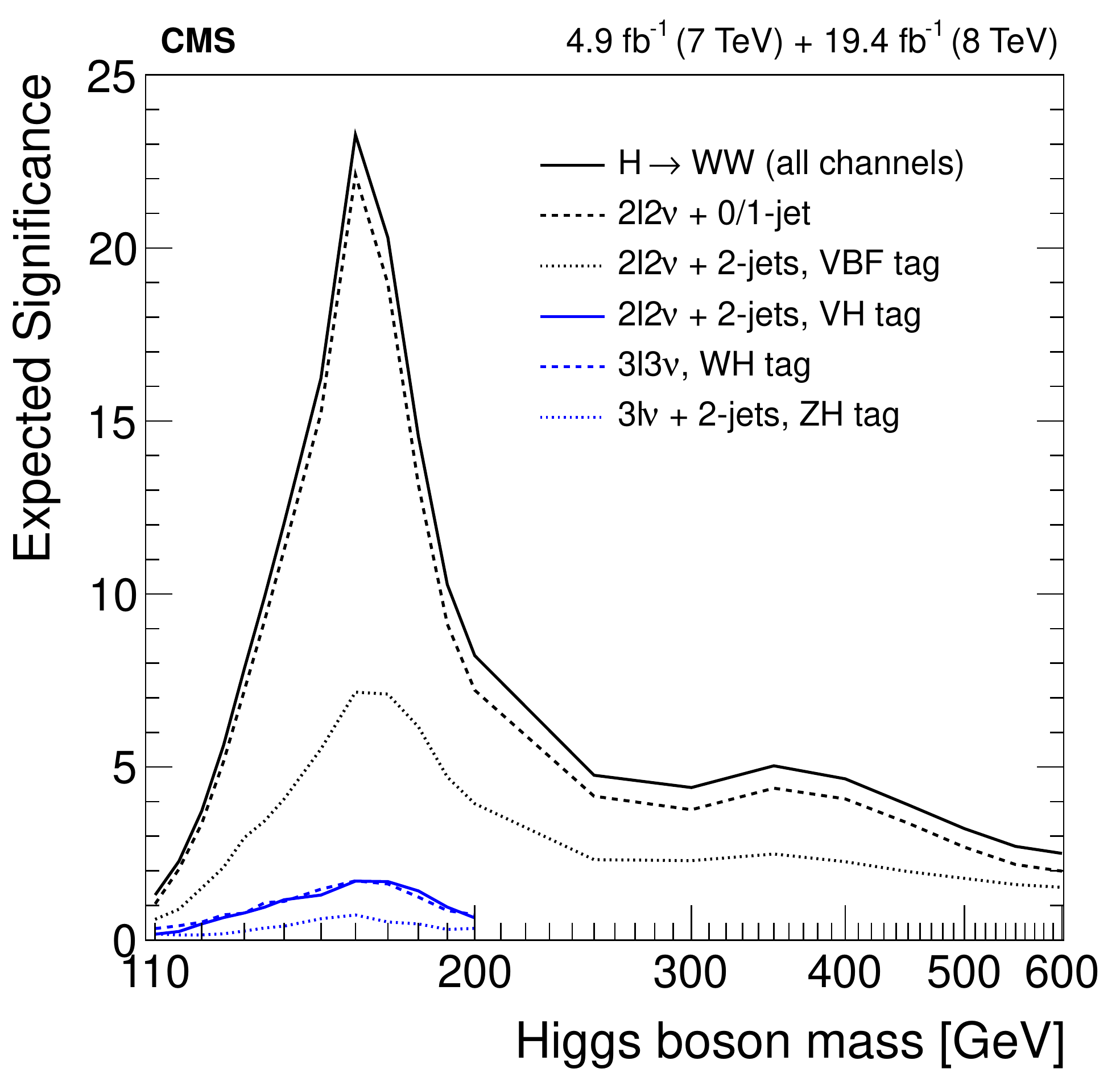}
\includegraphics[width=0.49\textwidth]{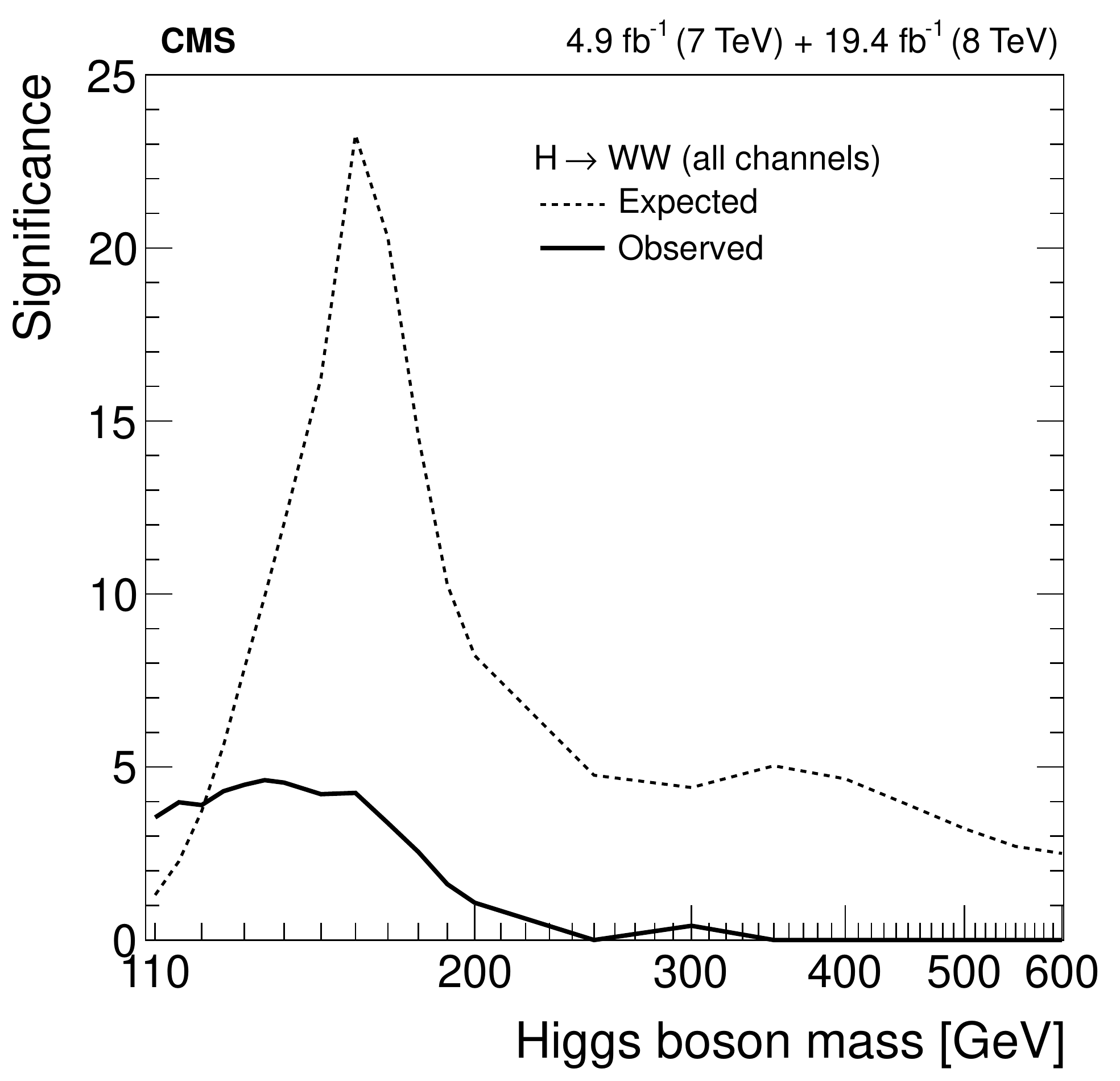}\\
\includegraphics[width=0.49\textwidth]{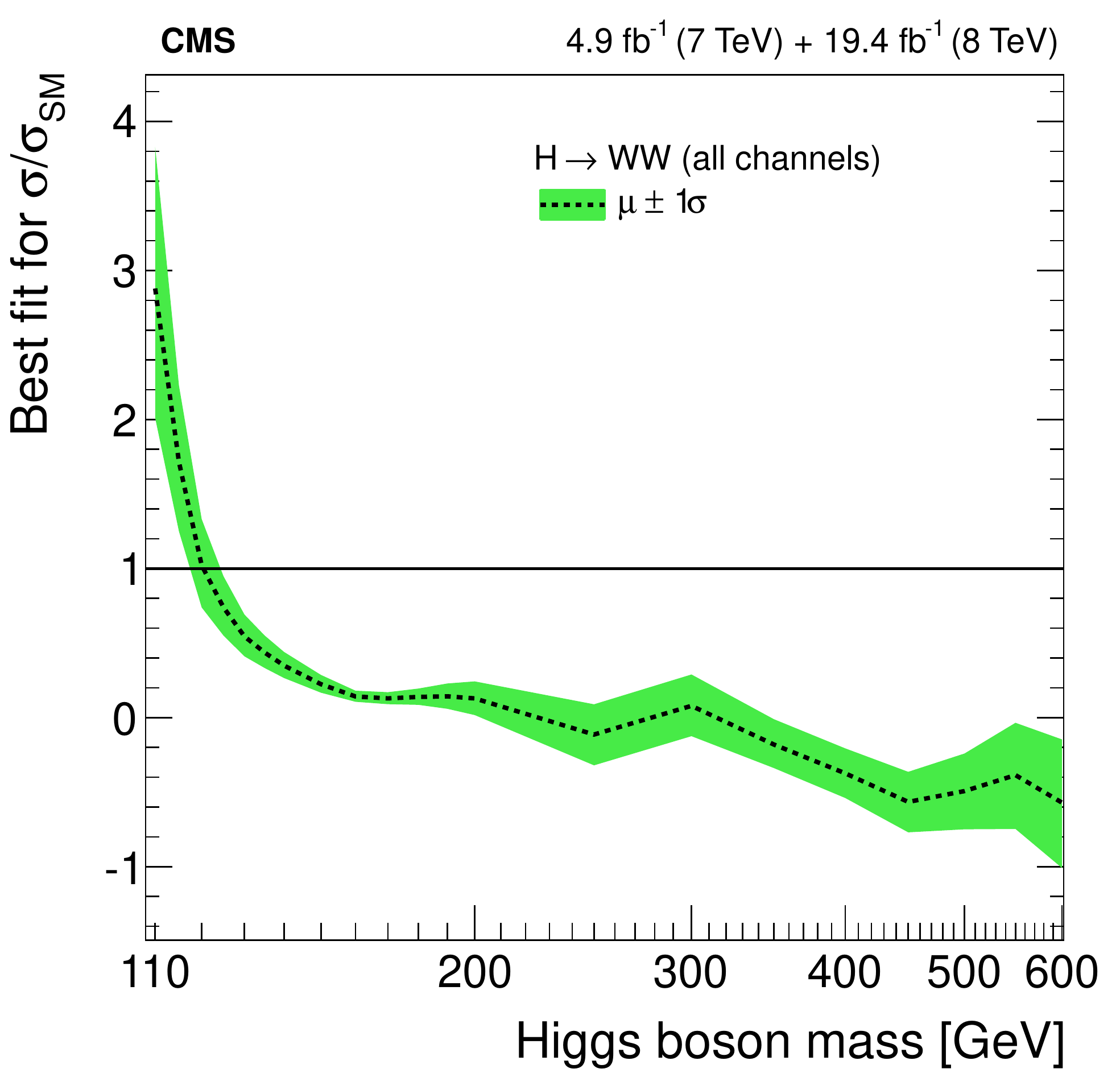}
\caption{
Expected significance as a function of the SM Higgs boson mass, individually for each
search category considered in the combination, and the combined result from all
categories (top left). Expected and observed significance (top right), and observed $\sigma/\sigma_\mathrm{SM}$ (bottom) as a function of the SM Higgs boson
mass for the combination of all $\PH \to \W\W$ categories. The very large expected significance
at $\mHi \sim 160\GeV$ is due to the branching fraction to $\W\W$ close to unity for those masses.
}
\label{fig:hww_signi}
\end{center}
\end{figure}

\begin{figure}[htbp]
  \begin{center}
    \includegraphics[width=0.65\textwidth]{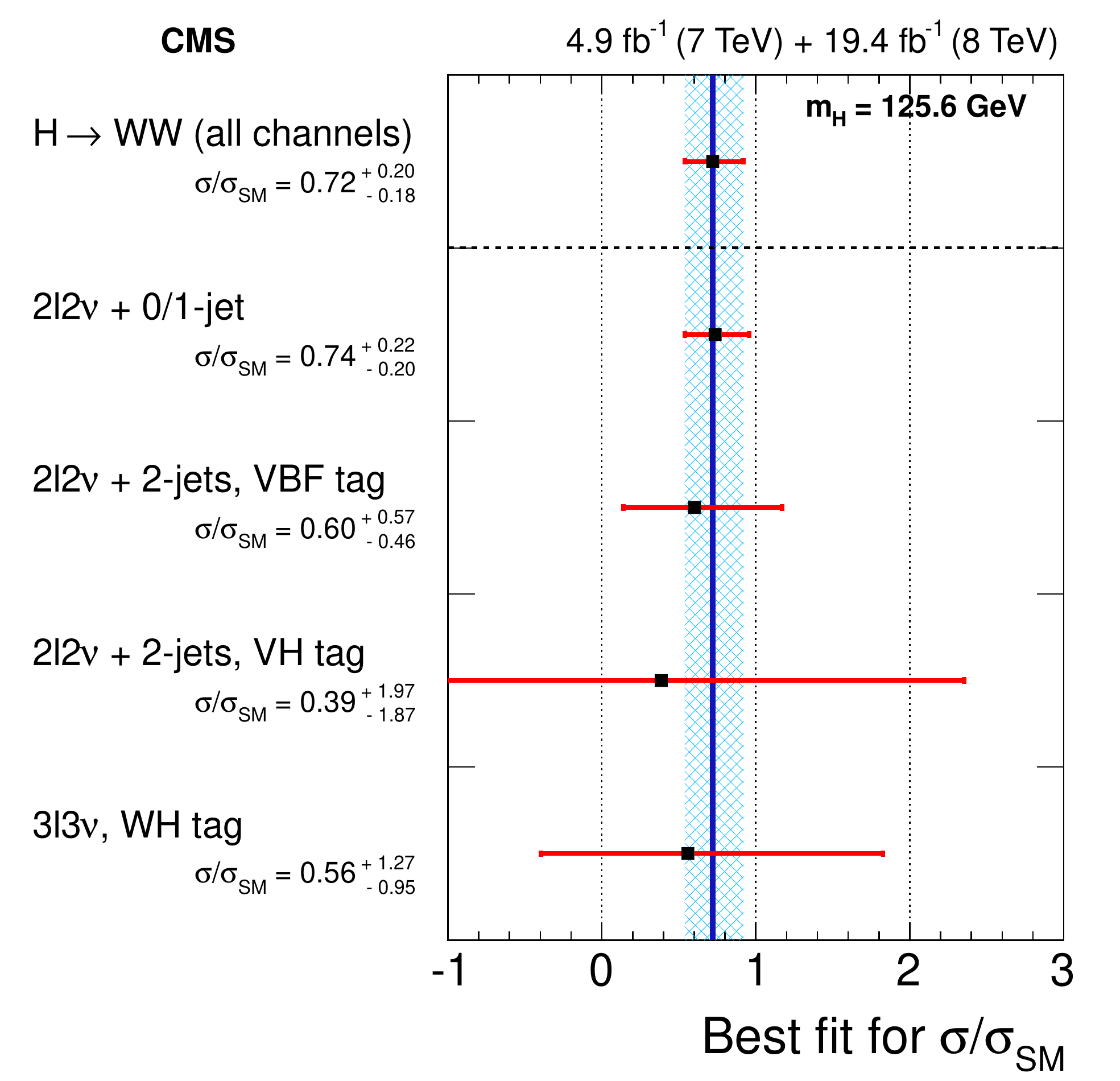}
  \caption{Observed $\sigma/\sigma_\mathrm{SM}$ for $\mHi = 125.6\GeV$ for each category used in the
  combination. The observed $\sigma/\sigma_\mathrm{SM}$ value in the $\Z\PH \to 3\ell \nu~\text{2 jets}$
  category is $6.41^{+7.43}_{-6.38}$. Given its relatively large uncertainty with respect
  to the other categories it is not shown individually, but it is used in the combination.}
  \label{fig:sig_mu_mh125}
  \end{center}
\end{figure}

The expected significance for the SM Higgs boson signal as a function of the mass hypothesis for each category
and for the combination is shown in Fig.~\ref{fig:hww_signi} (top left). The expected
and observed significances for the combination are shown in Fig.~\ref{fig:hww_signi} (top right).
The observed (expected) significance of the signal is 4.3 (5.8) standard deviations for
$\mHi = 125.6\GeV$. The observed $\sigma/\sigma_\mathrm{SM}$ as a function of the Higgs boson mass
is also shown in Fig.~\ref{fig:hww_signi} (bottom). The $\sigma/\sigma_\mathrm{SM}$
value for $\mHi = 125.6\GeV$ is
$0.72^{+0.20}_{-0.18}$ = $0.72^{+0.12}_{-0.12}\stat^{+0.12}_{-0.10}~\text{(th. syst.)}^{+0.10}_{-0.10}~\text{(exp. syst.)}$,
where the statistical, theoretical systematic, and experimental systematic
uncertainties are reported separately.
The statistical component is estimated by fixing all the nuisance parameters
to their best-fit values and recomputing the likelihood profile.
The most important systematic uncertainties are the theoretical uncertainties in
the signal, followed by those in the $\W\W$ background process. Other
important sources of systematic uncertainties are the lepton, $\met$, and jet
energy experimental uncertainties, as well as the limited knowledge of the $\Wjets$
and $\W\gamma^{(*)}$ background processes.
The observed $\sigma/\sigma_\mathrm{SM}$ for $\mHi = 125.6\GeV$
for each category used in the combination is shown in Fig.~\ref{fig:sig_mu_mh125}.
The results from all categories are consistent within the uncertainties.

\begin{figure}[htbp]
  \begin{center}
    \includegraphics[width=0.49\textwidth]{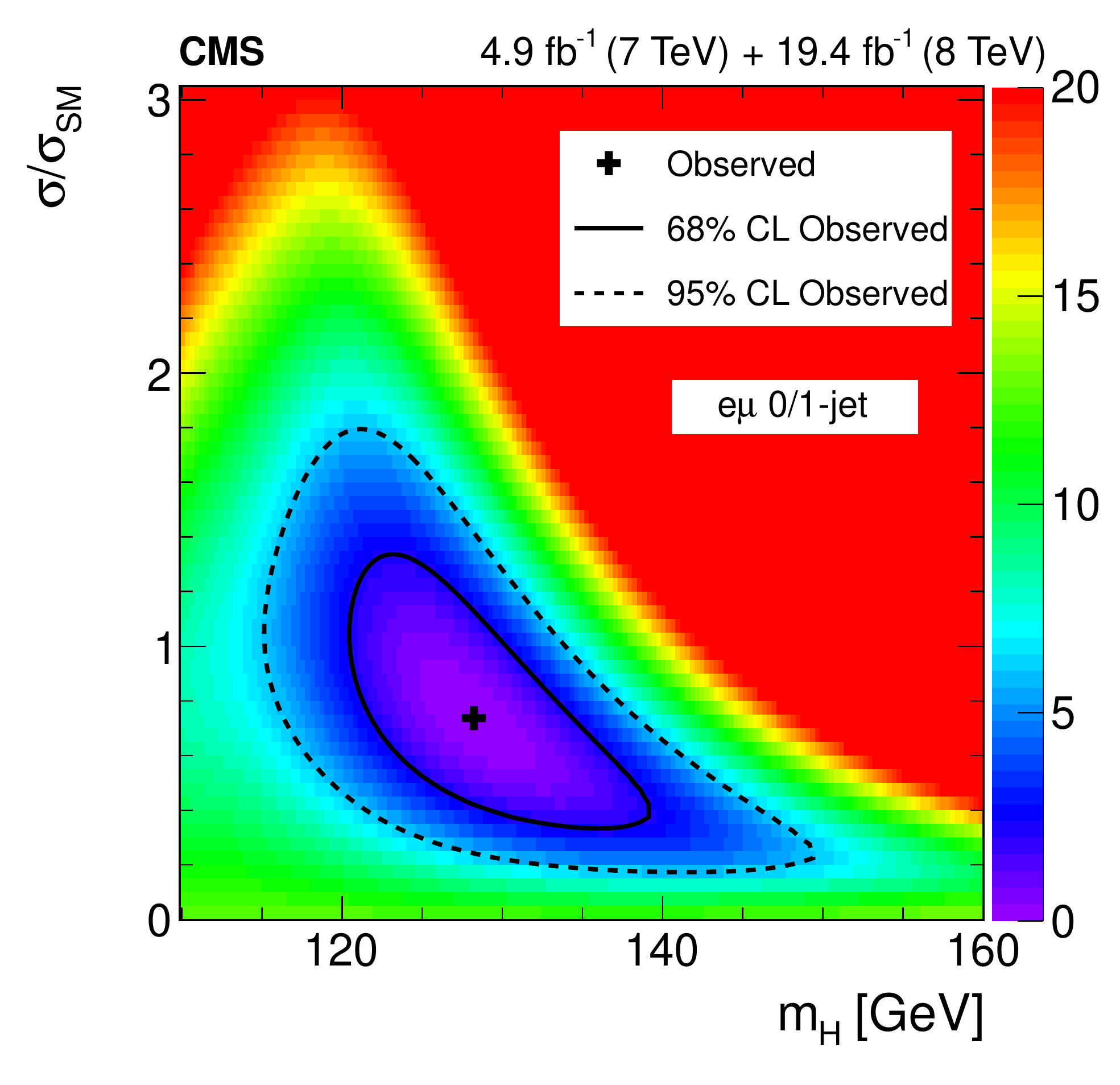}
    \includegraphics[width=0.49\textwidth]{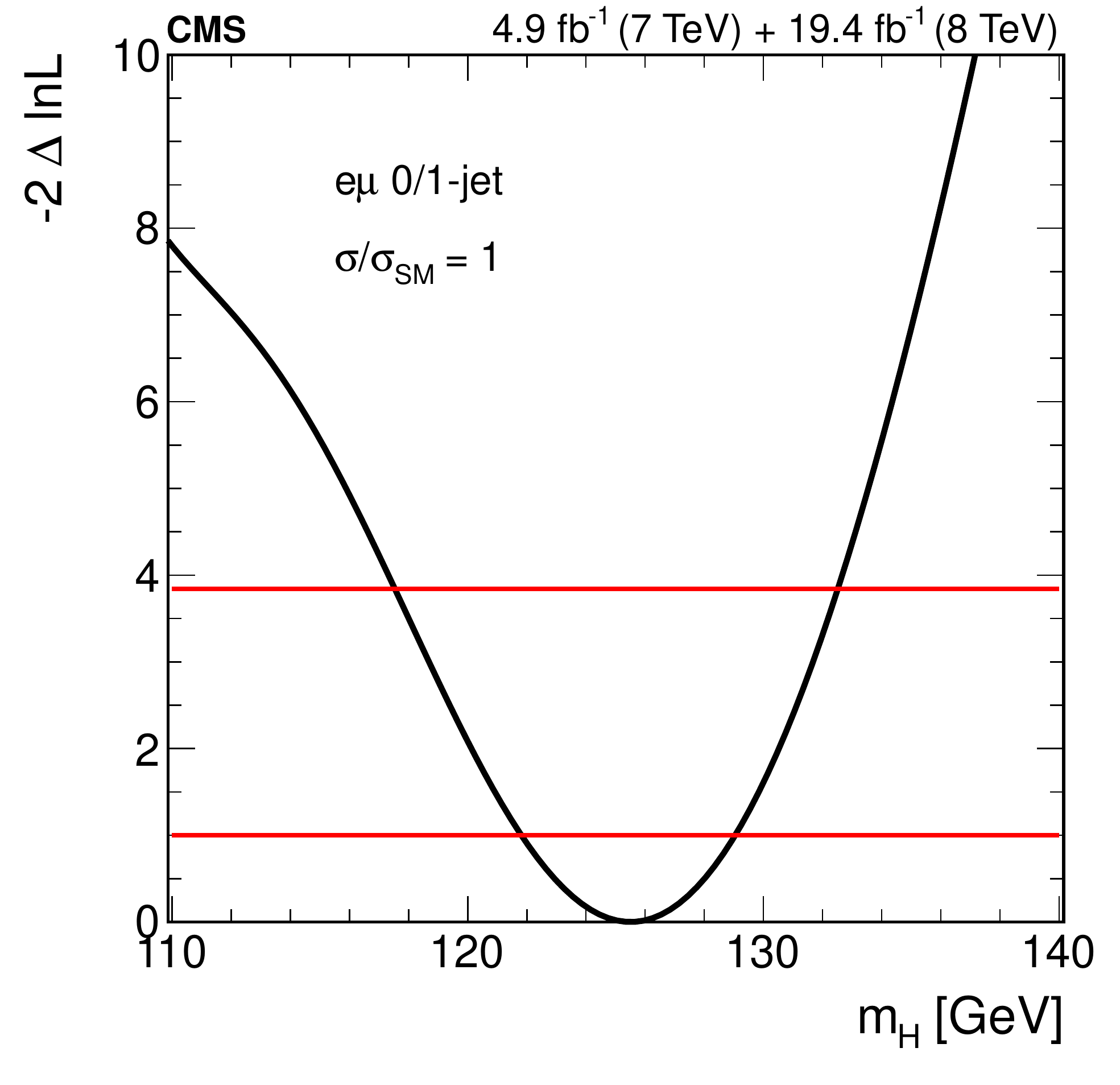}
  \caption{Confidence intervals in the ($\sigma/\sigma_\mathrm{SM}$, $\mH$) plane
           using the parametric unbinned fit in ($\mr$, $\delphir$) distribution (left)
	   for the 0-jet and 1-jet categories in the $\Pe\mu$ final states.
	   Solid and dashed lines indicate the 68\% and 95\% CL contours, respectively. On the right,
	   the one-dimensional likelihood profile for $\sigma/\sigma_\mathrm{SM}$=1 is shown.
           The crossings with the horizontal line at $-2\Delta \, \ln L$ = 1 (3.84) define the 68\% (95\%) CL interval.
	   The SM Higgs boson production cross section uncertainties are considered.}
  \label{fig:sig_mu_mh_mr}
  \end{center}
\end{figure}

Figure~\ref{fig:sig_mu_mh_mr} shows the confidence intervals in the
two-dimensional ($\sigma/\sigma_\mathrm{SM}$, $\mH$) plane and the one-dimensional
likelihood profile in $\mHi$ assuming the SM cross section and branching fraction, $\sigma/\sigma_\mathrm{SM}$=1,
where the SM Higgs boson uncertainties in the production cross section are considered.
The results are obtained with the analysis using a parametric fit to the
($\mr$, $\delphir$) distribution in the 0-jet and 1-jet categories of the
$\Pe\mu$ final state, as described in Section~\ref{sec:hww2l2n_01j}. The likelihood
curve at $\sigma/\sigma_\mathrm{SM}$=1 yields a best-fit mass of $125.5^{+3.6}_{-3.8}\GeV$.
Furthermore, without the constraint on $\sigma/\sigma_\mathrm{SM}$, the best-fit mass is at
$128.2^{+6.6}_{-5.3}\GeV$. The uncertainty on the best-fit mass value is consistent with
the expected resolution of the signal and the observed significance.

\subsection{Couplings}

\begin{figure}[htbp]
\begin{center}
\includegraphics[width=0.65\textwidth]{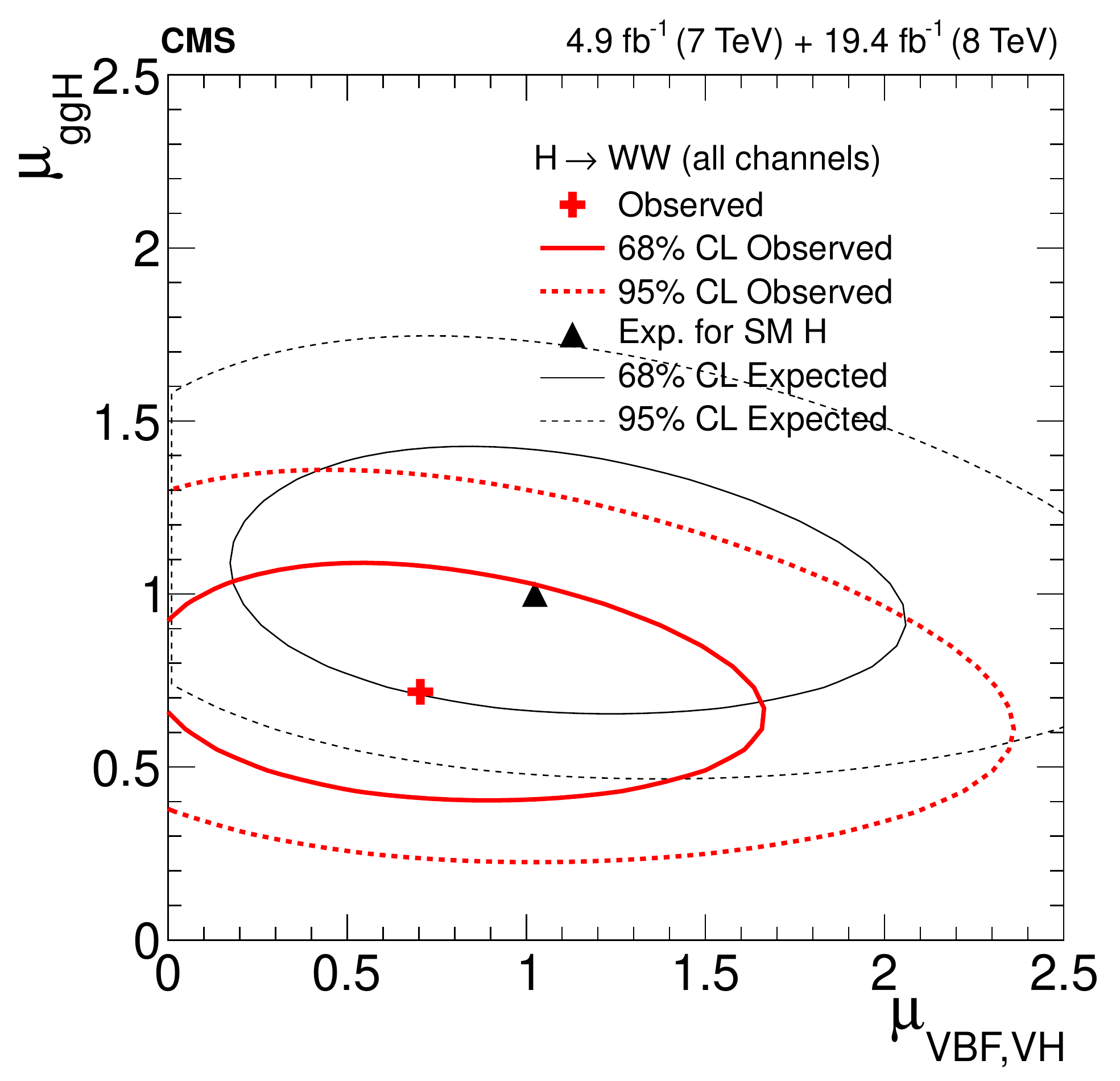}
 \caption{Likelihood profiles on $\mu_{\Pg\Pg\PH}$ and $\mu_{\mathrm{VBF},\V\PH}$ at 68\% (solid) and 95\% CL (dotted).
 The expected (black) and observed (red) distributions for $\mHi = 125.6\GeV$ are shown.
 \label{fig:rvrf}}
 \end{center}
\end{figure}

\begin{figure}[htbp]
\begin{center}
\includegraphics[width=0.49\textwidth]{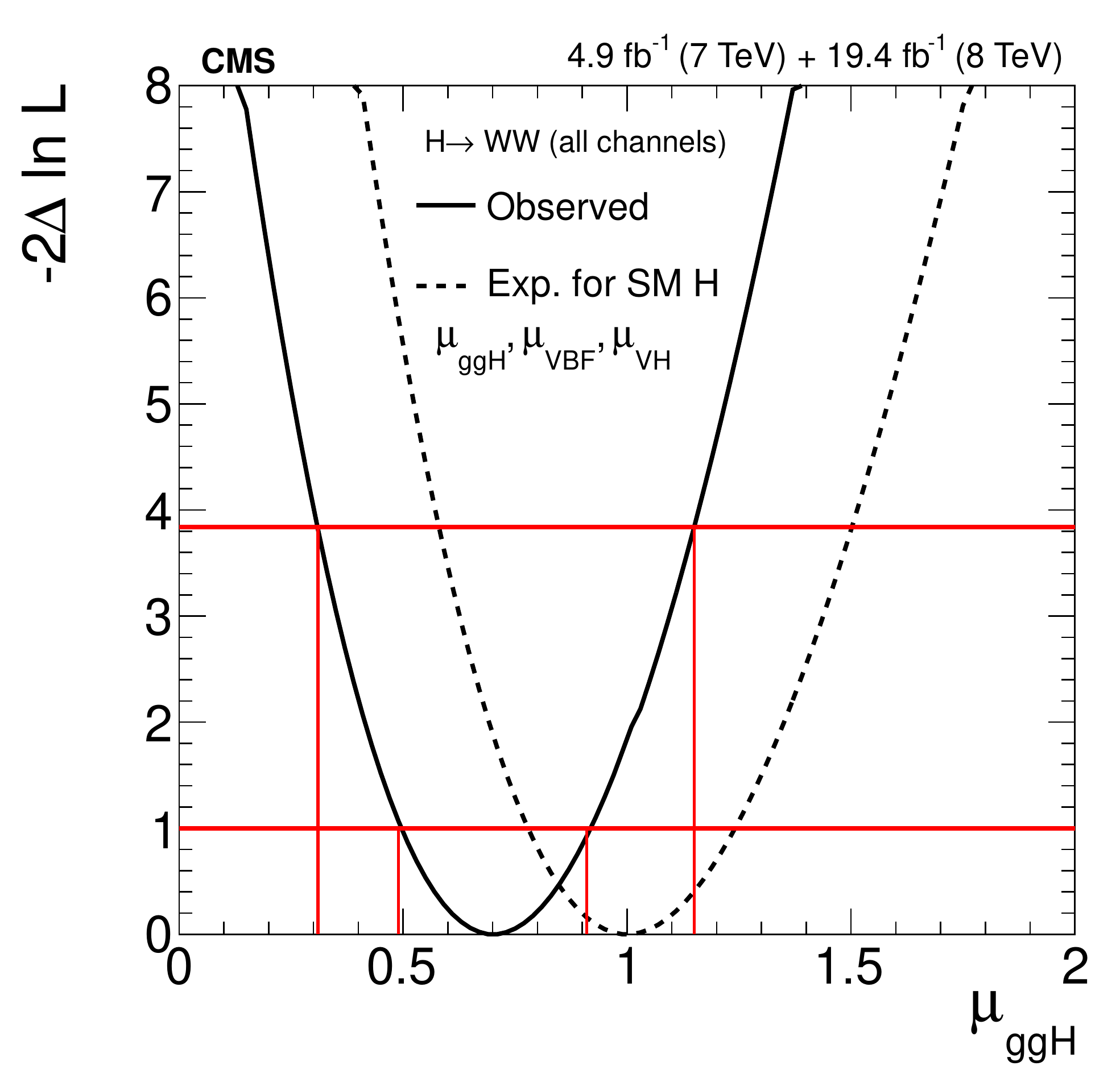}
\includegraphics[width=0.49\textwidth]{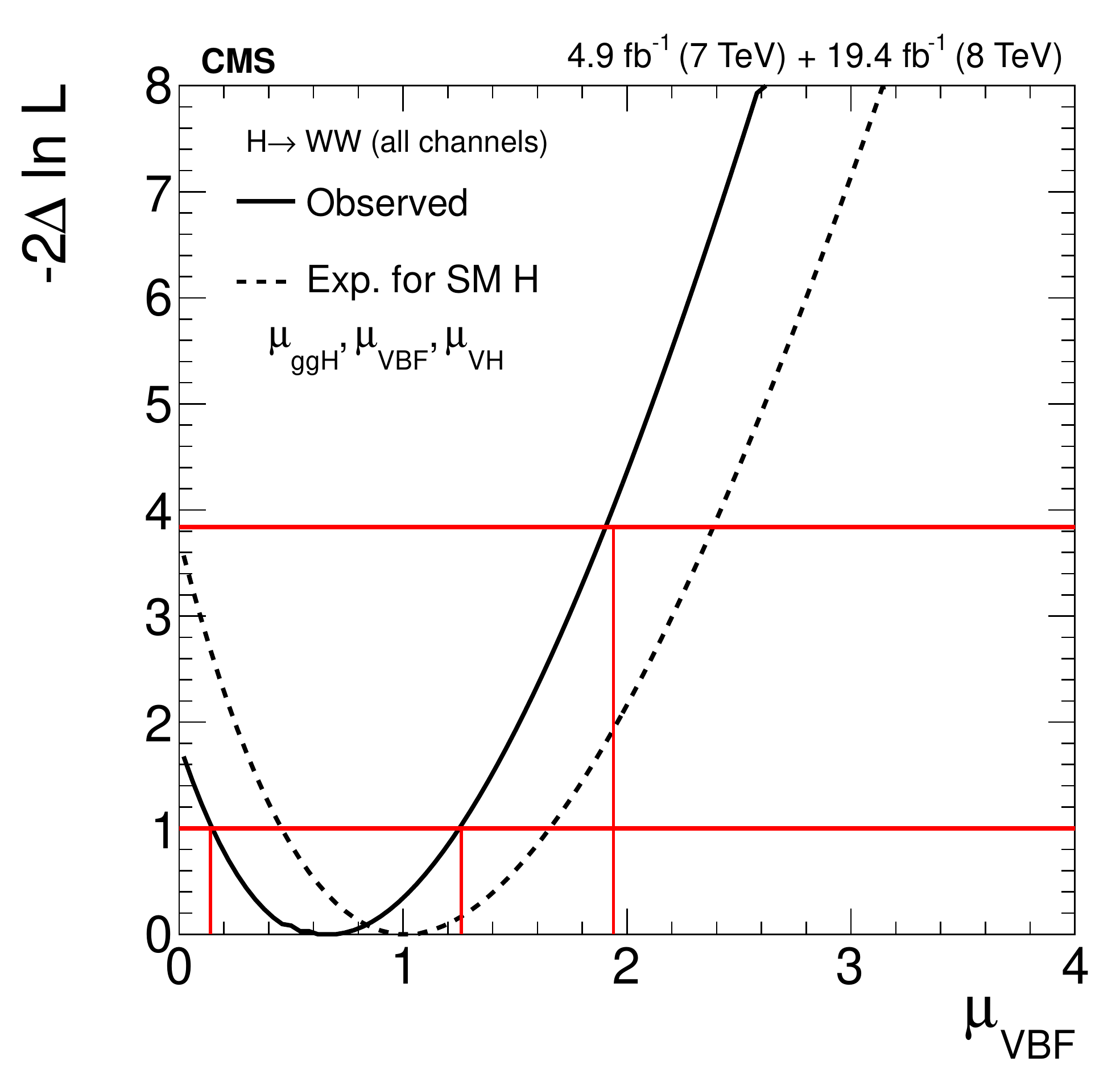}\\
\includegraphics[width=0.49\textwidth]{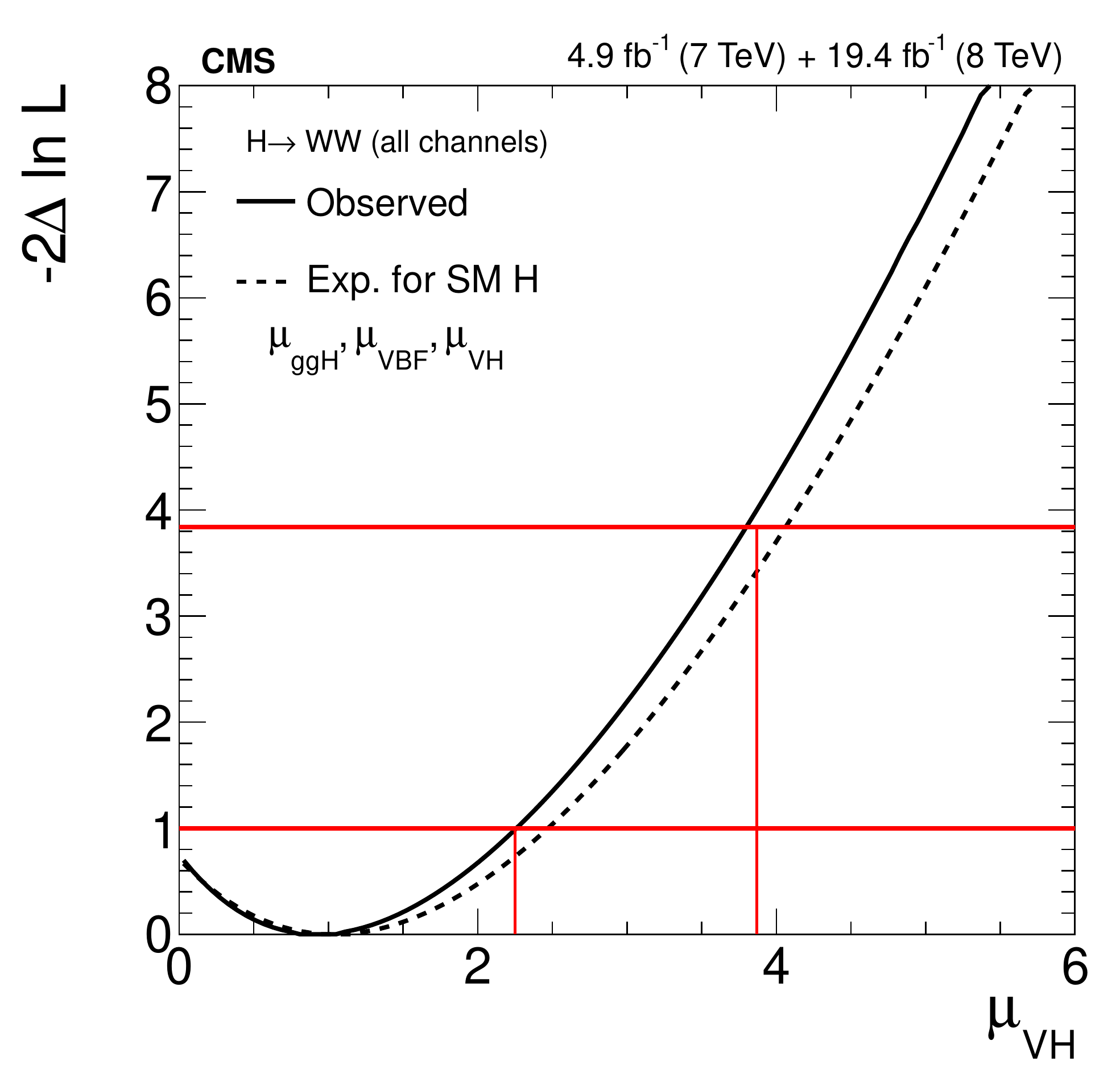}
 \caption{Expected and observed likelihood profiles for $\mHi = 125.6\GeV$ for the three
 production modes separately, $\Pg\Pg\PH$ (top left), VBF (top right), and $\V\PH$ (bottom). In each case,
 the modifiers for the other productions modes are profiled.
 The crossings with the horizontal line at $-2\Delta \, \ln L$ = 1 (3.84) define the 68\% (95\%) CL interval.
 \label{fig:mu_modes}}
 \end{center}
\end{figure}

The primary production mechanism contributing to the total cross section
for the SM Higgs boson is the $\Pg\Pg\PH$ process, with a smaller
fraction of the cross section coming from VBF and $\V\PH$ production.
Separating the $\Pg\Pg\PH$ process from the other contributions is particularly
relevant to explore the Higgs boson couplings, since in the first case the
coupling to the fermions of the virtual loop is involved, while in
the others tree-level couplings to vector bosons play a role.
The likelihood profiles for the signal strength modifiers associated with production
modes dominated by couplings to fermions ($\mu_{\Pg\Pg\PH}$) and vector
bosons ($\mu_{\mathrm{VBF},\V\PH}$) are shown at the 68\% and 95\% CL
in Fig.~\ref{fig:rvrf}. The expected and observed likelihood profiles
for $\mHi = 125.6\GeV$ for the three production modes, $\Pg\Pg\PH$, VBF, and $\V\PH$,
are shown separately in Fig.~\ref{fig:mu_modes}.

\begin{figure}[htbp]
\begin{center}
\includegraphics[width=0.49\textwidth]{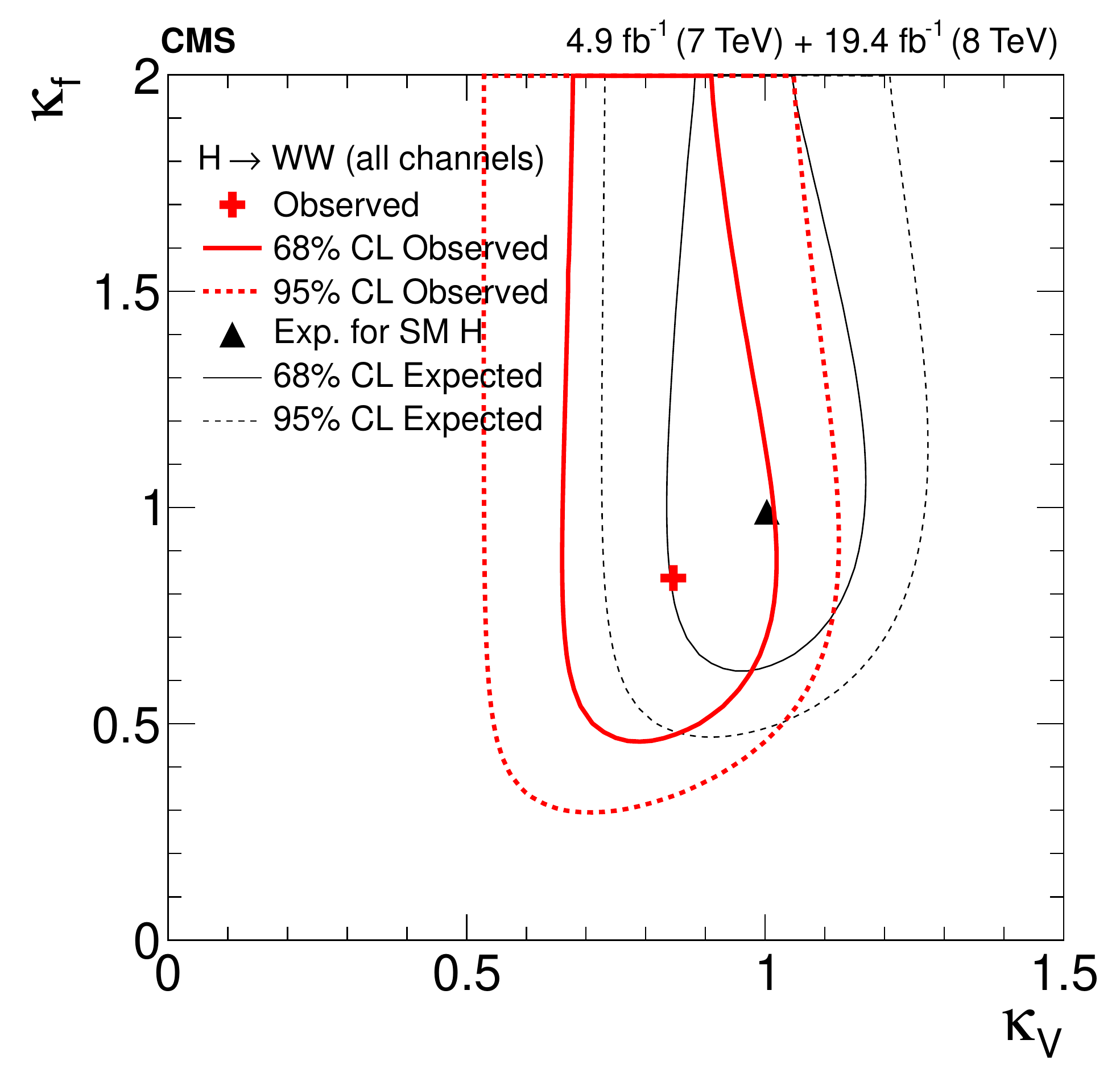}
\includegraphics[width=0.49\textwidth]{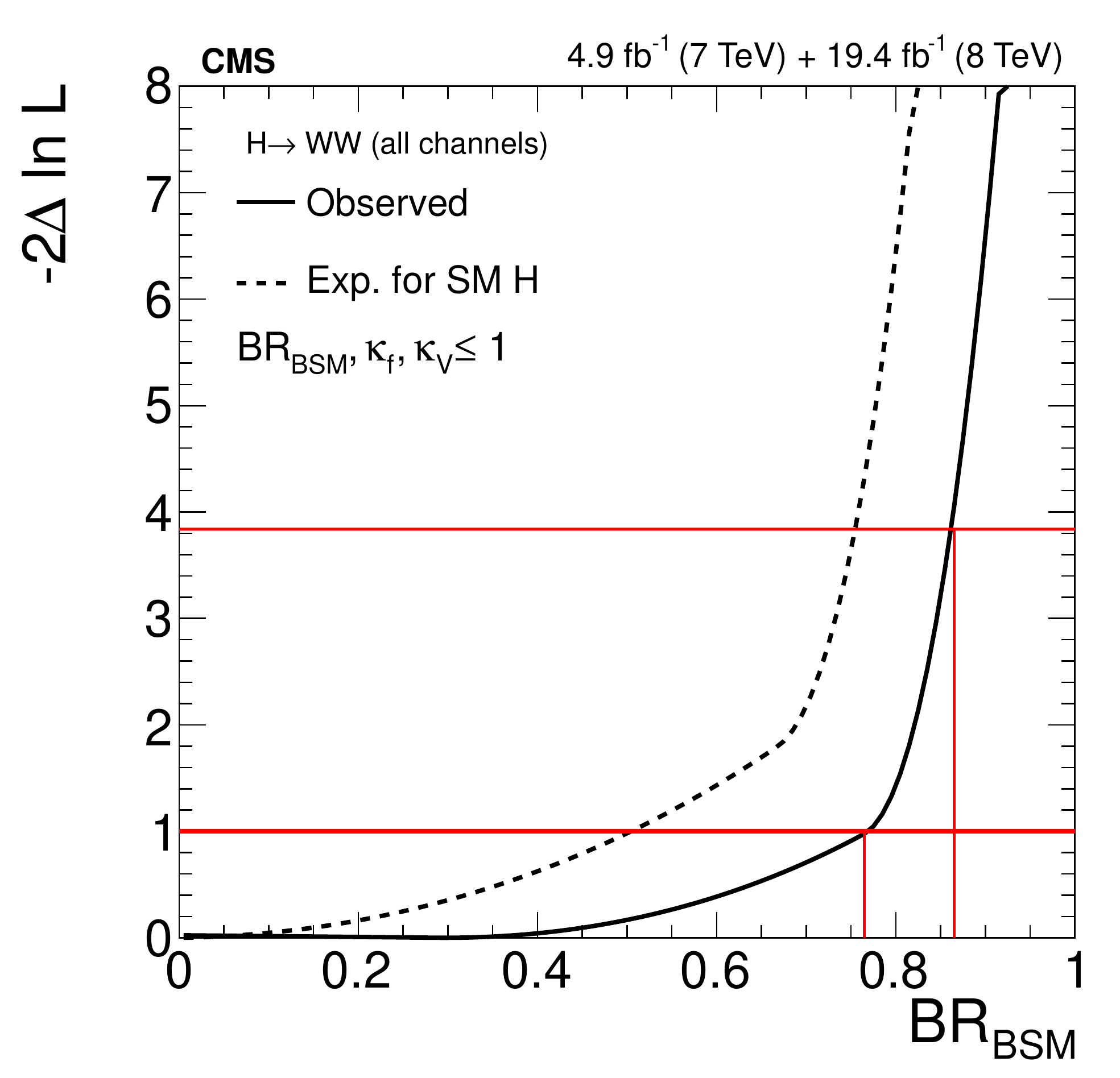}
 \caption{The two-dimensional likelihood of the $\kappa_{\mathrm V}$ and $\kappa_{\mathrm f}$ parameters (left). The observed
 value (red) and the SM expectation (black) are shown, together with the
 68\% (solid) and 95\% (dotted) CL contours. The likelihood scan
 versus $\mathrm{BR_{BSM}}$ (right) for the observed data (solid) and the
 expectation (dashed) in the presence of the SM Higgs boson with
 $\mHi = 125.6\GeV$ are shown.
 The crossing with the horizontal line at $-2\Delta \, \ln L$ = 1 (3.84) defines the 68\% (95\%) CL.
 The parameters $\kappa_{\mathrm V}$ and $\kappa_{\mathrm f}$ are profiled
 in the scan of $\mathrm{BR_{BSM}}$, with $\kappa_{\mathrm V} \leq 1$.
\label{fig:cvcf}}
 \end{center}
\end{figure}

A way to verify the theory prediction is to compare
the Higgs boson coupling constants to fermions and electroweak vector bosons with the SM
expectation~\cite{LHCHiggsCrossSectionWorkingGroup:2013tqa}. Two
coupling modifiers $\kappa_{\mathrm V}$ and $\kappa_{\mathrm f}$ are assigned to vector and
fermion vertices, respectively. They are then used to scale the
expected product of cross section and branching fraction to match the
observed signal yields in the data:

\begin{equation*}
\sigma \times \mathrm{BR}({\mathrm X} \to \PH\to \W\W)=\kappa_i^2\frac{\kappa_{\mathrm V}^2}{\kappa_{\mathrm H}^2} \sigma_{\mathrm{SM}} \times \mathrm{BR_{SM}}({\mathrm X} \to \PH \to \W\W),
\end{equation*}

where $\kappa_{\mathrm H}=\kappa_{\mathrm H}(\kappa_{\mathrm f},\kappa_{\mathrm V})$ is the total width modifier,
defined as a function of the two fit parameters $\kappa_{\mathrm V}$ and $\kappa_{\mathrm f}$.
The $\kappa_i$ modifier is $\kappa_\mathrm{f}$ for the $\Pg\Pg\PH$ process
and $\kappa_\mathrm{V}$ for the VBF and $\V\PH$ processes.
The assumption is made that only SM fields contribute to the total width.
In the context of this analysis the branching fraction is
always scaled by  $\kappa_{\mathrm V}^2/\kappa_{\mathrm H}^2$; the only direct coupling of the Higgs boson
to fermions occurs in the gluon fusion process, whose strength
is then parametrized by $\kappa_{\mathrm f}$. The two-dimensional likelihoods of the $\kappa_{\mathrm V}$ and $\kappa_{\mathrm f}$
parameters, for both the observed value and the SM expectation, are
shown in Fig.~\ref{fig:cvcf} (left).

An alternative general scenario can be obtained by allowing for non-vanishing
Higgs boson decays beyond the SM ($\mathrm{BR_{BSM}}$), while at the same time constraining
the fit to $\kappa_{\mathrm V} \leq 1$, which is well-motivated by the electroweak symmetry
breaking, with $\kappa_{\mathrm H}^2 = \kappa_{\mathrm H}^2(SM)/(1-\mathrm{BR_{BSM}})$.
The likelihood scan distribution versus $\mathrm{BR_{BSM}}$ is shown in Fig.~\ref{fig:cvcf} (right)
computed for this scenario. With these assumptions, an observed (expected) upper limit on
$\mathrm{BR_{BSM}}$ at the 95\% CL is set at 0.86 (0.75) using the
$\PH \to \W\W$ decay channel alone. This limit can be interpreted as, e.g., an
indirect limit on invisible Higgs boson decays.

\subsection{Spin and parity}
The different-flavor 0-jet and 1-jet categories are used to distinguish
between a $0^{+}$ boson like the SM Higgs boson and a $\spintwopmin$
boson or a pseudoscalar $0^{-}$ boson. The $\spintwopmin$ signal
templates for the $\Pg\Pg \to X$ and $\qqbar \to X$ processes, and the $0^-$
signal template for the $\Pg\Pg \to X$ process, are obtained from \textsc{jhugen}.

The results for the $\spintwopmin$ case are shown as a function of the
$\qqbar \to X$ component, $f_{\qqbar}$. The yields of the
$\Pg\Pg \to X$ and $\qqbar \to X$ processes are nominally taken from the simulated
samples assuming the SM Higgs boson cross section. A signal-plus-background model
is built for each hypothesis, based on two-dimensional templates in
$\mt$ and $\mll$, using the same bin widths
and data selection as for the low $\mHi$ case described in Section~\ref{sec:hww2l2n_01j}.
For the SM Higgs boson case, the signal templates derived from \POWHEG include
the  gluon fusion, VBF, and $\V\PH$ production modes. The background templates
are the same as in the SM Higgs boson search analysis.
The two-dimensional ($\mth$, $\mll$) distributions for the $0^{+}$ and $\spintwopmin$ hypotheses
are shown in Fig.~\ref{fig:hww01j_2D_0j} for the 0-jet category and in
Fig.~\ref{fig:hww01j_2D_1j} for the 1-jet category for the $8\TeV$ analysis. The distribution
of the two variables and the correlation between them clearly separates the two spin hypotheses, which
are related to the different $\ell\nu$ masses and $\ell\ell$ azimuthal angle distributions~\cite{JCPExpPaper2}.

For each hypothesis a binned maximum likelihood ($L$) fit is performed, to simultaneously
extract the signal strength and background contributions. This likelihood
fit model is the same as in the SM Higgs boson search. Fits are performed for both models,
and the likelihoods are calculated with the signal rates allowed to float independently for
each signal type. The test statistic,
$q = -2\,\ln(L_{J^P} /L_{0^+} )$, where $L_{0^+}$ and $L_{J^P}$
are the best-fit likelihood values for the SM Higgs boson and the alternative hypothesis
is then used to quantify the consistency of the two models with data. The expected separation between the
two hypotheses, defined as the median of $q$ expected under the $J^P$ hypothesis, is
quoted in two scenarios, when events are generated with a-priori expectation for the signal yields
($\sigma/\sigma_\mathrm{SM} \equiv 1$) and when the signal strength is determined from the fit to data
($\sigma/\sigma_\mathrm{SM} \approx 0.75$).

\begin{figure}[htbp]
\begin{center}
\includegraphics[width=0.49\textwidth]{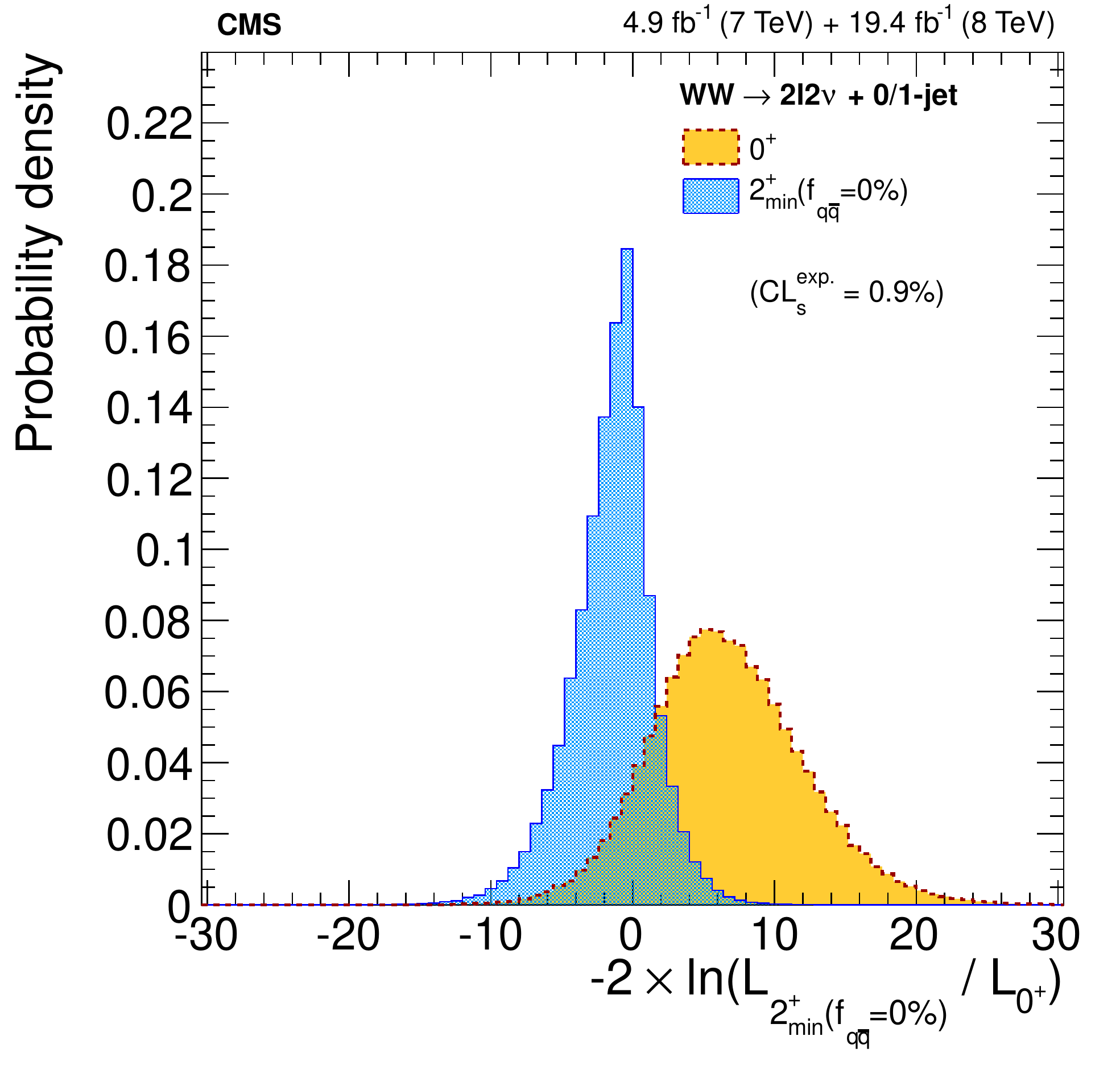}
\includegraphics[width=0.49\textwidth]{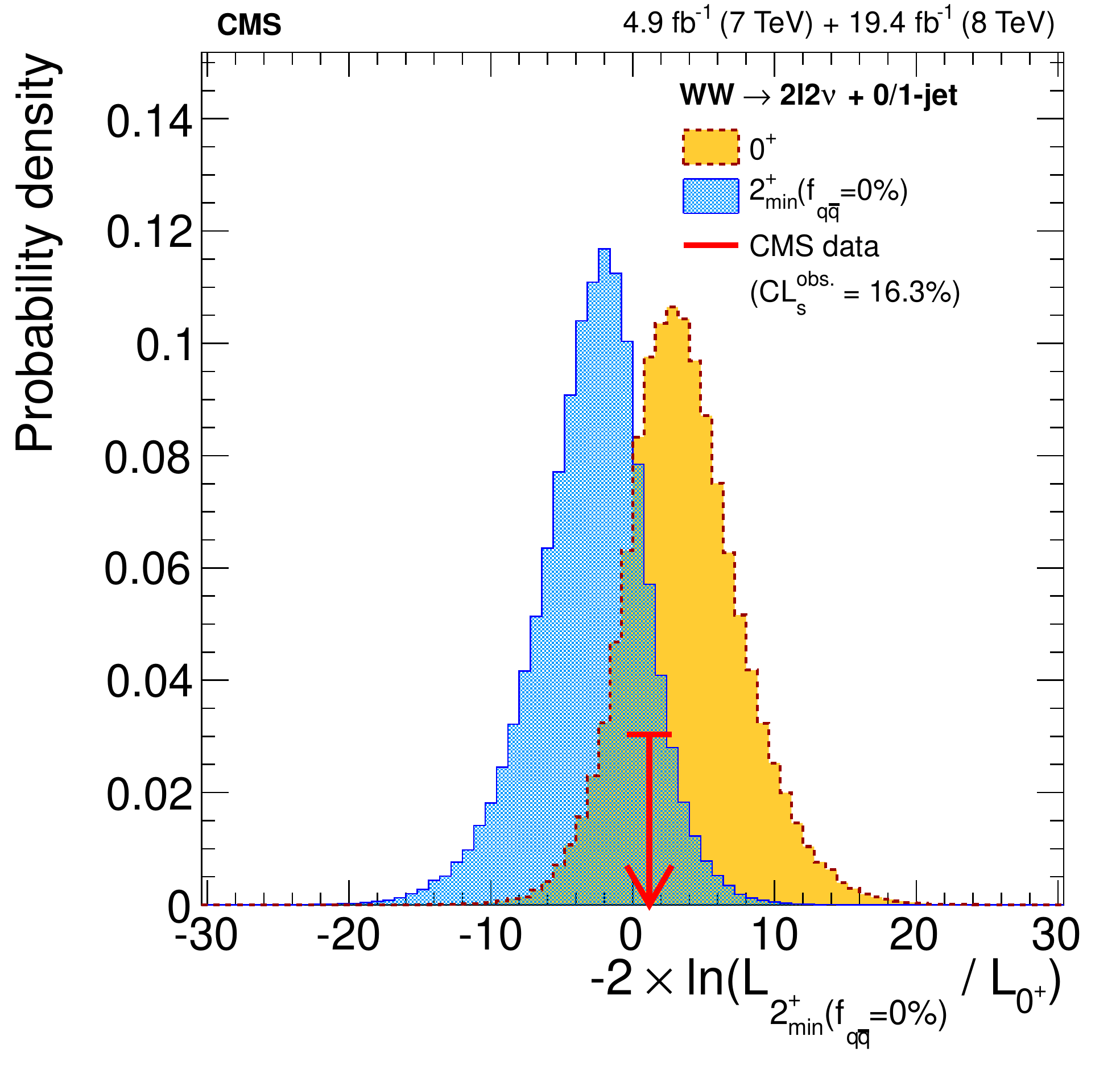}\\
\includegraphics[width=0.49\textwidth]{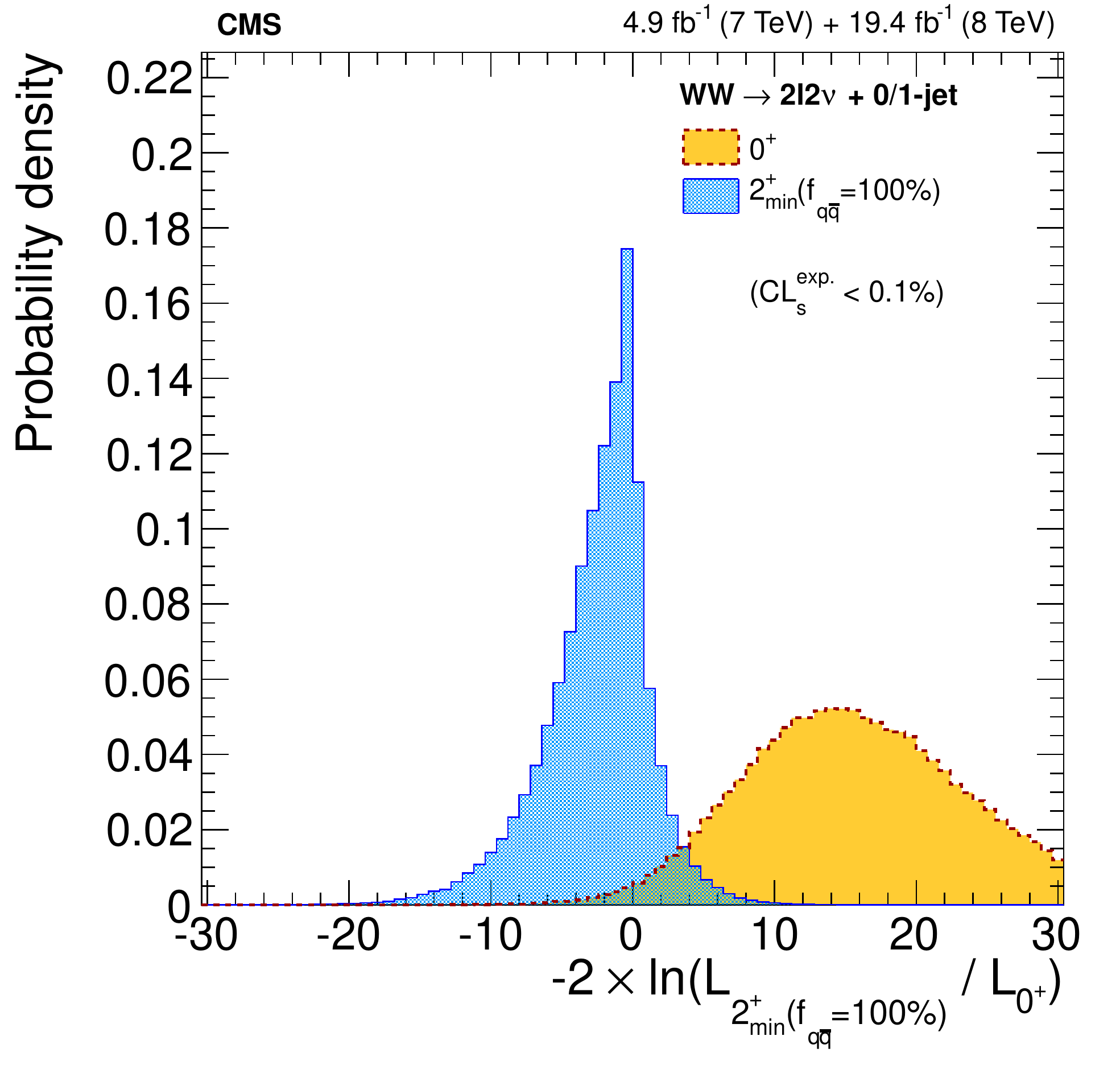}
\includegraphics[width=0.49\textwidth]{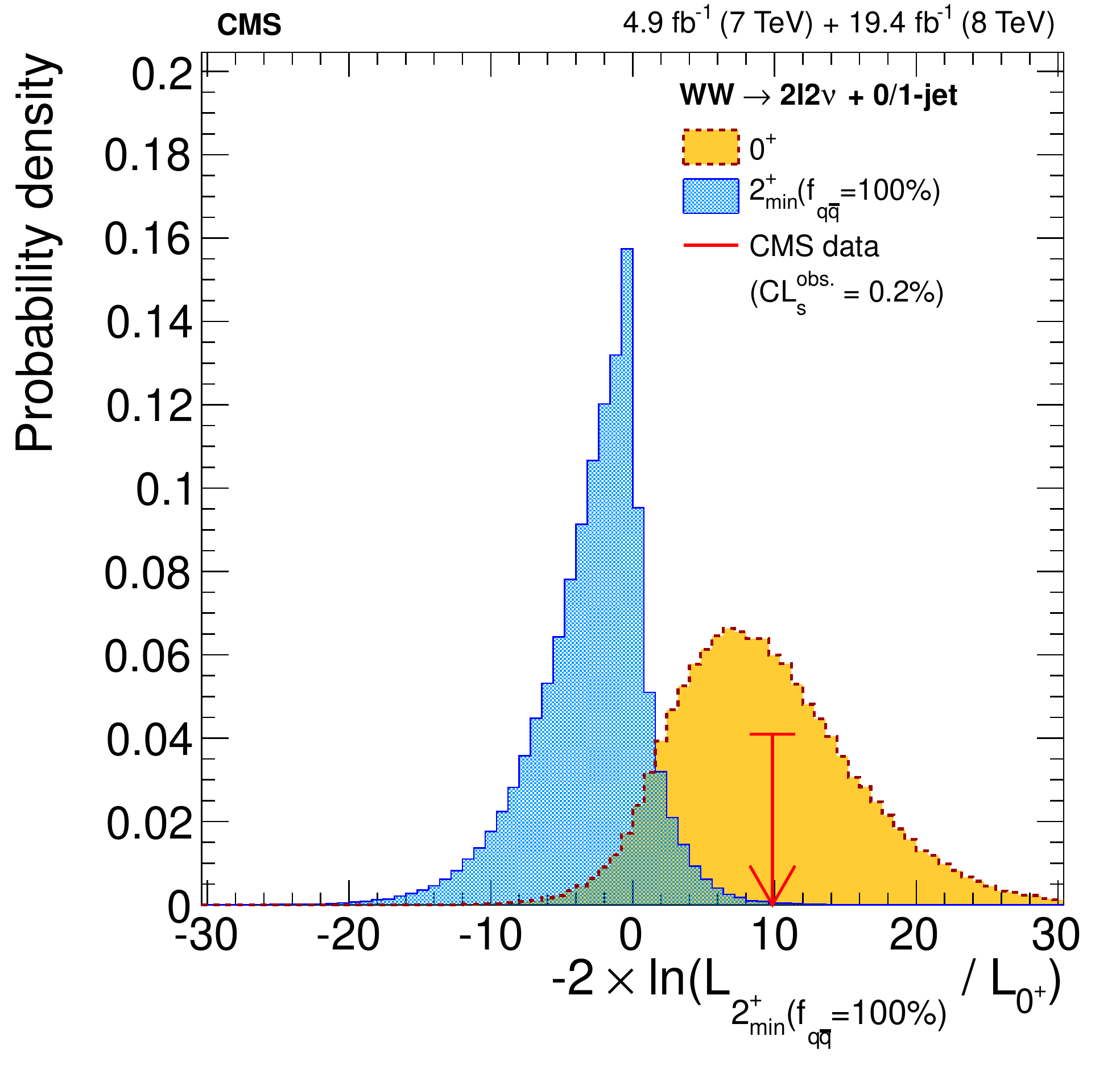}
 \caption{
   Distributions of $-2 \, \ln(L_{\spintwopmin}/L_{0^+})$, combining the 0-jet and 1-jet categories
   in the $\Pe\mu$ final state, for the $0^+$ and $\spintwopmin$ hypotheses at $\mHi=125.6\GeV$. The distributions
   are produced assuming $\sigma/\sigma_\mathrm{SM}$=1 (left) and using the $\sigma/\sigma_\mathrm{SM}$ value
   determined from the fit to data (right). The distributions are shown
   for the case $f_{\qqbar} = $0\% (top) and $f_{\qqbar} = $100\% (bottom).
   The observed value is indicated by the red arrow.
 \label{fig:JPCLimDist}}
 \end{center}
\end{figure}

\begin{figure}[htbp]
\begin{center}
\includegraphics[width=0.49\textwidth]{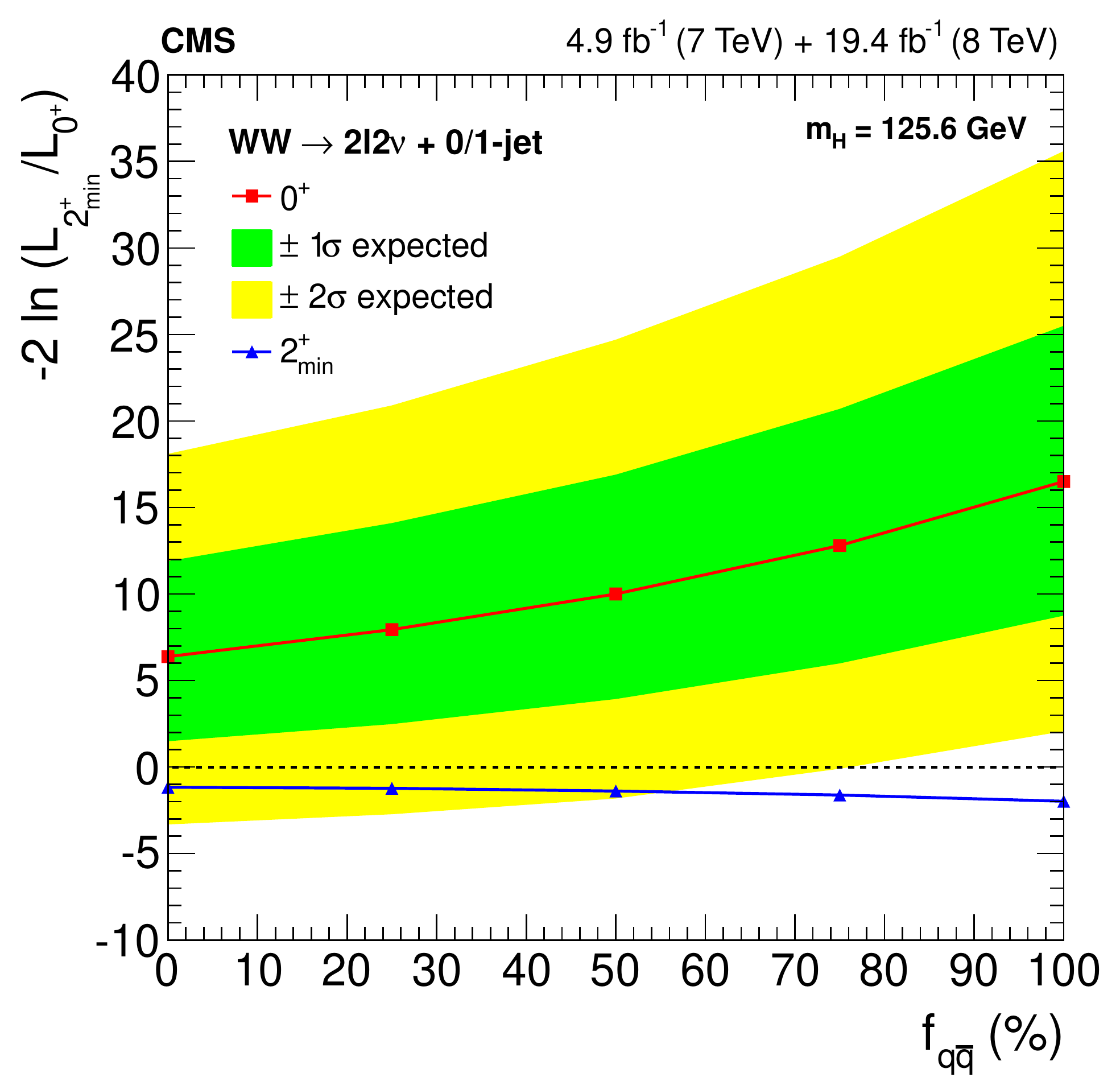}
\includegraphics[width=0.49\textwidth]{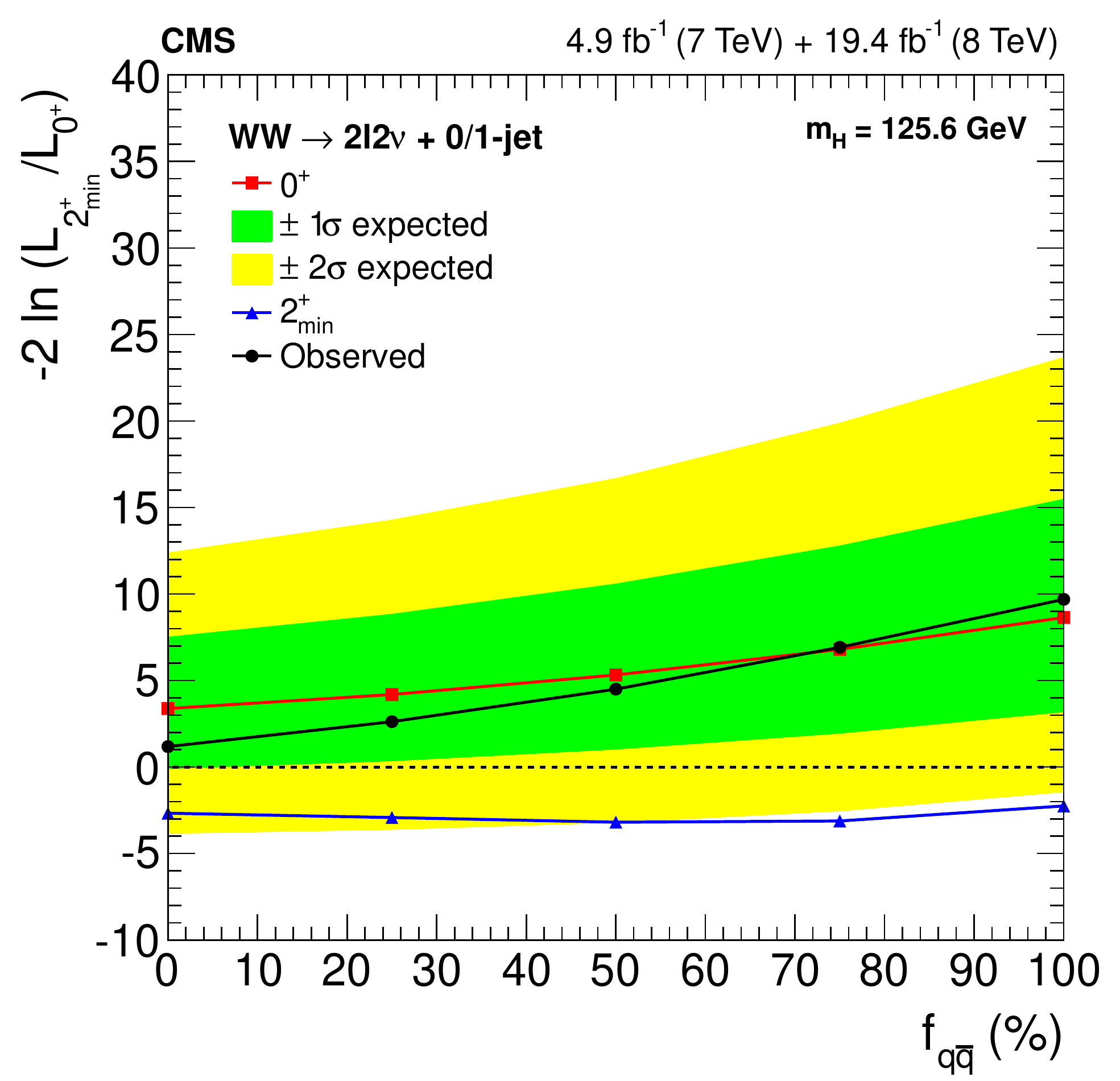}
 \caption{Median test statistic for the $0^+$ and $\spintwopmin$ hypotheses, as a function of $f_{\qqbar}$ of the
$\spintwopmin$ particle, assuming $\sigma/\sigma_\mathrm{SM}=1$ (left) and using the
$\sigma/\sigma_\mathrm{SM}$ value determined from the fit to data (right).
The observed values are also reported in the second case.
 \label{fig:JPCLim}}
 \end{center}
\end{figure}

The distributions of $q$ for the $0^+$ and $\spintwopmin$ hypotheses
at $\mHi=125.6\GeV$ for the two scenarios above and assuming
$f_{\qqbar} = $0\% or $f_{\qqbar} = $100\% are shown in Fig.~\ref{fig:JPCLimDist}.
Assuming $\sigma/\sigma_\mathrm{SM}=1$ for both hypotheses, the median test statistic
for the $0^+$ and $\spintwopmin$ hypotheses as well as its observed value,
as a function of $f_{\qqbar}$ of the $\spintwopmin$ particle is shown in
Fig.~\ref{fig:JPCLim} (left). The same results using the $\sigma/\sigma_\mathrm{SM}$
value determined from the fit to data are shown in Fig.~\ref{fig:JPCLim} (right).
In all cases the data favor the SM hypothesis with respect to the $\spintwopmin$ hypothesis.
The alternative hypothesis $\spintwopmin$ is excluded at a 83.7\%~(99.8\%) CL or higher
for $f_{\qqbar} = $ 0\%~(100\%) when the $\sigma/\sigma_\mathrm{SM}$ value determined from the fit to data is used.

\begin{figure}[htbp]
\begin{center}
\includegraphics[width=0.49\textwidth]{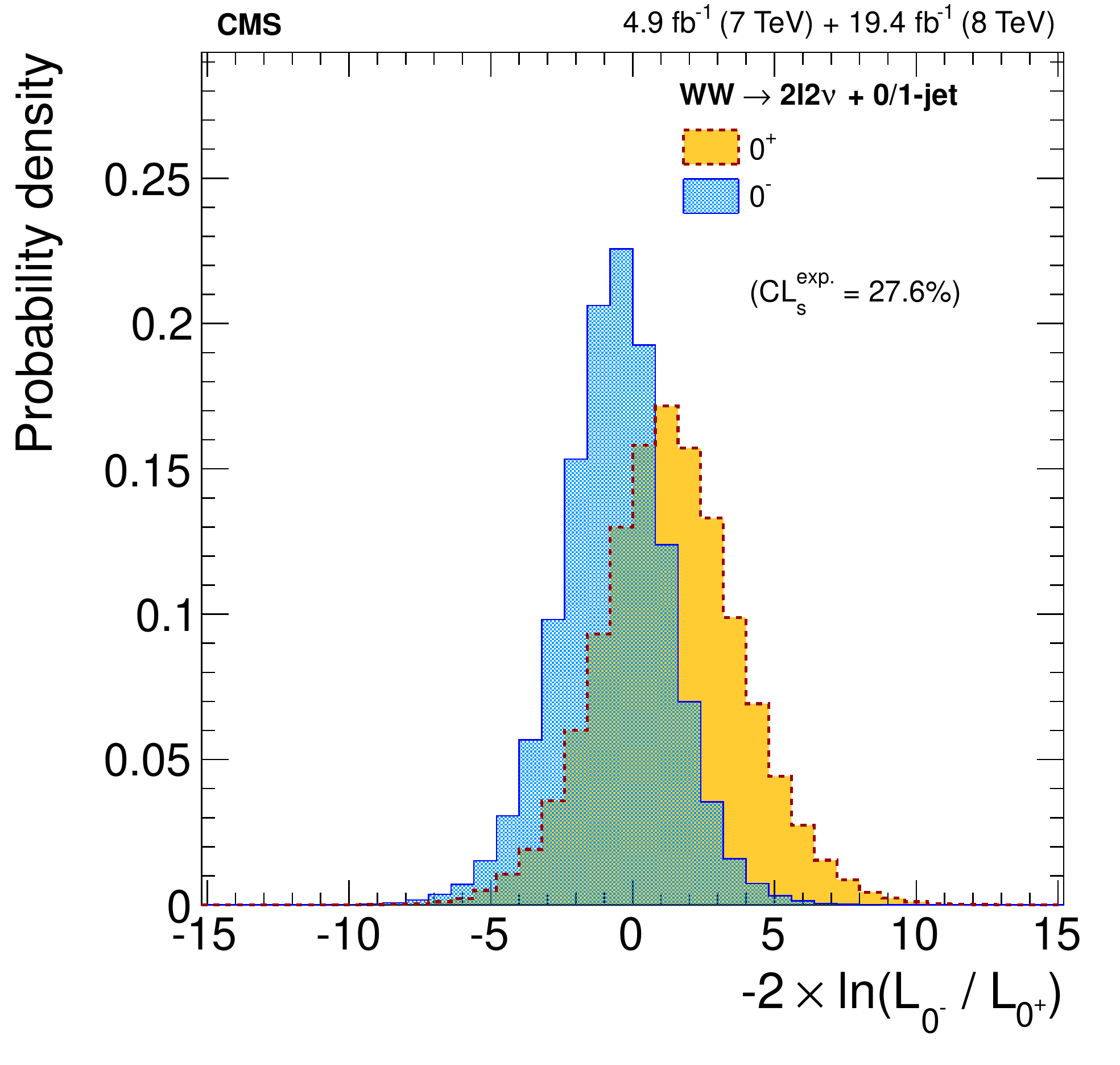}
\includegraphics[width=0.49\textwidth]{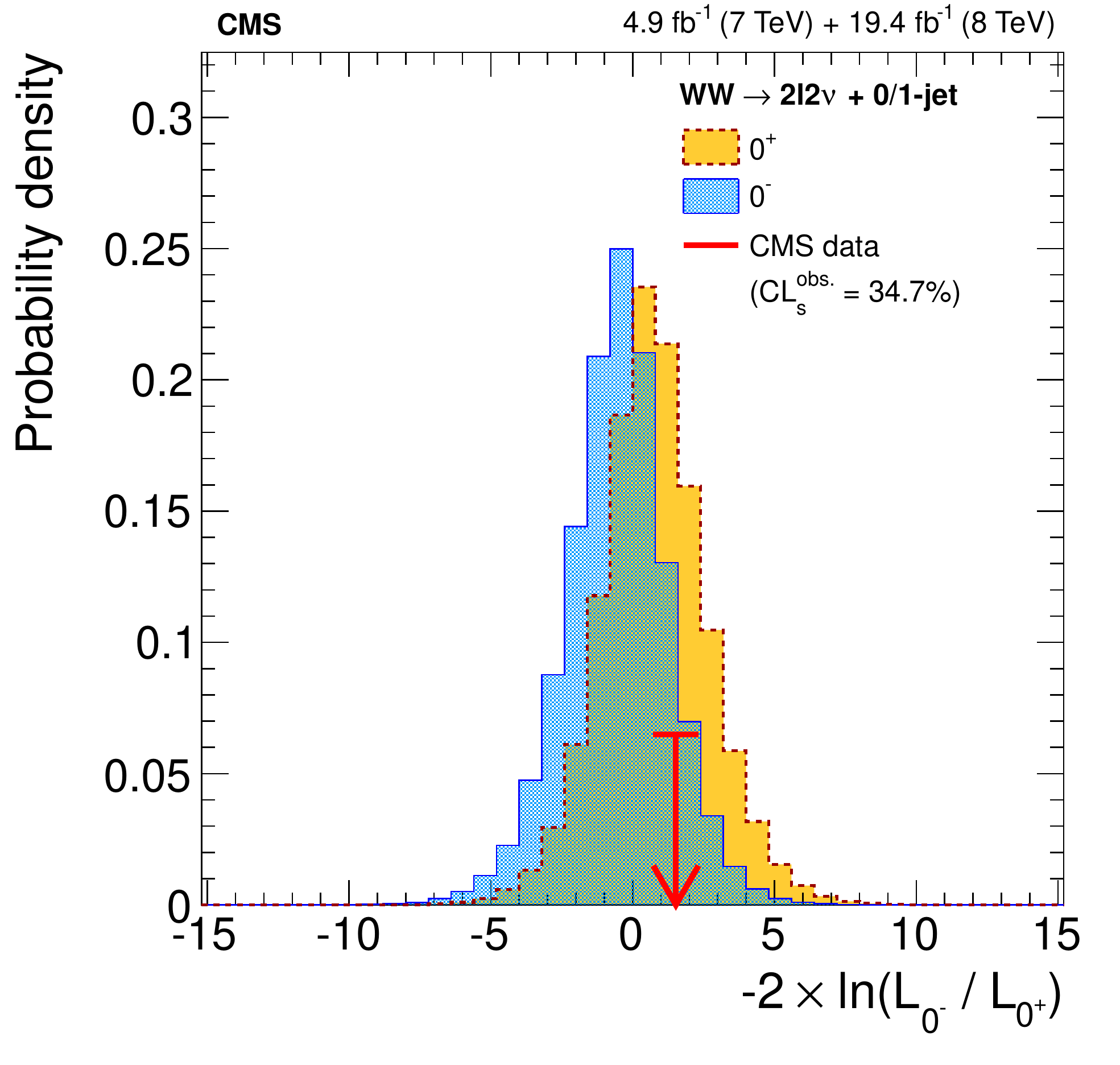}
 \caption{
   Distributions of $-2 \, \ln(L_{0^-}/L_{0^+})$, combining the 0-jet and 1-jet categories
   in the $\Pe\mu$ final state, for the $0^+$ and $0^-$ hypotheses at $\mHi=125.6\GeV$.
   The distributions are produced assuming $\sigma/\sigma_\mathrm{SM}$=1 (left) and using the signal strength
   determined from the fit to data (right).
   The observed value is indicated by the red arrow.
 \label{fig:JPCLimDist0}}
 \end{center}
\end{figure}

The same procedure described above is applied to perform a test of hypotheses between
a $0^{+}$ boson like the SM Higgs boson and a pseudoscalar $0^{-}$ boson. The average
separation between the two hypotheses is about one standard deviation, as shown in
Fig.~\ref{fig:JPCLimDist0}.
The alternative hypothesis $0^-$ is disfavored with a $\mathrm{CL_s}$ value of 34.7\%
when the $\sigma/\sigma_\mathrm{SM}$ value determined from the fit to data is used.
A summary of the list of models used in the analysis of the spin and parity hypotheses,
$J^P$, are shown in Table~\ref{tab:jpmodels} together with the expected and
observed separation $J^P/0^+$.

\begin{table}[htbp]
\centering
\topcaption{A summary of the models used in the analysis of the
spin and parity hypotheses. The expected
separation is quoted for two scenarios, where the value of $\sigma/\sigma_\mathrm{SM}$ for
each hypothesis is determined from the fit to data and where
events are generated with $\sigma/\sigma_\mathrm{SM}=1$.
The observed separation quotes consistency of the
observation with the $0^+$ model or $J^P$ model and corresponds to
the scenario where $\sigma/\sigma_\mathrm{SM}$ is determined from the fit to data.
The last column quotes the $\mathrm{CL_s}$ value that defines the minimum confidence level ($1 - \mathrm{CL_s}$) at which the $J^P$ model is excluded.
\label{tab:jpmodels}}
\begin{tabular}{cccccc}
\hline\hline
 $J^P$ model & $J^P$ production & Expected ($\sigma/\sigma_\mathrm{SM}=1$) &  obs. $0^+$  & obs. $J^P$ & $\mathrm{CL_s}$  \\
\hline
$\spintwopmin$     & $f_{\qqbar}$=0\%    &  1.8$\sigma$ (2.6$\sigma$)  & +0.6$\sigma$  & +1.2$\sigma$  &  16.3\% \\ 
$\spintwopmin$     & $f_{\qqbar}$=50\%   &  2.3$\sigma$ (3.2$\sigma$)  & +0.2$\sigma$  & +2.1$\sigma$  &   3.3\% \\ 
$\spintwopmin$     & $f_{\qqbar}$=100\%  &  2.9$\sigma$ (3.9$\sigma$)  & -0.2$\sigma$  & +3.1$\sigma$  &   0.2\% \\ 
$0^-$              & any                 &  0.8$\sigma$ (1.1$\sigma$)  & -0.5$\sigma$  & +1.2$\sigma$  &  34.7\% \\ 
\hline
\end{tabular}
\end{table}

\section{Summary}\label{sec:summary}
A search for the SM Higgs boson decaying to a $\W$-boson pair at the LHC has been reported.
The event samples used in the analysis correspond to
an integrated luminosity of $\usedLumiSeven$ and $\usedLumi$
collected by the CMS detector in pp collisions at $\sqrt{s} = 7$ and 8\TeV, respectively.
The $\WW$ candidates are selected in events with exactly two or three charged leptons.
The analysis has been performed in the Higgs boson mass range 110--600\GeV.
An excess of events is observed above background, consistent with
the expectations from the SM Higgs boson of mass around 125\GeV.
The probability to observe an excess equal or larger than the one seen, under the
background-only hypothesis, corresponds to a significance of 4.3 standard deviations
for $\ensuremath{m_\PH}$ = 125.6\GeV.
The observed $\sigma/\sigma_\mathrm{SM}$ value for $\mH = 125.6\GeV$ is $0.72^{+0.20}_{-0.18}$.
The spin-parity $J^P=0^+$ hypothesis is favored against a narrow resonance
with $J^P=2^+$ or $J^P=0^-$ that decays to a $\W$-boson pair.
This result provides strong evidence for a Higgs-like boson decaying
to a $\W$-boson pair.

\section*{Acknowledgements}
\hyphenation{Bundes-ministerium Forschungs-gemeinschaft Forschungs-zentren} We congratulate our colleagues
in the CERN accelerator departments for the excellent performance of the LHC and thank the technical and
administrative staffs at CERN and at other CMS institutes for their contributions to the success of the
CMS effort. In addition, we gratefully acknowledge the computing centres and personnel of the Worldwide
LHC Computing Grid for delivering so effectively the computing infrastructure essential to our analyses.
Finally, we acknowledge the enduring support for the construction and operation of the LHC and the CMS
detector provided by the following funding agencies: the Austrian Federal Ministry of Science and Research
and the Austrian Science Fund; the Belgian Fonds de la Recherche Scientifique, and Fonds voor Wetenschappelijk
Onderzoek; the Brazilian Funding Agencies (CNPq, CAPES, FAPERJ, and FAPESP); the Bulgarian Ministry of
Education and Science; CERN; the Chinese Academy of Sciences, Ministry of Science and Technology, and
National Natural Science Foundation of China; the Colombian Funding Agency (COLCIENCIAS); the Croatian
Ministry of Science, Education and Sport; the Research Promotion Foundation, Cyprus; the Ministry of
Education and Research, Recurrent financing contract SF0690030s09 and European Regional Development
Fund, Estonia; the Academy of Finland, Finnish Ministry of Education and Culture, and Helsinki Institute
of Physics; the Institut National de Physique Nucl\'eaire et de Physique des Particules~/~CNRS, and
Commissariat \`a l'\'Energie Atomique et aux \'Energies Alternatives~/~CEA, France; the
Bundesministerium f\"ur Bildung und Forschung, Deutsche Forschungsgemeinschaft, and Helmholtz-Gemeinschaft
Deutscher Forschungszentren, Ger\-ma\-ny; the General Secretariat for Research and Technology, Greece; the
National Scientific Research Foundation, and National Innovation Office, Hungary; the Department of
Atomic Energy and the Department of Science and Technology, India; the Institute for Studies in
Theoretical Physics and Mathematics, Iran; the Science Foundation, Ireland; the Istituto Nazionale di
Fisica Nucleare, Italy; the Korean Ministry of Education, Science and Technology and the World Class
University program of NRF, Republic of Korea; the Lithuanian Academy of Sciences; the Mexican Funding
Agencies (CINVESTAV, CONACYT, SEP, and UASLP-FAI); the Ministry of Business, Innovation and Employment,
New Zealand; the Pakistan Atomic Energy Commission; the Ministry of Science and Higher Education and
the National Science Centre, Poland; the Funda\c{c}\~ao para a Ci\^encia e a Tecnologia, Portugal;
JINR, Dubna; the Ministry of Education and Science of the Russian Federation, the Federal Agency
of Atomic Energy of the Russian Federation, Russian Academy of Sciences, and the Russian Foundation
for Basic Research; the Ministry of Education, Science and Technological Development of Serbia;
the Secretar\'{\i}a de Estado de Investigaci\'on, Desarrollo e Innovaci\'on and Programa
Consolider-Ingenio 2010, Spain; the Swiss Funding Agencies (ETH Board, ETH Zurich, PSI, SNF,
UniZH, Canton Zurich, and SER); the National Science Council, Taipei; the Thailand Center of
Excellence in Physics, the Institute for the Promotion of Teaching Science and Technology of Thailand,
Special Task Force for Activating Research and the National Science and Technology Development Agency
of Thailand; the Scientific and Technical Research Council of Turkey, and Turkish Atomic Energy Authority;
the Science and Technology Facilities Council, UK; the US Department of Energy, and
the US National Science Foundation.

Individuals have received support from the Marie-Curie programme and the European Research Council and
EPLANET (European Union); the Leventis Foundation; the A. P. Sloan Foundation; the Alexander von Humboldt
Foundation; the Belgian Federal Science Policy Office; the Fonds pour la Formation \`a la Recherche dans
l'Industrie et dans l'Agriculture (FRIA-Belgium); the Agentschap voor Innovatie door Wetenschap en
Technologie (IWT-Belgium); the Ministry of Education, Youth and Sports (MEYS) of Czech Republic; the
Council of Science and Industrial Research, India; the Compagnia di San Paolo (Torino); the HOMING
PLUS programme of Foundation for Polish Science, cofinanced by EU, Regional Development Fund; and
the Thalis and Aristeia programmes cofinanced by EU-ESF and the Greek NSRF.

\bibliography{auto_generated}   

\appendix

\section{Measurement of the \texorpdfstring{$\Wgstar$}{W gamma-star}~~cross~section scale factor}\label{app:wgstar}
The $\Wgstar$ electroweak process is included in standard CMS
simulations as a part of the $\WZ$ process using \MADGRAPH.
Nevertheless the low-mass dilepton region is not properly covered since the standard simulations
have a generator-level requirement at $m_{\gamma^*}>12\GeV$ and there
could be a significant rate of events below that threshold passing the selection criteria
described in Section~\ref{sec:objects}. Since the $\WZ$
and $\Wgstar$ processes may contribute as background to the Higgs boson signal whenever
one of the three leptons in the final state is not
selected, the low mass part of $\Wgstar$ background has been simulated using
\MADGRAPH, requiring two leptons each with $\pt>5\GeV$ and
no restrictions on the third one. Electron and muon masses have
been taken into account to properly simulate the kinematic cut-offs.
The key point is to observe the process in data and validate the simulation. In particular, the cross section of
the process needs to be measured to have a reliable prediction for the
background outside the control region.

The cases where the virtual photon decays into a pair of electrons or muons
have both been considered. The first is characterized by a cross section that is
about three times larger than the latter, since the production threshold,
defined by $m_{\Lep}$, is lower. In both cases, at
least one of the two leptons is soft, with an
average $\pt$ of $\sim$5\GeV. In the \ensuremath{\ell^{\pm}\Pep\Pem}
case the way of mimicking the signal is similar to that of the $\wgamma$
background, with the photon converting in the material close to the
interaction vertex, making the leptons look as though they were produced promptly.
For the \ensuremath{\ell^{\pm}\Pgmp\Pgmm} final state,
the low $\pt$ of the softest muon often prevents it from reaching the
muon detector and being correctly identified.

To measure the production rate of $\Wgstar$ in data, the
\ensuremath{\ell^{\pm}\Pgmp\Pgmm} final state has been studied, since
the large background from multijet production makes it difficult to extract the
$\Wgstar$ signal in the \ensuremath{\ell^{\pm}\Pep\Pem} case.
A region that has a high purity of $\Wgstar$ events is defined using the following
selection criteria:
\begin{itemize}
\item
the muons associated with the virtual photon need to have opposite signs.
In the $3\mu$ final state, the opposite-sign pair
with the lowest mass is assumed to originate from the $\gamma^{*}$;
\item
$m_{\mu^\pm\mu^\mp}<12$\GeV is required;
\item since events have two muons very close to each other,
the muon isolation is redefined to exclude muons from the isolation
energy calculation;
\item
to suppress the top-quark background, events with more than two reconstructed jets
are rejected, and events with at least one jet will be rejected if that jet is b-tagged;
\item
to suppress the multijet background, the minimum transverse mass of each lepton and $\vmet$
must be larger than 25\GeV, and the transverse mass of the lepton associated with the $\W$ boson
and $\vmet$ must be larger than 45\GeV;
\item the $\JPsi$ meson decays are rejected by requiring
$\abs{m_{\mu^\pm\mu^\mp}-m_{\JPsi}}> 0.1\GeV$. There is no need to apply
a requirement against Upsilon decays due to the very small cross section.
\end{itemize}

\begin{figure}[htbp]
  \begin{center}
  \includegraphics[width=0.49\textwidth]{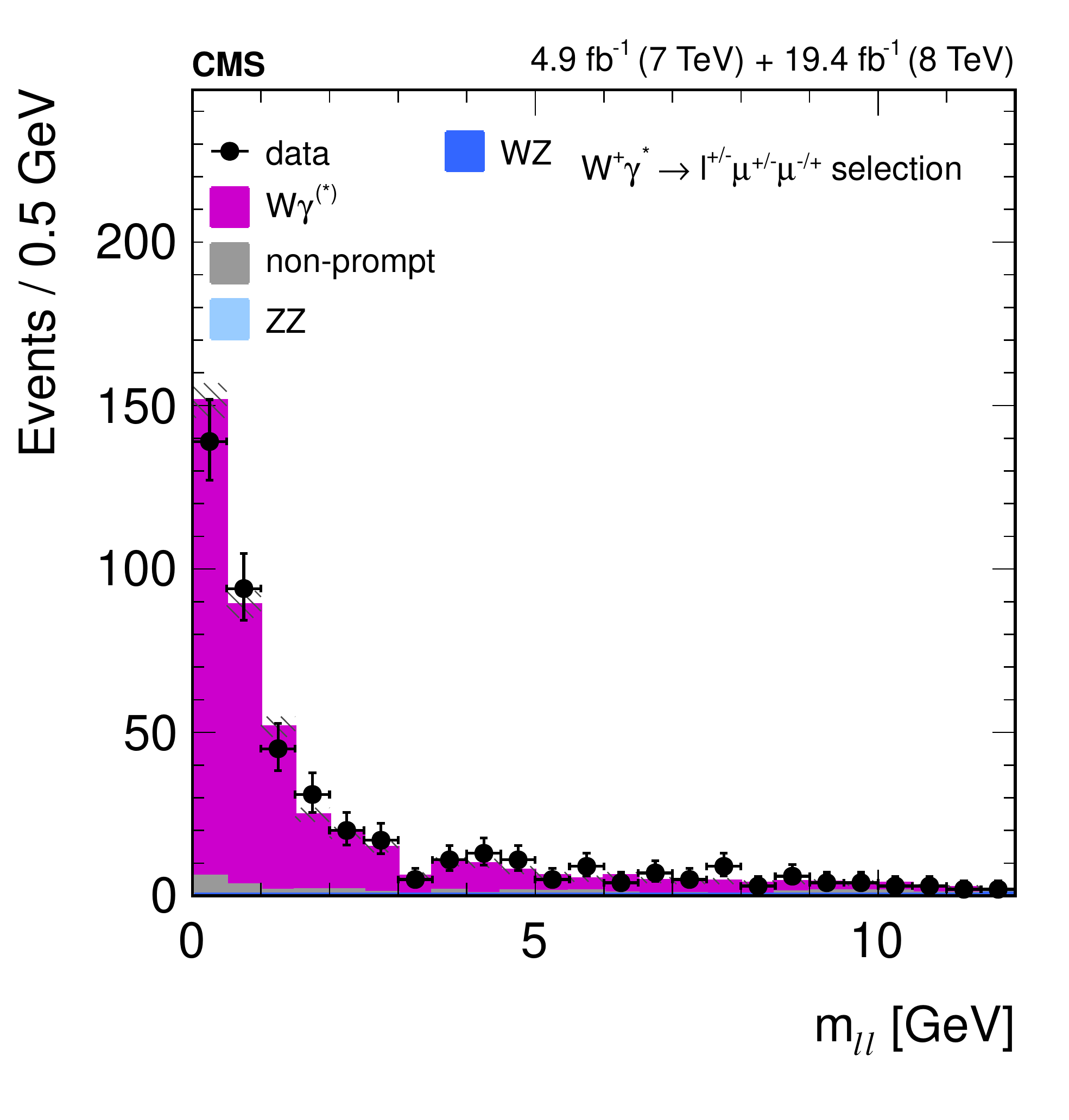}
    \caption{The $\mll$ mass distribution for opposite-sign muons after the $\Wgstar$ selection.
    The $\Wgstar$ contribution is normalized to match the data.}
    \label{fig:wg3l_mll}
  \end{center}
\end{figure}

The contribution from other background processes is rather small.
The only process which is not completely negligible is $\Wjets$,
as shown in Fig.~\ref{fig:wg3l_mll}.

The measured $K$-factor with respect to the LO cross section is around 1.5,
consistent with observations involving other electroweak processes computed at LO. This gives
further confidence on the accuracy of the simulation.
Some disagreement is observed between data and simulation in the virtual photon mass shape,
due to the mismodeling of the reconstruction efficiency of close-by muons at very low $\pt$.
To account for this difference in the normalization measurement,
the $K$-factor has been computed in different regions of the mass spectrum and compared to that
obtained from the full range. The same analysis is performed in four independent categories:
events with $m_{\mu^\pm\mu^\mp}<2\GeV$ and $2 \leq m_{\mu^\pm\mu^\mp}<12\GeV$, in both
$\ell^{\pm}\Pgmp\Pgmm$ final states. The average spread is taken as
systematic uncertainty, leading to a $K$-factor value of $1.5\pm0.5$.

\section{Estimation of the \texorpdfstring{$\wgamma$}{W gamma}~~background~template shapes}\label{app:wgamma}
\begin{figure}[htbp]
  \begin{center}
  \includegraphics[width=0.45\textwidth]{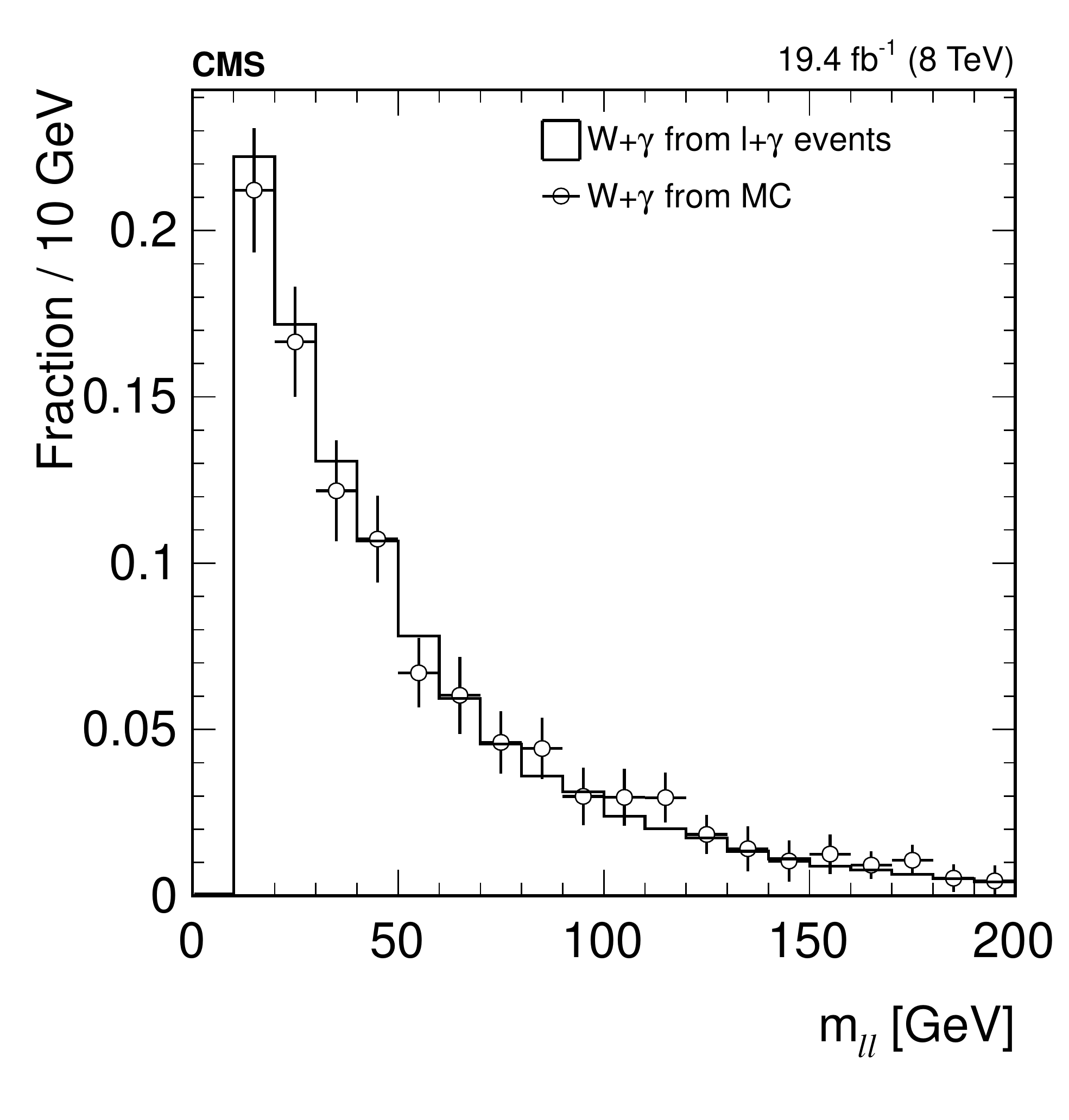}
  \includegraphics[width=0.45\textwidth]{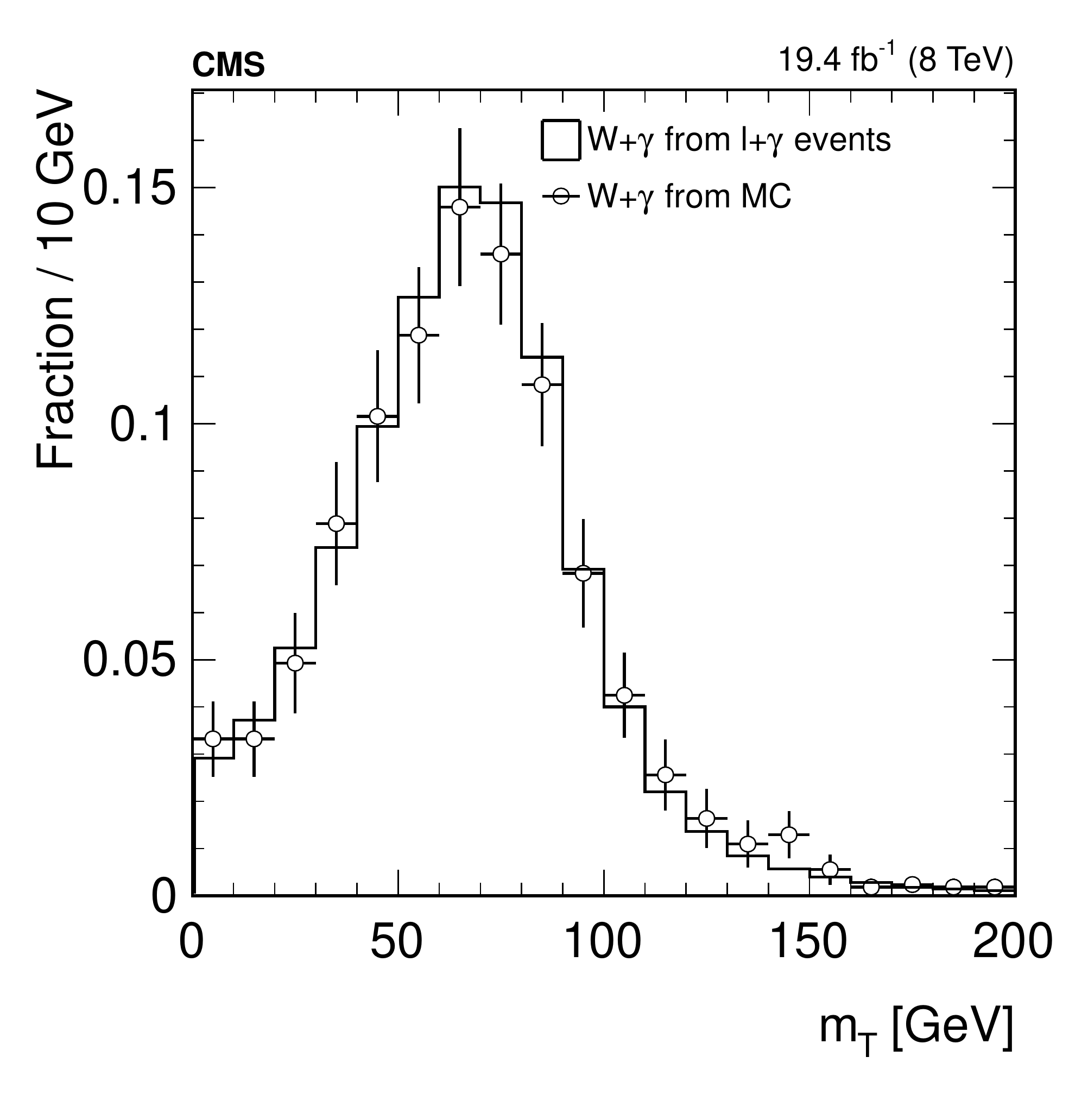}
    \caption{The $\mll$ (left ) and $\mt$ (right) distributions for the
    $\W+\gamma$ process in events passing the dilepton selection. The dots show the
    distribution from simulated events, while the histogram shows the distribution
    from a data sample with a lepton and a photon, which has about 200 times more events.}
    \label{fig:vg_mx}
  \end{center}
\end{figure}

In the dilepton final states, the $\wgamma$ background normalization
is taken from simulated samples, while the distributions of the final discriminant variables
are taken from data. To obtain the shapes, a sample of events with a lepton and an
identified photon is used. For the photon the same counting selection as applied in
ref.~\cite{CMSPaperCombination} is used.
The ratio of the photon-to-lepton identification efficiency as a function
of the photon $\eta$ and $\pt$ is used to properly weight the lepton-photon
event sample. The possible background contamination from non-prompt photons or
leptons shows a negligible effect on the shape of the distributions relevant for the
analysis. The $\mll$ and $\mt$ distributions for the $\W\gamma$ process in
events at the dilepton selection level as described in Section~\ref{sec:objects}
for simulated events and from a sample with a lepton and a photon
are shown in Fig.~\ref{fig:vg_mx}. The lepton-photon sample has about 200 times
more events than the simulated sample. Good agreement between the distributions
is observed.

\section{Estimation of the Drell--Yan background in the same-flavor dilepton final states}\label{app:dyll}
A method based on measurements in data is used to estimate the $\dyll$
contributions in the same-flavor $\ell^+\ell^-$ final states.
The expected contributions from $\dyll$ events outside a region around
the $\Z$ mass in data can be estimated by counting the number of events near
the $\Z$ mass region in data, subtracting from it the non-$\Z$
contributions, and scaling it by a ratio $R_{\text{out}/\text{in}}$ defined as the
fraction of events outside and inside the $\Z$ mass region in the
simulation. The $\Z$ mass region is defined as $\abs{\mll-m_{\Z}}<7.5\GeV$.
Such a tight window is chosen to reduce the non-$\Z$
contributions from top-quark and multi-boson backgrounds.
The non-$\Z$ contributions close to the $\Z$ mass region in
data are estimated from the number of events in the $e^\pm\mu^\mp$
final state $N_{\text{in}}^{\Pe\mu}$, applying a correction factor that
accounts for the difference in selection efficiency between
electrons and muons $k_{\Pe\Pe/\mu\mu}$.
The $R_{\text{out}/\text{in}}$ factor can be estimated both from simulated events and
data. In simulation it is defined as the ratio
$N_{\text{out}}^{\mathrm{MC}}/N_{\text{in}}^{\mathrm{MC}}$.

The number of Drell--Yan events in the signal region is therefore:

\begin{eqnarray*}
N_{\text{out}}^{\ell\ell,\text{exp}} = R_{\text{out}/\text{in}}^{\ell\ell}(N_{\text{in}}^{\ell\ell} - \frac{1}{2}N_{\text{in}}^{\Pe\mu}k_{\ell\ell}),
\label{eq:dyest}
\end{eqnarray*}

where $k_{\Pe\Pe} = \sqrt{\frac{N_{\text{in}}^{\Pe\Pe,\text{loose}}}{N_{\text{in}}^{\mu\mu,\text{loose}}}}$ for
$\dyee$ and $k_{\mu\mu} = \sqrt{\frac{N_{\text{in}}^{\mu\mu,\text{loose}}}{N_{\text{in}}^{\Pe\Pe,\text{loose}}}}$
for $\dymm$. The factor $\frac{1}{2}$ comes from the relative branching fraction between the
$\ell\ell$ and $\Pe\mu$ final states. In the $k_{\ell\ell}$ calculation, the selection on the missing
transverse energy is loosened to increase the available number of events under the $\Z$ peak. The
value of $k_{\Pe\Pe}$ is about 0.8, with a very loose dependence both on the
center-of-mass energy and jet category.

The $\Z\Z$ and $\W\Z$ ($\Z\V$) processes contribute to the events in the
$m_{\ell\ell}$ control region dominated by the Drell--Yan.
The contribution from $\Z\V$ becomes comparable to that of $\dyll$
after a tight $\pmet$ selection, since those events contain genuine
$\vmet$ for which the detector simulation is reliable. The expected
$\Z\V$ peaking contribution is subtracted from the yield in the $\Z$ peak using the
simulation. The $\Z\V$ events without $\met$ requirements are suppressed
by the same large factor as the Drell--Yan ones, and
therefore their contribution at the level of the final selection is as
negligible as it would be in the yield at the $\Z$ peak without $\met$ requirement.

When considering the full selection the Drell--Yan and $\Z\V$ components
allow for the extrapolation from control region to signal region to be different
for the two processes.

This $\dyll$ estimation method relies on the assumption that the dependence
of the ratio $R_{\text{out}/\text{in}}$ on the $\met$ requirement is relatively flat. On the other hand,
the value of $R_{\text{out}/\text{in}}$ changes as a consequence of the different kinematic
requirements applied to select the Higgs boson signal regions for different
Higgs boson mass hypotheses. Therefore $R_{\text{out}/\text{in}}$ is evaluated applying
selection requirements close to the full Higgs boson selections: all requirements
are applied except for variables depending on $\met$. As no
statistically significant difference is observed between the $\Pe\Pe$ and $\mu\mu$ final states,
both of them are combined.

\begin{figure}[htbp]
  \begin{center}
 \includegraphics[width=0.45\textwidth]{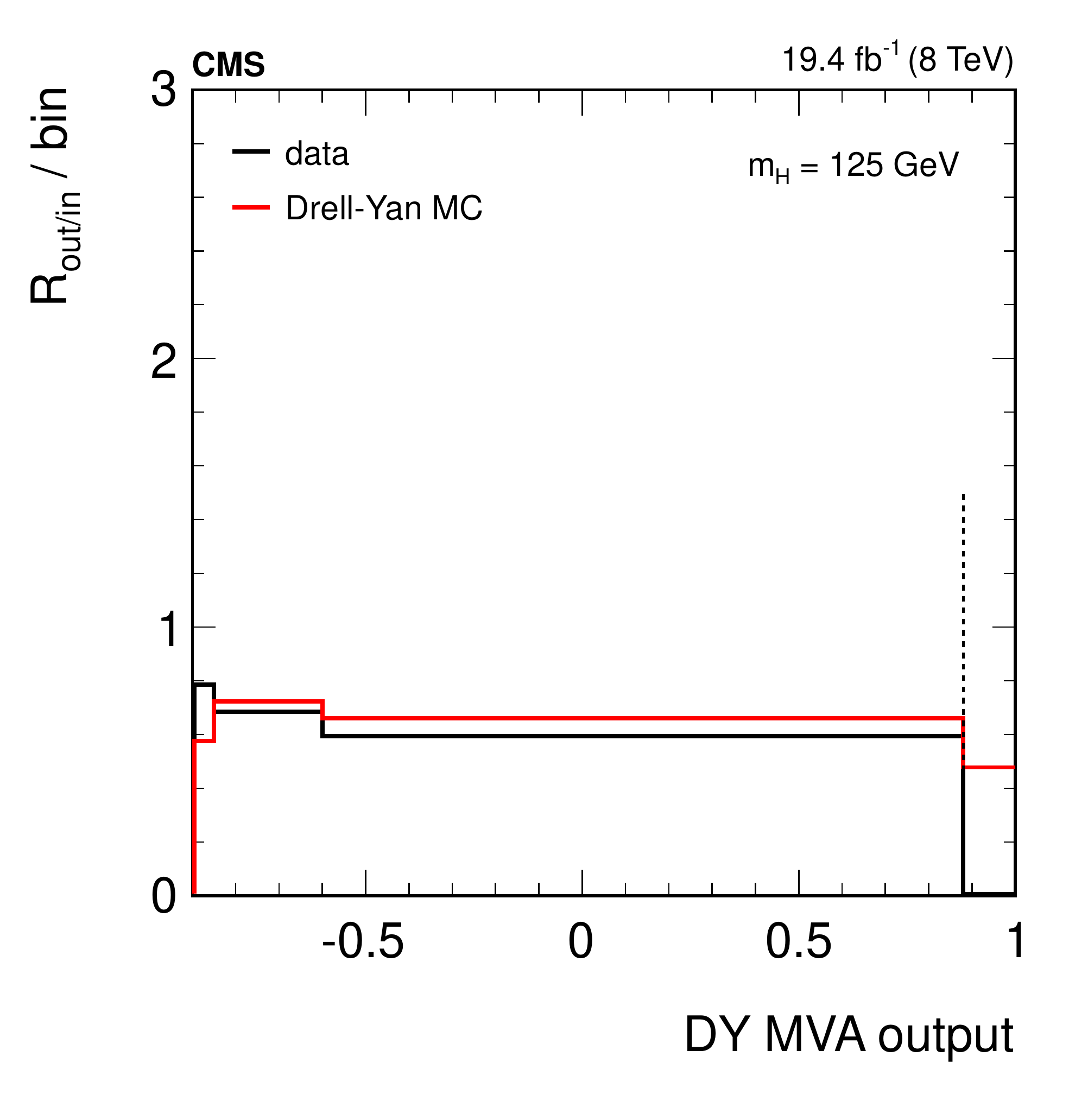}
 \includegraphics[width=0.45\textwidth]{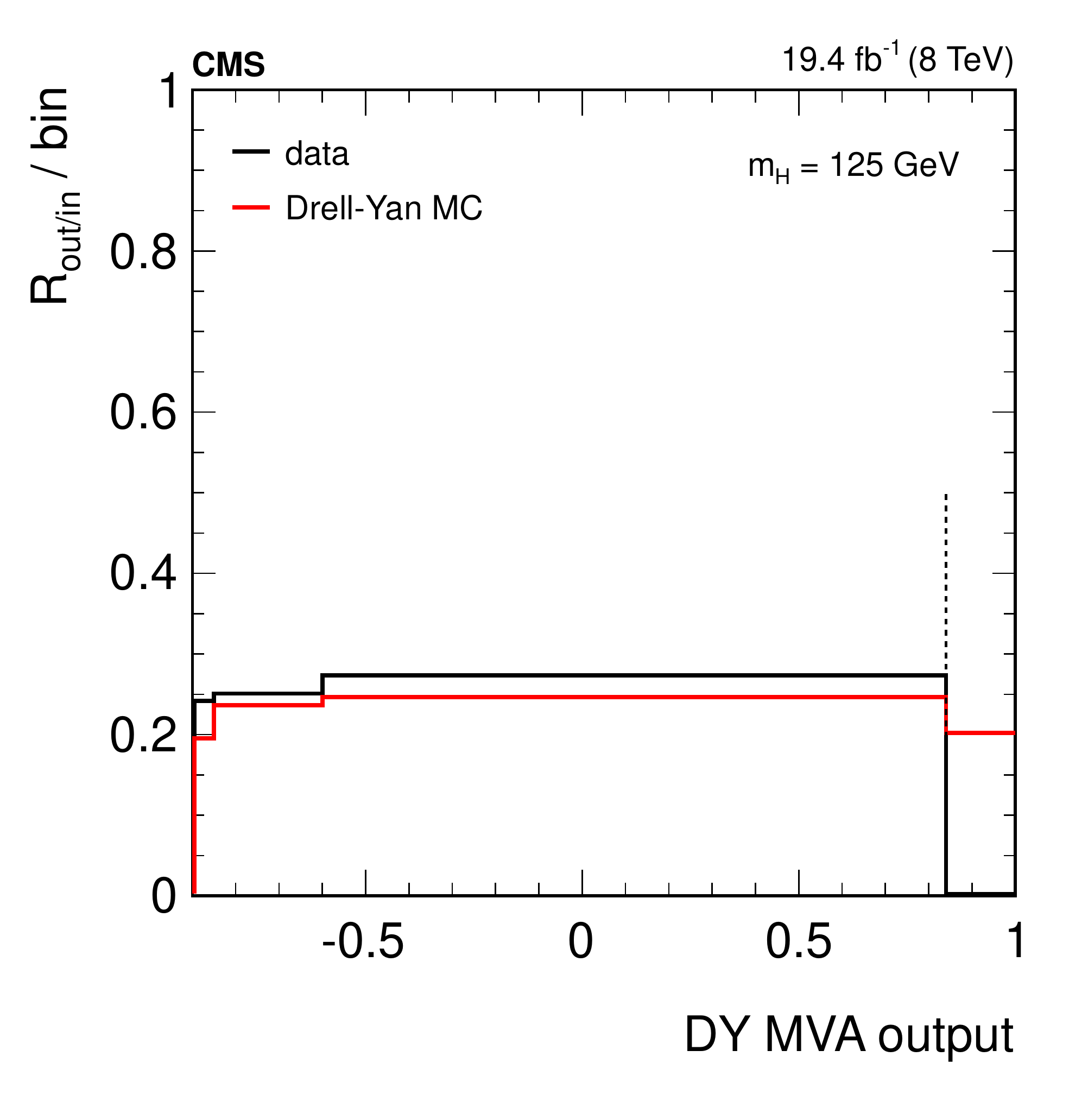}
    \caption{The $R_{\text{out}/\text{in}}$ values as a function of the multivariate Drell--Yan output variable in the 0-jet (left)
    and 1-jet (right) categories for the $\mHi= 125\GeV$ counting analysis at $\sqrt{s} = 8\TeV$. High output
    values are signal-like events, while low output values are more likely to be Drell--Yan events.
    The vertical dashed line indicates the minimum threshold on the discriminant
    value used to select events for the analysis, which is 0.88 for the 0-jet and
    0.84 for the 1-jet category. The dependence of the $R_{\text{out}/\text{in}}$
    ratio on the Drell--Yan discriminant value and the agreement between the data and
    the simulation are studied in the regions below this threshold.}
    \label{fig:Routin_0Jet_mH125_19467pb_dy}
  \end{center}
\end{figure}

The $R_{\text{out}/\text{in}}$ value is cross-checked in data as well. After the full selection, and after all efficiency corrections,
background processes contribute equally to $\Pe\Pe$, $\Pe\mu$, $\mu\Pe$, and $\mu\mu$ final states.
On the other hand, Drell--Yan only contributes to the $\Pe\Pe$ and $\mu\mu$ final states. Therefore the $\Pe\mu$ and $\mu\Pe$
contributions can be subtracted from the $\Pe\Pe$ and $\mu\mu$ samples to obtain an estimate of the
Drell--Yan background. The $R_{\text{out}/\text{in}}$ values as a function of the multivariate Drell--Yan output variable,
described in Section~\ref{sec:objects}, in the 0-jet and 1-jet categories for the $\mHi= 125\GeV$ counting analysis at $\sqrt{s} = 8\TeV$
are shown in Fig.~\ref{fig:Routin_0Jet_mH125_19467pb_dy}.

\section{Estimation of top-quark backgrounds in the dilepton final states}\label{app:top}
In the dilepton analysis, the top-quark-induced background originates from $\ttbar$ and
$\tw$ processes~\cite{wt7TeVPaper}, the latter being especially important
in the 0-jet category. A consistent theoretical description of the two
processes at higher orders is not straightforward to attain
as already at NLO some $\tw$ diagrams coincide with LO $\ttbar$ ones.
The simulated samples
used in the analysis exploit an approach recently
proposed, which addresses the overlap by
discarding the common diagrams from the $\tw$ process either at the
amplitude level (``diagram removal") or at the cross section level
(``diagram subtraction"). The former is considered the default
scheme, whereas the latter is used as a cross-check.

The top-quark background is estimated at the $\W\W$ selection level where
a common scale factor for the $\ttbar$ and $\tw$ simulated samples is
computed.  Once properly normalized, those samples are used to
predict the corresponding yields after the mass-dependent Higgs boson
selection requirements in the counting analyses and to produce the templates
in the shape-based analyses.

The procedure for top-quark background estimation can be summarized as follows.
The top-quark background is suppressed using a top-tagging veto. If the tagging
efficiency is known, the top-quark background can be estimated as:
\begin{equation*}
N_\text{not-tagged} = N_\text{tagged} \times (1-\epsilon_\text{top-tagged})/\epsilon_\text{top-tagged},
\end{equation*}

where $N_\text{not-tagged}$ is the estimated number of top-quark events in the signal region that pass the veto, $N_\text{tagged}$ is the
number of top-quark events that are top-tagged and $\epsilon_\text{top-tagged}$ is the top-tagging efficiency
as measured in a control region dominated by top-quark events.
For the evaluation of $N_\text{tagged}$ and $\epsilon_\text{top-tagged}$, non-top-quark backgrounds are properly
subtracted using the estimates depending on the jet category.
The systematic uncertainty in the top-quark background estimation is due to the
uncertainty in non-top-quark background contributions and the statistical uncertainty
in the efficiency measurement. The actual implementation of the estimation method
depends on the jet category, and is detailed below.

\subsection{Method for the 0-jet category}

Rejection for the top-quark background is achieved by top-tagging of events via the identification of
a low-$\pt$ b-tagged jet or a soft-muon as defined in
Section~\ref{sec:objects}. The estimation of this background
relies on the measurement of the top-tagging efficiency in data.

In the 0-jet category, the key ingredient for the top-quark background estimation is that $\ttbar$ events
are characterized by two b-jets with $\pt$ below 30\GeV, while $\tw$ events have one
low-$\pt$ b-jet.
Nevertheless a fraction $x$ of $\tw$ events contains two bottom-quark jets and these
events are effectively indistinguishable from $\ttbar$. The procedure described in
the following steps properly accounts for this feature:

\begin{itemize}

\item
First, the top-tagging efficiency for one ``top-taggable'' leg
($\epsilon_{\text{1-leg}}^{\text{data}}$) is computed.
A region enriched in top-quark background events is defined requiring exactly one b-tagged jet
with $\pt > 30\GeV$; this is the denominator. Events in this
sample but with an additional b-tagged jet with
$10\GeV < \pt < 30\GeV$ or one soft-muon define the
numerator. The ratio of the yields in the numerator and denominator
provides $\epsilon_{\text{1-leg}}^{\text{data}}$.
This efficiency is computed for $\ttbar$ only; i.e., non-top-quark
backgrounds and $\tw$ yields are subtracted from the measured data in
the control region. The $\tw$ yield is estimated from the simulation,
which is normalized accordingly, using the predictions previously evaluated
from the 1-jet category.

\item
The overall top-tagging efficiency, $\epsilon_{\text{top-tagged}}^{\text{data}}$, is
defined to account for the fraction $x$ of $\tw$ events that look
like $\ttbar$, that is with two top-taggable legs:

\begin{equation*} \label{eq:newTopTagEff}
\epsilon_{\text{top-tagged}}^{\text{data}} = \left[f_{\ttbar}^{\mathrm{MC}} + x(1-f_{\ttbar}^{\mathrm{MC}})\right]\left[1-(1-\epsilon_{\text{1-leg}}^{\text{data}})^2\right] + (1-f_{\ttbar}^{\mathrm{MC}})(1-x)\epsilon_{\text{1-leg}}^{\text{data}},
\end{equation*}

where the first term accounts for events with two taggable
legs and the second term for events with one taggable leg. The $f_{\ttbar}^{\mathrm{MC}}$ factor
represents the fraction of $\ttbar$ events with respect to the total $\ttbar+\tw$
and it is determined from simulation in the 0-jet category at the $\W\W$ selection level, without
applying the top-quark veto requirements.
The fraction $x$ matches the value of $\epsilon_{\text{1-leg}}$ estimated from
the $\tw$ simulation. This is considered a good approximation because
$\epsilon_{\text{1-leg}}$ is the fraction of events with one b-tagged jet
with $\pt$ larger than 30\GeV (the first ``top-taggable'' leg) out of all events with a
top-tagged leg (a b-tagged jet below 30\GeV or a soft-muon).

\item
Finally, a dedicated control region is defined in the 0-jet category by requiring
top-tagged events. The data yields in this region, corrected for the
contamination from other backgrounds, are then used together with the
top-tagging efficiency to predict the top-quark background:

\begin{equation*} \label{eq:topExtrapolation}
N^{\text{\text{top}}}_{\W\W~\text{region}}=N_{\text{top-tagged}}^{\text{top}}\frac{1-\epsilon_{\text{top-tagged}}^{\text{data}}}{\epsilon_{\text{top-tagged}}^{\text{data}}} =
(N_{\text{top-tagged}}^{\text{data}}-N_{\text{other-bkg.}}^{\text{data}})\frac{1-\epsilon_{\text{top-tagged}}^{\text{data}}}{\epsilon_{\text{top-tagged}}^{\text{data}}}.
\end{equation*}

\end{itemize}

\begin{figure}[htbp]
  \begin{center}
 \includegraphics[width=0.45\textwidth]{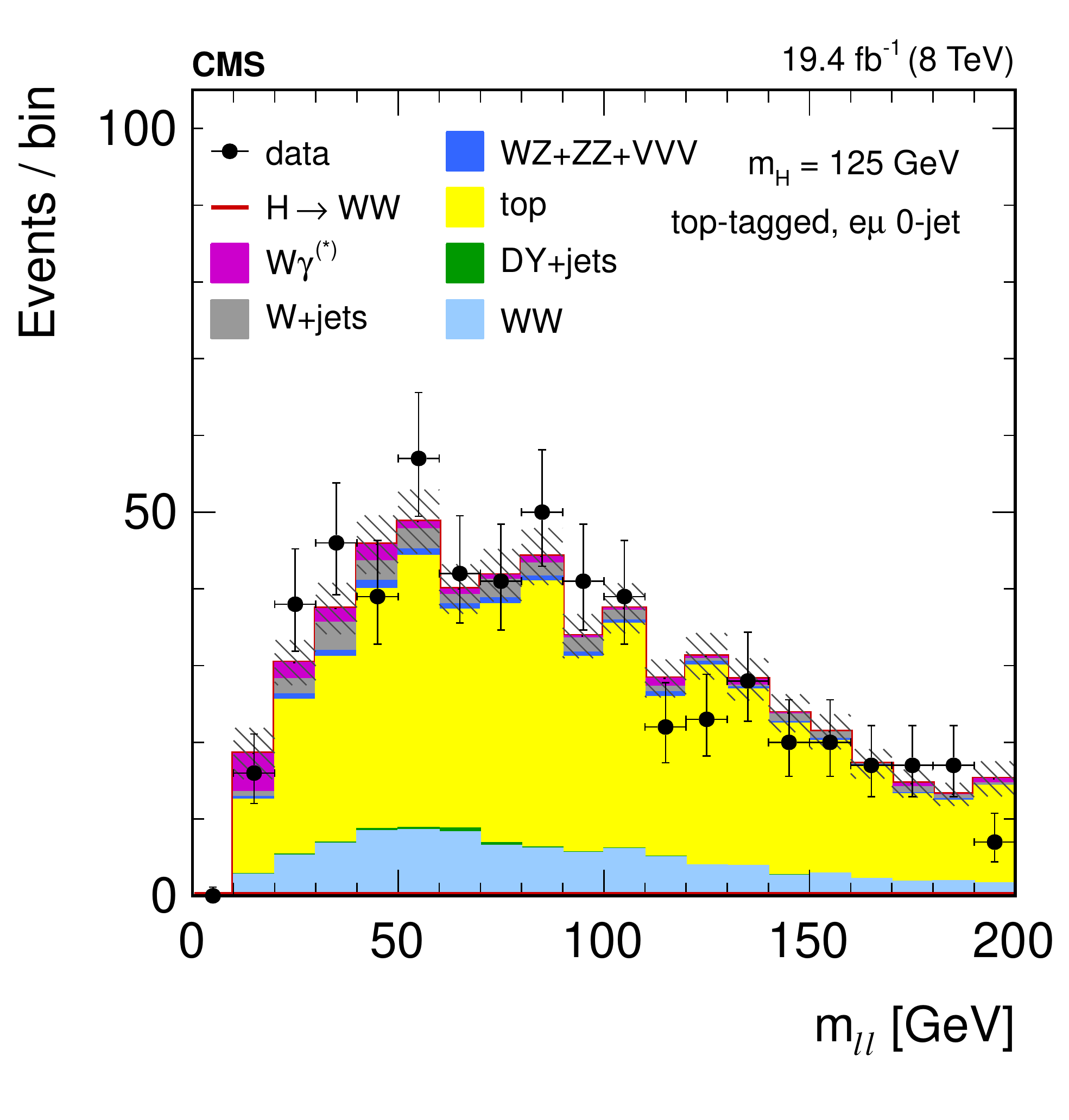}
 \includegraphics[width=0.45\textwidth]{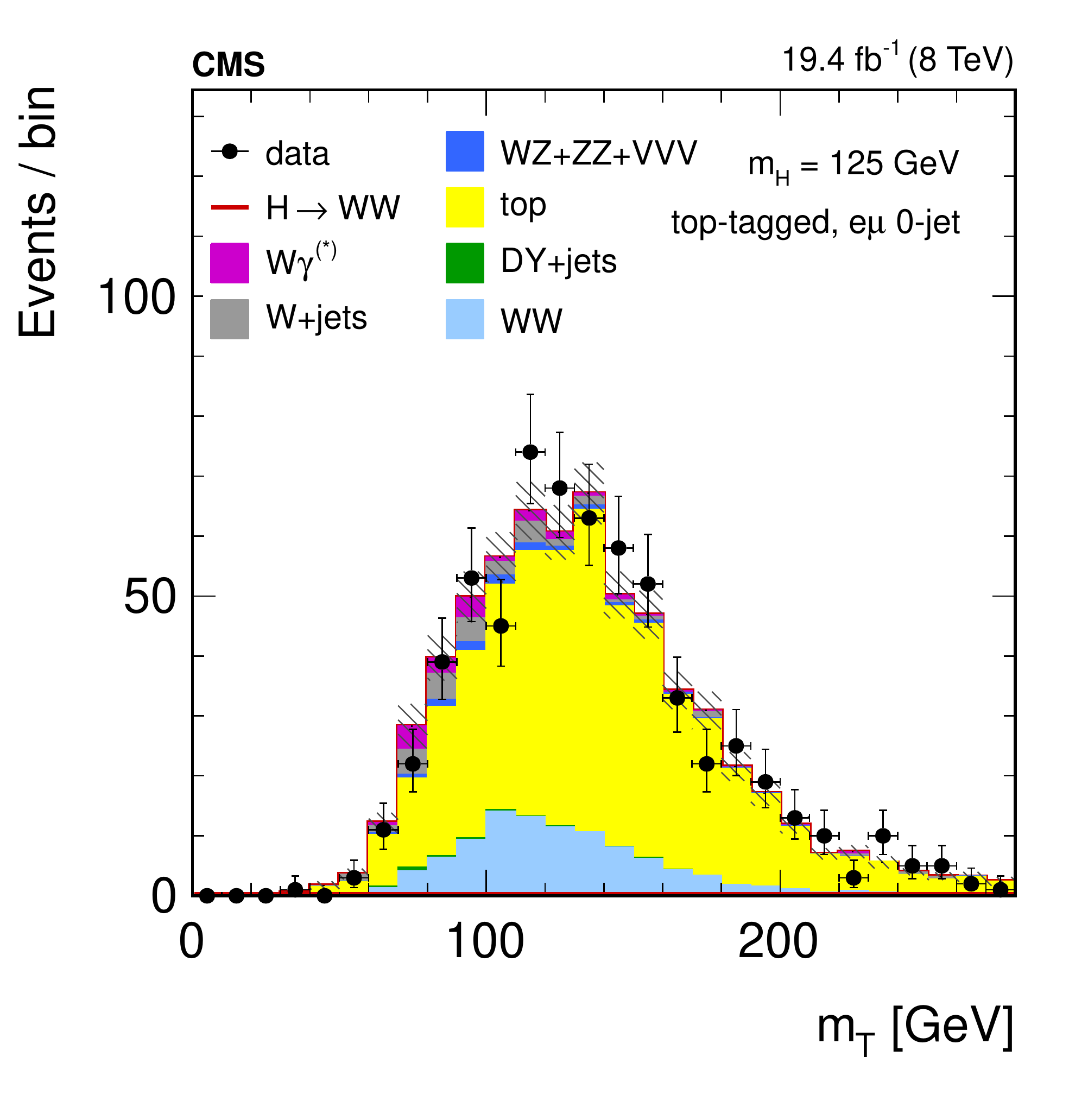}
    \caption{The $\mll$ (left) and $\mth$ (right) distributions in the 0-jet category for top-tagged events
    in the different-flavor final state at the $\W\W$ selection level for $\sqrt{s} = 8\TeV$ data sample. The uncertainty band includes
    the statistical and systematic uncertainty of all background processes.
       }
    \label{fig:topcontrol_of_0j}
  \end{center}
\end{figure}

The $\mll$ and $\mth$ distributions in the 0-jet category for top-tagged events
in the different-flavor final state at the $\W\W$ selection level for the
$\sqrt{s} = 8\TeV$ data sample are shown in Fig.~\ref{fig:topcontrol_of_0j}.

\subsection{Method for the 1-jet category}
To measure the top-tagging efficiency in the 1-jet category, top-quark events
with two reconstructed jets are used as the control sample.
The top-tagging efficiency for the highest $\pt$ jet is approximately
the same in the 1-jet and 2-jet categories. Therefore, the top-tagging efficiency
for the highest $\pt$ jet is used and it is measured in
the 2-jet category where, in order to increase the top-quark purity,
the second jet is required to be b-tagged.

The residual number of top-quark events in the 1-jet category is then given by,
$${N_{\text{non-tagged}}^{\text{1-jet}} = N_{\text{tagged}}^{\text{1-jet}} \times (1-\epsilon_{\text{highest-\pt-jet}})/\epsilon_{\text{highest-\pt-jet}}};$$
where $N_{\text{tagged}}^{\text{1-jet}}$ is the number of events where the counted jet is
tagged and none of the other non-counted jets are tagged, and $\epsilon_{\text{highest-\pt-jet}}$ is the
top-tagging efficiency for the highest $\pt$ jet measured from the 2-jet category.
The closure test, performed by comparing the estimate using this procedure in
simulated events, gives the same result to within $2\%$.

The scale factor is actually derived in a region that is slightly different from
the signal region, but then it is consistently applied to the yield from simulated samples in the signal
region. The difference is due to the soft-muon selection. In the signal region, events
with soft-muons are always rejected. Instead, in the 1-jet top-quark background estimation, soft-muons are
allowed inside the leading jet. This is also done in the top-veto region, in the
top-tag region and in the efficiency measurement. The reason is the correlation between soft-muons,
and b-tagging, since when a soft-muon is present in the jet, its b-tagging efficiency is
slightly higher. To avoid this correlation, the top-quark background is estimated without any requirement
on soft-muons close to the jet.

\begin{figure}[htbp]
  \begin{center}
 \includegraphics[width=0.45\textwidth]{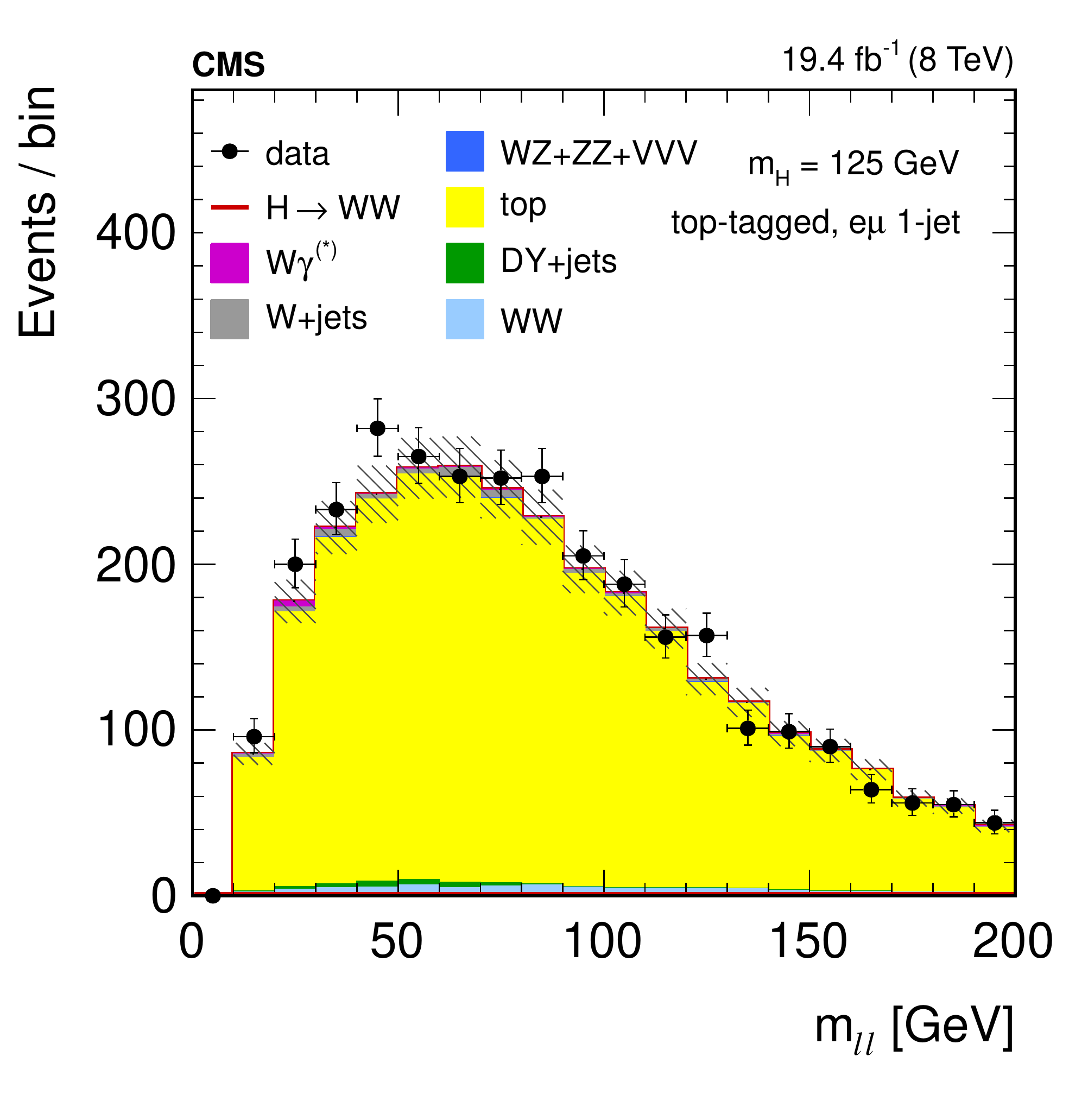}
 \includegraphics[width=0.45\textwidth]{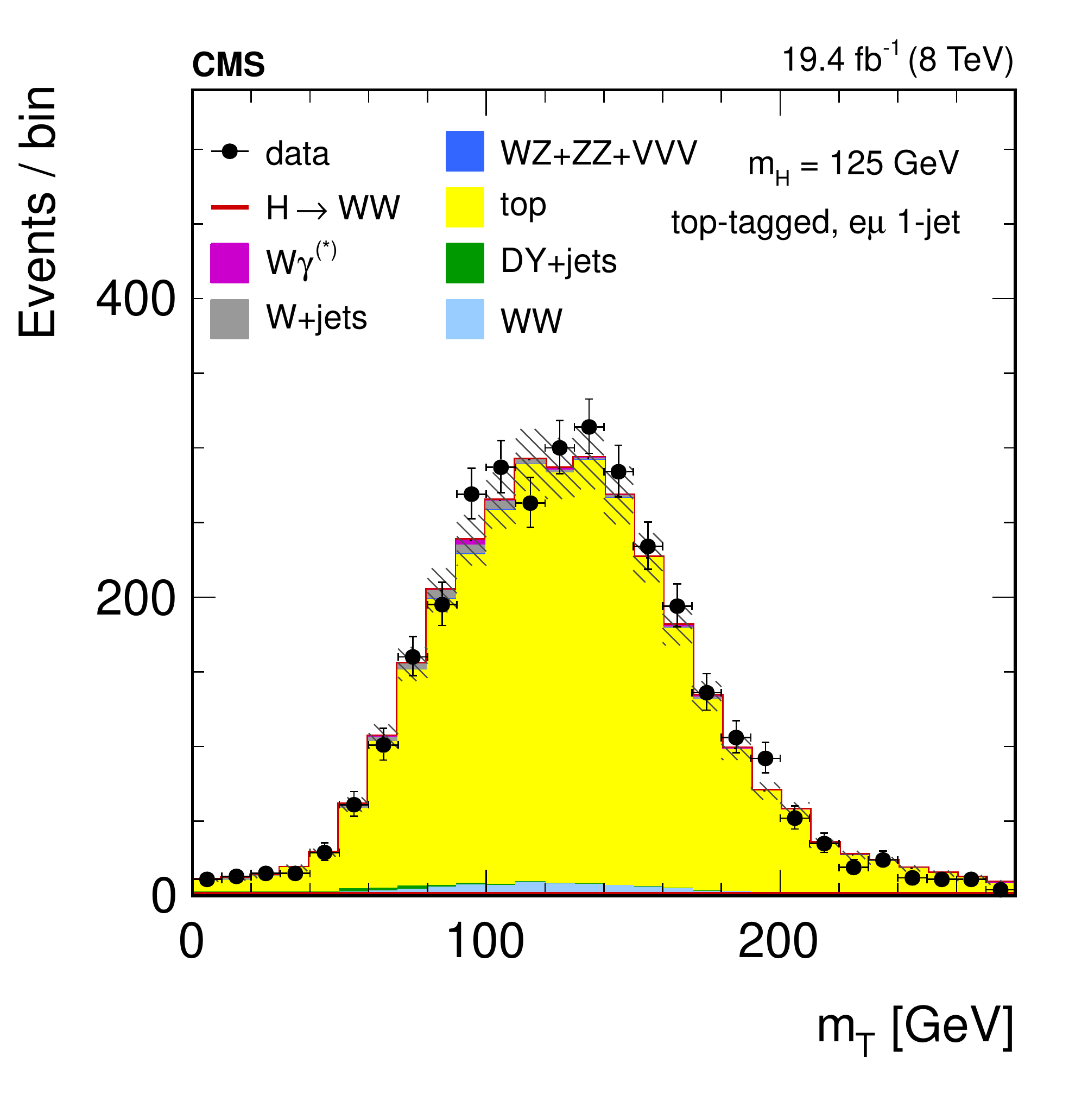}
    \caption{The $\mll$ (left) and $\mth$ (right) distributions in the 1-jet category for top-tagged events
    in the different-flavor final state at the $\W\W$ selection level for the $\sqrt{s} = 8\TeV$ data sample. The uncertainty band includes
    the statistical and systematic uncertainty of all background processes.
       }
    \label{fig:topcontrol_of_1j}
  \end{center}
\end{figure}

The $\mll$ and $\mth$ distributions in the 1-jet category for top-tagged events in the different-flavor final
state at the $\W\W$ selection level for the $\sqrt{s} = 8\TeV$ data sample are shown in
Fig.~\ref{fig:topcontrol_of_1j}.

\subsection{Method for the 2-jet category}
Estimation of the top-quark background in the 2-jet categories is complicated by the
additional requirements involved in tagging VBF and $\V\PH$ events since the
data sample is largely reduced.

The method employed measures the top-tagging efficiency for the most central
jet in the event as a function of its $\eta$ in an inclusive top-quark-enriched control
sample, and then applies that rate to fully selected events
where the most central jet is top-tagged. In this way the
possible kinematical differences between the control and signal regions
are taken into account.

Therefore, the residual number of top-quark events in the 2-jet category after
applying the selection is given by,
$${N_{\text{non-tagged}}^{\text{top}} = N_{\text{tagged}}^{\text{top}} \times (1-\epsilon_{\text{central-jet}})/\epsilon_{\text{central-jet}}},$$
where $N_{\text{non-tagged}}^{\text{top}}$ ($N_{\text{tagged}}^{\text{top}}$) is the number of
events where the most central jet is (not) top-tagged, and $\epsilon_{\text{central-jet}}$ is
the top-tagging efficiency as a function of $\eta$ of the jet.
A very small fraction of top-quark events has both jets outside the tracker acceptance and that fraction is
considered when estimating the systematic uncertainty of the method.

\begin{figure}[htbp]
  \begin{center}
 \includegraphics[width=0.45\textwidth]{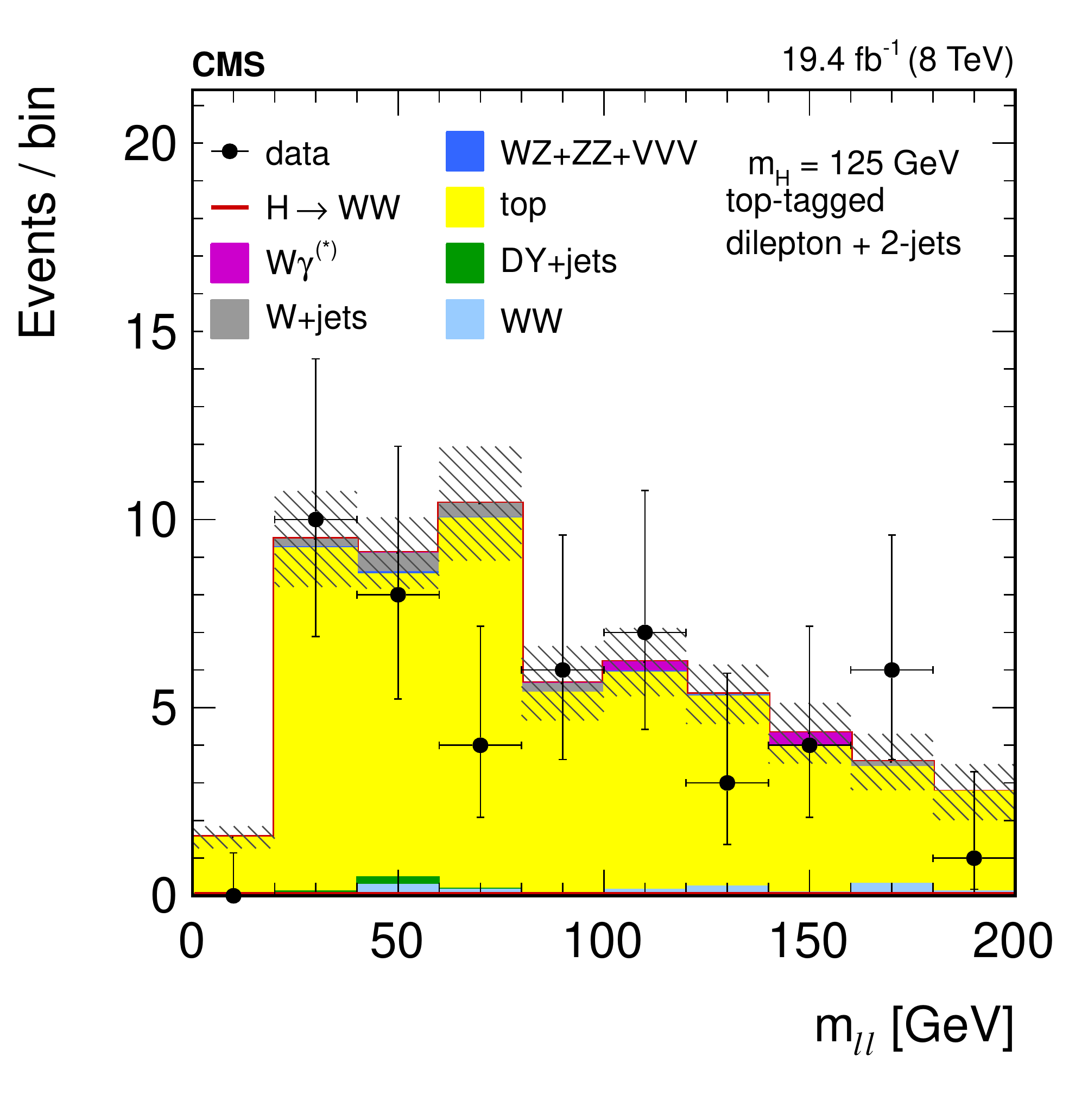}
 \includegraphics[width=0.45\textwidth]{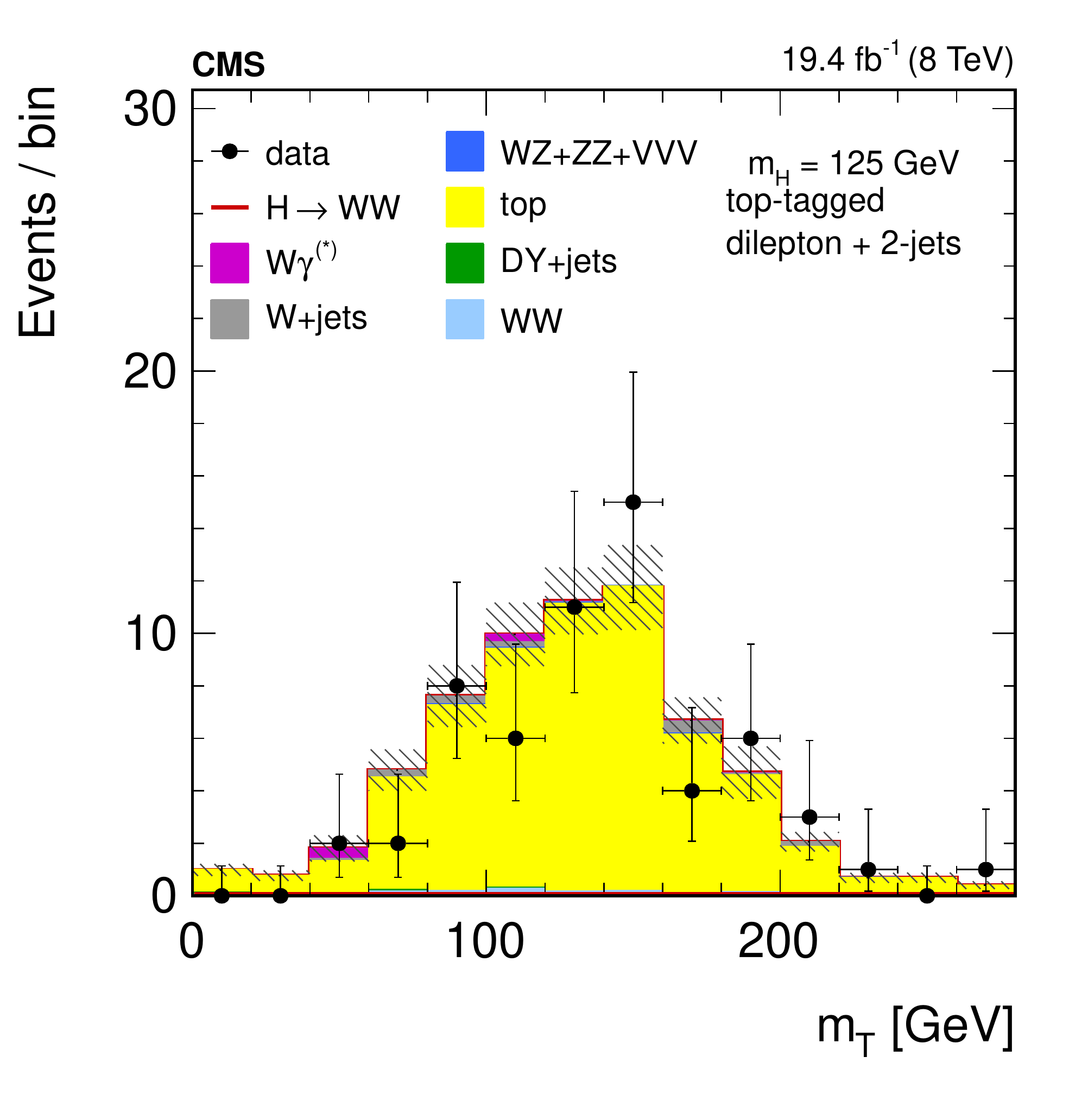}
    \caption{The $\mll$ (left) and $\mth$ (right) distributions in the 2-jet category for top-tagged events
    after applying the $\W\W$ and VBF-tag selections for the $\sqrt{s} = 8\TeV$ data sample. The uncertainty
    band includes the statistical and systematic uncertainty for all background processes.
       }
    \label{fig:topcontrol_incl_2j}
  \end{center}
\end{figure}

The $\mll$ and $\mth$ distributions in the 2-jet category for top-tagged events
after applying the dilepton 2-jet VBF tag selection for the $\sqrt{s} = 8\TeV$ data sample are
shown in Fig.~\ref{fig:topcontrol_incl_2j}.

\cleardoublepage \section{The CMS Collaboration \label{app:collab}}\begin{sloppypar}\hyphenpenalty=5000\widowpenalty=500\clubpenalty=5000\textbf{Yerevan Physics Institute,  Yerevan,  Armenia}\\*[0pt]
S.~Chatrchyan, V.~Khachatryan, A.M.~Sirunyan, A.~Tumasyan
\vskip\cmsinstskip
\textbf{Institut f\"{u}r Hochenergiephysik der OeAW,  Wien,  Austria}\\*[0pt]
W.~Adam, T.~Bergauer, M.~Dragicevic, J.~Er\"{o}, C.~Fabjan\cmsAuthorMark{1}, M.~Friedl, R.~Fr\"{u}hwirth\cmsAuthorMark{1}, V.M.~Ghete, C.~Hartl, N.~H\"{o}rmann, J.~Hrubec, M.~Jeitler\cmsAuthorMark{1}, W.~Kiesenhofer, V.~Kn\"{u}nz, M.~Krammer\cmsAuthorMark{1}, I.~Kr\"{a}tschmer, D.~Liko, I.~Mikulec, D.~Rabady\cmsAuthorMark{2}, B.~Rahbaran, H.~Rohringer, R.~Sch\"{o}fbeck, J.~Strauss, A.~Taurok, W.~Treberer-Treberspurg, W.~Waltenberger, C.-E.~Wulz\cmsAuthorMark{1}
% ...
\vskip\cmsinstskip
% ...
1:~~Also at Vienna University of Technology, Vienna, Austria\\
2:~~Also at CERN, European Organization for Nuclear Research, Geneva, Switzerland\\
% ...

\end{sloppypar}
\end{document}